\documentclass{JINST}
\pdfoutput=1
\usepackage{epsfig}
\usepackage{graphics}
\usepackage{url}

\title{CALICE Report to the DESY Physics Research Committee}

\author{\centering 
\LARGE\bf The CALICE Collaboration\footnote{The complete list of
  current CALICE authors can be found in Appendix 1.}}

\abstract{We present an overview of the CALICE activities on calorimeter development for 
a future linear collider. We report on test beam analysis results, the 
status of prototype development and future plans.}

\begin{document}

\section{Introduction}
The CALICE Collaboration pursues the development of highly
granular calorimeters for a future
e$^+$e$^-$ linear collider, based on the particle flow approach
for optimal overall detector performance.
The Collaboration consists of 57 institutes from 17 countries
in Africa, America, Asia and Europe and has about 350 physicists and
engineers as members. 

We are investigating several technological options for both
electromagnetic and hadronic calorimeters. Most of these are
candidates for both of the particle flow based ILC detector concepts,
ILD and SiD, and for a detector at a multi-TeV linear collider such as
CLIC.  Our aim is to cover the widest possible range of options with
prototypes and to test them using particle beams, thereby maximizing
the use of common infrastructure such as mechanical devices, ASIC
architecture or DAQ systems.  Wherever it is technically feasible and
expedient, we work within a common software and analysis framework.
This facilitates combination and comparison of test beam data while
reducing systematic effects, and allows us to achieve a common
understanding of the relative strengths and weaknesses of the options
under consideration.  This is one of the main strengths of a
broadly-based collaboration such as CALICE.

The major part of the effort is focused towards presenting realistic
proposals for the detector concept reports of the ILC Technological
Design Phase 1, which is due in 2012. This is closely co-ordinated
with the detector concept groups. Given the as yet uncertain schedule
and energy range that will be required of the future lepton collider,
we are also pursuing developments which will only reach a similar
level of maturity at a later stage.

The development of calorimeter prototypes generally proceeds in two
steps.  Physics prototypes provide a proof-of-principle of the
viability of a given technology in terms of construction, operation
and performance. In addition they are used to collect large data sets
for the study of hadronic shower evolution with high granularity.
These are invaluable for testing shower simulation programs, and for
the development of particle flow reconstruction algorithms with real
data.

 In contrast, technological prototypes address the issues of scaling, integration
 and cost optimization.  Due to the differing responses of the various
 active media used to the components of hadronic showers, the physics
 prototypes are necessary for each combination of active and passive
 materials under consideration. Technical prototypes are required for
 each technology, but the effort can be kept to an acceptable level by using
 common building blocks, and by addressing large area and multi-layer
 issues separately without instrumenting a full volume.

\subsection*{Synopsis} 
The completion of data taking with physics prototypes of a silicon
based ECAL and scintillator based ECAL and HCAL was reported in
2009. Since then, emphasis has shifted towards analysis of these data,
as illustraed by the new results reported here, including detailed
comparisons with Geant4- based simulations and a first test of
state-of-the art particle flow reconstruction with combined ECAL and
HCAL test beam data.

In the meantime, construction of the first gaseous digital HCAL prototype was 
finished, with front end electronics already integrated in the detector volume. 
Data taking started in 2010 at the Fermilab test beam facility, 
using the existing absorber structure and stage. We present first results on 
detector performance and the status of data taking, which is ongoing in 
conjunction with the silicon ECAL at the time of this review.

The year 2010 also saw the start o beam tests for a new tungsten HCAL
absorber structure, as proposed for the CLIC energy range. Here, the
well-understood active scintillator layers are being re-used. We
report on a first look into the time structure of the hadronic
response in a neutron-rich material.

Major progress was also made with technological prototypes with highly
integrated ultra-low power electronics. Final versions of
mixed-circuit read-out ASICs were mass-produced and successfully
tested, including power-pulsed operation in high magnetic fields.  The
full read-out chain of the second generation scalable DAQ was
established.

The biggest effort here is the construction of an RPC based
semi-digital cubic metre scale HCAL prototype for beam tests in 2011,
which will establish the electronics integration concept at large scale
and contribute to the study of hadronic interactions seen with highly 
granular gaseous read-out.

Smaller scale technical tests, some ranging up to square-metre sized active 
layers, were carried out with all considered HCAL read-out technologies, with 
GEMs, micromegas and scintillator, and are being prepared for the ECAL.

\section{Silicon-Tungsten ECAL: SiW}
\label{sec:SiWECAL}

\subsection{Introduction}

A high granularity ECAL is a central component of a particle-flow optimized
detector. It should efficiently identify photons and measure their
energy with reasonable precision, and should have excellent capabilities
(in conjunction with other subdetectors) to distinguish photonic depositions 
from those due to other particle types. 
A sampling calorimeter naturally gives the required granularity in one
dimension, while a highly segmented design of the readout layers provides 
it in the other two dimensions. The choice of tungsten as an absorber material
is motivated by its small Moli\`{e}re radius and radiation length and the
large difference between its radiation and interaction lengths, ensuring that
photonic showers are of small size, and with a significantly different size to 
hadronic showers. The use of matrices of PIN diodes in high resistivity
silicon allows an efficient and low-noise detection of charged particles,
and the possibility to efficiently implement highly segmented readout.

\subsection{Physics prototype}

\begin{figure}
\begin{center}
\includegraphics[width=0.5\textwidth]{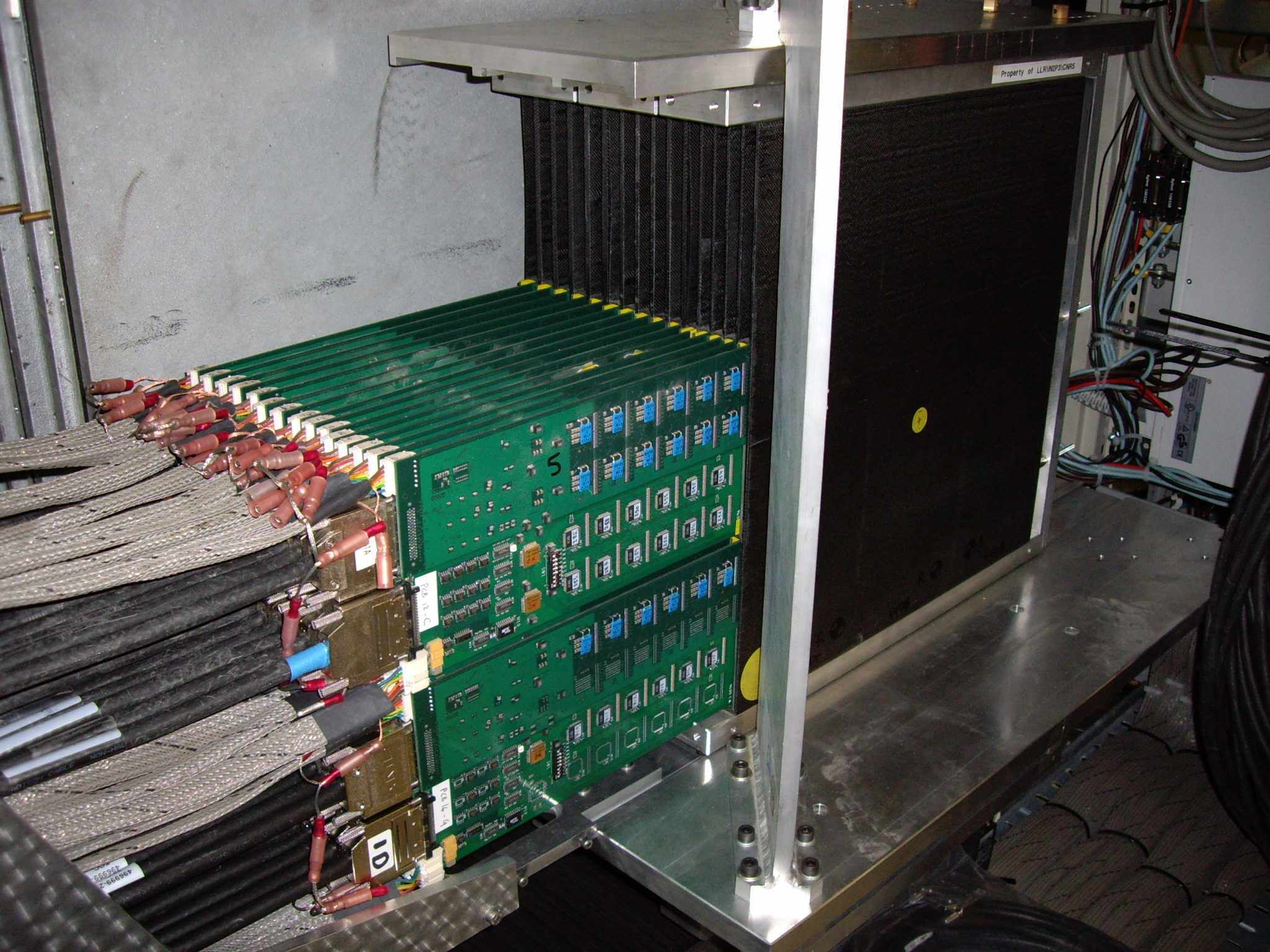}
\caption{\label{fig:physProto} Physics prototype installed at a test beam.}
\end{center}
\end{figure}

A first detector prototype, known as the ``physics prototype'' was
constructed during the years 2005--2007. It had an active area of $18
\times 18$~cm$^2$ and 30 sampling layers. It was extensively tested
in particle beams between 2006 and 2008 at CERN and FNAL (as seen in
Fig.~\ref{fig:physProto}). These tests were performed as part of the
combined CALICE test beam campaigns with the ECAL followed by a
(scintillator-based analogue readout) hadronic calorimeter, and a
tail-catcher system, and used beams of electrons, muons and pions with
energies between 2 and 180 GeV.

In 2011, the physics prototype will again be placed into test beams at FNAL, 
this time with a RPC-based digital readout hadronic calorimeter, allowing detailed
comparisons between the two HCAL approaches.

The measured performance of this prototype has demonstrated that
the silicon tungsten approach can satisfy the various requirements
for detector performance at a future lepton collider. The detector has run stably over
a period of five years without showing any signs of aging.

\subsection{Technological prototype}
The focus is now is on the development of the technologies necessary
for such a calorimeter to be integrated into a full detector, keeping
in mind that large scale, industrialised construction and quality control
will be necessary when a final detector is built.
Factors considered include a modular design using low power front end 
electronics integrated into the detector volume, 
realistic mechanical supporting structures, a compact leak-less cooling
system.

\begin{figure}
\begin{center}
\includegraphics[width=0.48\textwidth]{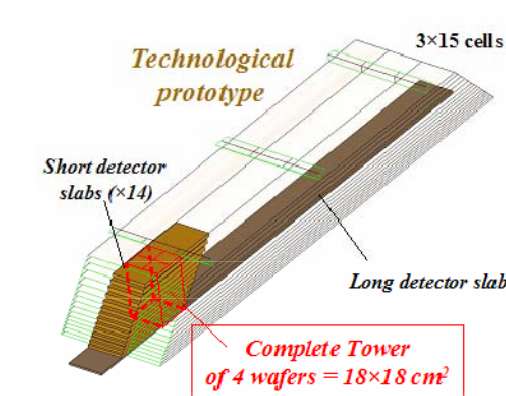}
\includegraphics[width=0.48\textwidth]{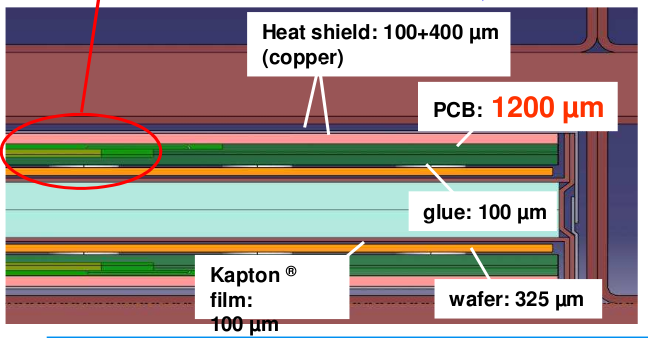}
\caption{\label{fig:techProto} 
Design of the technological prototype. The left figure shows an overview of the mechanical structure
and the instrumented region, while the right figure shows a section of a detector slab.}
\end{center}
\end{figure}

The general design of the module is shown in Fig.~\ref{fig:techProto}; details can be found in~\cite{eudet-memo}.
The composite-tungsten mechanical structure is a slightly scaled down version of
a barrel module envisaged for a linear collider detector, with a length of 1.5~m and a width
of around 55~cm, and a total weight of around 600~kg. It has a trapezoidal
shape and consists of $15 \times 3$ alveola into which sensitive detector slabs are inserted.

Each detector slab is built around a ``H'' structure with a tungsten core and composite walls.
Both sides of this structure are equipped with active elements consisting of the 
silicon sensors, a PCB which routes signals and commands between the interior and exterior
of the detector, and the front end ASICs embedded into the PCB thickness.
A Kapton cable provides the high voltage to the sensors, and a sheet of copper 
helps to extract the heat to a dedicated heat exchanger at the end of the module.
A small Detector InterFace board at the end of each slab is the first element of the 
common CALICE DAQ system, as described in more detail in Sect.~\ref{sec:daq}

For the technical prototype, a tower with an area of $18 \times 18$~cm$^2$ will be instrumented
in all 30 layers, and one long detector slab will be built to test the propagation
of signals over long distances within the calorimeter.

\subsubsection{Sensors}

\begin{figure}
\begin{center}
\includegraphics[width=0.5\textwidth]{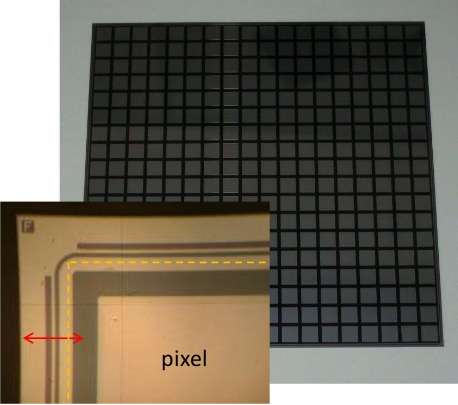}
\caption{\label{fig:sensor}
A photograph of a $9\times9$~cm$^2$ silicon sensor from Hamamatsu Photonics with 5.5~mm pixels.
The inset shows a microscopic view of the guard ring area: the length
of the red arrow corresponds to 750~$\mu$m.}
\end{center}
\end{figure}

The silicon sensors used in the ECAL are arrays of $5\times5$~mm$^2$ PIN diodes 
made in 300--500$\mu$m thick high resistivity silicon, and reverse biased at around 200~V. 
The sensors envisaged for the technical prototype have a total area of $9\times9$~cm$^2$. 
A guard ring structure at the edge of the sensor protects against breakdown. A photograph
of such a sensor is shown in Fig.~\ref{fig:sensor}.

Current studies include the minimisation of the guard ring region (an effectively dead
detector region), and the understanding of the cost drivers for mass production (in close 
collaboration with industrial partners), with a view to reducing the eventual cost.

Around 40 sensors have already been purchased from HPK, further sensors will
have different characteristics: relaxed quality control tolerances (a few dead or noisy channels
do not pose a serious problem in reconstruction), more aggressive
laser cutting of the wafer to minimise the edge area, 
and sensors sourced from other manufacturers.

\subsubsection{ASIC}

\begin{figure}
\begin{center}
\includegraphics[width=0.7\textwidth]{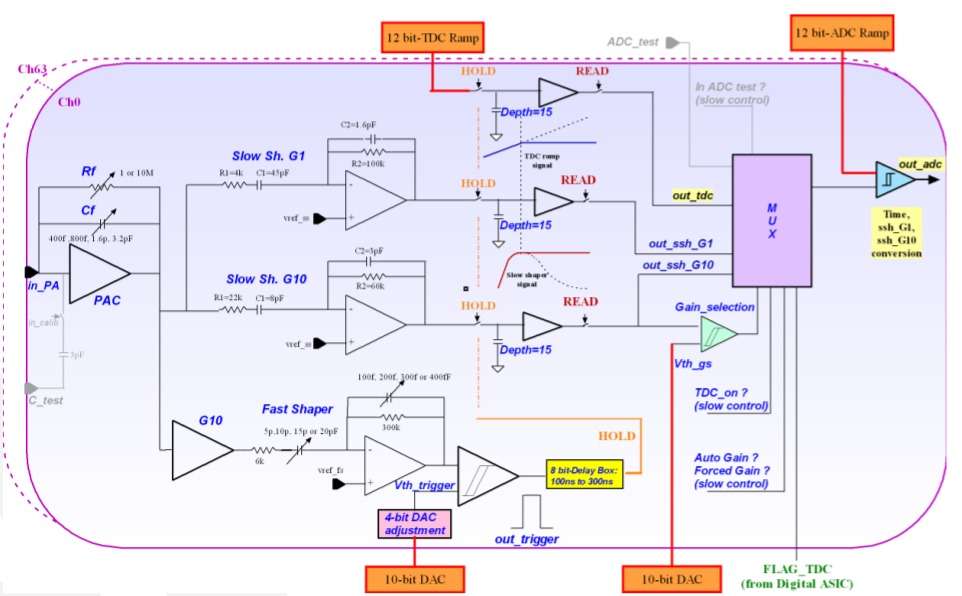}
\caption{\label{fig:skiroc2}
Schematic of the SKIROC2 ASIC.}
\end{center}
\end{figure}

A dedicated front end ASIC with 64 channels (SKIROC2) will be used to
read the signals from the PIN diodes.  It is designed to give a wide
dynamic range (from 0.5 to 2500 MIP signals in the sensor), low signal
to noise at the single MIP level (around 17 at 1 MIP), and low power
consumption, at the level of 25~$\mu$W/channel (using the power
pulsing technique). The schematic of the ASIC is shown in
Fig.~\ref{fig:skiroc2}.  More than 1000 such chips have been
produced, and are in the process of being tested.

\subsubsection{PCB}

The sensitive layers of the ECAL will be made up of several identical
modules, known as Active Sensor Units (ASUs).  An ASU is based around
a large and thin 8-layer PCB (thickness of around 1~mm, area of
$18\times18$~cm$^2$) which hosts 16 SKIROC2 ASICs on one side, and on
the other side of which are glued the silicon sensors.  Signals (both
for configuration and data) are routed along lines in the intermediate
layers.

Several prototypes of the PCB have been produced, for both packaged
and unpackaged ASICs.  The main obstacle at present is the planarity
of the PCB, particularly for thin models: since rather delicate
silicon sensors will be glued to the card, it should not be quite
flat.  Present prototypes have deviations of several millimetres from
perfect planarity over the full area of the card. Modified assembly
procedures (temperature/time allowed for cooling) and/or design may
result in a flatter PCB, and are under test.

\begin{figure}
\begin{center}
\includegraphics[width=0.48\textwidth]{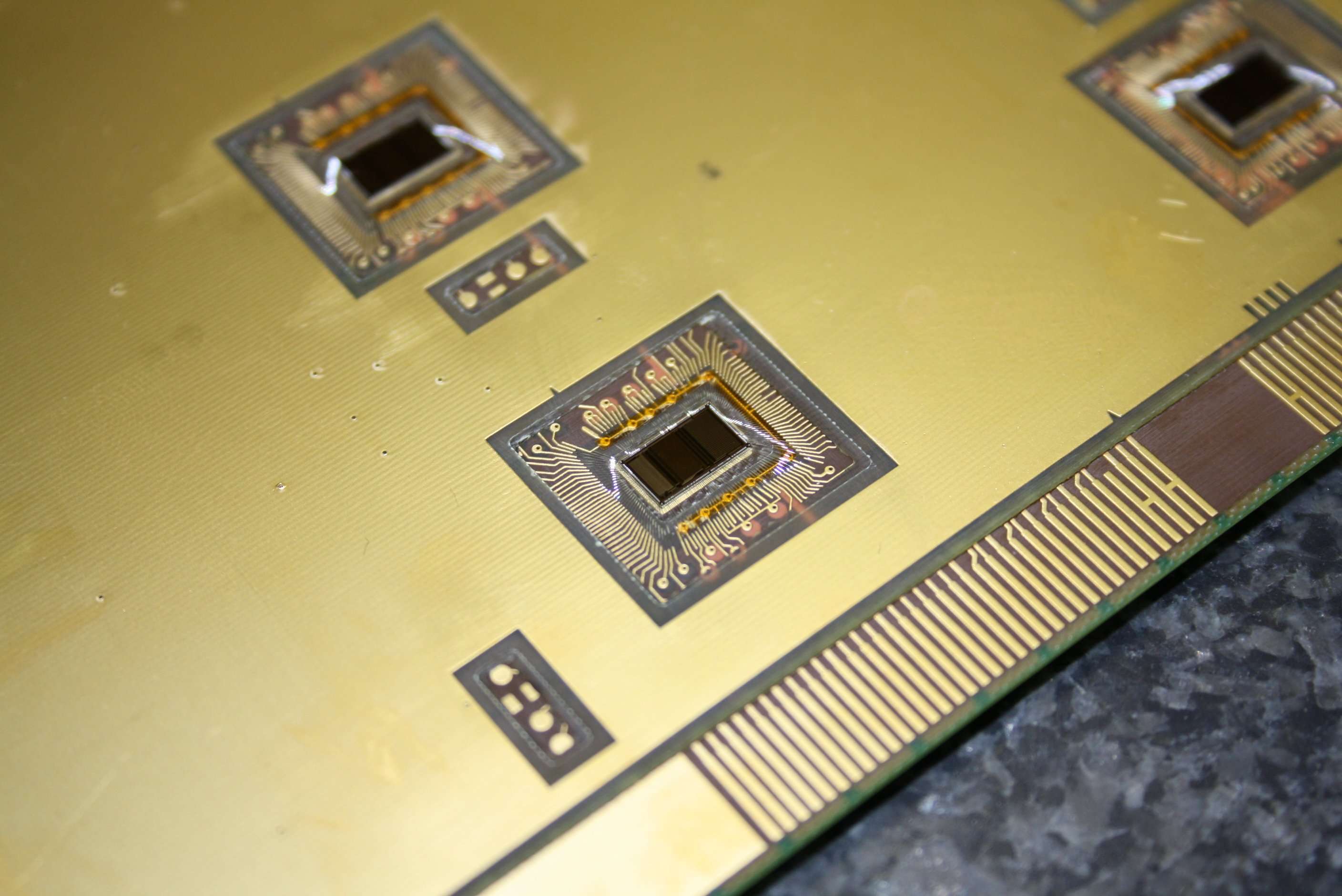}
\includegraphics[width=0.48\textwidth]{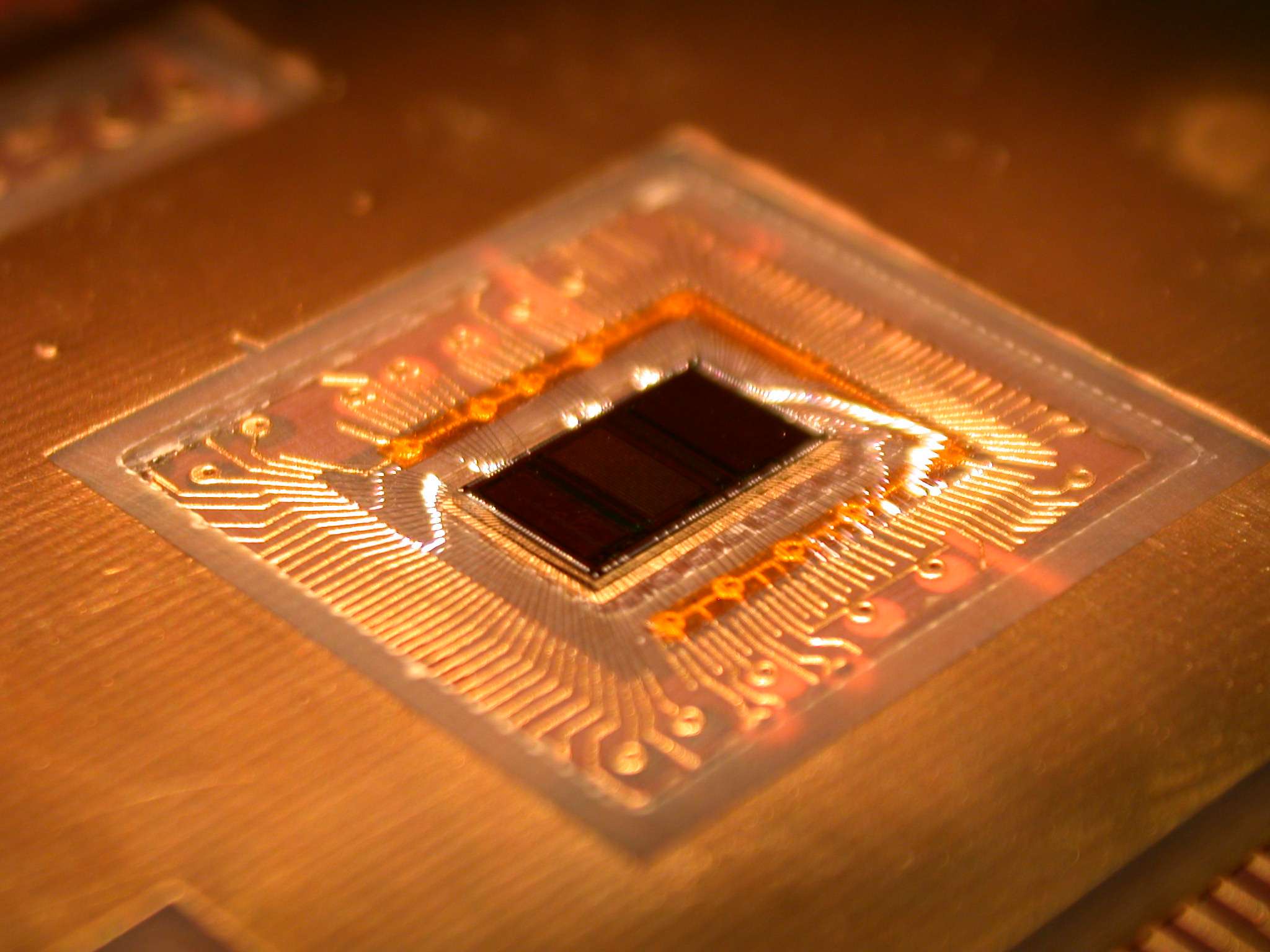}
\caption{\label{fig:asicBond}
View of the PCB with integrated front end ASIC.}
\end{center}
\end{figure}

In the case of unpackaged ASICs (as envisaged for the final
prototype), the ASICs will be contained within the volume of the PCB,
and wire bonded to pads on the PCB. A test of this procedure (results
shown in Fig.~\ref{fig:asicBond}) has already been carried out at the
CERN bonding lab.  Tests of an epoxy protection layer for the ASICs
and wire bonds are underway.

The bias voltage (200~V) will be supplied to the sensors via a flat
Kapton cable, prototypes of which have been successfully produced and
tested, and a $\sim 500$~$\mu$m thick copper layer is used to extract
the heat at the end of the slab.

\subsubsection{Integration}

The task of assembling several ASUs into a detector slab is also under
study. The connection between ASUs must be both mechanical and
electrical (since signals are routed along the entire length of a
detector slab).  A previously developed technique involved the
deposition of solder paste using a silk-screen and brief heating under
a halogen lamp. This technique works well, but is rather labour
intensive and requires great care: somewhat impractical for the future
industrialisation of the process.  A simpler process, based on the use
of anisotropic conductive film (ACF), seems to be a very promising
alternative. Tests have already been carried out in conjunction with
the 3M company, with good results.  The main question is whether the
fully equipped ASUs can withstand the pressure required to bond the
ACF without the fracture of the silicon sensors.

\begin{figure}
\begin{center}
\includegraphics[width=0.5\textwidth]{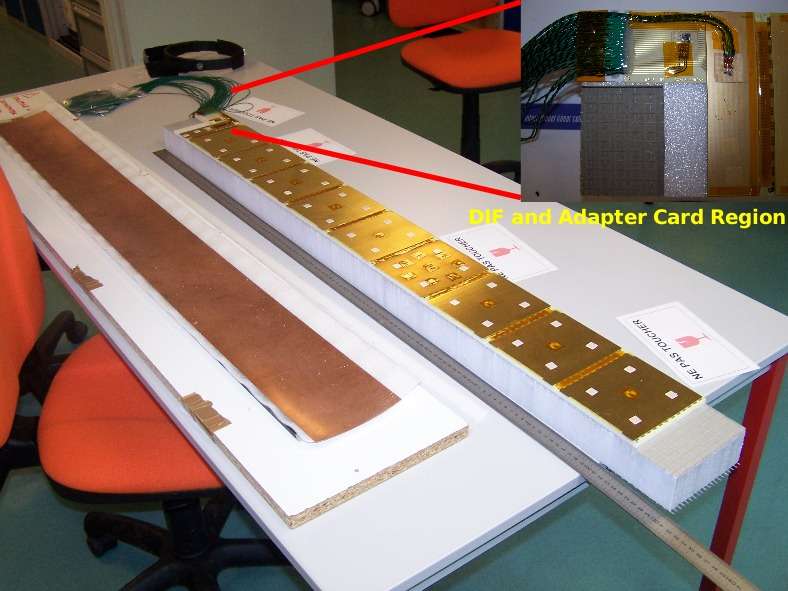}
\caption{\label{fig:thermalSlab} View of a ``thermal slab'', used to
  test integration techniques and to measure the thermal properties of
  the mechanical structure. Visible is a chain of interconnected test
  ASUs, and the continuous copper sheet which acts as a heat drain.}
\end{center}
\end{figure}

A dedicated assembly laboratory is being prepared at LAL, which will
have all necessary equipment to assemble the slabs in a suitable and
dedicated environment. A large-scale test of the integration procedure
has already been performed for the assembly of the ``thermal slabs'',
as shown in Fig.~\ref{fig:thermalSlab}.

\subsubsection{Data acquisition system}
\begin{figure}
\begin{center}
\includegraphics[width=0.5\textwidth]{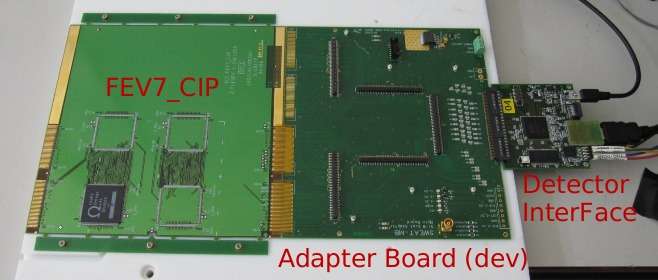}
\caption{\label{fig:ASU_DIF}
DAQ test hardware: prototype ASU - adapter board - Detector InterFace card.}
\end{center}
\end{figure}

The common CALICE data acquisition system is described in
Sect.~\ref{sec:daq}.  The integration of the system is continuing,
with recent progress including the configuring of the ECAL ASICs via
the complete DAQ chain. The various pieces of hardware of the system
have been produced, and good progress has been made of the development
of the necessary firmware and software.  Fig.~\ref{fig:ASU_DIF} shows
some of the hardware used to test the DAQ chain, with a prototype ECAL
ASU (left), an adapter board (centre), and the Detector InterFace card
on the right.  The full system will be tested using cosmics and particle
beams during 2011.

\subsubsection{Mechanical structure}
The mechanical supporting structure developed for the technical prototype
is close in shape and size to a barrel module envisaged for the barrel part
of a linear collider detector (e.g. ILD). It is based on
an alveolar structure made in carbon fibre composite material, and 
incorporates half of the tungsten absorber plates. This allows
a modular structure suitable for industrialisation, and minimises the
non-instrumented detector regions.

\begin{figure}
\begin{center}
\includegraphics[width=0.48\textwidth]{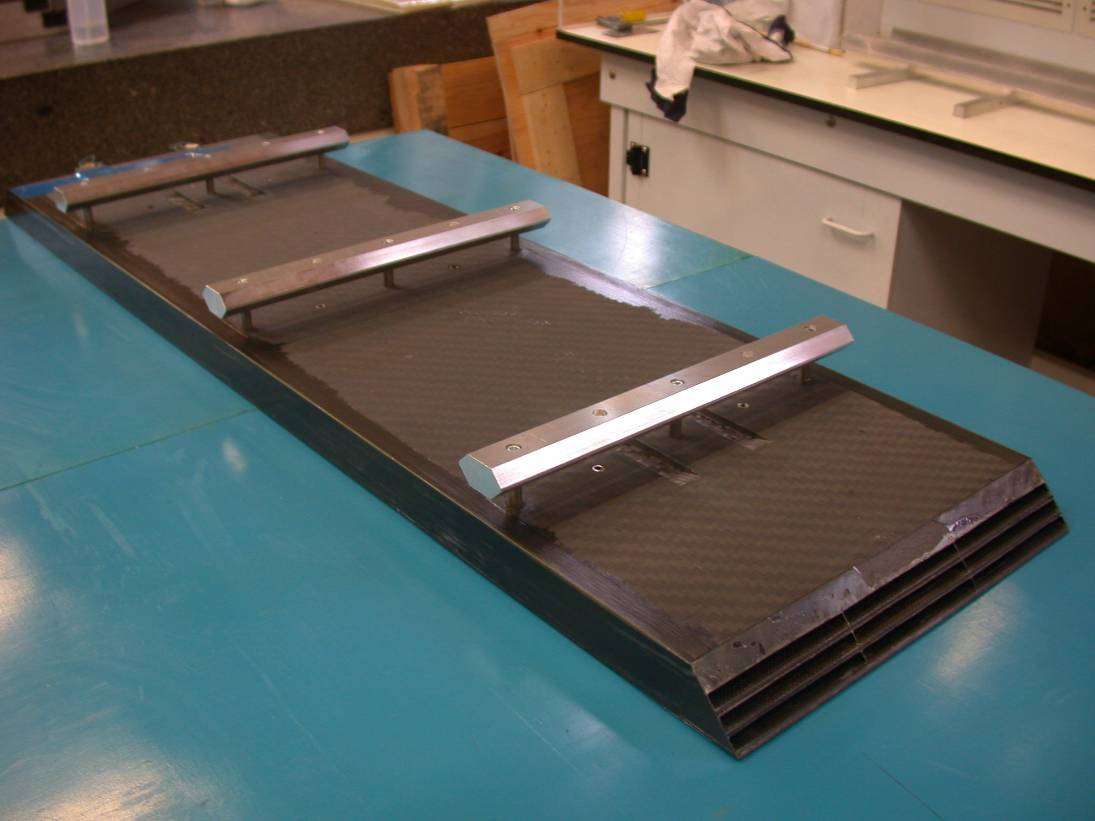}
\includegraphics[width=0.48\textwidth]{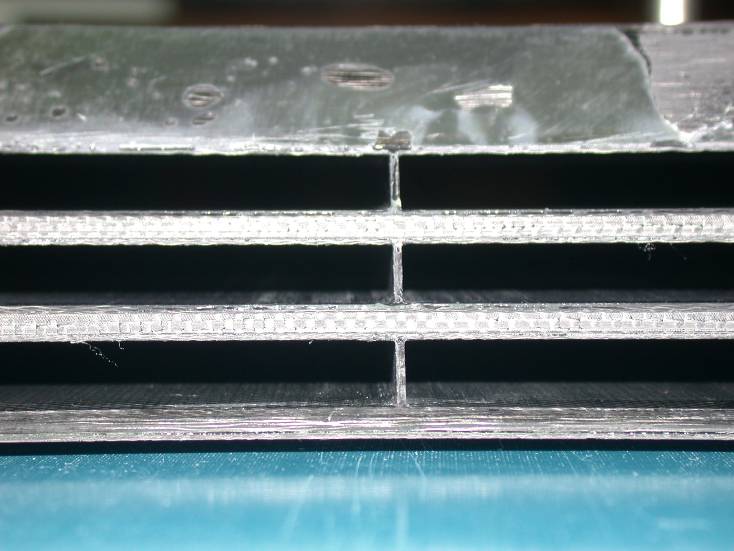}
\caption{\label{fig:demonstratorStruc} View of the composite-tungsten
  demonstrator structure used to validate the construction technique.}
\end{center}
\end{figure}

A small ``demonstrator'' structure, shown in
Fig.~\ref{fig:demonstratorStruc}, has already been built to validate
the assembly methods and materials. It was produced within the
required tolerances.

\begin{figure}
\begin{center}
\includegraphics[width=0.5\textwidth]{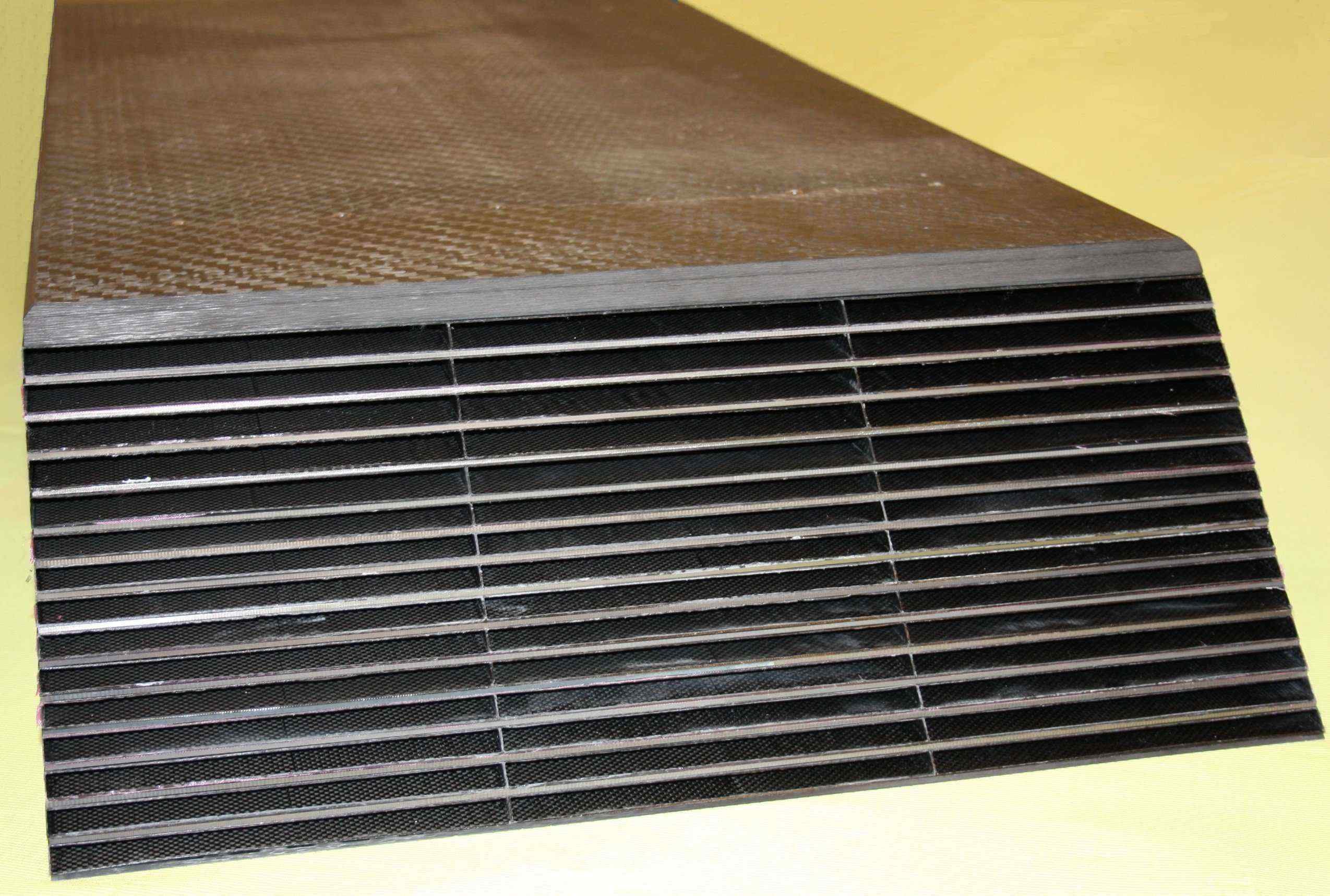}
\caption{\label{fig:mechanicalStruc}
Parts of the final mechanical structure awaiting the final assembly step.}
\end{center}
\end{figure}

The pieces for the final structure have been produced (as seen in
Fig.~\ref{fig:mechanicalStruc}), and an assembly mould has been
designed and constructed. Studies to better understand the final
thermal curing of the composite are underway in order to ensure
successful assembly (there is no possibility to re-do the final
curing, so care must be taken!).

The manufacture of the ``H'' structures is at present on hold, waiting
for a definite answer on the feasibility of a sufficiently thin and
flat PCB.

The production of longer alveolar structures, required for the endcap detectors
(up to 2.5~m long in the current design of ILD) are also under study. The additional 
length requires extra care to be taken in the manufacture to ensure that the
various pieces of the mould can be successfully extracted after the curing of the 
composite material.

\subsubsection{Cooling}

The ``demonstrator module'' has been used to understand the thermal
properties of the composite/tungsten alveolar structure. A number of
thermal detector slabs have been produced which produce heat at a
number of points in a controllable manner, and also measure the
temperature.  A cooling system has been developed based on cold water
circulation. It has been tested and has sufficient performance (with a
reasonably generous margin) to cope with the expected power
dissipation of the ECAL.  This results of the thermal tests have been
used to tune thermal simulations of the module's behaviour, and to
apply the same model parameters to simulation of the full technical
prototype module and also the entire ECAL system within the ILD
detector.

\begin{figure}
\begin{center}
\includegraphics[width=0.5\textwidth]{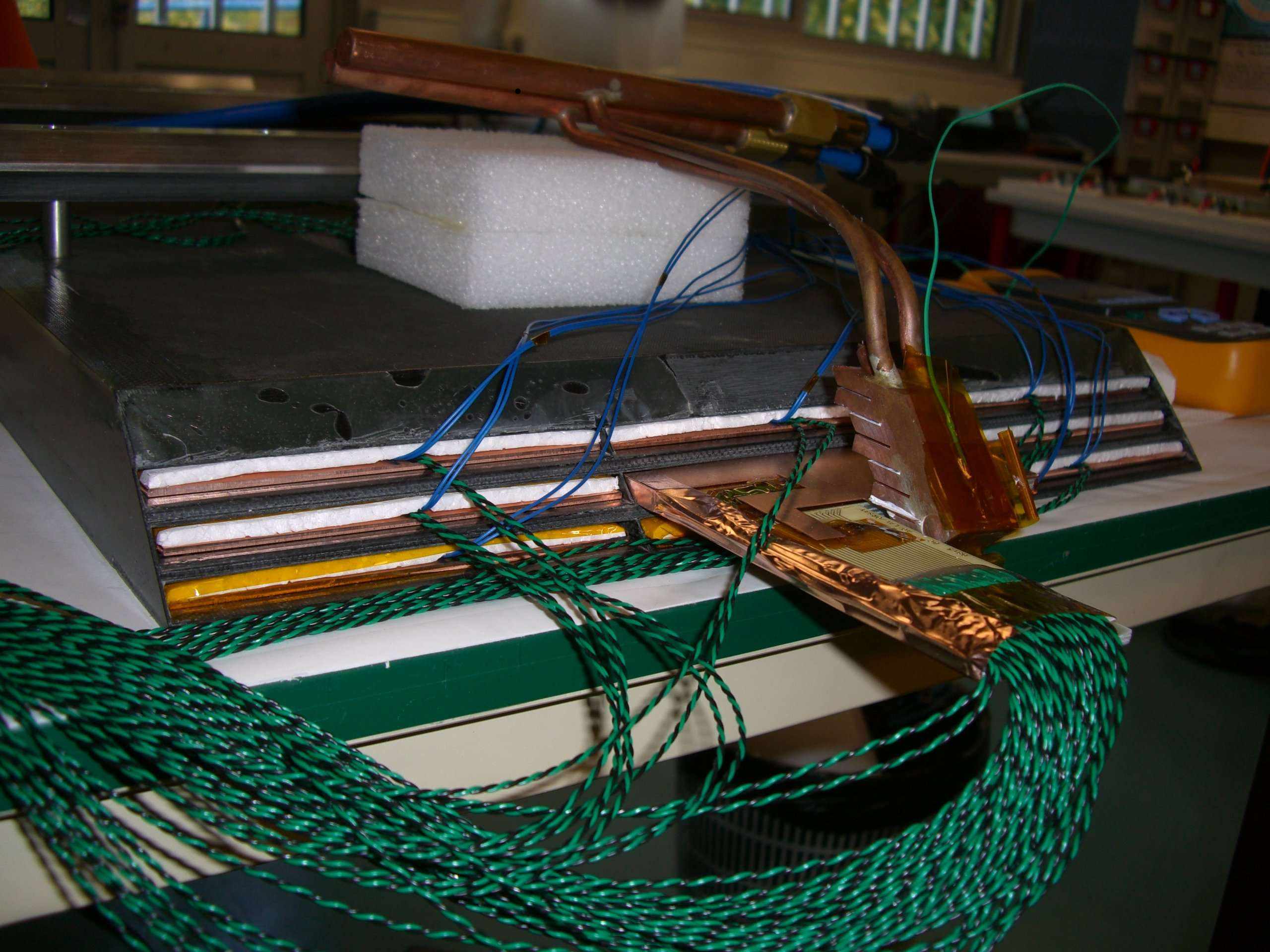}
\caption{\label{fig:cooling} Cooling tests using the demonstrator
  module. The ends of the nine thermal slabs are visible at the end of
  the mechanical structure, as is the copper heat exchanger connected
  to the slabs' thermal drain.}
\end{center}
\end{figure}

\subsection{Future Plans}
A first, partially equipped ASU readout using the full CALICE DAQ
system will be tested in cosmics and beams during 2011. This will be
equipped with a previous version of the ASIC (SPIROC2).  Once we are
confident that a sufficiently thin and flat PCB can be produced, the
``H'' structures can be produced. In the case that they cannot, we may
proceed with a temporary solution in the short term: a ``U'' structure
holding a single sensitive layer per alveola, which considerably
relaxes the constraint on the PCB thickness.

The structure will be gradually equipped, testing various different
approaches to the various technical challenges in order to understand
their respective strong and weak points.

Once a reasonable number of layers has been produced, they will be
tested within the mechanical structure using both cosmic rays and test
beams.

\section{Scintillator Strip ECAL: ScECAL} 
\label{sec:ScECAL}

\subsection{Operational experience with physics prototype}
\begin{figure}
\centering
    \includegraphics[angle=0,width=0.4\textwidth]{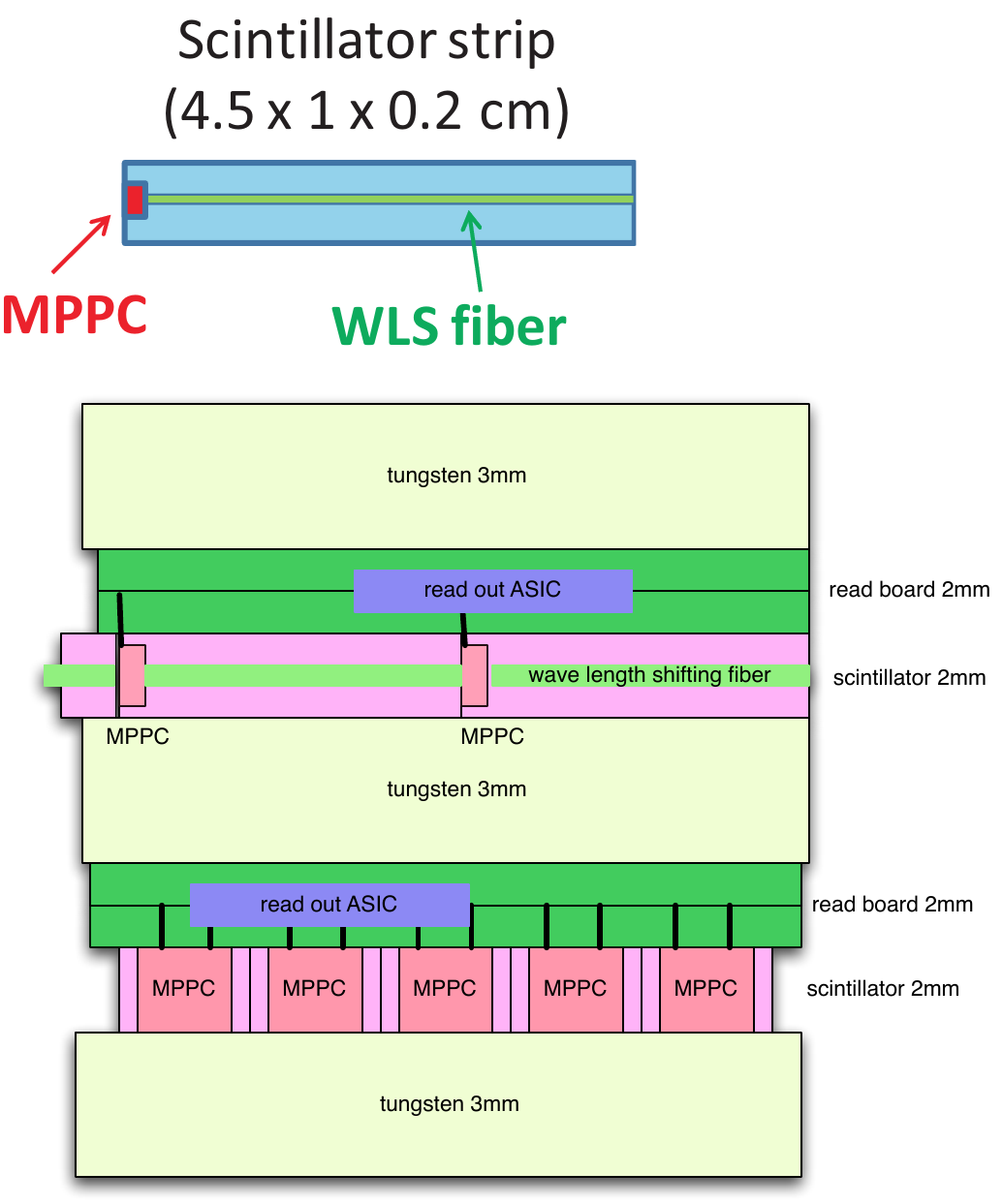}
\caption{\label{fig:scintecalxsect}\em Schematic cross section of the scintillator ECAL prototype.
}
\end{figure}

We have constructed and tested two prototypes. Both prototypes use
scintillator strips which are 10~mm wide, 45~mm long and 3~mm thick,
and wavelength shifting fibres read out by MPPCs, as shown in
Fig.~\ref{fig:scintecalxsect}. The original, smaller prototype was
constructed using 18 strips in each of 26 layers, giving a total of
468 strips, while the second prototype has 72 strips in each of 30
layers, giving a total of 2160 strips.
A good energy resolution performance as well as a reasonable
linearity was presented in the previous report.  For completeness, these results are
briefly reported in section~\ref{sec:analysis} below.
Those data were taken at the DESY beam
test line and FNAL beam test line MT6, in a combined effort from the CALICE
Collaboration together with HCAL and tail catcher as well.
The successful experience of those beam tests shows that the ScECAL
performs well. 

\begin{figure}
\centering
    \includegraphics[angle=0,width=0.4\textwidth]{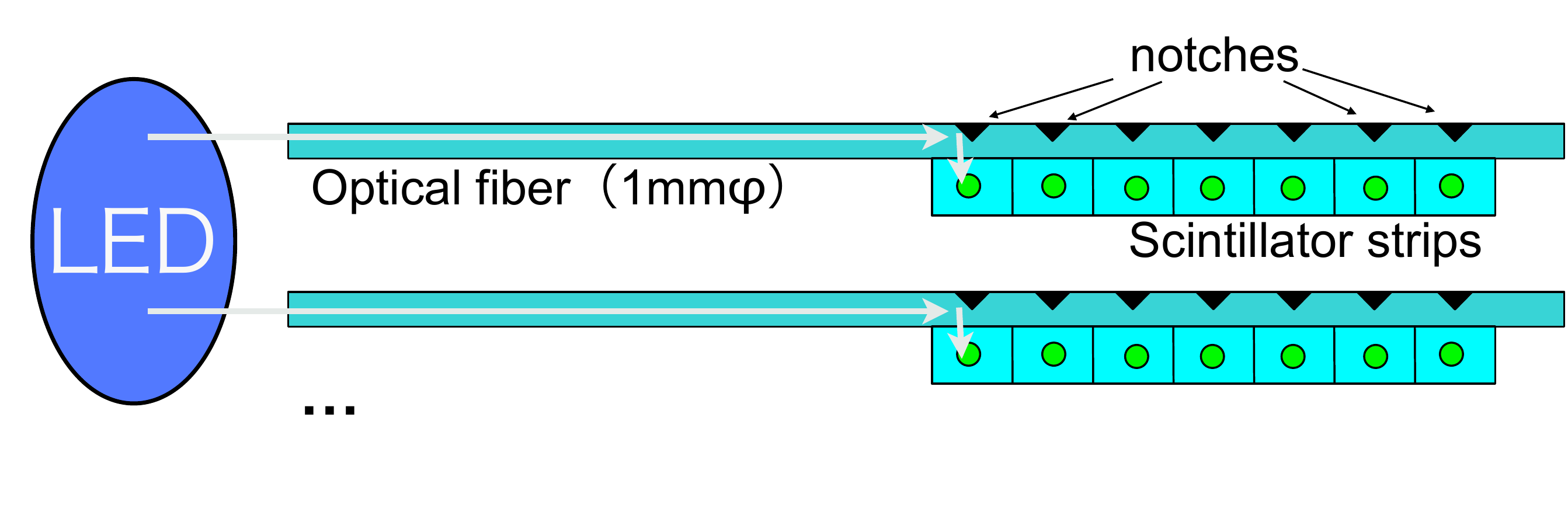}
        \includegraphics[angle=0,width=0.4\textwidth]{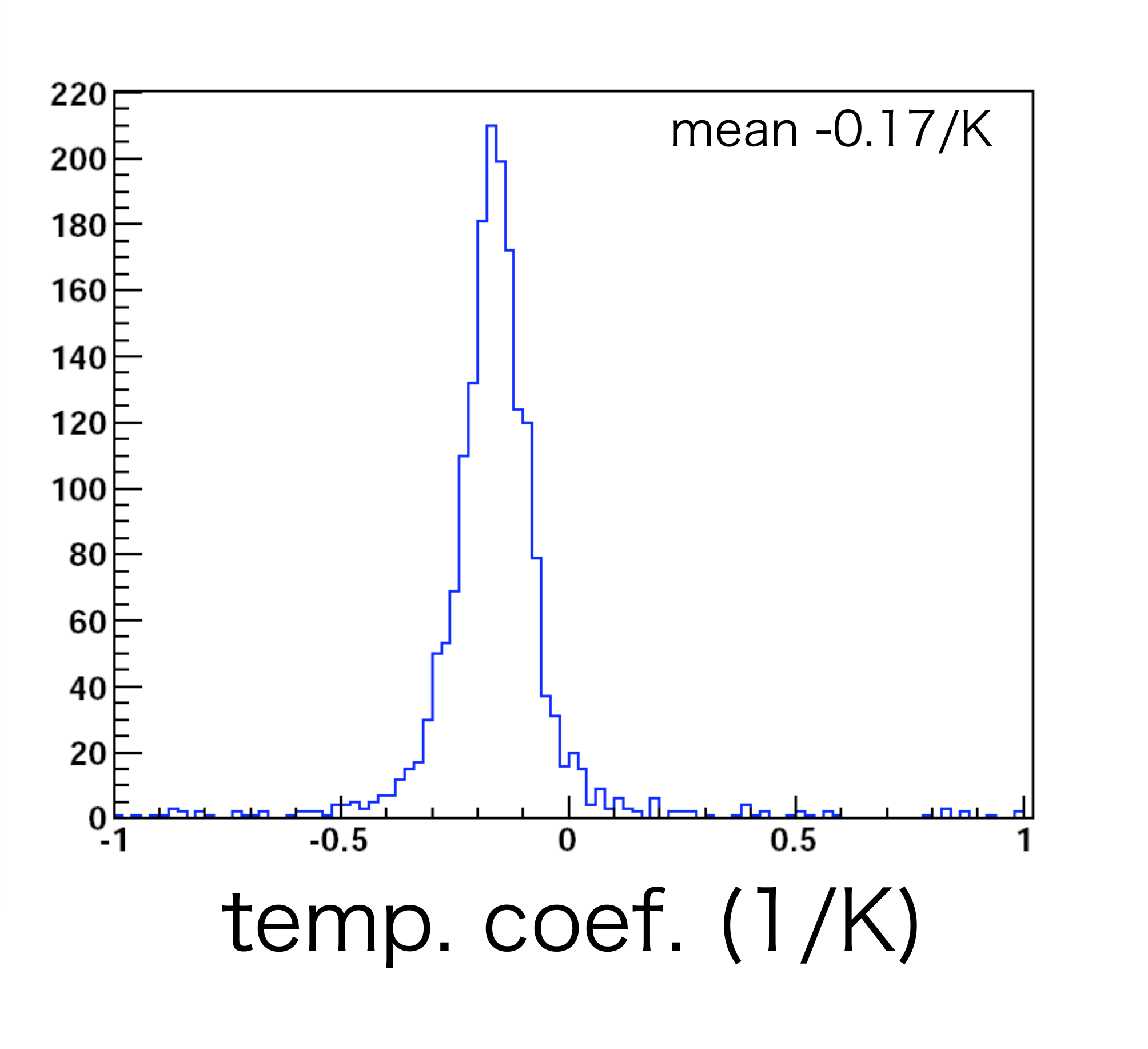}
\caption{\label{fig:scintecalled}\em  Scintillator ECAL LED monitoring system and temperature coefficients.
}
\end{figure}

 We have installed a calibration system in the prototype at FNAL, which
consists of a clear fibre with notches and a LED. Light from the
LED passes through a clear fibre and is reflected at the notches along the
fibre at the scintillator, then fed into a scintillator strip. The
number of lights can be counted and monitored for the stability of the
calorimeter system. The temperature coefficients are also extracted at
the FNAL beam test and shown in Fig.~\ref{fig:scintecalled}.

\subsection{Next generation engineering prototype and current status}

\begin{figure}
\centering
    \includegraphics[angle=0,width=0.4\textwidth]{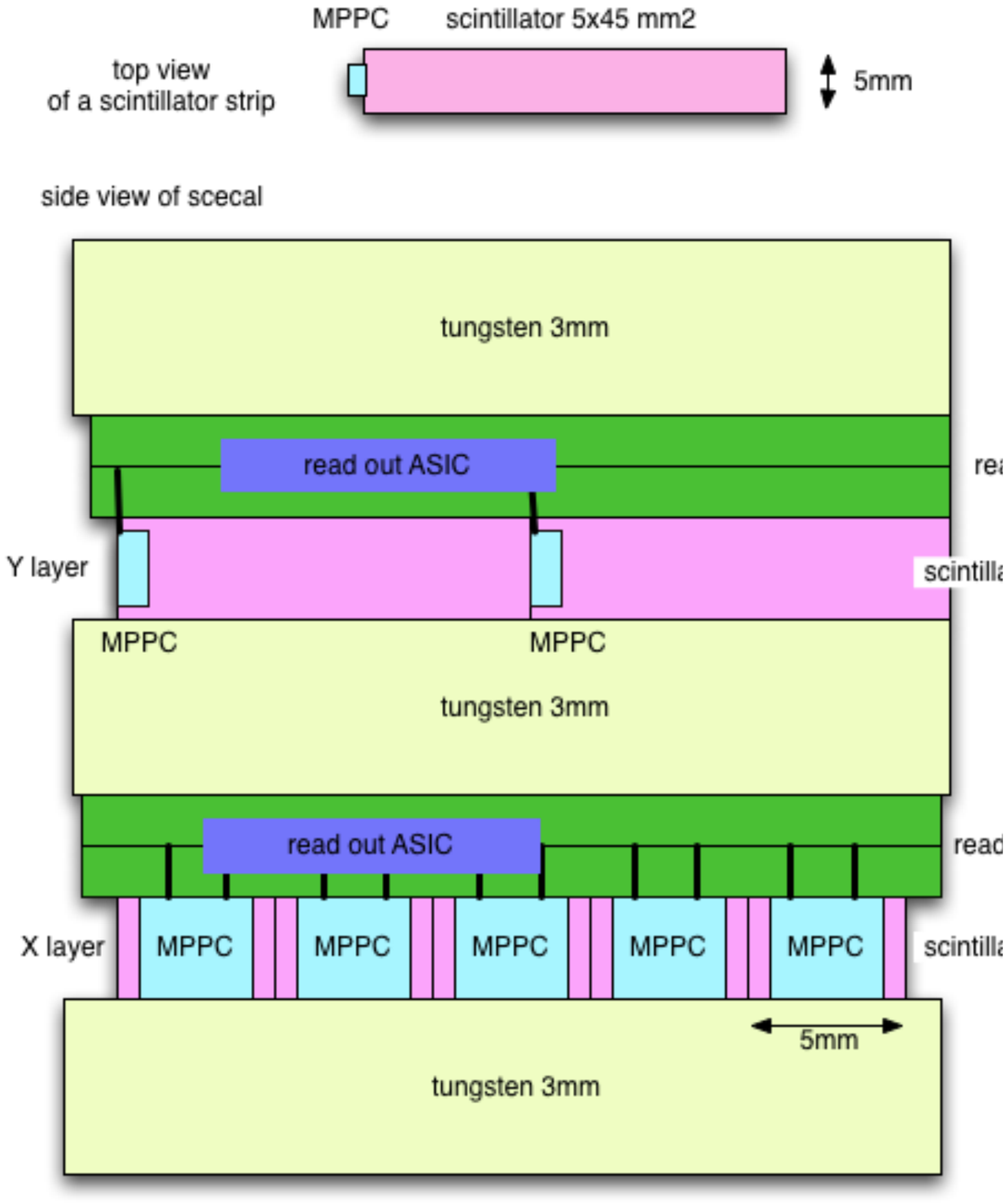}
\caption{\label{fig:scintecalxsect2}\em  Scintillator ECAL second generation prototype}
\end{figure}

According to the current understanding of PFA studies, a maximum transverse
segmentation of 5~mm is favoured, so that we are now aiming
to achieve a 5~mm wide strip. The ScECAL incorporating these narrower
scintillator strips will be constructed as in Fig.~\ref{fig:scintecalxsect2}.
A prototype should be produced and tested by 2012, according to the plan 
presented at the previous PRC review. 

 \subsubsection{Strip clustering algorithm development}
\begin{figure}
\centering
    \includegraphics[angle=0,width=0.4\textwidth]{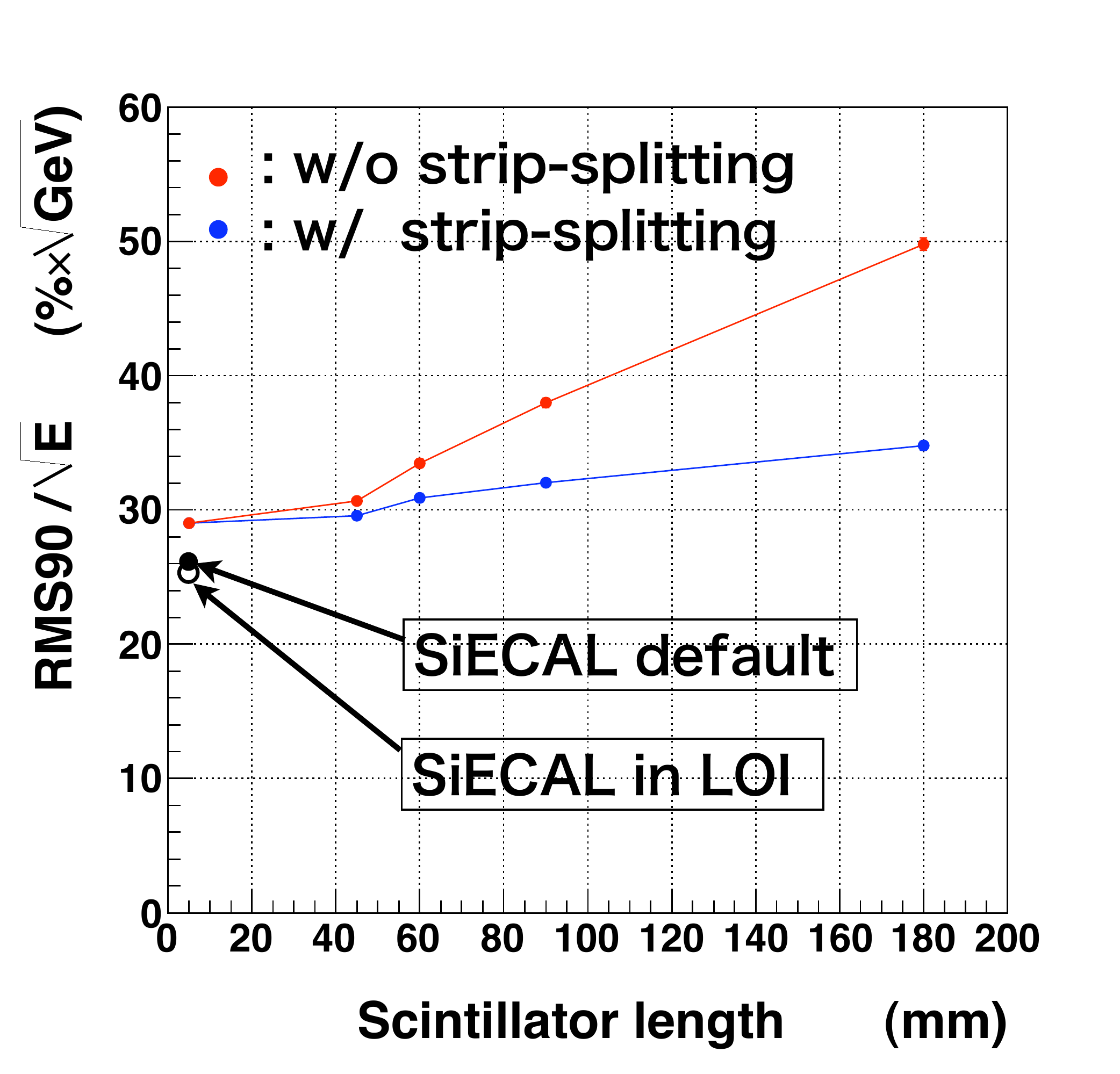}
\caption{\label{fig:scintecalstripclust}\em  Strip clustering perfromance: 
jet energy resolution vs.\ strip length.}
\end{figure}

 We have developed algorithms which will allow the energy desposited
 in the physical scintillator strips to be distributed among (virtual)
 square cells.  This is achieved by making use of the information of
 the previous and the next layers whose directions are orthogonal to
 the strip under consideration.  This algorithms is called
 Strip-Splitting and found to work well.  One of the results is shown
 in Fig.~\ref{fig:scintecalstripclust}, where the jet energy
 resolutions are plotted as a function of the length of the
 scintillator strip.

\subsubsection{Layer electronics development}
\begin{figure}
\centering
    \includegraphics[angle=0,width=0.4\textwidth]{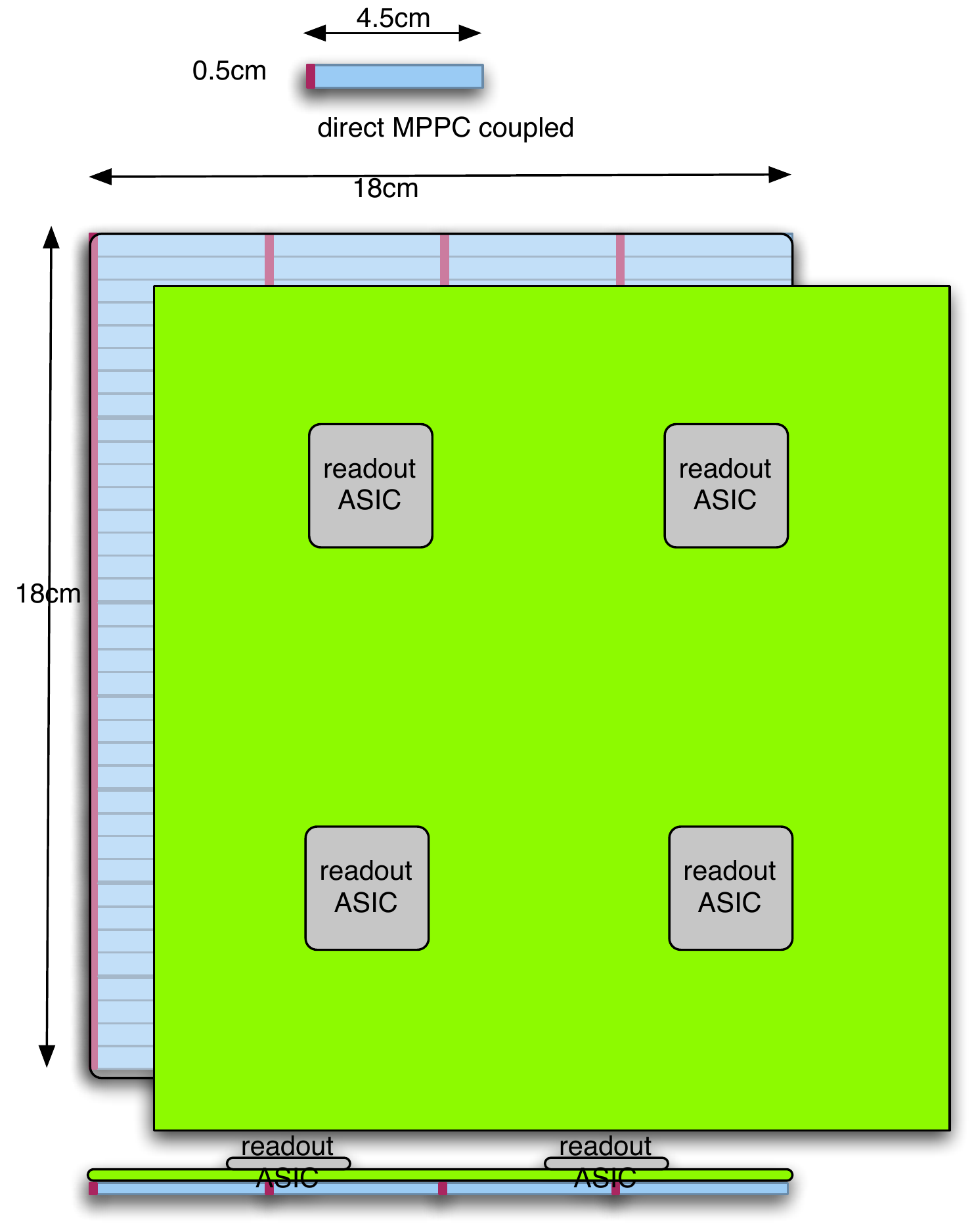}
\caption{\label{fig:scintecalebu}\em  Scintillator ECAL second generation 
integrated read-out electronics layer.}
\end{figure}

 A few layers are now under construction, as illustrated in
 Fig.~\ref{fig:scintecalebu}.  The scintillator layer will be
 constructed in Korea and Japan. The electronics layer, with SPIROC
 ASIC from France, is currently being designed by a DESY engineer and will be
 incorporated into layers in Japan.

\subsubsection{5~mm scintillator development}
\begin{figure}
\centering
    \includegraphics[angle=0,width=0.4\textwidth]{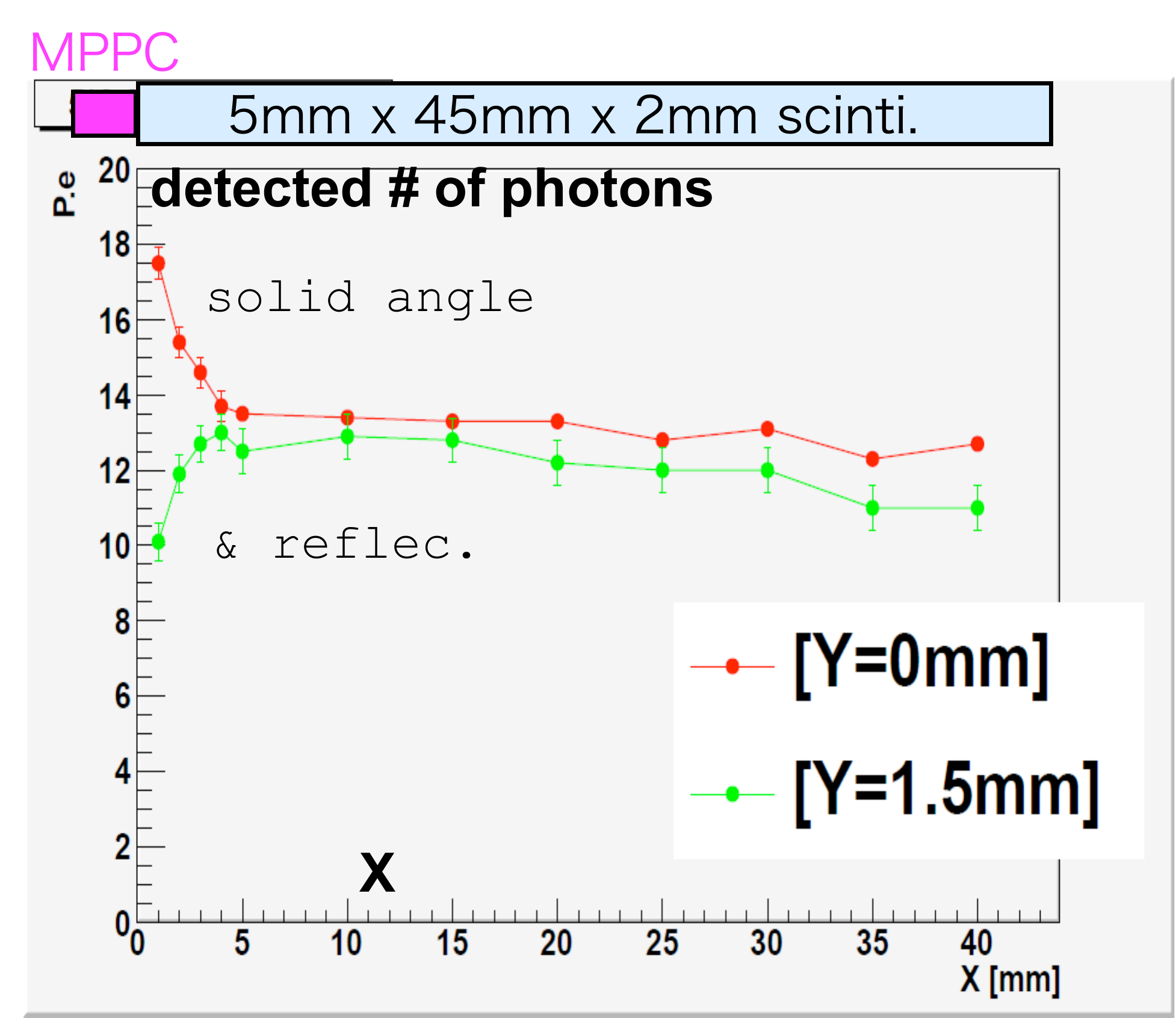}
\caption{\label{fig:scintecaluniformity}\em  Scintillator strip uniformity measured with 
a Sr-90 source.}
\end{figure}

 The width of a 5~mm scintillator is expected to be suitable without
 the use of wave length shifting fibre read out, though homogeneity should be
 confirmed. This has been tested using collimated $^{90}$Sr sources,
 with preliminary results shown in Fig.~\ref{fig:scintecaluniformity}.
Relatively good uniformity is measured at $x>5$~mm.  However, we find
non-uniformity due to the acceptance problem in the region near to the
photo-sensor. This indicates that further effort is required to
achieve optimal response, although in practise the overall energy
resolution for electrons and photons are little affected by this
non-uniformity.

\subsubsection{MPPC development}
We now have some MPPCs which consist of 2500 pixels of 20~$\mu$m$^2$
within a sensitive area of 1~mm$\times$1~mm. We have tested these
devices and found that the new MPPC exhibits an extended dynamic
range, and will continue to characterise their response further.

\subsection{Future R\&D}
 The integration of layers of scintillator sensors and readout
 electronics is the next key issue to be address for the current
 activity. Since this task is also a combination of several countr
 (Korea, Germany, France and Japan), international effort for the
 ILC/ILD will play an important role to achieve it.

\section{Digital ECAL: DECAL}
\label{sec:DECAL}


\subsection{Context}
   The studies of a digital ECAL (DECAL) continue in the UK, in spite
   of very significant funding difficulties.  In Dec.\ 2008, the STFC
   Executive recommended sufficient funding to allow the SPiDER
   Collaboration to construct a full physics prototype DECAL, as
   outlined in ~\cite{SPIDERPRC}.  By Dec.\ 2009, the funding for
   SPiDER had still not been issued and STFC informed the
   Collaboration that they would not do so.

   The UK groups in SPiDER have demonstrated that the INMAPS
   technology developed specifically for the DECAL application is
   viable in terms of basic pixel efficiency.  INMAPS is implemented
   as a 0.18$\mu m$ CMOS process in which a deep P-well implant stops
   signal charge from being absorbed in N-well circuits, and therefore
   allows the use of both NMOS and PMOS within the pixel, as well as
   (optionally) high resistivity silicon in the thin epitaxial layer
   to reduce the charge collection time.

\subsection{Testbeams in 2010}
   Following a successful test beam run at CERN in Sep.\
   2009 using 120~GeV pions, two further data taking runs have been
   carried out.  The first of these was at DESY in
   Mar.\ 2010, for which the primary goal was to quantify the peak
   electromagnetic shower density observed downstream of specific
   absorber materials.  A secondary goal was to make further pixel
   efficiency measurements.
\begin{figure}[htbp]
\centering
    \includegraphics[width=0.8\textwidth]{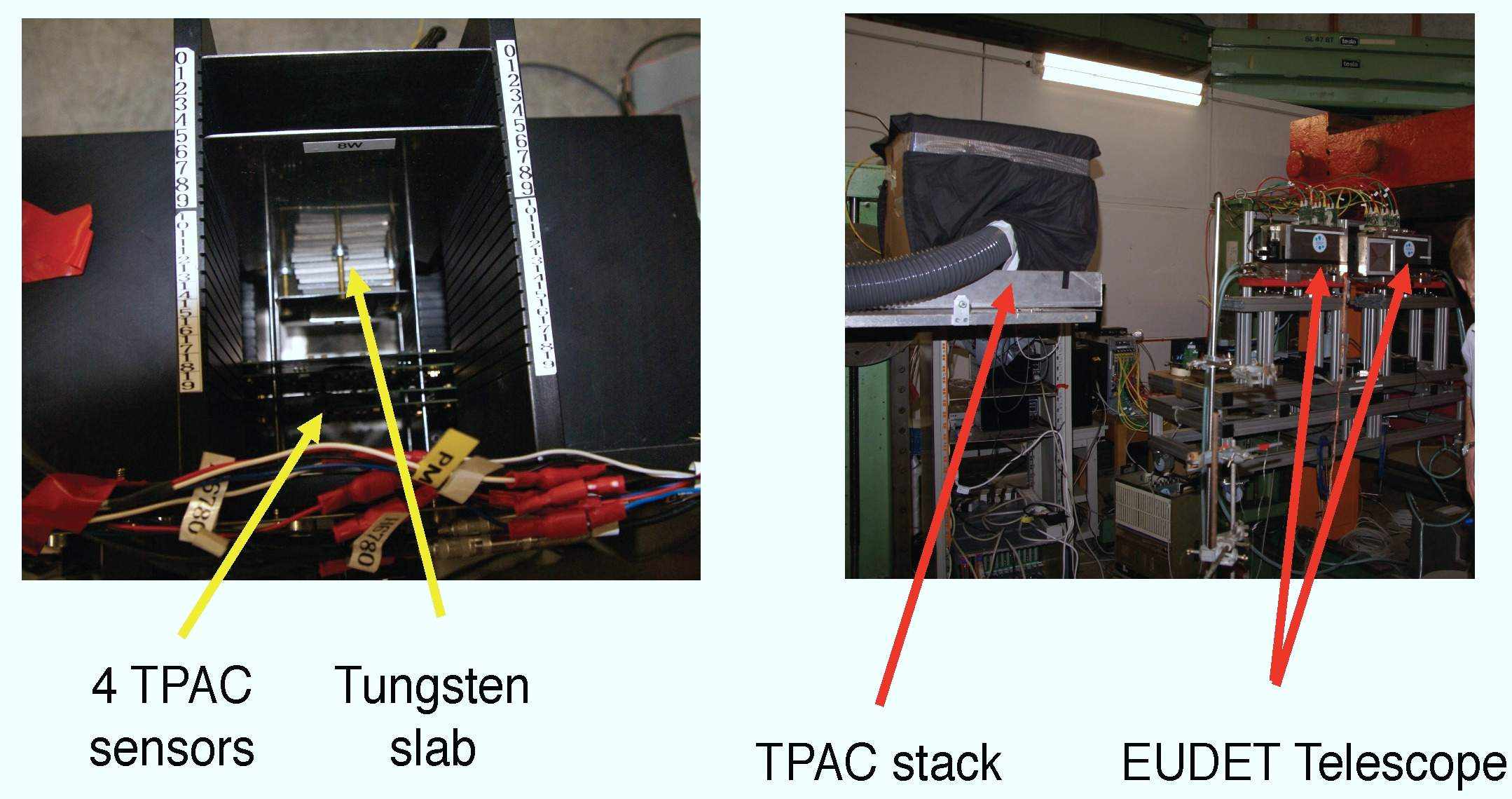}
\caption{\label{fig:decal_desy2010}
\em The DESY 2010 testbeam: (left) test stand, showing four upstream
sensors and tungsten absorber, and (right) test stand with cooling
system and EUDET telescrope upstream.}
\end{figure}
   Data were recorded with the 1--5~GeV electron beam, using a
   configuration in which four TPAC 1.2 sensors were aligned precisely
   along the beam direction using the same custom-built mechanical
   frame as at CERN.  Absorber material (W, Fe, Cu) was placed
   downstream of these, followed immediately by a further pair of TPAC
   sensors, to study the shower density.  The EUDET telescope was
   located upstream of the DECAL test stand, as shown in
   Figure~\ref{fig:decal_desy2010}.

   To complement the DESY run, similar, additional data was recorded
   at CERN in Sep.\ 2010, using the EUDET telescope alone as it has
   finer pitch than the TPAC sensor, with positrons between 10 and
   100~GeV.  Similar absorber materials and thicknesses to those at
   DESY were used. Analysis of these data is ongoing, with the aim of
   having first results to present at TIPP'11 in June.

\subsection{Pixel efficiency results}
\begin{figure}[htbp]
\centering
\includegraphics[width=0.45\textwidth]{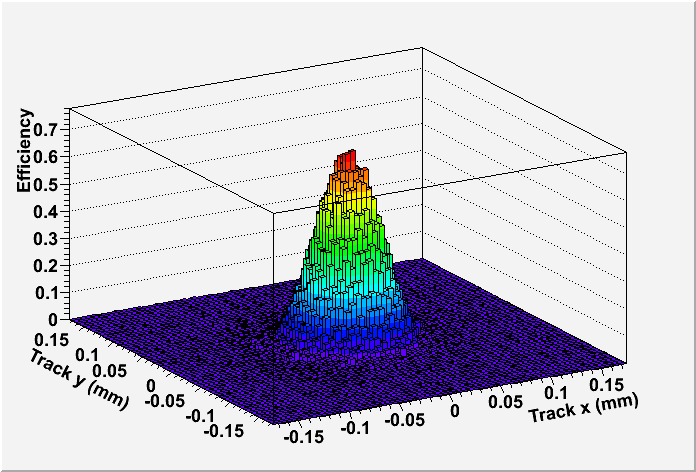}
\includegraphics[width=0.51\textwidth]{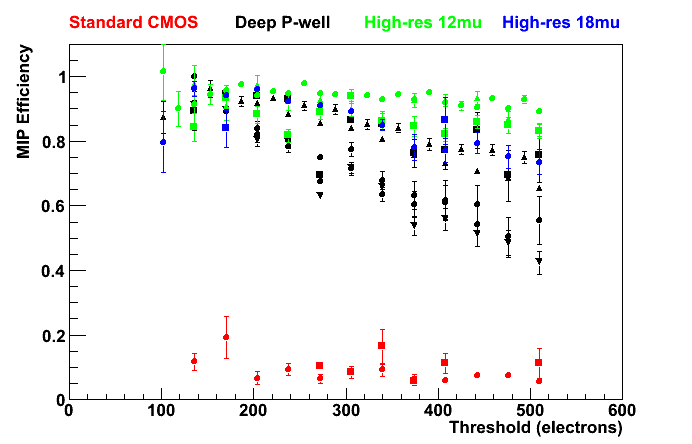}
\caption{\label{fig:decal_pixel_efficiency}
\em (left) Distribution of the probability of a pixel registering a hit
in response to a MIP, as a function of distance to the projected
track, and (right) MIP efficiency as a function of the sensor digital
threshold, for all four sensor variants studied.}
\end{figure}

   The studies of pixel efficiency from CERN 2009 testbeam and DESY
   were performed using a set of six TPAC 1.2 sensors aligned along
   the beam direction, in which the outer four sensors served as a
   beam telescope, while the two innermost sensors were considered as
   the devices under test.  The trajectory of the beam particle was
   projected onto the plane of both of these pixels, and each pixel of
   the test sensors was examined for the presence of hits as a
   function of the distance from the projected track.  The MIP hit
   efficiency was determined by fitting the distribution of hit
   probability to a flat top function, convoluted with a gaussion of
   the appropriate resolution to allow for finite tracking
   performance.  This efficiency, folded for all pixels together, is
   illustrated in Figure~\ref{fig:decal_pixel_efficiency}.

  The MIP efficiency was determined per pixel for both the DESY and
  CERN data, and for each of the four pixel variants tested.
  The variants (and corresponding marker colour in
  Figure~\ref{fig:decal_pixel_efficiency}) are:
\begin{enumerate}
 \item (red) in 12$\mu m$ standard (non-INMAPS) CMOS;
 \item (black) 12$\mu m$ deep P-well CMOS;
 \item (green) deep P-well within a 12$\mu m$ high resistivity epitaxial layer;
 \item (blue) deep P-well within an 18$\mu m$ high resistivity epitaxial layer.
\end{enumerate}

  The results \cite{decal:dauncey_ichep2010} are summarised in
  Figure~\ref{fig:decal_pixel_efficiency}, for a range of the sensor
  digital thresholds representative of the signal levels expected in
  DECAL pixels due to charge spreading.  (A typical MIP signal in a
  12$\mu m$ epitaxial layer of silicon is 1200 electrons and a single
  absorbs at most 50\% of this due to charge spreading.)

  From the results shown in the figure, it is observed that the
  standard, non-INMAPS sensors have markedly low efficiencies, which
  is attributed to signal charge being absorbed by in-pixel PMOS
  transistors.  In contrast, the use of the deep P-well reduces the
  absorption of signal charge by N-wells in the circuitry, improving
  very substantially the pixel efficiency by a factor of $\sim5$.  The
  addition of the high resistivity epitaxial layer further improves
  the pixel efficiency to $\sim100$\%.

\subsection{Future plans}
   It is no longer an option to plan for a physics prototype DECAL and
   the short-term future of the DECAL project is extremely uncertain
   at present.  A programme of radiation hardness testing is in
   progress during 2011, using X-ray sources at RAL, and may be
   extended to include neutrons and protons.  This is in part to
   understand how the TPAC sensor would satisfy the requirements of
   SuperB and ALICE ITS.  The technology development will continue
   while this is still possible.  The studies which have been carried
   out so far are in the process of being finalised, and a series of
   papers, e.g. \cite{decal:tpac_paper}, are in preparation to document
   what has been achieved.

\section{Analogue HCAL Technological Prototype}
\label{sec:AHCAL}
A technological prototype of a highly granular scintillator based 
AHCAL is under design and construction to demonstrated the feasibility of 
this approach in a realistic linear collider (LC) detector. 
The challenge is the high level of integration to maintain maximum 
compactness and hermeticity of the final detector, 
once the AHCAL design is realized as a whole barrel detector for a LC experiment.

\begin{figure}
\centering
    \includegraphics[angle=0,width=0.7\textwidth]{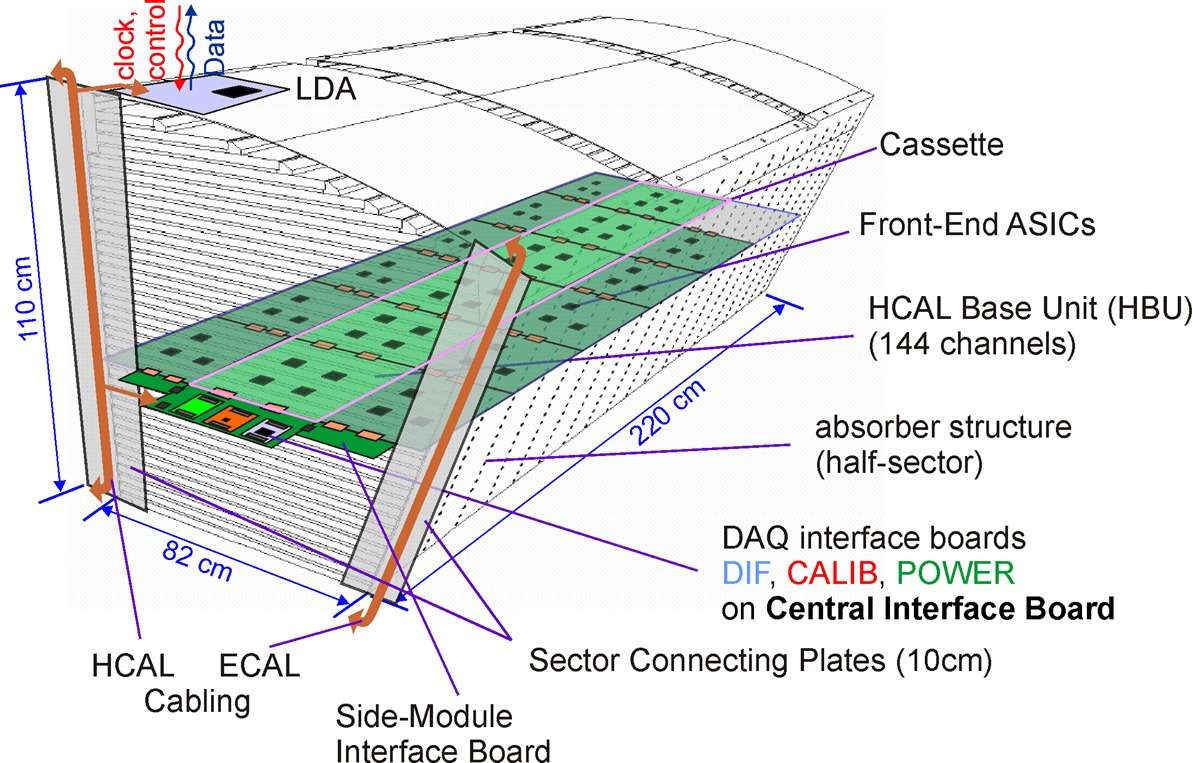}
\caption{\label{fig:ahcalmodule}\em Electronics integration
architecture for the technological AHCAL prototype.
}
\end{figure}

The envisaged detector architecture~\cite{EUDET_memo_2,EUDET_memo_23} is sketched in 
Fig.~\ref{fig:ahcalmodule}.
It is inspired by one variant of the ILD detector concept, but is very 
similar to those envisaged for SiD or CLIC. 

The barrel of the AHCAL has a cylindrical structure and will be placed
outside the electromagnetic calorimeter, while it is surrounded by the
magnet. The cylindrical structure is divided into 16 segments with 48 detector
layers each. 
The figure shows one sector of a barrel subdivided only once along the beam axis. 
This layout provides access to electronics and service interfaces once the 
detector end-cap is opened, but poses tight space constraints to the barrel end-cap
transition region.

One active layer consists of three parallel slabs. 
Each slab is again subdivided in six HCAL basic units (HBU) 
and the middle slab is connected to the DAQ via the Central Interface Board (CIB). 
The side slabs are in turn connected to the CIB via the Side 
Interface Boards (SIBs). 

\begin{figure}[!b]
\centering
\includegraphics[width=0.45\textwidth]{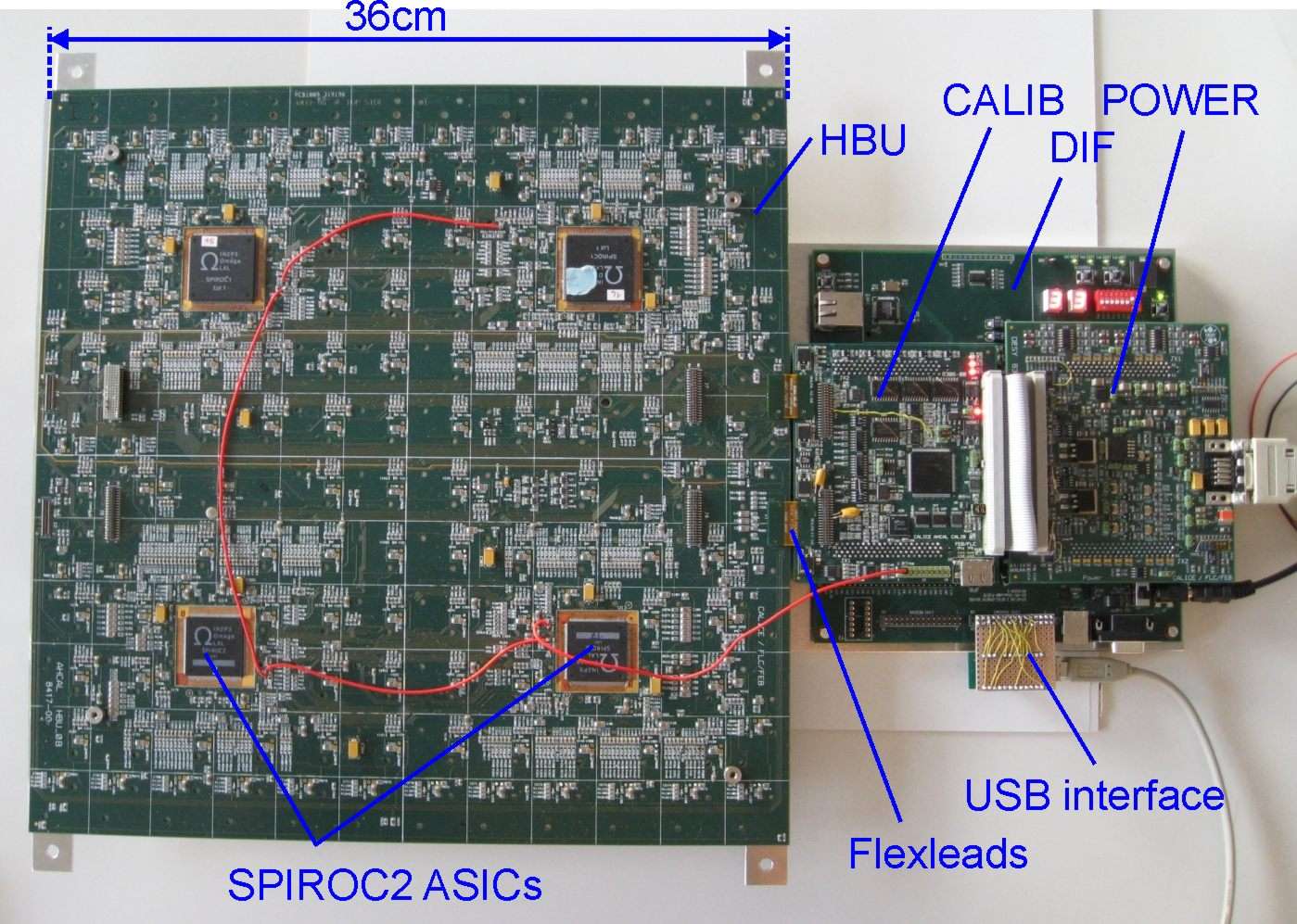}
\includegraphics[width=0.45\textwidth]{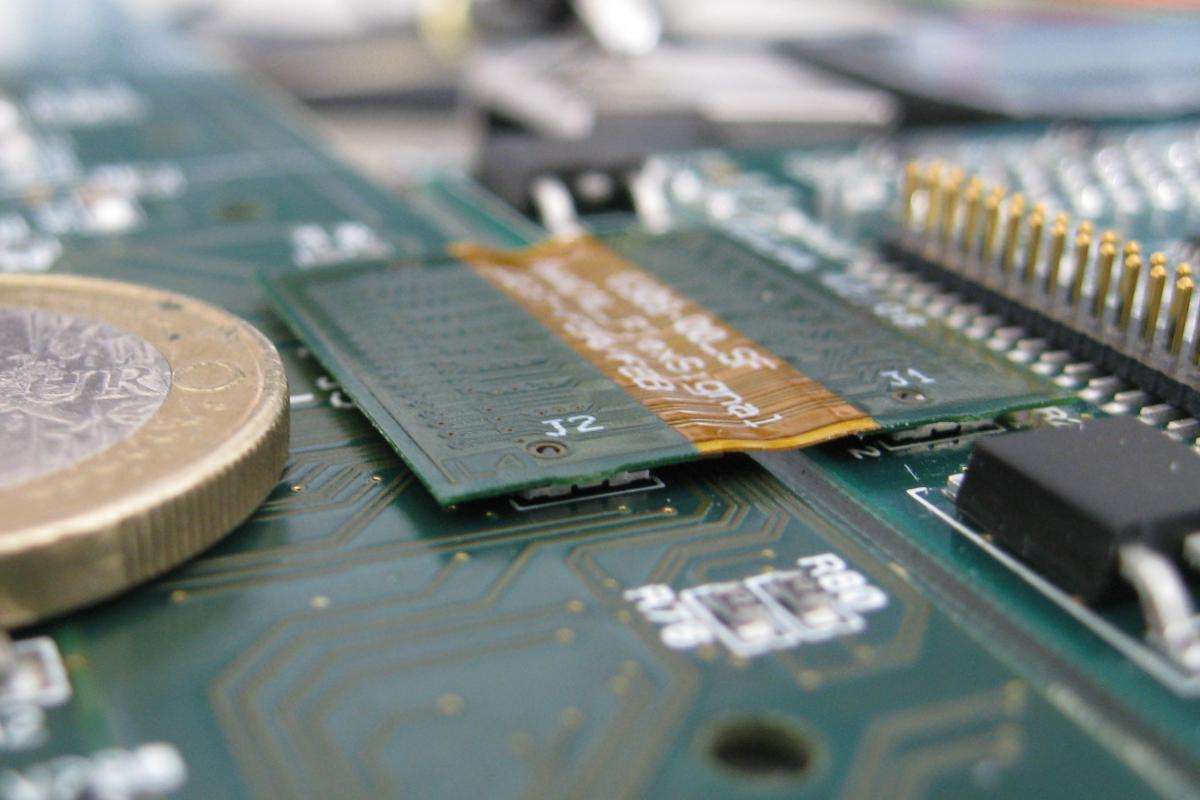}
\caption{Left) Current setup of an HBU as used in the DESY test measurements. Right) HBU interconnection via flexleads and ultra-thin connectors. }
\label{HBU}
\end{figure}

The first HBU module, along with the interface modules, is shown in
Fig.~\ref{HBU} left, as it is used in the DESY test setups. In the final
design the HBUs are interconnected by flexleads and ultra-thin
connectors with a stacking height of 0.8\,mm (see Fig.~\ref{HBU} right),
which are also used to connect the HBUs to the CIB.


\subsection{Tiles and ASICs}

\begin{figure}
\centering
\includegraphics[width=0.45\textwidth]{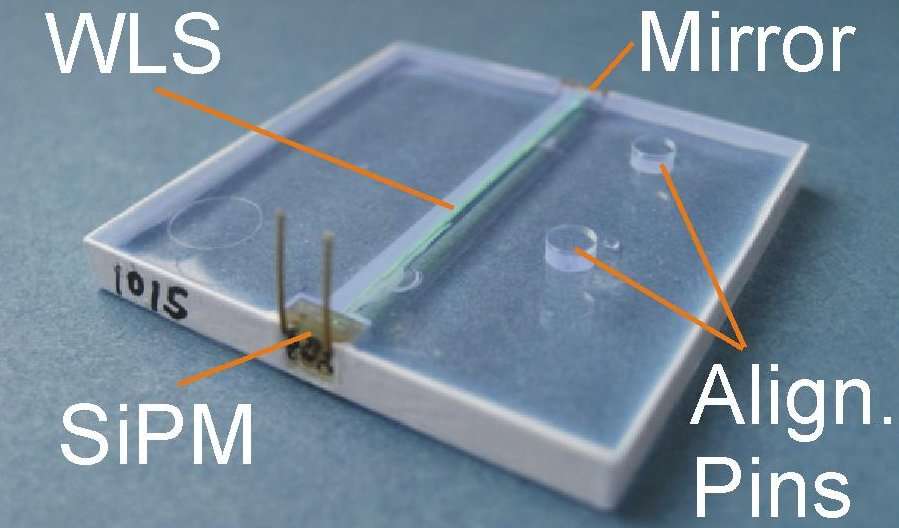}
\includegraphics[width=0.45\textwidth]{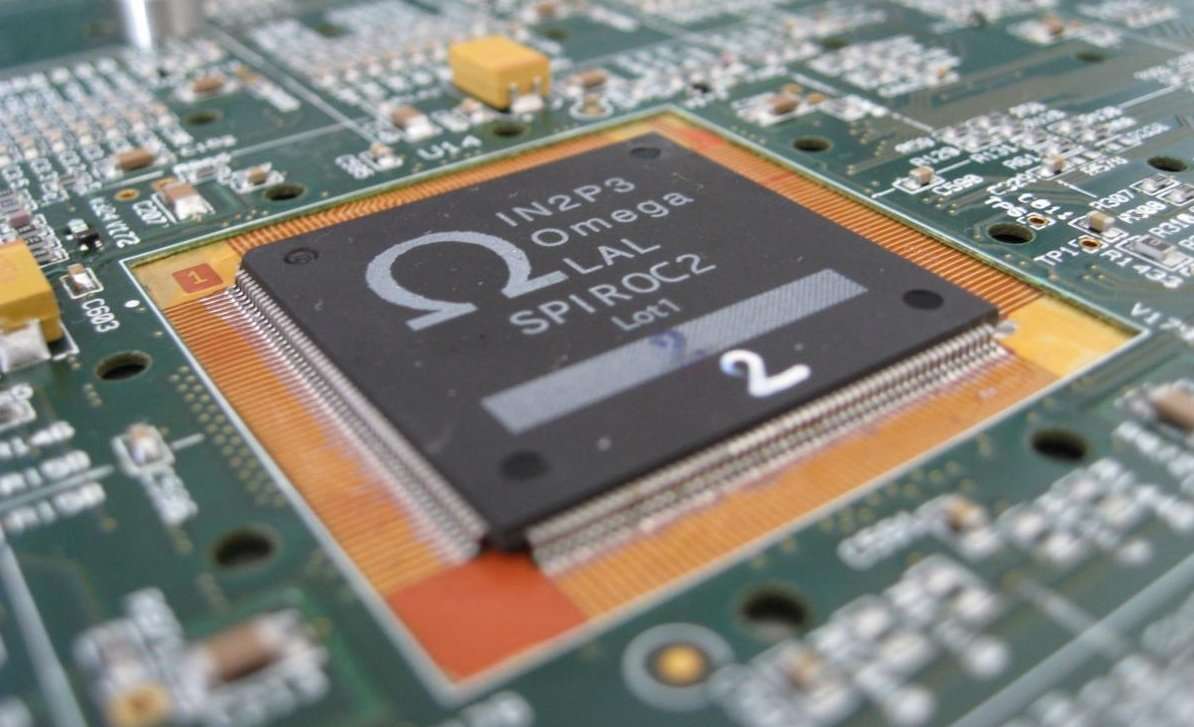}
\caption{Left) Scintillating tile with embedded wavelength shifting fiber,
  SiPM, mirror and alignment pins. Right) Integration of the SPIROC ASIC into the PCB.}
\label{Tile}
\end{figure}

The signal that is detected by the SiPMs is produced by scintillating
tiles with a size of $3\times 3\times 0.3$\,cm$^3$, as shown in 
Fig.~\ref{Tile} left. The new design differs from the design used in the
physics prototype~\cite{PPT} and includes a straight wavelength
shifting fiber coupled to a SiPM with a size of 1.27\,mm$^2$ on one
side and to a mirror on the other side. The SiPM comprises 796 pixels
with a gain of $\sim 10^6$. Two alignment pins are used to connect the
tiles to the HBU's printed circuit board (PCB) by plugging them into
holes in the PCB. The nominal tile distance is 100\,$\mu$m.


For each HBU the analog signals from the SiPMs are read out by four
36-channel ASICs equipped with 5\,V DACs for a channel-wise bias
voltage adjustment. They provide two gain modes, which leads to a
dynamic range of 1 to 2000 photo electrons. The chips are designed to operate with
pulsed power supply for minimized heat dissipation. The foreseen power
consumption amounts to 25\,$\mu$W per channel for the final LC
operation. The main new features of the ASICs compared to the physics
prototype are the integration of the digitization step (12-bit ADC and
12-bit TDC for charge and time measurements) and the self-triggering
capability with an adjustable threshold, which acts as a on-detector 
zero suppression. To reduce the height of the
active layers the ASICs are lowered into the PCB by $\sim
500$\,$\mu$m. This leads to a total reduction of the AHCAL diameter of
48\,mm. 
A picture of an ASIC as it is embedded into the PCB is shown in Fig.~\ref{Tile} right.

\subsection{Detector/DAQ interface}

\begin{figure}[!t]
\centering
\includegraphics[width=3.5in]{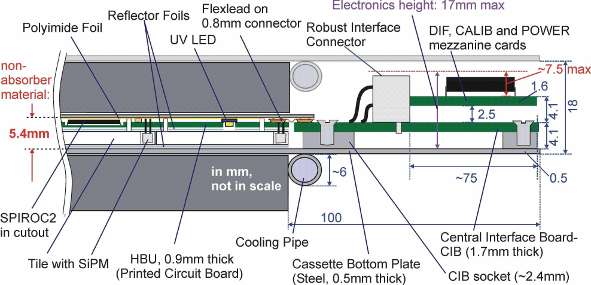}
\caption{Cross section of one HCAL layer, including the absorbers,
  tiles, SiPMs, PCBs, ASICs and the detector/DAQ interface modules.
  All dimensions are given in mm.}
\label{HCAL_cross_endface}
\end{figure}

Fig.~\ref{HCAL_cross_endface} shows the cross section of one AHCAL
layer including the dimensions of the single components. The active
layers, including the tiles, SiPMs, PCBs and ASICs, are shown as they
are placed between two layers of absorber material and connected to
the CIB. The total height of the detector/DAQ interface modules hosted
by the CIB has to be very small ($\sim$~18\,mm in case of a tungsten
absorber) in order to fit between two layers.

\begin{figure}[!t]
\centering
\includegraphics[width=3.5in]{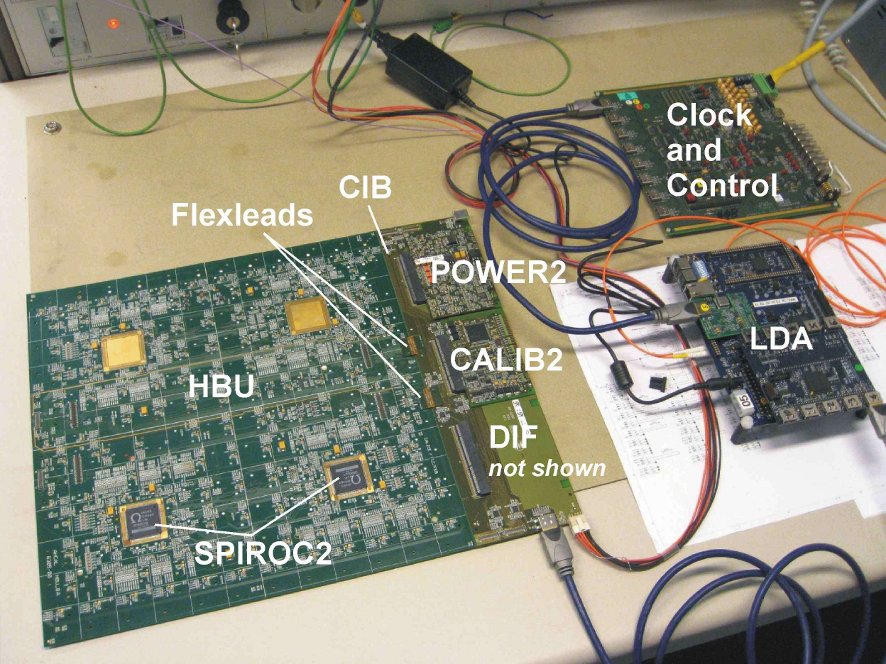}
\caption{Electronics setup of the inner detector electronics (HBU) connected
to the DAQ interface modules (DIF, CALIB2 and POWER2) via the flexleads and
the CIB module, and together with the final CALICE DAQ modules LDA and C\&C.
The setup is not yet in operation, but still in commissioning phase.}
\label{CIB}
\end{figure}

In Fig.~\ref{CIB} the electronics setup is shown in the final assembly stage: The
DAQ interface modules DIF, CALIB2 and POWER2 are realized as mezzanine
cards on top of the central-interface board (CIB). All modules fulfill the
stringent space requirements for an arrangement in the pitch of the
absorber plates. 
All DAQ interface modules have been realized and
are in commissioning stage at the moment. The module DIF has been realized
by the Northern Illinois University (NIU). It serves as the interface
between the inner detector module (HBU) and the back-end DAQ. The CALIB2
module controls the light calibration and gain monitoring system for the
SiPMs, which is based on the usage of ultraviolet LEDs. The POWER2 module
provides all necessary supply voltages to the inner detector electronics,
including the SiPM bias voltages. Additionally, it enables the
power-cycling of the HBU electronics in the scheme of the ILC bunch-trains.
The final HBU module with the newest generation of readout ASICs SPIROC2b
is currently in production.

Up to 10 CIB modules can be connected to one Link and Data Aggregator (LDA) module of the final CALICE DAQ. The LDA  collects the parallel
incoming data streams of the DIFs on the CIB modules, and serializes the
data to a single, duplex optical line, which connects to the backend DAQ
(ODR module, not shown in Fig.~\ref{CIB}). Timing critical signals as the front-end
clock, which is used by all front-end modules for a synchronous operation
of all detector parts, are provided from the central experiment through the
Clock-and-Control module (C\&C). In testbeam setups, the C\&C module can be
used for the distribution of a fast trigger as well. The setup as shown in
Fig.~\ref{CIB} is still in the beginning of the commissioning phase, the prototype
has been operated so far by a preliminary Labview DAQ via an USB connection
between a PC and the DIF-module.


\subsection{Light calibration system}


Since the SiPM response shows a strong dependence on the temperature
and bias voltage and saturates due to the limited number of pixels, a
gain-calibration and saturation-monitoring system with a high dynamic
range is needed. In the calibration mode of the ASICs a very low light
intensity is needed to measure the gain as the distance between the
peaks in a single-pixel spectrum, while at high light intensities
(corresponding to $\sim$100 minimum-ionizing particles (MIPs)) the
SiPM shows saturation behavior. Currently there are two concepts
under investigation:
\begin{itemize}
\item One LED per tile that is integrated into the detector gap.
  This system is used in the HBUs in the DESY test setups.
\item One strong LED outside the detector, while the light is
  distributed to each tile via notched fibers (see~\cite{Ivo}).
\end{itemize}


Both options have been successfully tested on the DESY test setups in
the laboratory and under testbeam conditions. 
The measured cross-talk is purely optical and is of the order of 2.5\%. 
The dynamic range of the system redesigned for the
construction of the engineering prototype is currently under
investigation. The channel uniformity is also an open issue, since for
the first LED system the individual LEDs have a large spread of the
emitted light intensity, while for the second system the light
coupling from the fiber to the tile and its mechanical integration in
a full prototype is unsolved.

\subsection{Measurements and results}


\begin{figure}[!t]
\centering
\includegraphics[width=0.45\textwidth]{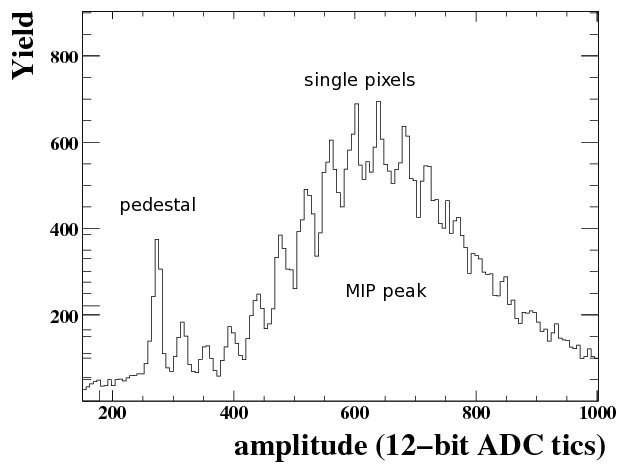}
\includegraphics[width=0.45\textwidth]{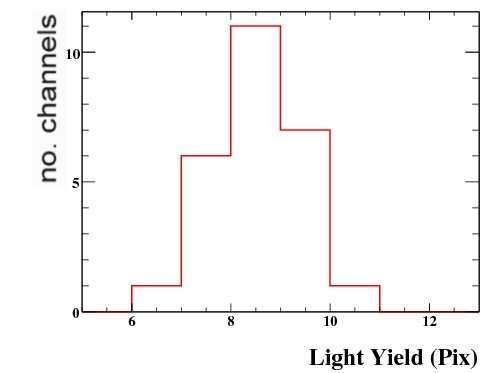}
\caption{Left) Measurement result of a typical MIP spectrum obtained with
  the HBU and the DAQ interface modules using the newly developed
  Labview/USB DAQ in the 6\,GeV DESY electron testbeam. Right) Light yield as measured in the DESY testbeam facility for multiple channels of the HBU.}
\label{MIP}
\end{figure}

The main task of the current characterization is to prove the
suitability of the realized detector-module concept for the
larger-scale prototype with 2500 channels and the final length of
2.2\,m. Two HBUs are in operation, one in
the DESY 6\,GeV electron testbeam facility (the 2-6\,GeV electrons
that have been used are MIPs in the scintillating tiles) and the
second in a laboratory environment.  

After the investigation of the fundamental properties of the SPIROC chip 
like noise behavior and signal delays~\cite{Riccardo}, measurements in the
laboratory using the LED calibration system and a charge injection
setup, as well as testbeam measurements have been performed to
investigate the uniformity of the tile/SiPM response for multiple channels. 
Fig.~\ref{MIP} left shows a
typical MIP spectrum measured with the electron testbeam. It can be
seen that single pixel peaks are clearly distinguishable for more than
10 peaks. The first peak is the pedestal peak and the maximum of the
spectrum is at 9 pixels. The distribution of the light yields, defined
as the most probable number of active pixels for a MIP signal, is
plotted in Fig.~\ref{MIP} right.


\begin{figure}[!t]
\centering
\includegraphics[width=0.45\textwidth]{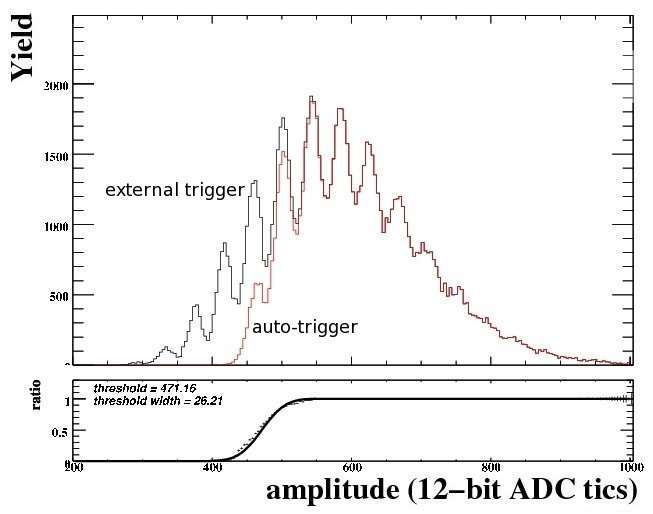}
\includegraphics[width=0.45\textwidth]{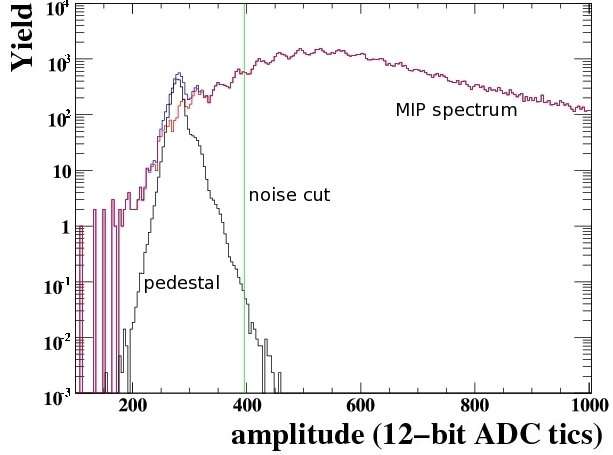}
\caption{Left) Comparison of a single-pixel spectrum produced with LED light
 for external- and auto-triggering with a given threshold. Right) Calculated auto-trigger threshold for having a noise to signal ratio of smaller than 10$^{-4}$. 
 An independent pedestal measurement is compared to a MIP spectrum.}
\label{AT_LED_MIP}
\end{figure}

The auto-triggering function of the SPIROC chip has been tested~\cite{Jeremy}. 
Fig.~\ref{AT_LED_MIP} left shows two single-pixel spectra
measured with LED light and external trigger (black histogram) and
auto-trigger (red histogram), respectively. The ratio of the two histograms 
is also shown and gives an
impression of the width of the trigger turn-on curve. 
After the turn-on the trigger efficiency is 100\%. 




The threshold of the auto-trigger will be adjusted in order to
minimize the noise hits and simultaneously maximize the efficiency for
measuring a MIP. 

For a threshold requirements of less then than 10$^{-4}$ noise hits per event, one gets a MIP efficiency of around 95\% as illustrated in Fig.~\ref{AT_LED_MIP} right.

Here the noise threshold  is fixed with respect to the pedestal distribution for one given channel (black histogram) and the MIP detection efficiency is calculated from the MIP distribution (red histogram) of the same channels.



\subsection{Future plans}

In 2011, the system will be further commissioned, including also the timing functionality.In parallel,
final versions of the read-out and interface boards are in production, With this, all components are in
hand for a multi-module layer tests, using the tile sand new sensors under production at ITEP. 

For 2012 a multi-layer test is planned where several options are possible. The HBUs can be arranged in
a flexible manner, such that one can instrument an electromagnetic section if the stainless steel
module (a tower of 12 HBUs), or produce a few larger area layers (two by tow HBUs) to start exploring
time-resolved 4-dimensional shower measurements in the tungsten HCAL. Provided sufficient funding, the
second generation prototype will be extended to 40 hadronic layers.

\section{Tungsten Analogue HCAL: W-AHCAL}
\label{sec:WAHCAL}

\subsection{Motivation}

A hadron calorimeter required for the multi-TeV range of the CLIC linear
collider will have to cope with increased jet energies. Particle separation will
become more difficult and confusion will be of increased importance in the
detector resolution when particle flow algorithm is used. But also leakage will
be an important contribution to the energy resolution. In the design of the
detectors proposed for CLIC a depth of the hadron calorimeter of 7.5~$\lambda_I$ is
required. The calorimeter has also to be placed inside the solenoid to achieve
optimum resolution. Iron ($\lambda_I=16.77$~cm) as absorber would yield then to a larger radius for the
solenoid, which would be costly and difficult to realize. The choice of an
heavier absorber material like tungsten ($\lambda_I=9.95$~cm) leads to a substantially smaller solenoid
diameter.

However, tungsten needs to be validated as absorber material. A particular
question is the influence of delayed neutrons produced by spallation of the
tungsten on the shower development. Also the simulation of the hadronic shower
development in tungsten has not yet been validated to the same precision as the
one in iron as absorber. Therefore a sampling calorimeter has been built using a
tungsten absorber and scintillator tiles as readout with a total depth of 4.9~$\lambda_I$.
First experimental data have been taken using a calorimeter depth of 3.9~$\lambda_I$.

\subsection{Experimental prototype}

\begin{figure}[!ht]
  \centering{
  \includegraphics[width=0.45\textwidth]{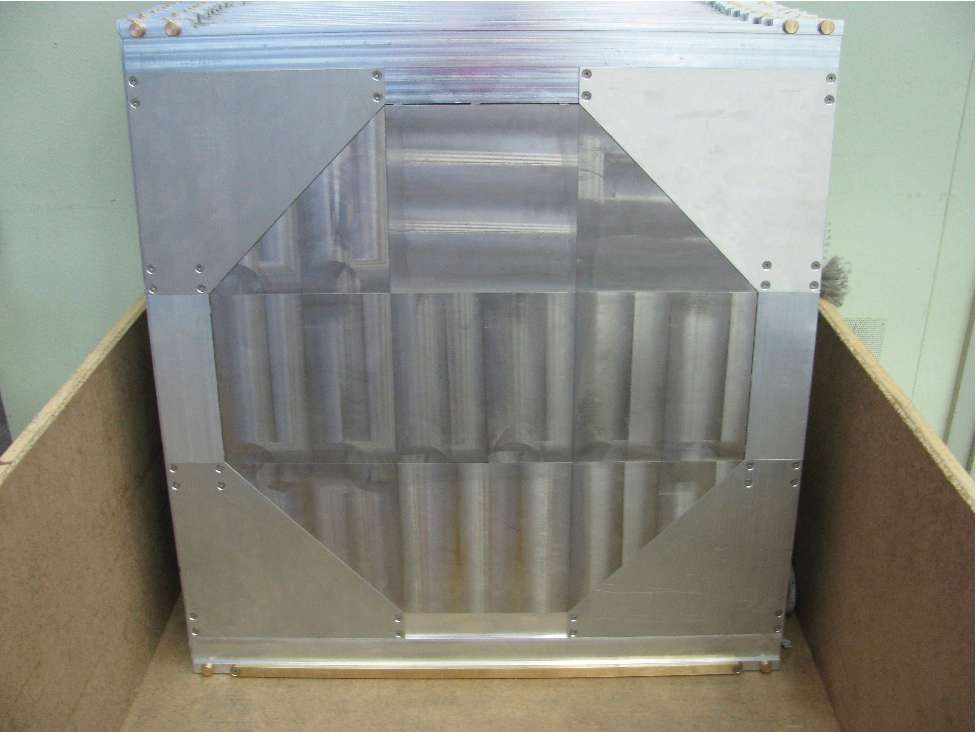}
   \hspace{0.2cm}
  \includegraphics[width=0.45\textwidth]{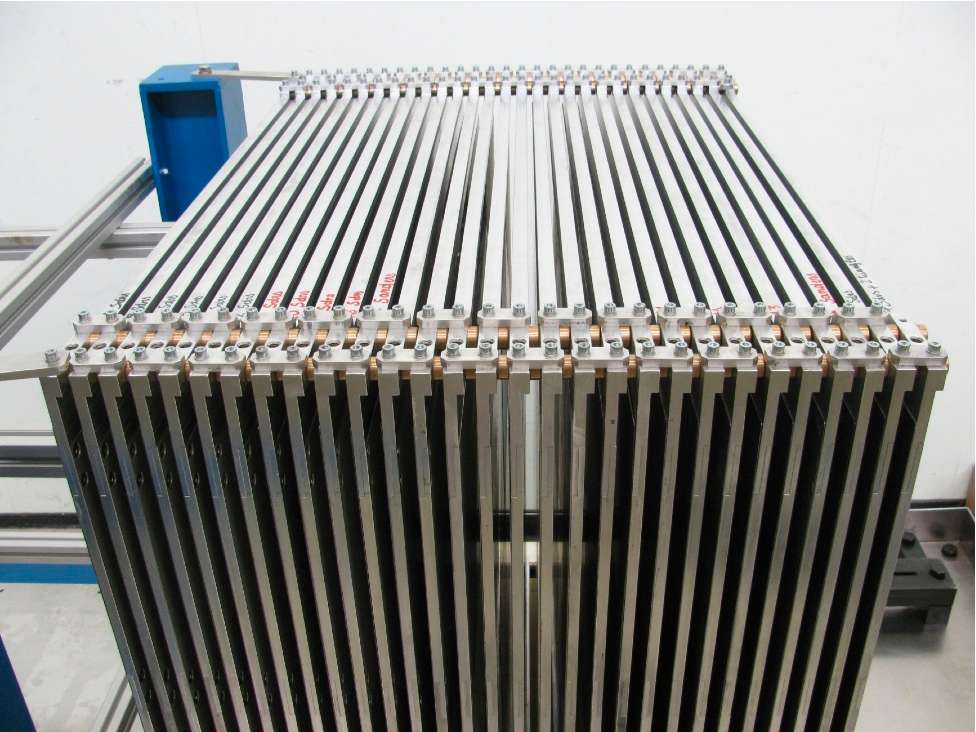}
  }
  \caption{View of W-HCAL absorber plates (left) and stack during assembly(right).}
  \label{fig:WAHCAL-tungstenPlate} 
\end{figure}

The W-HCAL prototype has been designed in order to use the scintillator
cassettes built for the AHCAL physics prototype~\cite{PPT}.
The material used
in the absorber was a sintered alloy formed by 93\% W, 5.3\% Ni and 1.7\% Cu. This
alloy allows easy machining and handling since it is much less brittle then pure
tungsten. Each absorber layer has been assembled from 5 square
tungsten plates 
of $27\times 27 \times 1 \; \textrm{cm}^3$ size and 4 triangular plates with corresponding size to form an
octagon with 81~cm diameter (see Fig.~\ref{fig:WAHCAL-tungstenPlate}, left).

\begin{figure}[!ht]
  \centering{
  \includegraphics[width=0.4\textwidth]{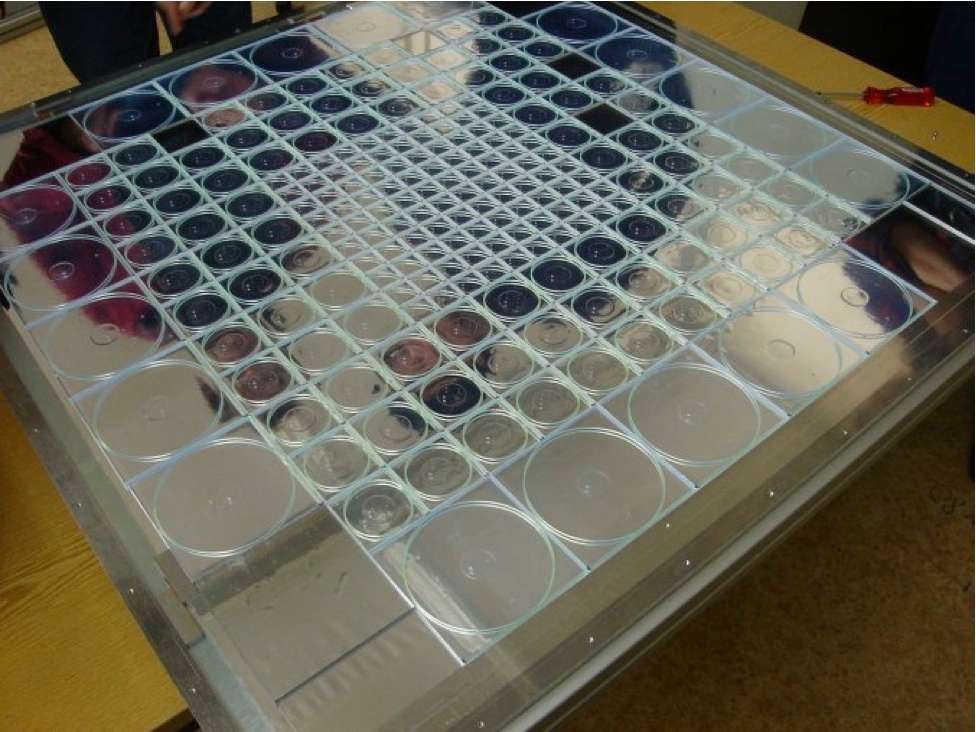}
   \hspace{0.2cm}
  \includegraphics[width=0.55\textwidth]{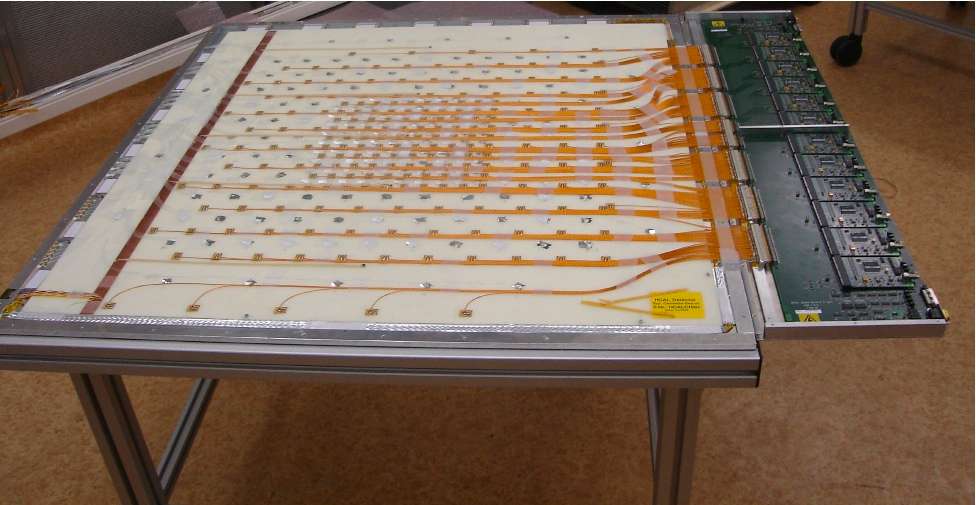}
  }
  \caption{Scintillator tiles layer (left), assembled module with front end electronics (right).}
  \label{fig:WAHCAL-electronics} 
\end{figure}

This octagon has been glued into an aluminum frame of a size of $1
\times 1$~m~$^2$. For stability reasons the assembly has been
glued to a 0.5~mm thick and 1~m$^2$ sized stainless steel
plate. Forty layers with an absorber thickness of 10~mm each have in
total been assembled. Thirty and of these layers have been constructed
in 2010 and another ten in 2011. The first thirty layers have been
assembled in a stack with an interleaving space of 14~mm in between
successive absorber layers leaving room for scintillator tile
cassettes as active modules (see
Fig.~\ref{fig:WAHCAL-tungstenPlate}, right).  Each cassette contains
216 scintillator tiles with a thickness of 5~mm: In the central core
of the detector are situated 100 tiles with a size of $3\times
3\;\textrm{cm}^2$, surrounded by 96 tiles sized $6 \times
6\;\textrm{cm}^2$.  As can be seen in
Fig.~\ref{fig:WAHCAL-electronics}, left, cells with a size of $12
\times 12\;\textrm{cm}^2$ are arranged on the very outside of the
cassette.

Each tile is read out individually via a wavelength shifting fibre
coupled to a silicon photomultiplier (SiPM) mounted on the tile. The
SiPM is a multi-pixel avalanche photodiode operated in Geiger mode and
provides a gain of more than 105. The scintillator tile layers and the
read-out are the same used already in the AHCAL tests with steel
absorber in the previous test beams at DESY (2006), CERN (2007) and
FNAL (2008, 2009).

The front end electronics is mounted on one side of the cassettes (see
Fig.~\ref{fig:WAHCAL-electronics}, right).  It is based on 16
channel ASICs which are read out by the standard CALICE DAQ
system. For the calibration and equalization of all the detector
channel, test bench characterizations as well as test beam data are
used. The detailed procedure is described elsewhere \cite{PPT}.

Minimum ionizing muon beams were used to equalize the cell
response. All the cells are equipped with an LED illumination to
monitor the gain of each of the SiPMs. In addition the SiPM
temperature and SiPM bias voltage have been monitored.  Detailed
studies of the applied temperature offset corrections were performed.

\begin{figure}[!h]
  \centering
  \includegraphics[width=0.90\textwidth]{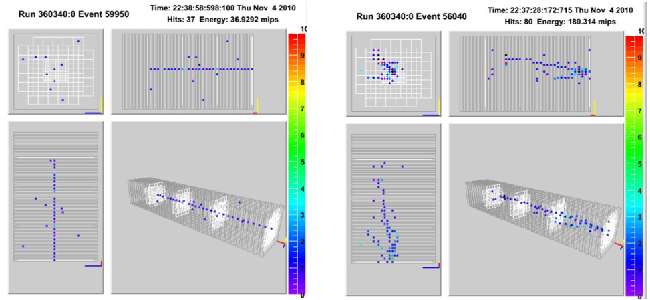}
  \caption{Example of muon (left) and pion (right) event displays in
    the W-HCAL, for a beam energy of 8~GeV.}
  \label{fig:WAHCAL-eventDisplay} 
\end{figure}

The prototype, equipped with thirty layers of tungsten absorber, has
been put into the CERN-PS.  About 28 million triggers have been taken
using $e^{\pm}$, $\mu^{\pm}$, $\pi^{\pm}$ and protons in the energy
range of 1 to 10~GeV. The beam passed through 2 threshold Cherenkov
counters placed upstream in the beam which were used for offline
particle identification.  Typical events of muon and pions are shown
in Fig.~\ref{fig:WAHCAL-eventDisplay}.

\begin{figure}[!h]
  \centering{
  \includegraphics[width=0.45\textwidth]{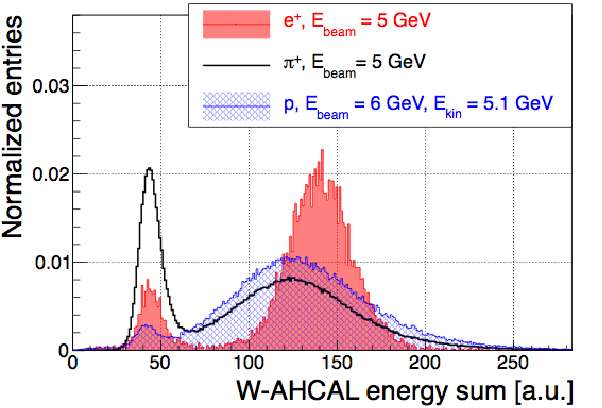}
   \hspace{0.2cm}
   \includegraphics[width=0.5\textwidth]{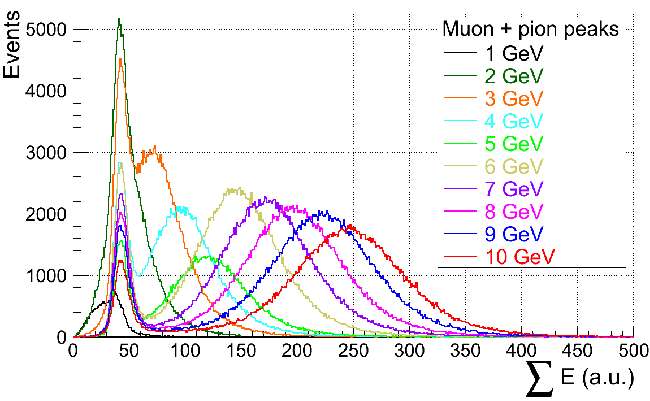}
  }
  \caption{Total energy deposited in the W-HCAL: (left) 5~GeV
    positrons, pions and protons; (right) Muon and pion peaks for beam
    energies from 1 to 10~GeV.}
  \label{fig:WAHCAL-response} 
\end{figure}

In Fig.~\ref{fig:WAHCAL-response}, left, the energy sum of the total
calorimeter is plotted for positrons, muons, pion and protons at
5~GeV. The beam is always contaminated with muons which behave as
minimum ionizing particles. Therefore the muon peak is well
distinguished for beam energies $E_{beam} \geq 3$~GeV.  In
Fig.~\ref{fig:WAHCAL-response}, right, are plotted the calorimeter
reponse of muons and pions for energies varying from 1 to 10~GeV.

Note that the presented results still use old calibration data (from
CERN 2007). However, a first look at the new MIP calibrations
indicates very similar calibration values.

We foresee to extend the test in 2011 to an energy range of up to
300~GeV at the \mbox{CERN-SPS} accelerator. From energies of
approximately 60~GeV onwards, transversal leakage will become an
important contribution to the observed resolution. In order to cope
with this, it is planned to add an instrumented tail catcher with a
thickness of 6~$\lambda_I$ behind the calorimeter.

 \subsection{T3B: Time Structure of Hadronic Showers in the W-AHCAL}
 \label{sec:T3B}
 For calorimeters in CLIC detectors, the time stamping capabilities are
of significant importance because of the high bunch crossing frequency
of 2~GHz and the high hadronic background from $\gamma\gamma
\rightarrow \mathrm{hadrons}$ processes. For hadronic showers, the
possible time resolution is not only given by the active detector
elements, but may well be limited by the intrinsic time structure of
the showers themselves. For absorbers with heavy nuclei, such as
tungsten, a particularly complex time structure is expected, calling
for experimental verification of this aspect of the detector
simulations.

To provide first experimental input, a specialized experiment, the
Tungsten Timing Test Beam (T3B) detector, was developed and installed
in the tungsten analogue HCAL (WHCAL) prototype. A first data taking
period at the CERN PS in November 2010 was successfully completed,
providing first analysis results \cite{CAN-033} that constrain Geant4
physics lists.

\subsubsection{T3B: setup and data analysis}

The T3B setup consists of fifteen $3\times3$ cm$^2$ scintillator tiles
with a thickness of 5~mm, directly read out with 1~mm$^2$ Hamamatsu
MPPC50P SiPMs with four hundred $50\times50$ $\mu$m$^2$ pixels. The
scintillator tiles have a ``dimple'' drilled into the side face at the
SiPM coupling position to achieve a uniform response over the full
active area \cite{Simon:2010hf}. At nominal operation, they provide a
signal of approximately 27 photoelectrons (p.e.) for minimum ionizing
particles, including afterpulses of the photon sensor.

The photon sensors are read out with 4-channel USB
oscilloscopes\footnote{PicoTech PicoScope 6403
  (http://www.picotech.com/)} with 1.25~GS per second, using long
acquisition windows of 2.4 $\mu$s per event to record the time
structure of the energy deposits in the scintillator in detail.  Each
SiPM was connected to a preamplifier board, which then feeds the
signal to the oscilloscope via coaxial cable. The preamplifier boards
with packaged scintillator cells were mounted on a 2~mm thick aluminum
plate and protected by a 1~mm thick aluminum top cover, forming a
robust cassette.

\begin{figure}[!t]
\centering
\includegraphics[width=0.95\textwidth]{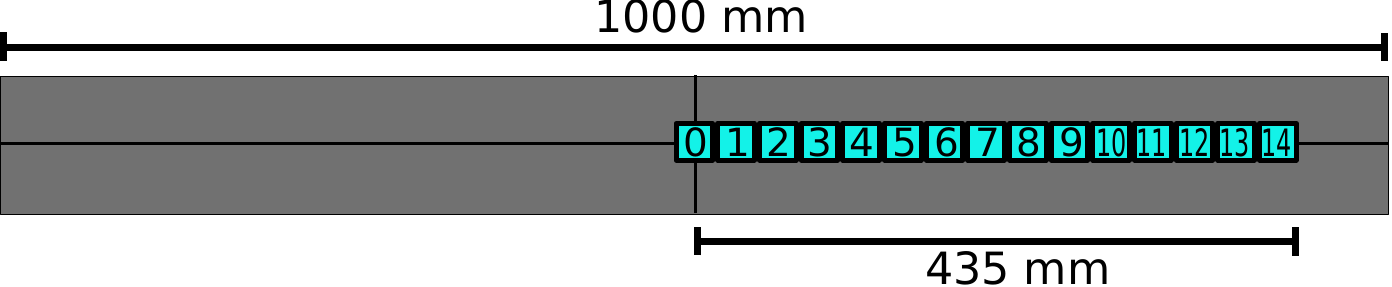}
\caption{Layout of the T3B scintillator tiles. From the nominal beam
  axis, the setup extends by 15~mm to one and 435~mm to the other
  side.}
\label{fig:T3BLayout}
\end{figure}

The T3B scintillator tiles are arranged in one row extending from the
center of the calorimeter layer out to one side of the detector. The
first tile is centered on the nominal beam position, thus the setup
extends 15~mm beyond the nominal beam center on one and 435~mm on the
other side, as shown in Figure \ref{fig:T3BLayout}. This permits the
measurement of a full radial timing profile of the hadronic shower at
the position of T3B, given sufficient statistics. The limited coverage
however only allows averages over many events to be measured, and is
not suitable for the study of the time evolution on an event by event
basis.

A first analysis of 645\,000 10 GeV $\pi^-$ events was performed,
using T3B in standalone mode without attempting to correlate the
events with CALICE WHCAL events to obtain additional information about
the showers. The data were analyzed on a cell by cell level. As a
first step, zero suppression based on pedestals determined on a
spill-by-spill basis was applied. Then, waveforms with an integral
above 0.3~MIP were decomposed into individual photon equivalents to
provide precise information on the arrival time of photons at the
light sensor. This was done by consecutively subtracting single photon
signals from local maxima detected in the waveform, until no maxima
above approximately 0.5~p.e. remained. The single photon signals were
obtained from noise events taken between spills and are determined for
each tile separately. This results in an implicit gain calibration,
since possible cell-to-cell gain differences lead to corresponding
differences in the average single photon signals used in the
analysis. The resulting number of photons is thus independent of the
SiPM gain. This reference signal was refreshed every 10 spills,
typically corresponding to time intervals of less than 5 minutes. This
provided continuous automatic corrections of gain variations due to
temperature changes.

\begin{figure}
\centering
\includegraphics[width=0.8\textwidth]{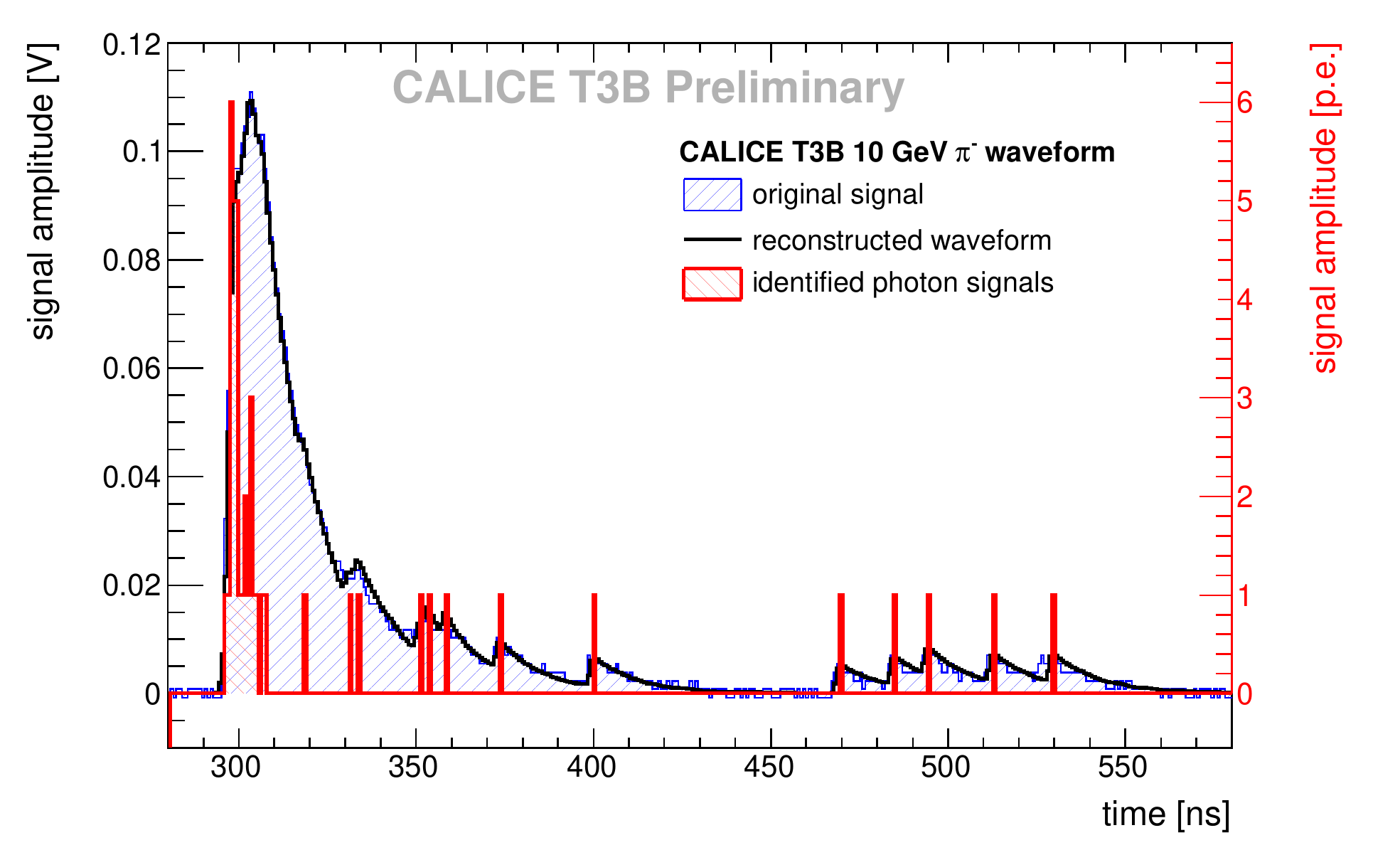}
\caption{Typical waveform with a high initial signal, decomposed into
  individual photon signals during the data analysis. Very good
  agreement of the original waveform and the reconstructed signal from
  standard single photo-electron distributions is observed. }
\label{fig:Waveform}
\end{figure}

Figure \ref{fig:Waveform} shows one example of a waveform decomposed
using this reconstruction technique. To check the quality of this
analysis, a waveform based on the identified photon signals was built
up with the reference single photon signals and compared to the
original waveform. The very good agreement between measurement and the
reconstructed waveform demonstrates the quality of reconstruction.
First results using this reconstruction are outlined in Sect.~\ref{sec:analysis} below.



\section{Digital HCAL: DHCAL}
The Digital Hadron Calorimeter Project trades the typical tower
structure of past hadron calorimeters and their high-resolution
readout with large number of finely segmented active elements, read
out individually with a single bit resolution.

 \subsection{RPC-based DHCAL}
 \label{sec:DHCAL}
 
\subsubsection{Description of the project}
 A collaboration of institutes (Argonne, Boston, FNAL,
IHEP Beijing, Iowa, McGill, Northwestern, and Texas at Arlington) is
developing such a novel calorimeter with Resistive Plate Chambers
(RPCs) as active media. Currently the readout is segmented into
$1\times 1$~cm$^2$ pads or 10\,000 per square metre. 

The project progressed in several stages. In a first stage, different
designs of RPCs were developed and tested with a high-resolution
readout system~\cite{dhcal1}. In parallel to this activity a 1-bit readout
system capable of handling large numbers of channels in a
cost-effective way was developed. The second stage put the two
together in a small prototype calorimeter, here named the Vertical
Slice Test (VST), and included detailed tests with both cosmic rays and
in the Fermilab test beam. For the first time within the CALICE
collaboration the VST utilized a readout system with the digitization
taking place directly on the front{}-end boards. Based on the
successful experience with the VST~\cite{dhcal2, dhcal3, dhcal4, dhcal5, dhcal6} and after a further round of
R\&D, the third stage consisted of the construction of a large
technical prototype hadron calorimeter (the DHCAL) with, close to
350\,000 readout channels. The active elements were inserted into the
CALICE hadron calorimeter absorber structure. The calorimeter is now
undergoing tests in the Fermilab test beam. 

Located behind the hadron calorimeter is the CALICE Tail Catcher and
Muon Tracker (TCMT) with 16 active layers. Over the past six month its
Scintillator layers have been gradually replaced with RPC layers,
identical to the one's in the DHCAL. This brought the total number of
readout channels of the combined DHCAL + TCMT system to approximately
480\,000. 

Following the tests in the Fermilab test beam, the DHCAL group will
return to R\&D to tackle the remaining technical issues in preparation
of the construction of a so-called Module 0.

Additional information pertaining to this project can be obtained from~\cite{dhcal7}.

\subsubsection{Past achievements since the last review}
The DHCAL group has been very active since the last review: publishing
papers, completing the R\&D necessary for the construction of the
DHCAL, constructing the DHCAL and the RPC-TCMT, installing the active
layers into the CALICE structures in the Fermilab test beam, taking
data with various beam configurations, and last but not least analyzing
the collected data. In the following we provide a few additional
details on these activities:

\paragraph{Completion of the analysis of the VST data.}
The 5$^{th}$ and last paper based on data from the VST was
published in JINST and documented the environmental dependence of the
performance of RPCs.

\paragraph{R\&D in preparation of the construction of the DHCAL}
Moving from the VST to the DHCAL represented an increase in size and
channel count of roughly a factor of 200. Techniques to spray the glass
plates with resistive paint providing the required surface resistivity
and homogeneity were developed. Fixtures for the preparation of the
rims of the RPCs were designed, built and optimized. Three identical
fixtures for the assembly of RPCs were built. In order to provide the
required uniformity of the gas gap, the fixtures were machined with a
precision better than 0.1~mm.

The application of high voltage posed a significant problem, as an
attempt was made to minimize the inactive rim around the edge of the
chambers. After several set{}-backs we developed a technique which
allowed us to extend the resistive paint up to within a couple of mm
from the edge, without the risk of a high voltage break down.

The readout boards consist of a pad board (each with 1536 pads of
$1\times 1$~cm$^2$) and the corresponding Front-end board 
(with 24 DCAL ASICs), thus avoiding the need for costly blind vias. The
two boards are mated to together by applying drops of conductive glue
on glue pads on the back side of the pad board. A gluing fixture was
designed and built to apply the 1536 glue dots in a timely fashion. 

Several prototype iterations were necessary before embarking on the mass
production of the front-end boards. The challenge was to minimize the
cross-talk between the digital activity of the board and the analog
front-ends.

Finally, a viable design of the detector cassette needed to be
developed. The role of the cassettes is to: (i) provide a mechanical
structure to hold the 3 RPCs in a given layer together, and (ii) to
provide a surface to cool the front-end ASICs. Our cassette design
exerts slight pressure on the front-end boards against the underlying
RPC. This is required to minimize the distance between the two and
therefore to reduce the average pad multiplicity for single Minimum
Ionizing Particles (MIPs).

\paragraph{Construction of the DHCAL and RPC-TCMT}

The construction of the DHCAL and RPC-TCMT took approximately two
years and involved up to 15 people at a given time.

Over 700 sheets of glass were sprayed with resistive paint. The
procedure was never completely under our control and the efficiency was
only about 60\%. \ We made no effort in controlling the environment and
abrupt changes in the weather typically required a tedious re{}-tuning
of the various parameters of the spraying gun.

205 RPCs (with the dimensions of $32 \times 96$~cm$^2$) were
produced in three parallel assembly lines. The gap size was maintained
at a very uniform level, with slightly larger values ($<$100
${\mu}$m) at the four corners. The chambers were tested for gas
tightness and the high voltage connections were added. The chambers
typically operate at 6.3~kV, but were tested overnight at 7.0~kV.

Over 300 Front-end boards and pad boards were produced. The boards
were checked out thoroughly in three test stations, working in
parallel. Each test took between 3--6 hours. Faulty boards, with
e.g. more than 4 dead channels, were repaired and retested.

Cassettes consisting of a 2~mm copper matched by a 2~mm steel cover, a
top and a bottom bar were produced by an outside company. The assembly
of the cassettes was relatively straightforward and could be
accomplished in less than one hour per cassette. 

Of the order of 50 Data collector modules were built and tested at
Boston University. Again a dedicated test station was needed to check
out the boards. Twelve Timing and Trigger Modules were fabricated and
assembled. The check out was done at Fermilab. 

New gas mixing and distribution racks were built and commissioned. The
RPCs require a mixture of three gases (Freon, Isobutan and SF$_{6}$). 
The distribution rack features 28 outputs with
individual controls. 

The Front-end boards require +5 V. A low voltage power supply system
was built which uses commercial Wiener power supplies and custom made
distribution boxes. The system provides low voltage to all 306
front{}-end boards of the combined DHCAL and TCMT system. Each line can
be individually controlled and is separately fused.

The high voltage system is on loan from CERN and Fermilab. Due to its
relative old age (from the early 1980s), the system is prone to
frequent break downs. The group is considering replacing the system
with a commercially available system.

\begin{figure}[h]
\begin{center}
\includegraphics[width=0.48\textwidth]{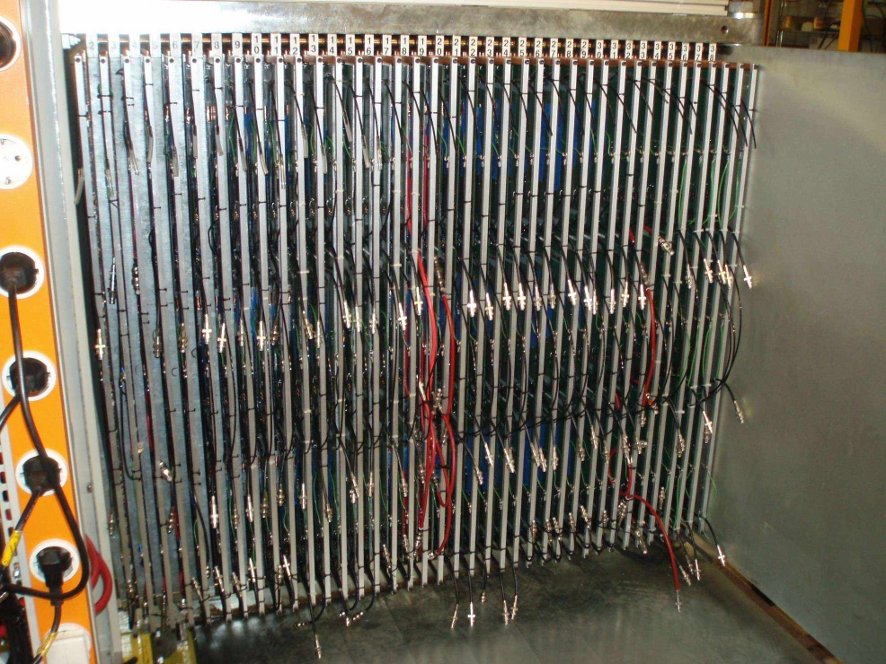}%
\includegraphics[width=0.48\textwidth]{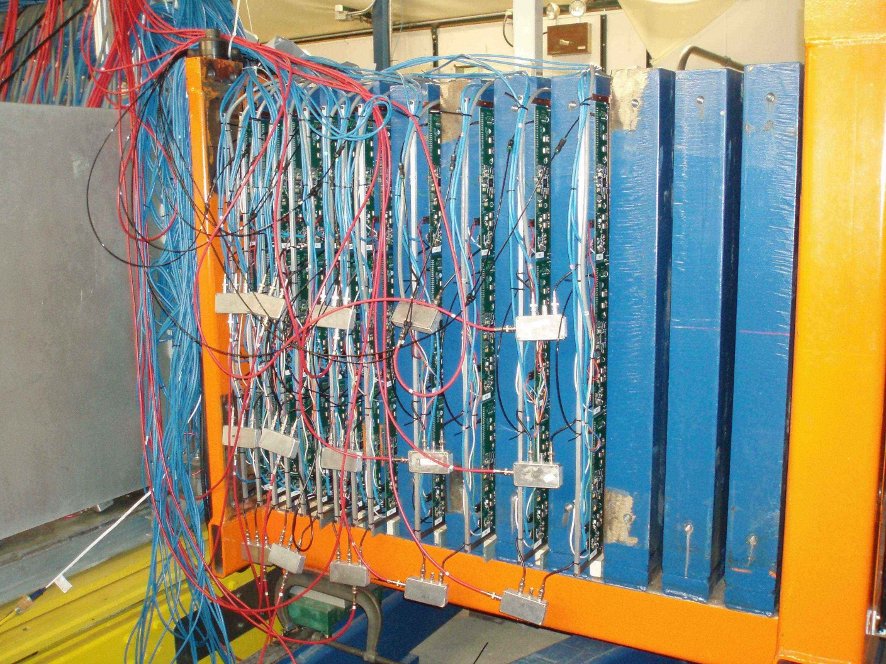}
\caption{{\sl Photograph of the DHCAL (left) and the RPC-TCMT
(right). The photos were taken before (after) cabling of the DHCAL
(RPC-TCMT).}}
\label{fig:dhcal_fig1}
\end{center}
\end{figure}

The completed cassettes were transported from Argonne to Fermilab in a
specially devised transport structure, which minimized the impact of
shocks due to bumps on the road. The cassettes were inserted into the
CALICE Hadron Calorimeter absorber structure and into the CALICE TCMT.
Fig.~\ref{fig:dhcal_fig1} shows photographs of the two structures after installation of
the RPC layers.

\paragraph{Data taking at Fermilab}

Testing of the DHCAL in the Fermilab test beam started in October 2010.
Table I summarizes the data taken since. In general the data are of
very high quality. The broadband muons were obtained with the 32 GeV/c
secondary beam and a 3 meter long iron beam blocker.

As the January run progressed we gradually replaced the Scintillator
TCMT layers with RPC layers identical to the ones in the DHCAL. In
Tab.~\ref{tab:dhcal_tab1}. the number of RPC -- TCMT layers is indicated in
parenthesis. 

In its current and final configuration the system counts 480\,000 readout
channels. In April 2011 the CALICE Silicon-Tungsten electromagnetic
calorimeter was installed in front of the DHCAL, bringing the total
number of readout channels to 490\,000.

Due to the rate limitation of RPCs, pion and positron events are
collected simultaneously. Separate data sample are generated offline
using the Cerenkov signal. 

\begin{table}[tbp]
\begin{centering}
\begin{tabular}{|p{0.5816598in}|p{2.5858598in}|p{1.8587599in}|p{0.74065983in}|}
\hline 
\bf{Date}
&
\bf{Configuration}
&
\bf{Beam}
&
\bf{Number} 

\bf{events}
\\\hline
Oct'10
&
DHCAL+SCINT\_TCMT
&
\multicolumn{2}{p{2.67816in}}{\hspace*{-\tabcolsep}\begin{tabular}{|p{1.8587599in}|p{0.74065983in}|}
\hline
Broadband muons
&
1,405 k
\\\hline
Secondary beam at 

2,4,8,10,12,16,20,25,32 GeV/c
&
1,524 k
\\\hline
\end{tabular}\hspace*{-\tabcolsep}
}\\\cline{1-2}
Jan'11
&
DHCAL+RPC\_TCMT (4-13)
&
\multicolumn{2}{p{2.67816in}}{\hspace*{-\tabcolsep}\begin{tabular}{|p{1.8587599in}|p{0.74065983in}|}
\hline
Broadband muons
&
1,591 k
\\\hline
Secondary beam at

2,4,6,8,10,60 GeV/c
&
3,619 k
\\\hline
\end{tabular}\hspace*{-\tabcolsep}
}\\\cline{1-2}\cline{4-4}
Apr'11
&
ECAL+DHCAL+RPC\_TCMT (14)
&
\multicolumn{1}{p{1.8587599in}|}{\hspace*{-\tabcolsep}\begin{tabular}{|p{1.8587599in}}
\hline
Broadband muons
\\\hline
Secondary beam
\\\hline
\end{tabular}\hspace*{-\tabcolsep}
}&
Ongoing
\\\cline{1-2}\cline{4-4}
\end{tabular}
\caption{\sl Summary of the DHCAL data taking at the Fermilab test beam
facility.}
\label{tab:dhcal_tab1}
\end{centering}
\end{table}

\subsubsection{First DHCAL results}

A rigorous effort is underway to analyze the DHCAL data in a timely
manner. Due to the fact that the data have been collected quite
recently, the results shown here are still preliminary. In order to
provide a flavor for the ongoing activities, in the following, we
sample results from different analysis efforts. Additional results can
be obtained from the DHCAL CALICE notes~\cite{CAN-031, CAN-030, CAN-032}.

\begin{figure}[h]
\begin{center}
\includegraphics[width=0.6\textwidth]{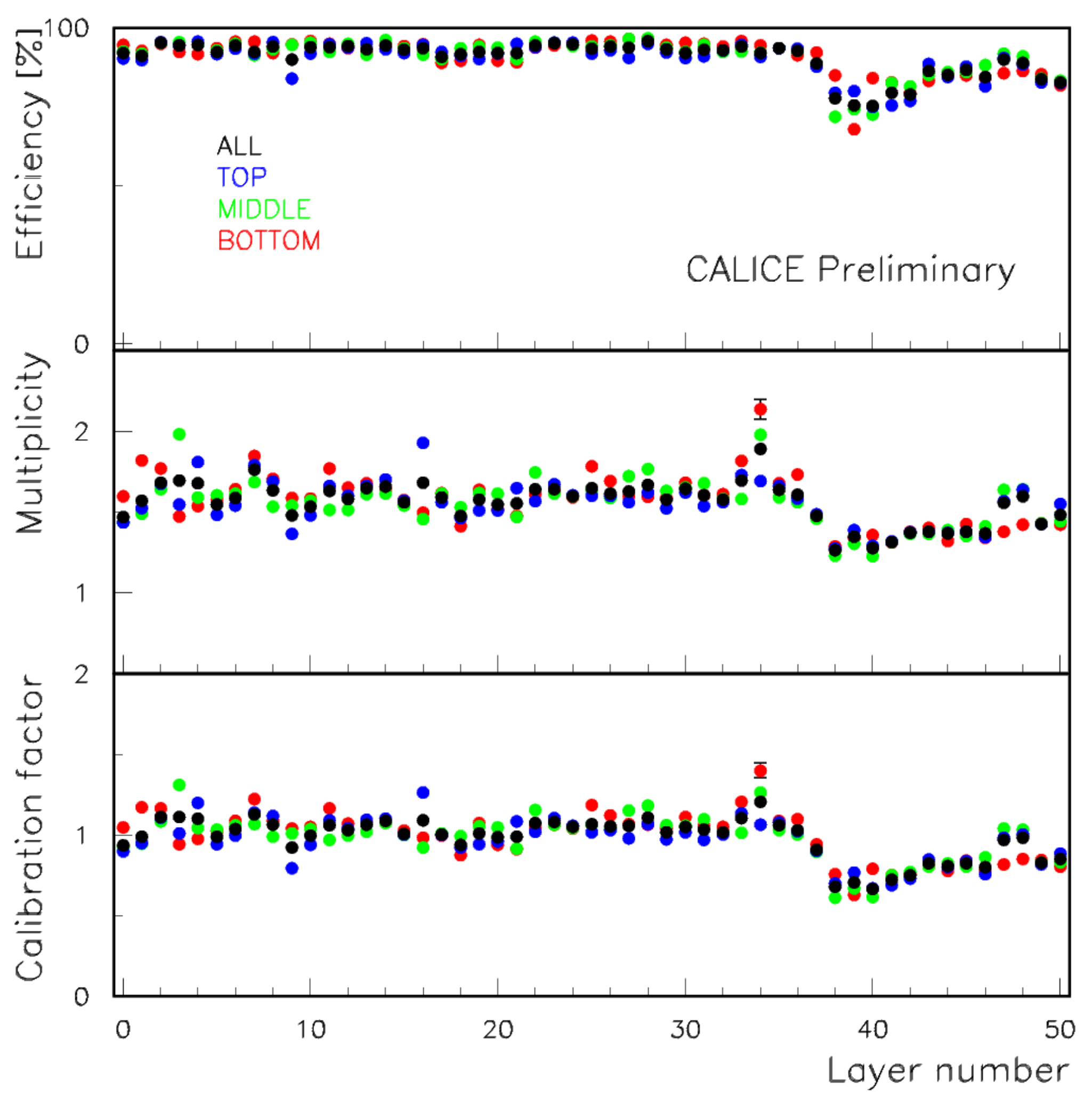}
\caption{{\sl 
MIP detection efficiency, average pad multiplicity
and calibration factors as function of detector layer as measured with
broadband muons.
}}
\label{fig:dhcal_fig4}
\end{center}
\end{figure}

\begin{figure}[h]
\begin{center}
\includegraphics[width=0.55\textwidth]{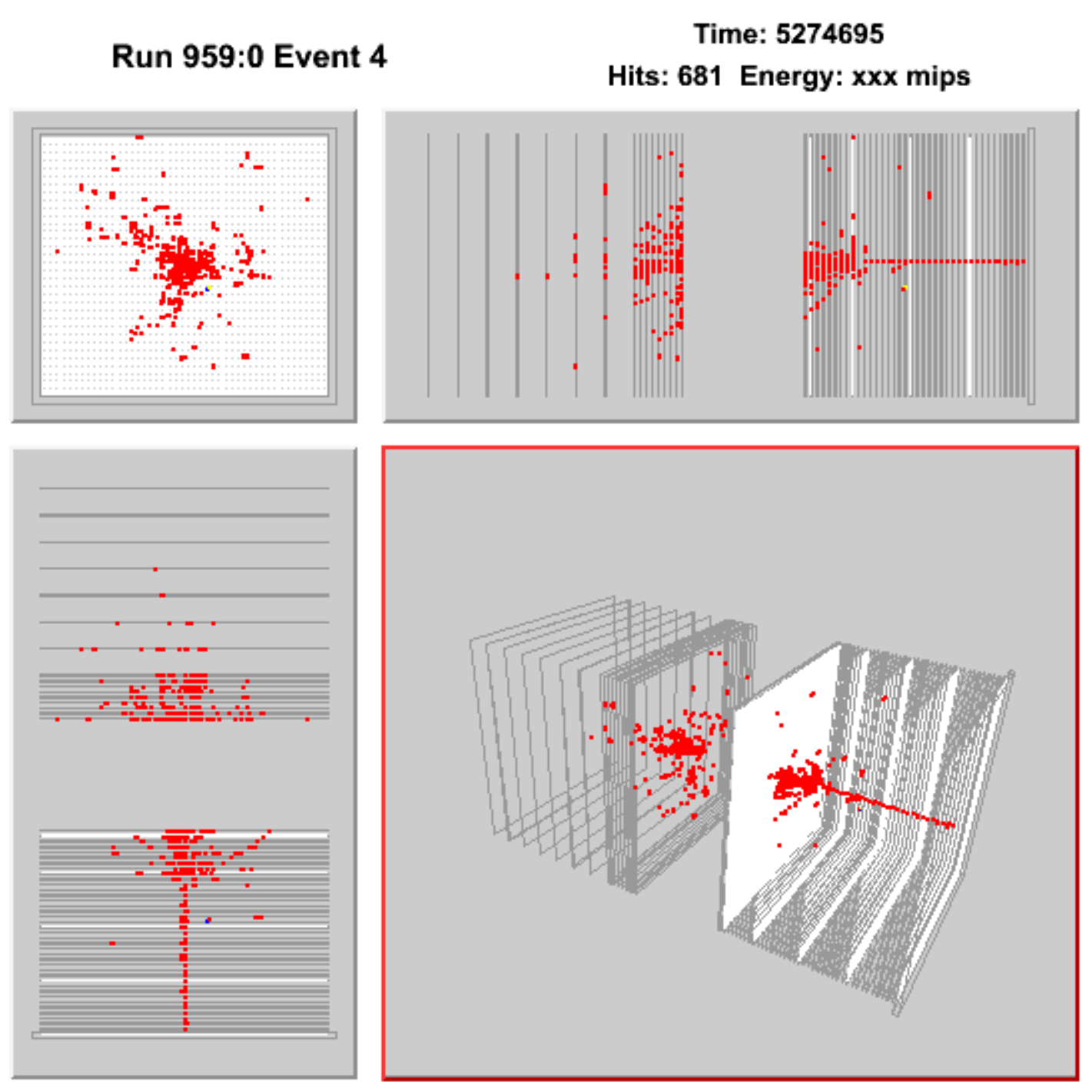}
\caption{{\sl 
Display of a 60 GeV pion event with significant
leakage into the RPC-TCMT. The observed isolated hits are part of the
hadronic shower and are not due to noise.}}
\label{fig:dhcal_fig6}
\end{center}
\end{figure}

\begin{enumerate}
\item \textbf{Analysis of cosmic ray data.} The completed chambers were
inserted into a cosmic ray test stand and tested with cosmic rays. The
stand could accommodate up to nine chambers at a given time. As an
example, Fig.~\ref{fig:dhcal_fig2} shows the distribution of cosmic rays, the efficiency
and the pad multiplicity as a function of the dip angle of the cosmic
rays.
\item \textbf{Measurement of the noise rate. }The noise rate is measured two
ways: i) in trigger{}-less operation, where any hit in the detector is
recorded and ii) in triggered operation during off{}-beam times, where
the trigger is provided by a pulse generator. The two methods are seen
to provide consistent results~\cite{CAN-031}. Fig.~\ref{fig:dhcal_fig3} shows the noise rate as
function of layer number or of RPC number. The average noise rate of
0.6~Hz/cm$^2$ is somewhat elevated for this type of
detector and due to the elevated temperature in the stack.
Nevertheless, the measured noise rate contributes in average only 0.06
hits or 4 MeV per triggered event. 

\begin{figure}[h]
\begin{center}
\includegraphics[width=0.7\textwidth]{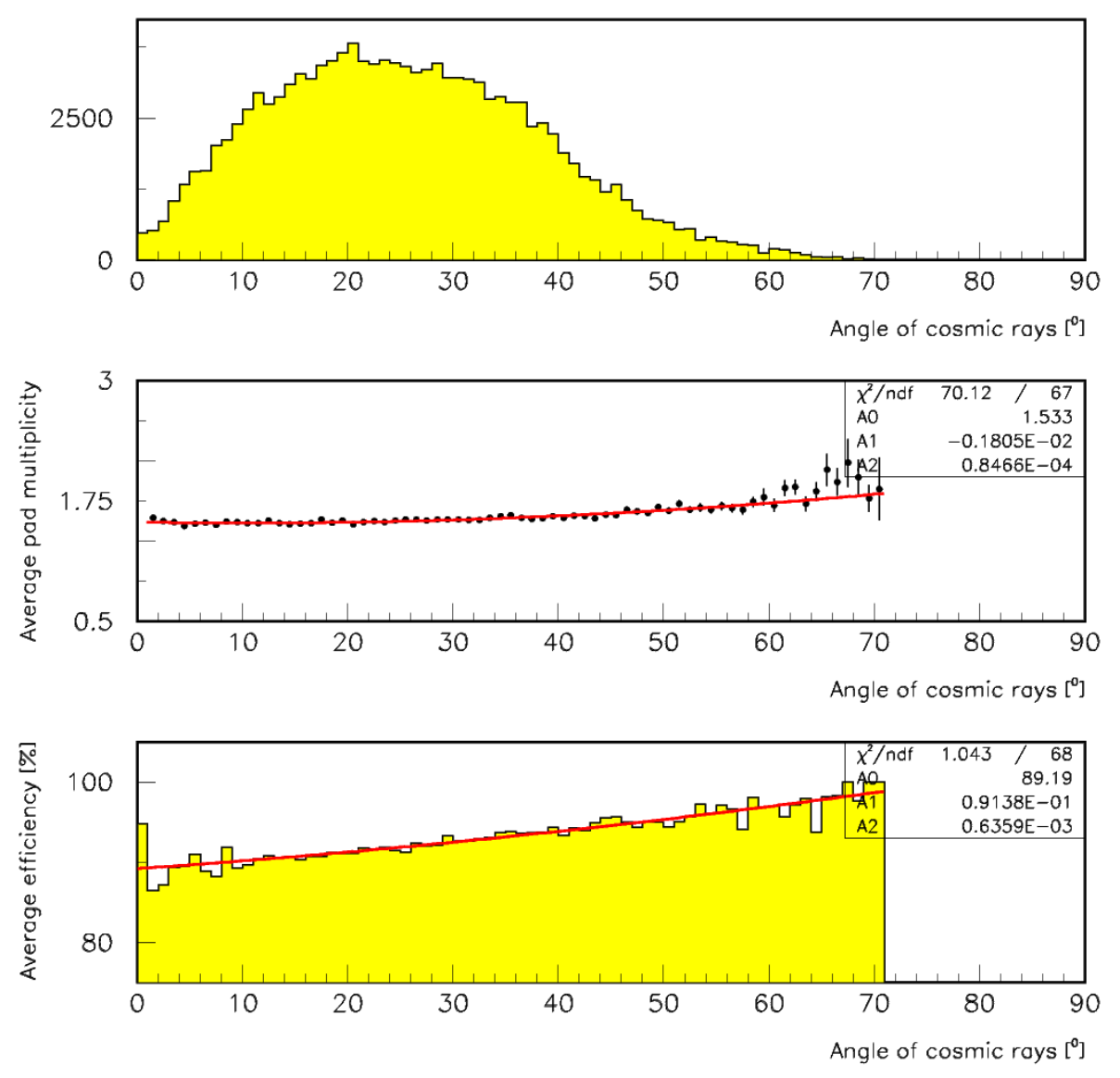}
\caption{{\sl 
Distribution of cosmic rays, average pad multiplicity
and MIP detection efficiency as function of the dip angle of cosmic
rays.}}
\label{fig:dhcal_fig2}
\end{center}
\end{figure}

\begin{figure}[h]
\begin{center}
\includegraphics[width=0.7\textwidth]{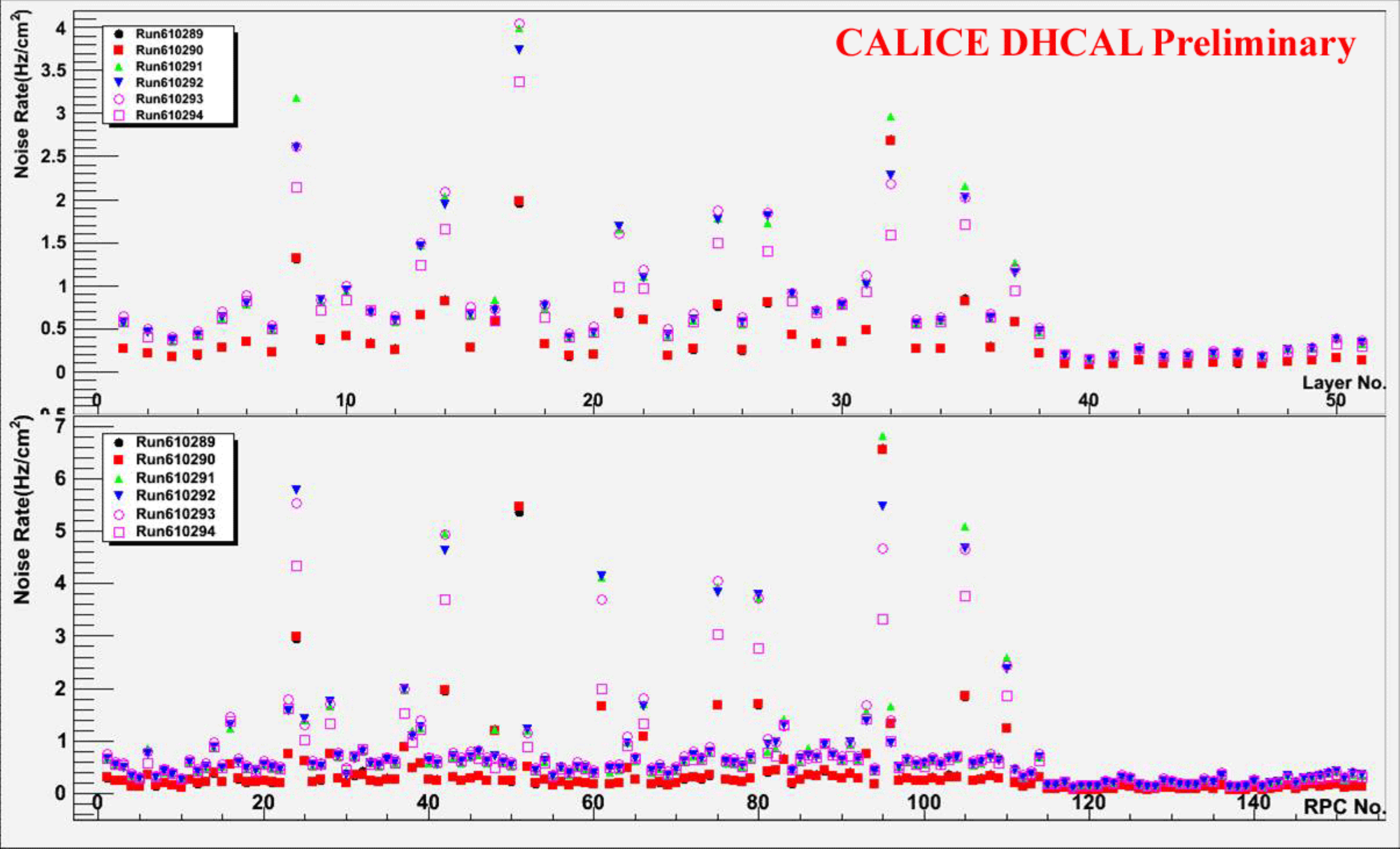}
\caption{{\sl Noise rate in Hz/cm\textsuperscript{2} as function of
layer number and as function of RPC number. The different colors
correspond to different runs taken at different times.}}
\label{fig:dhcal_fig3}
\end{center}
\end{figure}

\item\textbf{Analysis of muons.} Muons are used to geometrically align the
layers in the stack and to measure and monitor the performance of the
calorimeter~\cite{CAN-030}. Fig.~\ref{fig:dhcal_fig4} shows the MIP detection efficiency, the
average pad multiplicity and the calibration factors as function of
layer number. 
\item\textbf{Analysis of secondary beam data. }  Fig.~\ref{fig:dhcal_fig6} shows the event
display of a 60 GeV pion with significant leakage into the RPC-TCMT.
Note that the observed isolated hits are not originating from noise,
but are generated by the hadronic shower.
Some first results based on events like this are outlined in Sect.~\ref{sec:analysis} below.
\end{enumerate}

\subsubsection{Future plans}

In parallel to the large effort of assembling the technical prototype
the collaboration has initiated R\&D related to the remaining technical
issues of an RPC{}-based hadron calorimeter. Tab.~\ref{tab:dhcal_tab2} summarizes the
various ongoing and planned activities.

\begin{table}[tbp]
\begin{center}
\begin{tabular}{|p{1.2462599in}|p{1.0559598in}|p{3.41846in}|}
\hline
R\&D topic
&
Funds
&
Comment
\\\hline
Thin RPC
&
Applied for
&
Further investigation of 1-glass design
\\\hline
Large area RPCs
&
Currently not 

pursued
&
Areas of several m$^2$ needed
\\\hline
Gas system
&
Funded
&
Exploration of new gas mixtures, recycling, gas distribution
\\\hline
High Voltage distribution
&
Funded
&
System capable of supplying HV to all layers of a module individually
\\\hline
Low Voltage distribution
&
Currently not 

pursued
&
System capable of supplying LV to all layers of a module individually
\\\hline
Wedge shape
&
Currently not 

pursued
&
Develop concept to accommodate wedge shaped module designs
\\\hline
Pad/FE-board
&
Currently not 

pursued
&
Develop new design which minimizes thickness
\\\hline
Front-end ASIC
&
(Funded)
&
Develop next iteration with reduced power consumption, token ring
passing, and redundancy for reliability
\\\hline
Data concentrator
&
Currently not 

pursued
&
Develop new system which minimizes space requirement and provides high
reliability
\\\hline
Mechanical structure 
&
Currently not 

pursued
&
Develop cassette structure which can be oriented which ever way, develop
module structure which accommodates all supplies and data lines
\\\hline
Magnetic field
&
Currently not 

pursued
&
Tests of all subsystems in magnetic field
\\\hline
\end{tabular}
\caption{\sl  Summary of R\&D topics beyond the construction of the
technical prototype.}
\label{tab:dhcal_tab2}
\end{center}
\end{table}

 \subsection{GEM-based DHCAL}
 \label{sec:GEM}

\subsubsection{GEM 30~cm $\times$ 30~cm prototype tests with KPiX readout system}
A new GEM chamber, with greater ease of assembly and dis-/re-assembly
and improved gas flow with new spacers, was designed. In order to use
the prototype chamber for beam tests and for integration of KPiX
Analog and DCAL digital readout systems, we constructed three of these
chambers. After completing one of these new chambers, with the new
gas-transparent G10 spacer from CERN and the updated KPiX readout
board, we performed source tests using $^{55}$Fe and $^{106}$Ru
radioactive sources. Since $^{55}$Fe has characteristic peaks from
5.9~keV and 4~keV X-rays losing their entire energy in the chamber, it
provides an excellent test for chamber performance.

\begin{figure}[!t]
\centering
\includegraphics[width=0.95\textwidth]{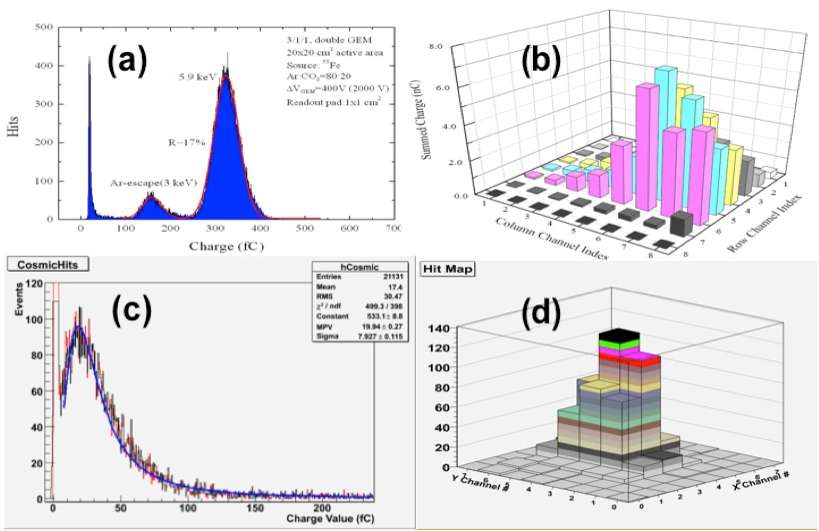}
\caption{(a) Pulse height distributions from $^{55}$Fe source, showing
  characteristic peaks from 5.9~keV and 4~keV X-rays (b) Lego plot of
  hits in all active channels with radioactive source (c) Pulse height
  distribution from cosmic ray muons, conforming to a Landau
  distribution (d) Lego plot of hits from cosmic ray muons which
  conforms to the trigger coverage area}
\label{fig:GEMFig1}
\end{figure}

Fig.~\ref{fig:GEMFig1} (a) shows two distinct peaks from $^{55}$Fe
X-rays in channel 49.  The source particles were narrowly collimated
to ensure that the particles traversed a normal incidence path rather
than with large incident angles. After confirming the minimum ionizing
particle pulse height distributions from $^{106}$Ru $\beta$ source
which shows the typical MIP Landau distributions, we moved into two
dimensional measurement by removing the collimator. The source
particles are then able to go through various channels demonstrating
two dimensional profile distributions in Fig.~\ref{fig:GEMFig1}
(b). Fig.~\ref{fig:GEMFig1} (c) shows the characteristic Landau
distributions from cosmic ray muons obtained through the 2~cm $\times$
2~cm coincidence trigger coverage and plotting the charges from the
highest charge pad.  Fig.~\ref{fig:GEMFig1} (d) shows the lego plot
of the hits from the cosmic ray muons, clearly showing the area
covered by the trigger paddles.

 We are now in the process of investigating uniformity of the chamber
 responses by performing response measurements in many different
 channels.  We do see some implication of differences in channel gains
 a factor of 3.5, we believe this is caused by incorrect application
 of electronic gains correction in KPiX software.  We believe we will
 resolve this issue shortly working with the SLAC team.  Once these
 issues are resolved, we will be able to determine chamber gains using
 sources and will take cosmic ray data for further MIP studies and
 efficiency studies.  We will then, as described in detail in the next
 section, take the chamber to a particle beam for high statistics
 chamber characterization.

\subsubsection{Pressure dependence corrections}

Due to the recirculation setup of the Ar-CO$_2$ gas, our GEM chamber
gains depend on the atmospheric pressure.  We have observed the
changes in chamber results over a period in which the atmospheric
pressure was known to be changing. Therefore, it is of critical
importance for chamber characterization, the chamber gains are
corrected for pressure dependence.  This section describes the
pressure dependence correction.

The measurement was performed using a  30~cm $\times$ 30~cm GEM
prototype chamber read out through the 64 channel KPiX7 chip. The
electronics was comprised of two low voltage power supplies, both set
at 7~A; one for the FPGA board and another located for the
interface board. The GEM detector is directly connected to a high
voltage power supply, which is used to set the potentials between the
GEM foils. To study pressure dependence, a $^{55}$Fe source was
elevated a distance above the GEM window and is centered in relation
to the readout pads. The pressure dependence of the chamber was
studied by performing hour-long runs. The pressure of the local area
(City of Arlington, Texas) can be retrieved from national weather
services and for each run; and thus subsequent pressure is recorded.

Each hour-long measurement is then analyzed and fitted with a Gaussian
distribution where its peak position (peak charge position) is
recorded. After multiple data is obtained, a plot of peak position
vs. pressure can be made and a linear-regression line made.

The fitted equation from the regression line is obtained where the
dependent variable is gain and the independent variable is
pressure. It is used to correct for pressure at 1 atmosphere (atm),
which is theoretical gain ($g_{theo}$) from the fit. The ratio of
theoretical gain, at all the pressures where data was taken, over the
measured gain at 1 atm is used as the correction factor ($C_f$). Using
the theoretical gain equation we can input the pressure of the elapsed
time for each hour interval and combined with the theoretical gain for
that interval and obtain the correction factor, as follows:

\begin{equation}
g_{theo} =  -303.9 \, p \ + \ 35509 \ \rightarrow \ C_f \, = \, \frac{g_{theo}}{g_{1atm}} 
\end{equation}

\subsubsection{Evolution to large area chamber}

\begin{figure}[!t]
\centering
\includegraphics[width=0.95\textwidth]{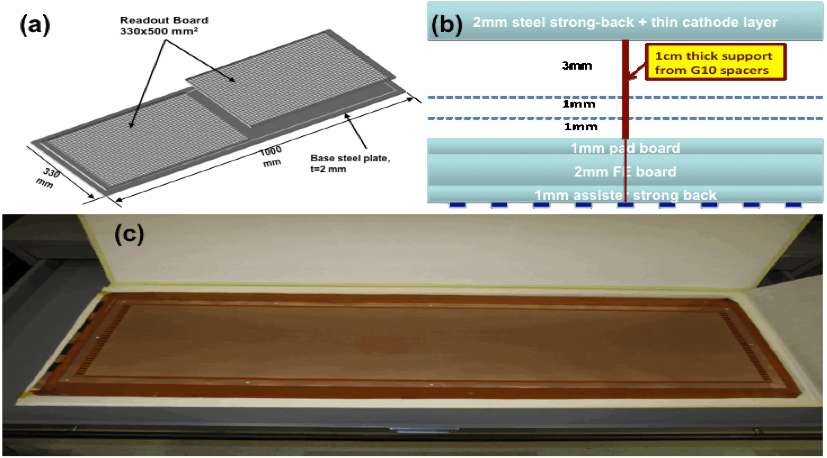}
\caption{(a) A schematic layout of readout boards for a  Unit Chamber. (b) Proposed layer and support structure for large GEM chambers.  (c) The first  30 cm $\times$ 100 cm GEM foils received from CERN.}
\label{fig:GEMFig2}
\end{figure}

The next step in the evolution of GEM-DHCAL layers is to construct and
test a chamber as an element of future planes within a full test beam
stack. The team has been working with the CERN GDD workshop in
developing GEM foils, the largest foils so far produced to date. The
foils have been made using a new single-sided-etching approach, which
eliminates some of the manufacturing problems associated with
side-to-side registration in the standard double-sided-etching
process. The first set of five foils has been produced and delivered
to UTA. A procedure for testing and qualification of these foils has
been developed based primarily on electrical properties, and all foils
have been qualified in three categories.  Two of the foils have been
identified and ready to be used in the first 33~cm $\times$ 100~cm.

We have been working on the design of the first chamber and a
schematic is shown in Fig.~\ref{fig:GEMFig2} (a) In order to avoid
thick side walls for the chamber resulting from tensioning the large
foils, we will use a 2 mm steel strong-back on which to mount the
chamber. In a final calorimeter configuration, this would be part of
the absorber structure between two active layers. The required
separation of the cathode, foils and anode layers is achieved by the
use of thin spacers, also made for us by the CERN GDD Workshop as
shown in Fig.~\ref{fig:GEMFig2} (b).  Once we have assembled,
tested, and successfully operated a 30~cm $\times$ 100~cm chamber, we
will proceed to construct a full 1~m $\times$ 1~m plane.

\subsubsection{Progress on thick-GEM alternative approach}

Since the invention of the standard GEM foil, it has been widely used
in various applications and is a promising candidate for use in a
DHCAL as discussed above. However, the 50~$\mu$m thickness can make
handling larger foils quite difficult, and there can be issues with
dust getting trapped in the 70~$\mu$m standard hole size. As an effort
to overcome such possible drawbacks, another technique has been
developed to produce GEMs. In this context we have been collaborating
with Weizmann Institute (with travel support from a U.S. Israel
Bi-National Science Foundation grant) on thick GEM (THGEM) production
and testing using a 0.4~mm thick PCB plate as the base material. Holes
with diameter of 0.5~mm are perforated with pitch of 1~mm. For initial
tests, two 10 $\times$ 10~cm$^2$ THGEMs were used in the amplification
stage and the KPiX readout electronics were used as the data
acquisition system. The pressure of the Ne/CH$_4$ = 95/5 gas was kept
constant using a containment vessel. The measurement was carried out
with gain of 2000, and the measured energy resolution was about 22\%.
The next step was to try various configurations of single THGEM to
achieve a useful level of MIP signal with stable chamber
performance. Unfortunately, with the basic THGEM setup read out with
KPiX, sparking occurred and two successive chips were found to be
dead.  We plan to return to the CERN test beam in Summer 2011 and
test the ``well'' and other configurations with later versions of
KPiX.

\subsubsection{DCAL digital readout integration}

In preparation for its digital readout, we have started the effort for
integrating the ANL-FNAL developed DCAL readout chip.  We spent two
weeks at ANL in early 2010 as the initial attempt for integration.
While we were able to readout the noise of the chamber and the
electronics via the DCAL readout board and back-end DAQ system at ANL,
we did not observe a clear signal for detector responses at that time.
The timing also did not allow us to continue beyond the allotted two
weeks since the ANL RPC-DHCAL team had to prepare for their beam test
runs at FNAL.  Given this progress and with the help of ARRA funds, we
have decided to purchase the back-end DAQ equipment, identical to that
of the ANL's cosmic ray test stand and proceed with two week
integration efforts in early 2011.  The goals are to fully integrate
the DCAL readout board with the prototype 30~cm $\times$ 30~cm GEM
chamber, clearly understand the chamber properties with the digital
readout, establish a fully functioning back-end DCAL DAQ system for
continued operations and characterization of the chamber with DCAL at
UTA.  We will construct 2 to 3 chambers fully integrated with DCAL
readout system in preparation for the beam tests at FTBF.

\subsubsection{Future plans}
Through late 2011, the team will complete the characterization of the
30~cm $\times$ 30~cm chambers, complete the development of 33~cm
$\times$ 100~cm large GEM foils and complete the design of the 33~cm
$\times$ 100~cm foil stretching and gluing stations in preparation for
prototype construction.  
In late 2011 through early 2013 time scale,
the team will work on assembly technique for 33 cm $\times$ 100~cm
unit chambers, characterization of 1024 channel KPiXA chips, using
30~cm $\times$ 30~cm chambers with the intent to use them in 33~cm 
$\times$ 100 cm unit chamber, complete understanding chamber behavior
with DCAL chips and begin construction of 33 cm $\times$ 100 cm unit
chambers. In early 2013 through mid 2014 time scale, the team will
characterize a 33~cm $\times$ 100~cm unit chamber built with DCAL chip
based anode board on the bench and in test beam, construct a 100~cm
$\times$ 100~cm plane using three 33~cm $\times$ 100~cm unit chambers
read out through DCAL chip and build 6 additional 33~cm $\times$ 100
cm unit chambers for two 100~cm $\times$ 100~cm planes.  Finally in
mid 2014 through 2015 time scale, the team will complete constructing
three additional 100~cm $\times$ 100~cm planes and expose to particle
beams as a calorimeter in the existing CALICE stack.

\section{Semi-digital HCAL: SDHCAL}
\label{sec:sDHCAL}
Members of the CALICE Collaboration, including Belgian, Chinese,
French, Russian and Spanish groups, are pursuing a new development
aimed at constructing a highly granular gaseous hadronic calorimeter
prototype based on a semi-digital readout and a transverse
segmentation of 1~cm$^2$.  In addition to the tracking capability it
offers, a semi-digital readout HCAL can provide very good energy
resolution which can be, according to simulations, as good as that of
an analogue calorimeter with an appropriate choice of threshold
values.

 The semi-digital HCAL prototype is intended to come as close as
 possible to the hadronic calorimeters of the future ILC experiments
 in terms of resolution, efficiency and compactness.  For instance,
 they should have negligible dead zone in order to keep the tracking
 capability as high as possible. They should also be very thin to
 reduce the cost of the whole experiment.  Indeed, as the HCAL will be
 placed inside the magnetic field, reducing the thickness of the
 sensitive media is essential to avoid an excessively large radius of
 the magnet coil, which would have very serious cost implications for
 the overall detector.

Two kinds of gaseous detectors are being investigated as candidates to
become the sensitive medium of such a SDHCAL: glass RPCs (GRPC) and
MICROMEGAS. New readout electronics satisfying the ILC constraints was
developed and successfully tested on small GRPC detectors and more
recently on a 1 m$^2$ detector.

 \subsection{RPC-based SDHCAL}
 


 The use of gaseous detectors such as the GRPC ensures
 excellent efficiency and good homogeneity of the sensitive medium at
 a low cost.
In contrast to the other options, the high granularity of this gaseous
HCAL option is provided by the readout electronics system. The signal
created by the passage of charged particles in GRPC detector is
collected thanks to 1~cm$^2$ pads etched on one of the two faces of
the electronics board and put in contact with the GRPC.  To keep the
the sensitive medium as compact as possible, the readout ASICs need to
be tiny, embedded on the detector and connected with each using the
electronics board itself thanks to a DAISY chain scheme. This leads to
a limited number of connections coming out of the sensitive medium.
With such a granularity more than 50 million electronics channels are
needed for the HCAL of the ILD experiment.  This renders the analog
readout prohibitive from the acquisition point of view.

Aa a trade-off, a 2-bit readout is proposed.   The choice of this
semi-digital scheme rather than the binary one was motivated by
simulation studies which show that better performance can be obtained
by the semi-digital scheme at high energy.  This result can be
explained by the fact that at high energy many particles might go
through one pad especially in the centre of the hadronic shower. Since
the extension of the avalanche created by charged particle in GRPC is
of the order of few  square mm \cite{chinese}, the semi-digital readout
with its three thresholds can help to distinguish among scenarios with
few, many and too many particles going through one pad. This improves
 the energy  resolution by providing a better estimate of the number
of charged particles produced in the hadronic shower, and also in 
separating the electromagnetic and the hadronic contributions.

The technology mentioned above was successfully tested with small
GRPCs~\cite{small}.  Large GRPCs of 1~m$^2$ size were then developed
using new concepts with the goal to keep the detector as homogenous as
possible. Electronics boards were also designed to cope with the large
detector surface and a compact cassette was conceived to tie them
together.  The successful tests of two of those cassettes in cosmic
ray test benches, as well as in test beam at CERN, paved the way to start
the construction of a technological prototype of 1~m$^3$.  It is
intended that this will be completed in May 2011.

\subsubsection{Detector design}
As can be shown in Fig.~\ref{scheme} the GRPC is made of two glass
plates of 0.7~mm and 1.1~mm thickness.  The thinner is used to form
the anode while the the thicker forms the cathode.  68 Ceramic balls
of diameter 1.2~mm are used as spacers to separate the glass
plates. The balls are glued on only one  of the glass plates.  In
addition to those balls,  13 cylindrical buttons of 4mm diameter made
of fiber glass are also used.  Contrary to the Ceramic balls the
buttons  are glued to both plates ensuring thus a robust structure.

The distance of the spacers (10~cm) and hence their number was fixed
so that the deviation of the gap distance between the two plates under
the glass weight and the electric force when the high voltage is
applied on the two electrodes does not exceed 45$\mu m$.  The choice
of these spacers rather than the fishing lines was intended to reduce
the dead zones (0.1\% ) rather than the few percents in the case of
the fishing lines. It was also aimed at reducing the noise
contribution observed along the fishing lines in the small GRPC
chambers.

 The gas volume is closed by a glass fiber frame of 1.2~mm thick and
 3~mm wide glued on both glass plates. The glue used for both the
 frame and the spacers was chosen for its chemical passivity and long
 term performance. The resistive coating on the glass plates which is
 used to apply the high voltage and thus to create the electric field
 in the gas volume was found to play important role in the pad
 multiplicity associated to a MIP.  To find the best coating for our
 chambers many products were tested.  Finally two products were
 identified, both of which are based on colloids containing graphite
 and both can be applied using the silk screen print method which
 ensures very uniform surface quality.  Both were found to provide
 stable resistivity in the range 0.5-2~M$\Omega$/Square. Repeatability
 tests were conducted to validate the use of such coating for a large
 amount of detectors. Both products can be painted using the silk
 screen printing method which guarantees very good
 homogeneity. Another important aspect of this development concerns
 the gas distribution within the GRPC. In order to improve on the gas
 distribution in large chambers taking into account the requirement
 that both gas outlets should be on the same side of the detector to
 satisfy all possible mechanical structures proposed for ILD hadronic
 calorimeter, new schemes were studied.  The one we finally adopted is
 based on channeling the gas along one side of the chamber and
 releasing it into the main gas volume at regular intervals thanks to
 a PEEK tubes of 1.2~mm diameter fixed at 5~cm from the chamber side. A
 similar system is used to collect the gas at the other side of the
 chamber.  A finite element model has been established to verify the
 gas distribution \cite{tipp09}. The simulation confirms that the gas
 speed is reasonably uniform over most of the chamber area.

\subsubsection{Electronic readout}
To read out the 1~m$^2$ GRPC detector, an electronic board with the same
size is needed.  This electronic board is an important piece in the
present design. It hosts both the pick-up pads and the ASICs in
addition to the connections linking the pads to the ASICs as well as
those  among the different ASICs.   To ensure good transmission
qualities and low cross-talk,  8-layer ASU (Active Sensor Unit) was
conceived. Feasibility constraints, make the tasks of circuit
production, components soldering, testing and handling of the
assemblies, exceedingly difficult in the case of a single board of one
square meter.   The solution of dividing that circuit into 6 smaller
but more manageable ASU boards was adopted.

Each of these small ASUs hosts 24 Hardroc chips to read out $48\times
32$ pads of  1~cm$^2$ each.  The routing of each input signal from
its own pad up to the HARDROC pin has been carefully optimized to
reduce the cross-talk.  All input signals are laid out in the same
analog signal layer  which is sandwiched between two GND layers. Great
care has been taken keeping the routing of digital signals well
separated from the vias connecting signals from one  layer to another.

\begin{figure}[ht]
\begin{minipage}[t]{.6\textwidth}
\begin{center}
\includegraphics*[width=\textwidth,keepaspectratio]{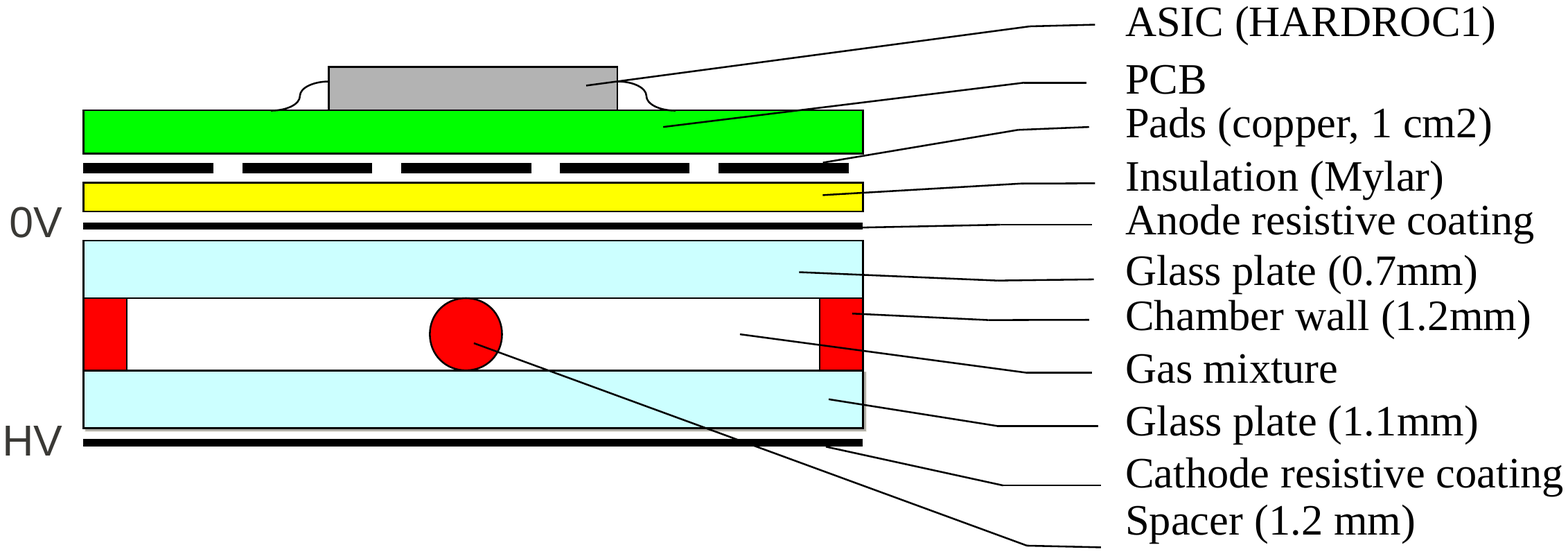}
\caption{Schematic view of a glass RPC\label{scheme}}
\end{center}
 \end{minipage}
\begin{minipage}[t]{.4\textwidth}
    \begin{center}
        \includegraphics*[width=\textwidth,keepaspectratio]{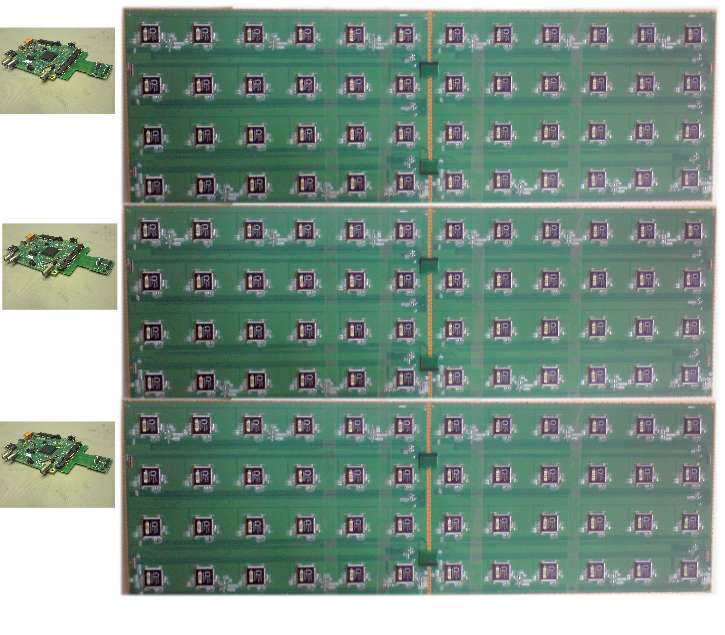}
  	\caption{ An electronic board of 1~m$^2$ made of 6 ASUs \label{board}}
      \end{center}
 \end{minipage}
 \end{figure}

The HARDROC base pattern is then replicated 24 times in the
$33.33\times 50$~cm$^2$ small board following a $4 \times 6 $ form
factor.  The rooting was conceived so two of the ASUs can be
associated to form one slab hosting 48 ASICS.  Each slab is then
connected to one DIF (Detector InterFace board). The connection
between the DIF and the slab as well as the connection of the two ASUs
is performed thanks to tiny connectors allowing the different clocks,
signals as well as the power to circulate between the two ASUs.  Three
slabs are then assembled to form the required electronics board. To
ensure the same electric reference level for the six ASUs, the GND
layer of the six ASUs is connected thanks to a copper gasket on all
the common sides.  The thickness of the final electronic board taking
into account that of the HARDROC chip with its TQFP160 package is less
than 3~mm. In Fig.~\ref{board} a picture of the electronic board of 1~m$^2$.

\subsubsection{Data acquisition}
To communicate with the different ASICs of one slab a board called
Detector InterFace (DIF) was conceived. The DIF hosts an FPGA  device
which by distributing the DAQ commands to the ASICs and by
transmitting the collected data to the DAQ system constitutes the
master piece of the DIF.
 
In addition to the FPGA the DIF hosts the connectors allowing the
communication with the slab on the one hand and with the DAQ system on
the other hand. Two kinds of connections with the DAQ system are
available: A USB-based and and HDMI-based ones.  Furthermore, The DIF
is used to power the slab.  The first phase of the acquisition is the
transmission of the configuration parameters to the 48 ASICs of one
slab. Indeed $872\times 48$ configuration parameters are transmitted
through the FPGA. Once the ASICs are configured, the second phase can
start. In this phase the communication between the DAQ and the ASICs
can be summarized in two steps: an acquisition phase and a readout
one.  The second phase starts after the end of the first one due to an
external signal which can be triggered by a full-memory state
(RAMFULL).  The readout of each ASIC is performed after receiving a
StartReadout and ends by emitting an EndReadout commands. The latter
is transmitted to the next ASIC which interprets it as a StartReadout
command. The readout phase of each ASIC can last up to 4~ms in the
worst case when the 127-frame memory is full.  As was mentioned
before, to read out large GRPC detectors with a 1~cm$^2$ lateral
segmentation many ASICs need to be assembled.  To keep small the
number of connections between the ASICs and the DAQ system the ASIC
digital readout signals were conceived to be linked to an open
collector bus and hence daisy chained leading to only one
connection. This is also the case for what concerns the assignment of
configuration parameters to all the ASICs of one board.  The situation
is different with respect to the acquisition where all the ASICs
should start this phase at the same moment. To synchronize the
previous operation among the ASICs of one slab as well as that of the
three slabs, the DIF can be connected with each other and the commands
to stop and start the acquisition is transmitted simultaneously to the
144 ASICs. This is also true in the case of a RAMFULL.  This scheme
was used for validation tests. It will be soon replaced with the
CALICE second generation acquisition system is under development and
will be used in the hadronic calorimeter prototype with its 144
DIFs. The acquisition system will be based on the Xdaq software
developed for the CMS tracker acquisition.  More information on this
can be found in the CALICE acquisition section.

\subsubsection{Cassette}
As mentioned before, the GRPC and its associated electronics are
housed in a special cassette which protects the chamber and ensures
that the readout board is in intimate contact with the anode
glass. The cassette is a thin box consisting of 2.5~mm thick
stainless steel plates separated by 6~mm wide stainless steel spacers
which form the walls of the box. These spacers are precision machined
so no space left between the two cassette plates and the sensitive
medium.  The dimensions of the plates are such that a space of 2~mm
wide is left between the GRPC and the cassette's walls. This space is
filled with insulating product to protect against sparks. One of the
two plates is 20~cm larger than the other. This allows one to fix the
three DIFs as well as the gas outlets and the high voltage connector.

The electronics board is assembled thanks to a polycarbonate spacer
which is also used to fill the gaps between the readout chips and to
improve the overall rigidity of the detector. The board is then fixed
on the small plate thanks to tiny screws and the new set is fixed on
the other plate which hosts the detector and the spacers. The whole
width of the cassette is 11 mm with only 6 of them corresponding to
the sensitive medium including the GRPC detector and the readout
electronics.

\subsubsection{Tests and results}
The cassette including the GRPC and the associated electronics
described in the previous section was first tested in a cosmic rays
bench. The bench was equipped with a triggering system  made of  two
scintillator-PMs. One is placed above the cassette and the other
beneath.  In order to evaluate the homogeneity of our detector
response  we equipped the bench with a movable structure with two arms
on which we fixed the scintillators.  The scintillators position can
be chosen easily. We have however selected three positions
corresponding to a geometrical areas read out by three different slabs
(and hence three different DIFs).  To ensure the absence of fake
triggers due to the scintillator noise and to have a rather precise
information on the particle impact in the GRPC, the same setup that
was used to study the small GRPCs as described in reference
\cite{small} was used. This setup made of 5 small GRPCs ($33.2\times
8.3$~cm$^2$) equipped with the same readout electronics was fixed on the
aforementioned  structure that supports the scintillators and used as
a tracker.  Both the large GRPC as well those of the tracker system
were operated with the following gas mixture:  Tetrafluoroethane (TFE,
93\%), Isobutane (5\%) and SF6 (2\%).   Since the  large GRPC and the
small ones are not equipped with the same acquisition a DIM-based
protocol was used. The protocol allows to accept triggers only if none
of the GRPCs readout electronics is busy. This ensures that the two
systems are stooped simultaneously with the occurrence of an external
trigger and their memories are read before to start again recording
data as described previously. This procedure allows to associate the
hits due to the same trigger in both the large and small chambers. 

The polarization voltage of the small GRPC chambers was fixed at
7.4~kV.  This was found to provide the best efficiency as described in
\cite{small}. The efficiency and the pad multiplicity of
the large chamber was then studied by varying its polarization voltage
in the range 6.2--8~kV.  The hits belonging to the small
chambers were then used to build tracks following the same method
described in \cite{small}. The coordinates of the particle
impact on the large chamber are then deduced.  An uncertainty of less
than 1 mm in both direction is expected from the straight line fit
applied on the small chambers hits. The hits of the large chamber
belonging to the same time window are then looked for and their
coordinates are compared with the that of the particle impact.  If at
least one hit is found in the vicinity of the impact particle (a zone
of 2~cm radius around the particle impact) then the hit is considered
to belong to the track. The number of hits found with a distance less
than 5~cm with respect to the impact coordinates provides the pad
multiplicity associated to this track.  The efficiency obtained for
the three different zones was found to be almost identical and the pad
multiplicity in the range of 1.6 to 2.3 pads/MIP.  The homogeneity
observed in the results obtained in the three zones indicates that
both the detector and the readout electronics are homogeneous.  The
noise was also measured and found to be about 1 Hz/cm$^2$.  The cassette
was then exposed to a pion beam of 10~GeV at the the T9 beam line of
the CERN-PS.

  A triggering system using four scintillators-PM was used with two of
  them were placed upstream and the two others downstream. The
  scintillators position was chosen so a particle passing through the
  four of them passes necessarily in the GRPC chamber.  The cassette
  was placed on a movable stage with X and Y movement.  Due to a space
  limitation only two thirds of the cassette could be exposed to the
  beam. Contrary to the cosmic-rays -bench test the small GRPC setups
  was not used here. In addition and because of CERN safety norms the
  isobutane was replaced in the gas mixture by CO$_2$ with the
  percentage.

To check the GRPC behavior in a beam and in the vertical position in
opposition to the horizontal one of the cosmic-rays-bench test, the
central zone behavior was studied by varying the polarization voltage.
Although the gas mixture is not identical we observe the same
efficiency behavior as the one obtained with the cosmic rays test was
observed. The same statement is valid for the pad multiplicity.

\begin{figure}[ht]
\begin{minipage}[t]{.2\textwidth}
    \begin{center}
        \includegraphics*[width=\textwidth,keepaspectratio]{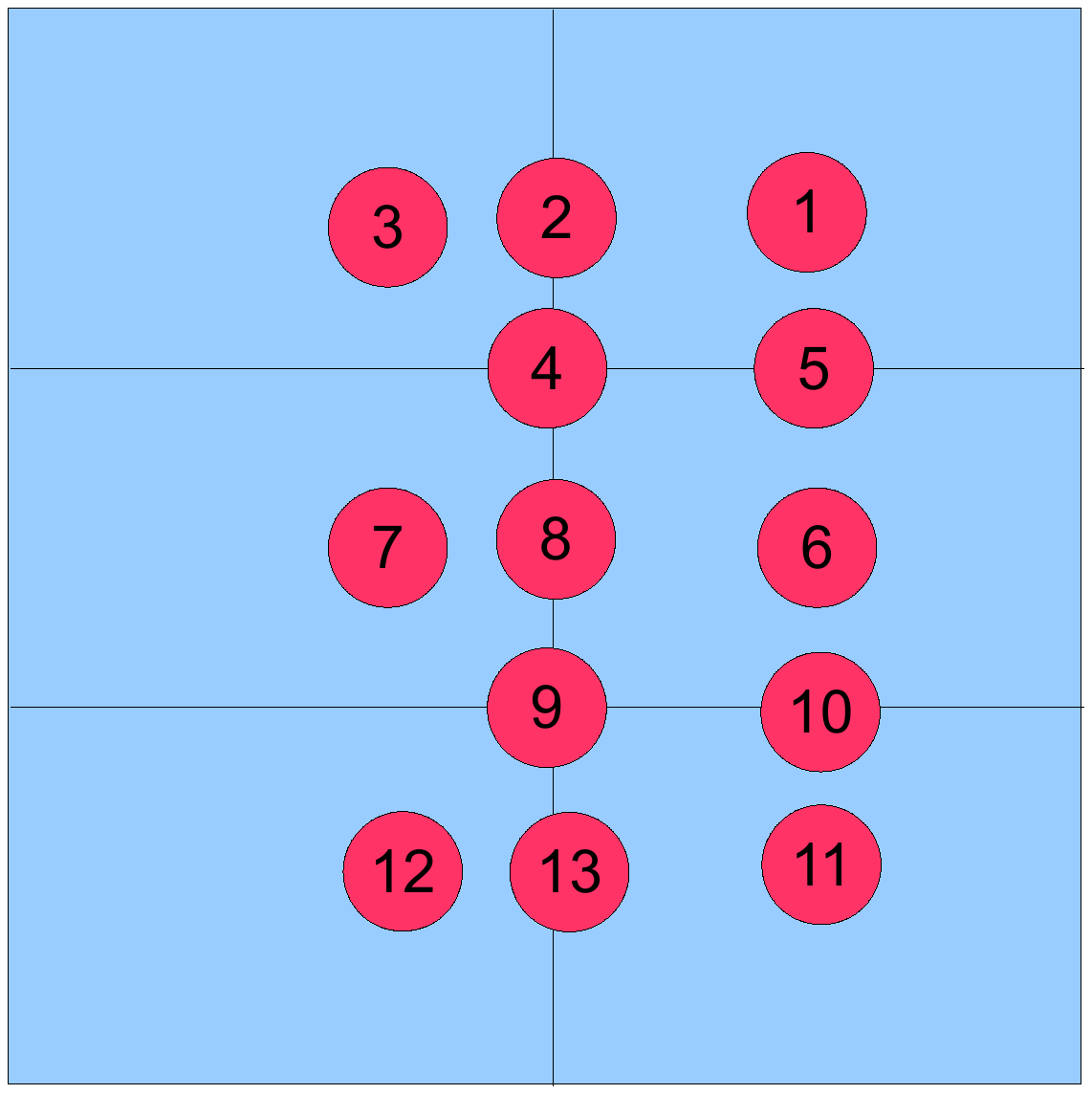}  
      \end{center}
  \end{minipage}
\begin{minipage}[t]{.3\textwidth}
      \begin{center}
        \includegraphics*[width=\textwidth,keepaspectratio]{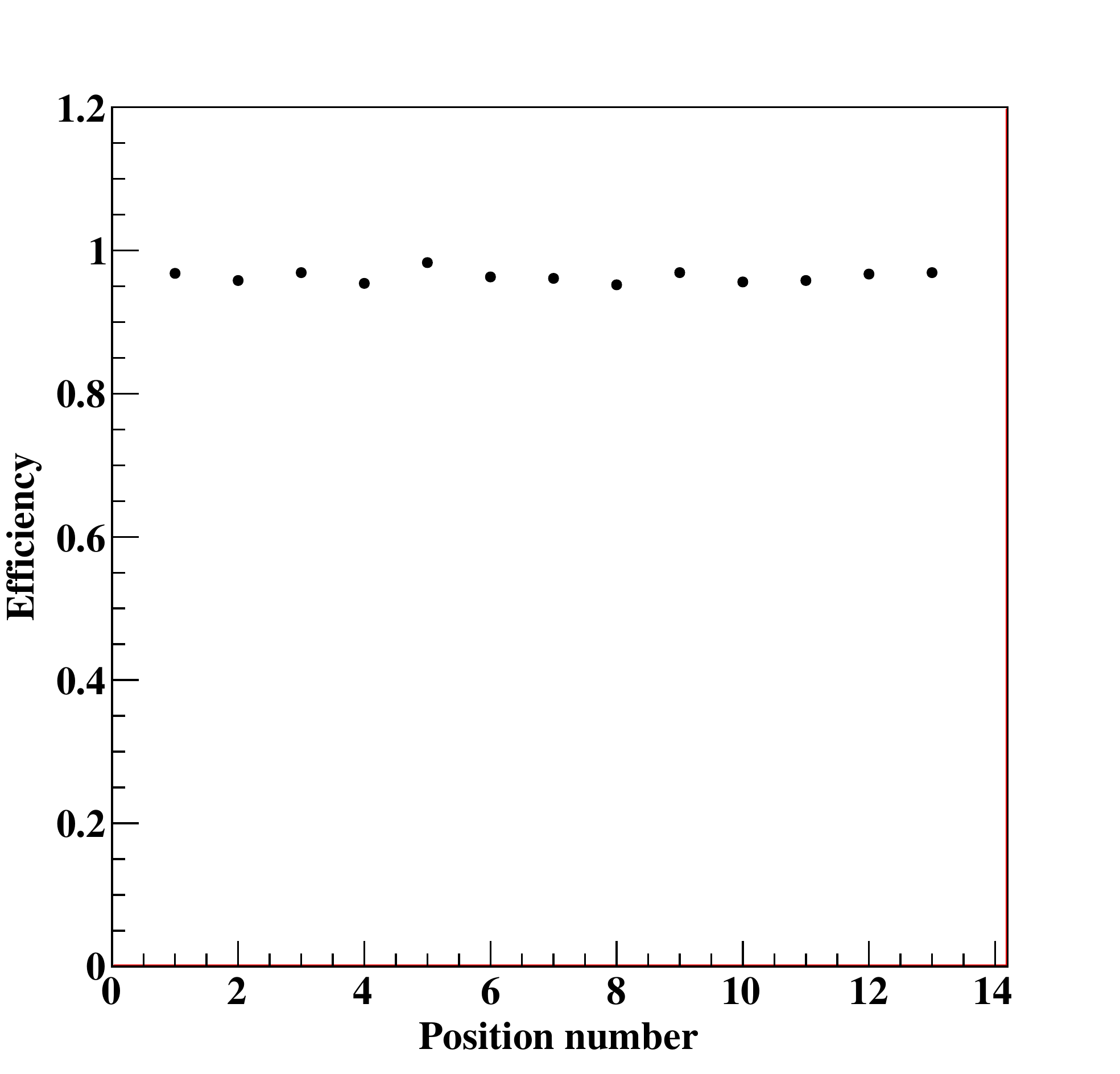}
     \end{center}
  \end{minipage}
  \begin{minipage}[t]{.3\textwidth}
      \begin{center}
        \includegraphics*[width=\textwidth,keepaspectratio]{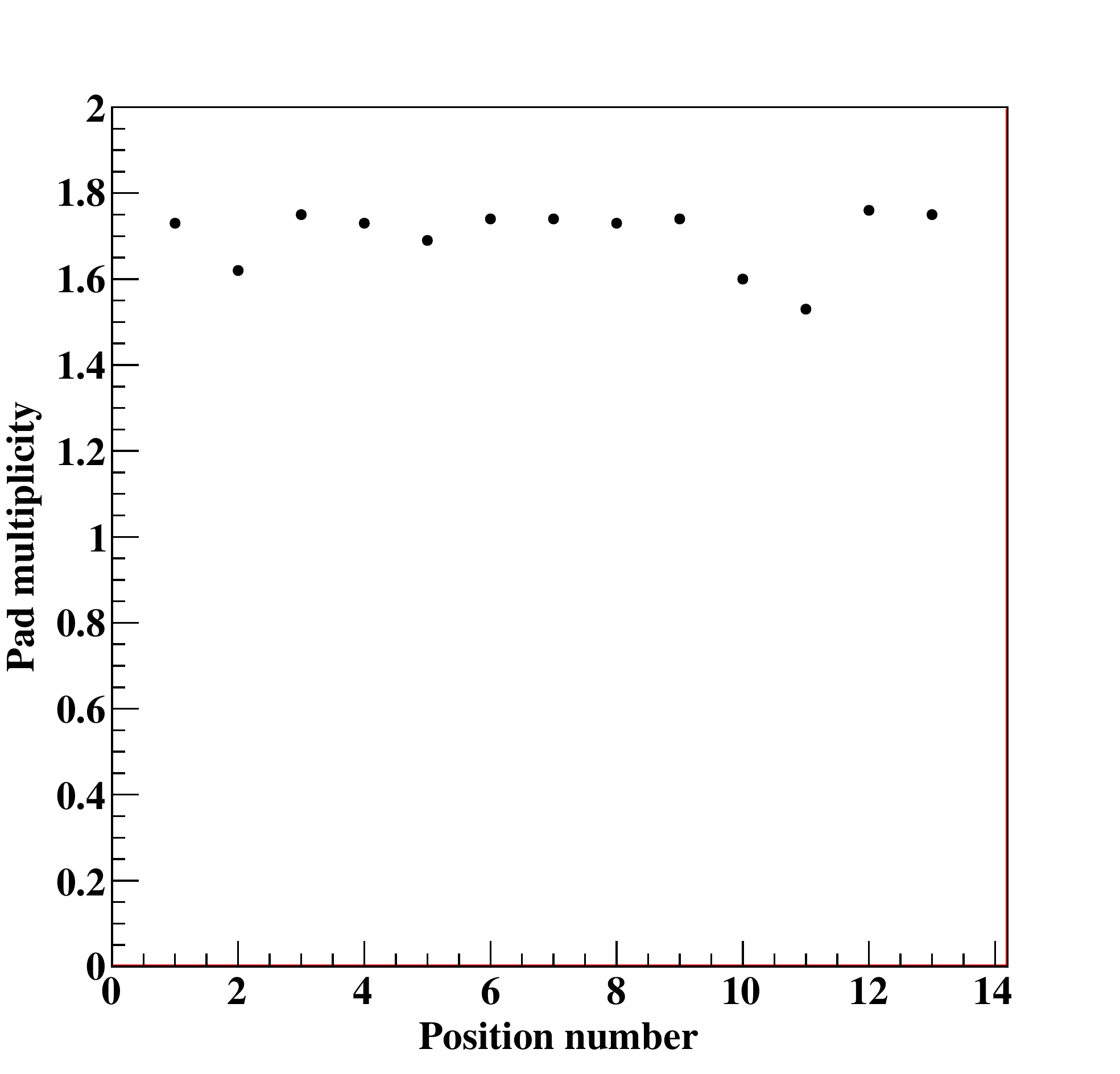}
	     \end{center}
	     
  \end{minipage}
  \caption{Positions, efficiency and pad multiplicity in the 13 selected zones. \label{homogen}}

\end{figure}

The second step was to confirm the homogeneity of the detector. 13
zones corresponding to different positions of the electronics board
and the detector were chosen as shown in Fig.~\ref{homogen}.  For all
these positions the GRPC was operated at the same polarization voltage
of 7.4~kV.  The results obtained for all those positions are shown in
Fig.~\ref{homogen}. They confirm those found in cosmic-rays-bench
test. These results with the vertical position show that the cassette
we conceived provides enough rigidity to maintain the electronics
board in contact with the GRPC detector.

An important test to validate the power-pulsing scheme in a magnetic
field of 3~T was performed in the H2 magnet of the SPS-CERN in
2010. The results obtained with a fully equipped detector of
$33.33\times 50$~cm$^2$ showed that the power -pulsing scheme which
allows to reduce the power consumption considerably when applied to
the ILC duty cycle of 1~ms of cross-benching every 200~ms is functional
and no loss of efficiency was observed.

\subsubsection{Mechanical structure}
A self-supporting structure was conceived to host the 48 cassettes
needed for the technological prototype.  It is made of 48 stainless
steel plates of 1.5~cm thickness each.  The plates should satisfy
stringent conditions on flatness and thickness resolutions (250~$\mu$m
and 100~$\mu$m respectively). They are assembled together thanks to
spacers of ($13 \times 17$~mm$^2$) section precisely machined.  Three
spaces are indeed used to assemble two consecutive plates thanks to
well distributed bolts. This leaves a 13~mm to insert the previous
cassette.  The deformation of such structure was carefully study in
different positions and found negligible on the HCAL performance.  The
mechanical structure is being indeed built and the first tests show
that the results are in agreement with expectations. The whole
structure will be completed in May 2011. In addition to this structure
a rotation tool allowing to use the prototype in vertical and
horizontal position has already been built.

\subsubsection{Prototype construction}
The technological prototype construction has started in late 2010.
More than 30 GRPC satisfying the requested quality control were
already built. The cassettes described previously were all
produced. The electronics board are being produced and cabled with the
HARDROC ASICs which have been all tested thanks to a dedicated
robot. The electronics boards production as well as the assembling
process are expected to be completed in May 2011.

After a period of 1 month commissioning the prototype will be exposed
to particle beams at CERN first in late June and then in October. The
first period will be used essentially for calibration study and
optimization procedure while the second will be dedicated to physics
study.

 \subsection{MICROMEGAS-based SDHCAL}

Invented in 1996, MICROMEGAS is today widely used in several running
experiments (COMPASS, KABES, T2K) and is also considered for LHC
upgrades (SUPER-ATLAS) and at future electron colliders (ILD TPC).
The rapid progress recently made on Micro Pattern Gas Detectors (MPGD)
makes MICROMEGAS a viable choice for a semi-digital hadronic
calorimeter. With the Bulk technique developed by the CERN technical
service MICROMEGAS becomes a robust detector which satisfies the
requirements of thickness, efficiency and large area. In addition, it
has working voltages below 500~V and is insensitive to neutrons if
used in hydrogen free gas mixtures. Finally, it delivers signals
proportional to the deposited energy and demonstrates very high rate
capability (up to some GHz/mm$^{2}$): MICROMEGAS is thus well suited
for a multi-threshold readout and could instrument both the barrel and
forward regions of a linear collider experiment (ILC or CLIC).

\subsubsection{Small prototypes}
\begin{figure}[h!]
\begin{centering}
\includegraphics[width=0.8\textwidth]{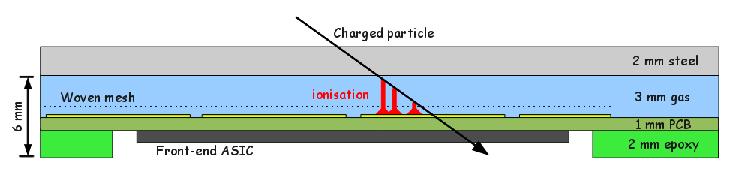}
\caption{\em Cross-section drawing of a MICROMEGAS chamber for an sDHCAL.}
\label{micromegas_dhcal}
\end{centering}
\end{figure}

Baseline parameters of MICROMEGAS chambers are 3~mm of argon-based gas
mixture (with \textit{i}C$_{4}$H$_{10}$ or CO$_{2}$) and square pads
of 1~cm$^{2}$ (Fig.~\ref{micromegas_dhcal}). First prototypes were
small chambers with dimensions up to 32$\times$12~cm$^{2}$ equipped
with analogue readout electronics based on GASSIPLEX ASICs. At a
detector gas gain of 10$^{4}$, the measured most probable value of the
charge from MIPs is 25~fC with 10~\% variation over each chamber
pads. Most importantly, the efficiency to MIPs reaches 97~\% (at
1.5~fC readout threshold) with 1~\% absolute variation per
chamber. Due to very little transverse diffusion in the gas volume,
the fired pad multiplicity lies below 1.15 even at the lowest workable
thresholds \cite{MM_ref}.

Small prototypes have been first used for characterisation
(2008--2009) \cite{MM_beam,MM_technote1,MM_technote2} and more
recently (2010) as a telescope during the test in a beam of the first
MICROMEGAS chamber of 1~m$^{2}$. Over the three years, they showed
constant performance and resistance to gas discharges thanks to
dedicated protections between the pads and the ASIC inputs.

\subsubsection{Active sensor units of 48$\times$32~cm$^{2}$}
\begin{figure}[h!]
\begin{centering}
\includegraphics[width=0.7\textwidth]{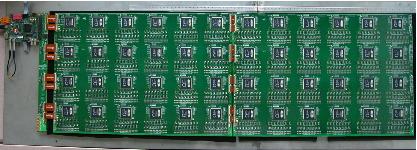}
\caption{\em Photograph of 2 ASUs of 48$\times$32~cm$^{2}$ with 24
  HARDROC2 chips chained with flexible cables, readout boards (DIF and
  inter-DIF) appear in the top left corner.} 
\label{micromegas_asu}
\end{centering}
\end{figure}

The basic units composing the 1~m$^{2}$ MICROMEGAS chamber are 1~mm
thick Printed Circuit Boards (PCB) of 48$\times$32~cm$^{2}$ with anode
pads and Bulk mesh on the top side and embedded readout ASICs and
discharge protections (passive components) on the opposite side
(Fig.~\ref{micromegas_asu}). They are called Active Sensor Units
(ASU). The fabrication proceeds first with soldering the ASICs and
protections to the PCB. In order to laminate the Bulk mesh on the pad
side, a mask is then glued on the ASIC side; the final ASU thickness
is 3~mm. This together with the 3~mm drift gas gap yield a minimal
thickness of a Micromegas chamber of 6~mm (baseplate and cover, being
part of the absorber, are ignored).

A detailed calibration procedure of the ASICs has been established to measure the pedestals and noise levels of each channel (1536/ASU). Eventually this information is used to correct for the channel to channel non-uniformity by tuning the preamplifier gains or to lower the individual detection thresholds by aligning the pedestals to a central value.

\subsubsection{First prototype of 1~m$^{2}$}
The design of the m$^{2}$ chamber consists of 6 ASUs placed next to
each other inside a common gas volume (Fig.~\ref{micromegas_h4}
left).  Spacers are inserted in the 1~mm gap between boards to define
the 3~mm drift gap. Together with 2~mm wide insulating lines on the
ASU edges to support the Bulk meshes, they are responsible for dead
regions of about 2~\% of the total area inside the chamber.  The first
m$^{2}$ prototype was assembled in May 2010 from 5 ASU each equipped
with 24 HARDROC2 ASICs. Although the fast HARDROC2 shaper is not
suitable for the 150--200~ns long MICROMEGAS signals, no other chip
was available in required quantity at that time. It was hence decided
to build and test the prototype in order to validate the design as
well as several technical choices (materials, assembly procedure, gas
distribution, feedthrough...).

\subsubsection{Beam test}
\begin{figure}[h!]
\begin{centering}
\includegraphics[width=0.33\textwidth]{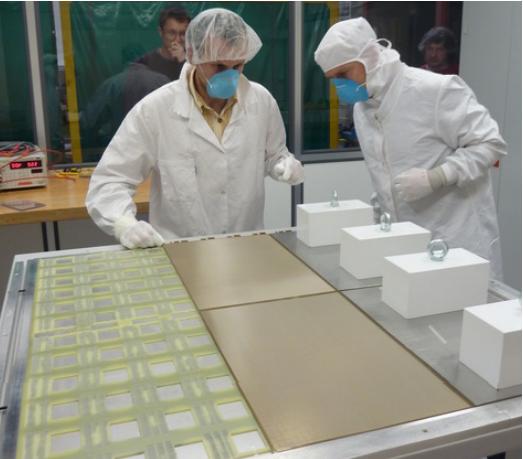}
\includegraphics[width=0.29\textwidth]{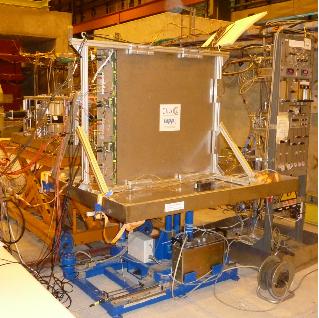}
\includegraphics[width=0.31\textwidth]{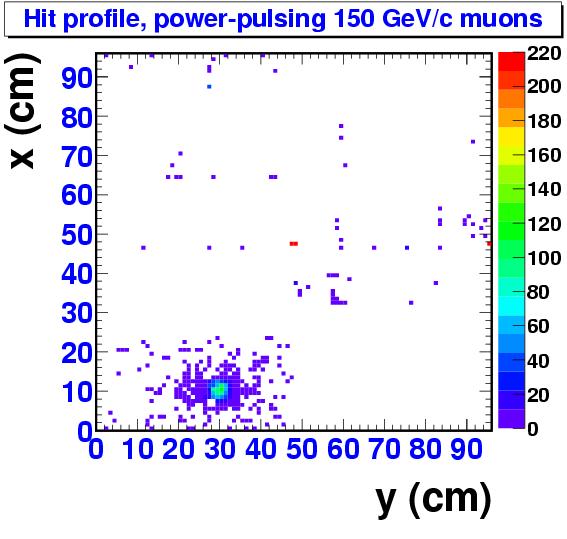}
\caption{\em From left to right: assembly of the 1 m$^{2}$ MICROMEGAS
  prototype, test setup in CERN/SPS/H4 line in June 2010 and recorded
  muon beam profile.} 
\label{micromegas_h4}
\end{centering}
\end{figure}

The m$^{2}$ prototype was tested in a high energy muon beam in
June/July 2010 at the CERN SPS (Fig.~\ref{micromegas_h4}).  At the
nominal gas gain of 10$^{4}$ a pad efficiency of 50~\% was reached.
This value was expected from the fast shaping of the HARDROC2 chip
which makes use of roughly 10~\% of the detector signal only.  A pad
to pad study performed with a telescope revealed no significant
spatial dependence of the detector response.  The prototype was also
tested under power-pulsing of the ASIC analogue parts (2/10~ms
ON/OFF).  Efficiency measurements with and without pulsing yielded
similar results.

\begin{figure}[h!]
\begin{centering}
\includegraphics[width=0.33\textwidth]{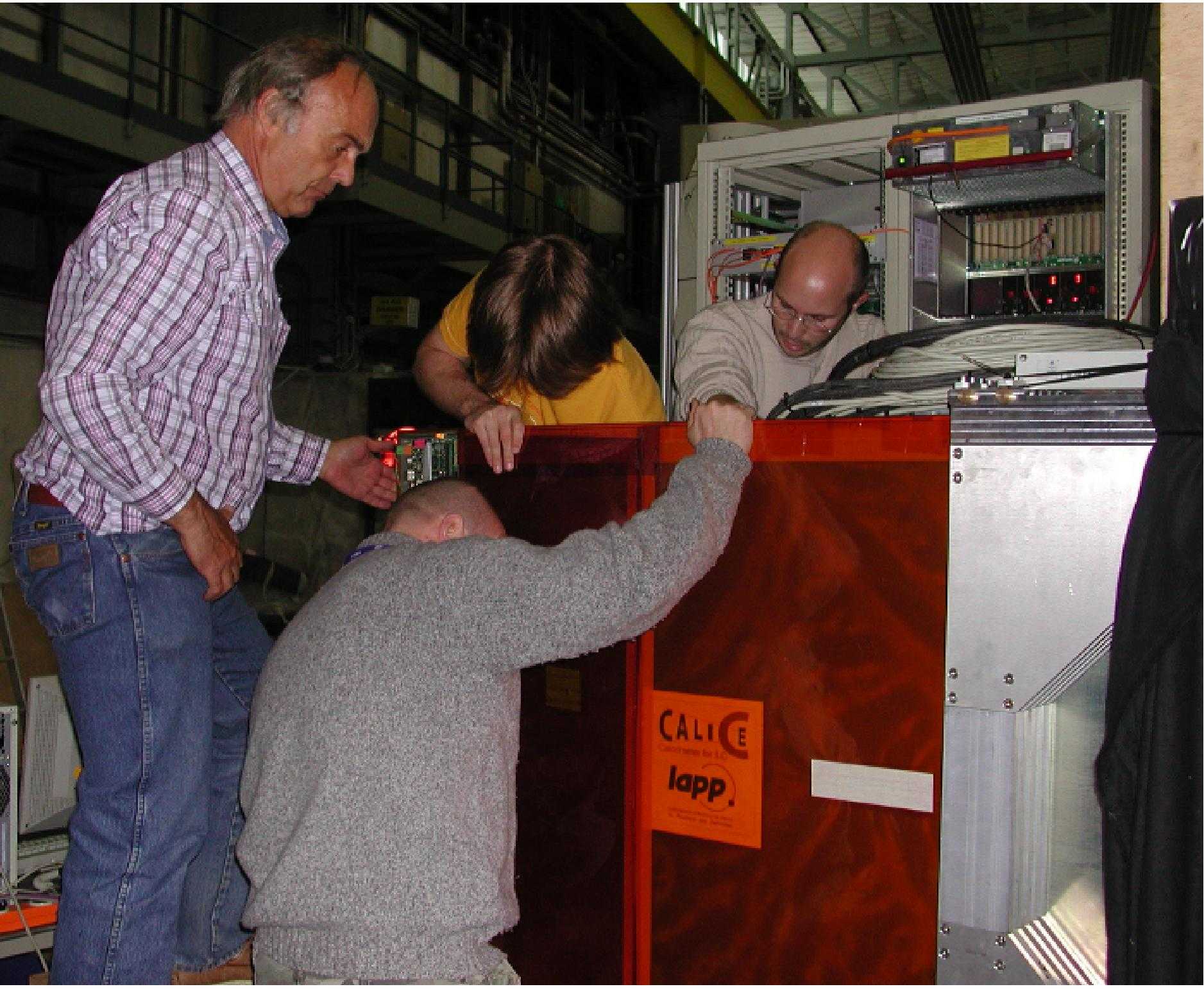}
\includegraphics[width=0.31\textwidth]{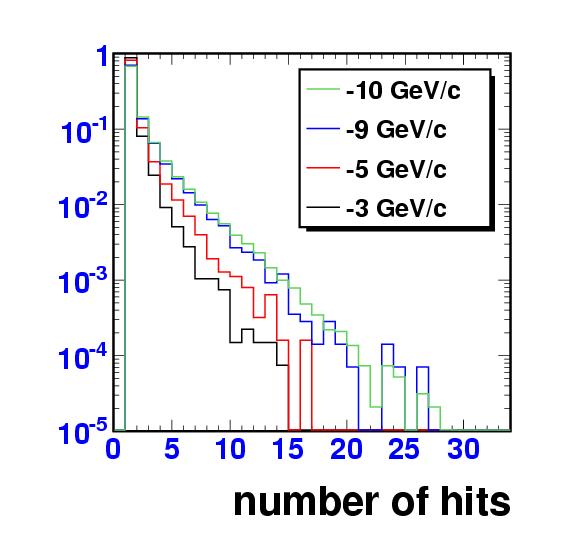}
\includegraphics[width=0.31\textwidth]{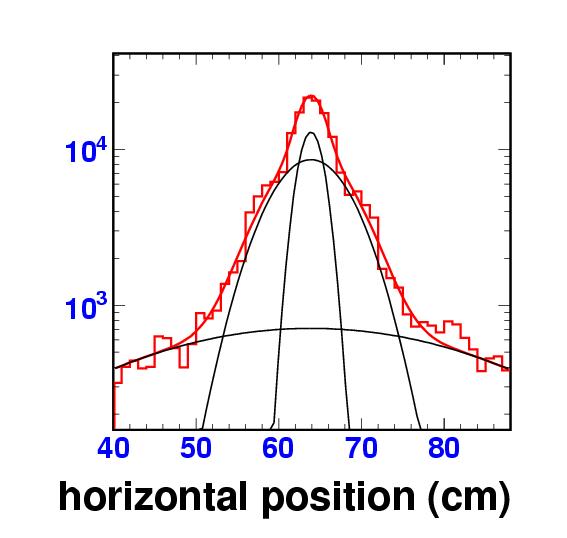}
\caption{\em From left to right: Insertion of the m$^{2}$ MICROMEGAS
  prototype inside the last slot of the AHCAL-tungsten structure in
  CERN/PS/T9 line in November 2010. Number of hits measured at various
  beam energies and projection of the hits recorded at -10~GeV/c along
  the horizontal direction showing the contributions from beam muons,
  electromagnetic shower core and hadronic halo.} 
\label{micromegas_t9}
\end{centering}
\end{figure}

At the end of 2010, the MICROMEGAS group joined the test of the
tungsten analogue hadron calorimeter (W-AHCAL) in a beam of low energy
hadrons at the CERN/PS (Fig.~\ref{micromegas_t9}).  The m$^{2}$
prototype acted as the last layer and recorded hits from hadrons
showering deep in the calorimeter.  A synchronisation between the two
acquisitions could be set up which should allow the reconstruction of
events recorded by both detectors.  The analysis of the data is
on-going though preliminary results of standalone reconstruction have
been obtained.

\subsubsection{ASIC development}
In collaboration with the LAL/Omega group and based on the experience
obtained with the HARDROC2 and DIRAC circuitry \cite{MM_asic}, LAPP is
developing a new ASIC. Optimised for the detection of MICROMEGAS
signals (and also GEM/THGEM signals), this new chip is called MICROROC
and has 64 channels with individual pedestal settings and three
thresholds common to all channels. Oriented for use at a future linear
collider, it has a 127 event depth memory and power pulsing
capability.

A batch of 350 chips has been fabricated in 2010 and caracterised on a
test board. First measurement of the noise level (around 0.24 fC)
allows the detection of input charges down to 1~fC. Systematic
calibrations are now on-going on the full batch and already half of
the chips have been tested and mounted on 7 ASU. These will compose a
new 1~m$^{2}$ prototype which should be ready for test in July
2011. The construction of a second prototype with MICROROC should be
completed by fall.

\subsubsection{Future plans}
The short term plan is to test at CERN the first MICROROC prototype of
1~m$^{2}$.  A test in a muon beam should be carried out in August
2011, followed in fall by a test inside steel or tungsten structures
of one or two planes. In the latter case, the MICROMEGAS group intends
to join the test of the Glass RPC Fe-DHCAL and/or the scintillator
W-AHCAL.

Longer term plans are the fabrication of more MICROROC ASICs and
prototypes.  The number of planes to be produced in 2012 and later
critically depends on financial ressources and is thus not precisely
known today.  Shower development and containment of low energy
electrons and hadrons can already be studied with a small number of
planes. Future test in beams should therefore be conducted inside
steel or tungsten structures, preferably at the CERN PS facility.

\section{Front-End Electronics}
\label{sec:fee}

A second generation of readout ASICs have been developed to
read out the technological prototypes defined in EUDET and
CALICE. These are based on the first generation of chips that were used
for the physics prototype for the analogue front-end part but add
several essential features:

\begin{itemize}
\item[$\bullet$] Auto trigger to reduce the data volume
\item[$\bullet$] Internal digitization to allow purely digital data output
\item[$\bullet$] Integrated readout sequence and common interface to
  the second generation data acquisition to minimize the number of
  lines between chips
\item[$\bullet$] Power-pulsing to reduce the power dissipation by a
  factor of 100
\end{itemize}

Three chips have been designed, following the EUDET milestones:
\begin{itemize}
\item[$\bullet$] HARDROC for the digital Hadronic Calorimeter, for
  RPCs or Micromegas chambers. A new ASIC, MICROROC, has also been
  designed and submitted in June 2010 for 1m$^2$ MICROMEGAS detectors,
  which require HV spark robustness for the electronics and very low
  noise performance to detect signals down to 2~fC.
\item[$\bullet$] SPIROC for analog Hadronic Calorimeter.
\item[$\bullet$] SKIROC for the the Si-W Electromagnetic Calorimeter, submitted in March 2010.
\end{itemize}

\begin{figure}[h]
\centering
\includegraphics[width=0.58\textwidth]{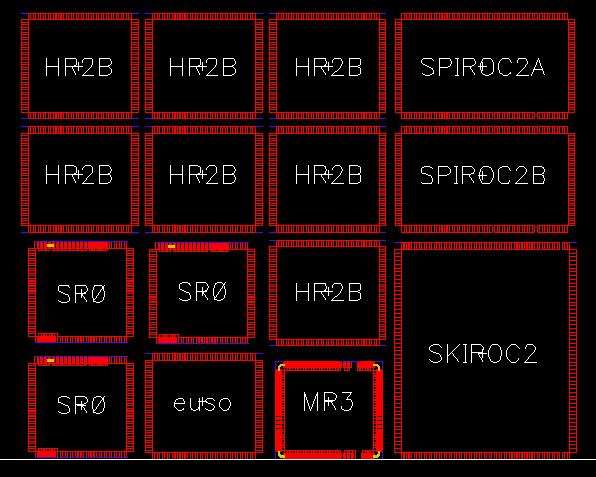}
\caption{Reticle of the production run.}
\label{fig:fee_prod_reticle}
\end{figure}

\begin{figure}[h]
\centering
\includegraphics[width=0.38\textwidth]{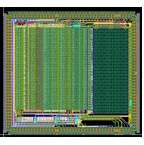}
\includegraphics[width=0.20\textwidth]{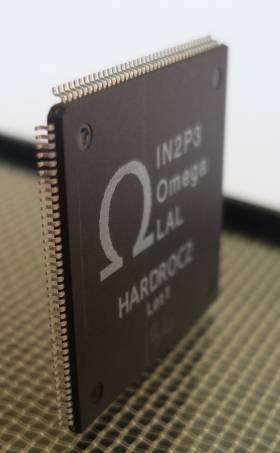}
\includegraphics[width=0.38\textwidth]{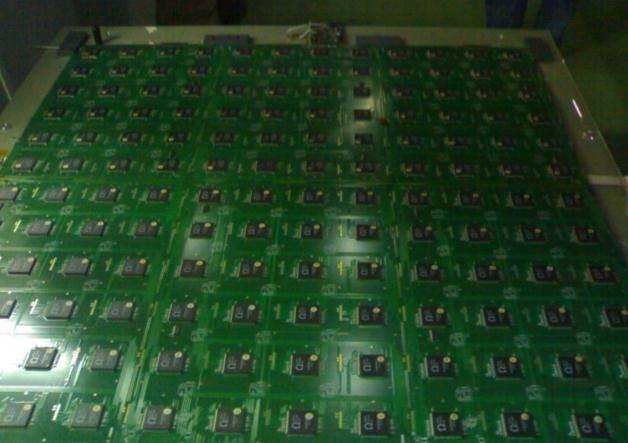}
\caption{(left) Layout of HARDROC2, (centre) view of the chip packaged
  in TQFP160, and (right) square metre prototype of RPC DHCAL with 144 HARDROC.}
\label{fig:fee_hardro2c_layout}
\end{figure}

In March 2010, these ASICs (except MICROROC) have been produced
(Fig.~\ref{fig:fee_prod_reticle} in a dedicated run of AMS SiGe 0.35~$\mu$m technology. 25
wafers, each with 500 HARDROC2 (see Fig.~\ref{fig:fee_hardro2c_layout}), 70 SPIROC2A, 70 SPIROC2B, 70 SKIROC2
have yielded a few thousand chips that can equip respectively
DHCAL, AHCAL and ECAL modules. The chips have been packaged
commercially (I2A company, USA) in an ultra-flat TQFP package to be embedded inside the
detector modules with a minimal thickness.

\subsection{DHCAL technological prototype}

\subsubsection{HARDROC ASIC}
The HARDROC readout is a semi-digital readout with three thresholds (2
bits readout) which allows both good tracking and coarse energy
measurement, and also integrates on chip data storage. The chip
integrates 64 channels of fast, low-impedance current preamplifiers with
6 bits variable gain (tuneable between 0 and 2), followed by a fast
shaper (15~ns) and low offset discriminators. The discriminators feed a
128-deep digital memory to store the 2*64 discriminator outputs and
bunch crossing identification coded over 24 bits counter. Each is then
readout sequentially during the readout period.

A first version was been fabricated in AMS SiGe 0.35$\mu$m technology
in September 2006 and met design specifications. A second version was
produced in June 2008 to fit in a smaller low-height package (TWFP160)
which necessitated changing the double-row bonding pad ring into a
single row, rerouting all of the inputs and removing many pads. A
possibility for a third threshold was added at that time, also
separating more widely the three thresholds (typically 0.1--1--10~pC)
and the ``off'' power dissipation was brought down to a few $\mu$W for
the whole chip.

\begin{figure}[h]
\centering
\includegraphics[width=0.52\textwidth]{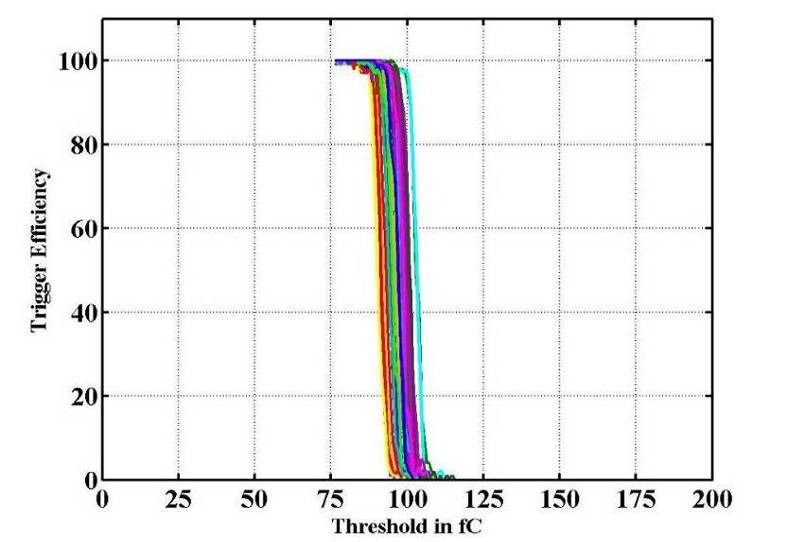}
\includegraphics[width=0.46\textwidth]{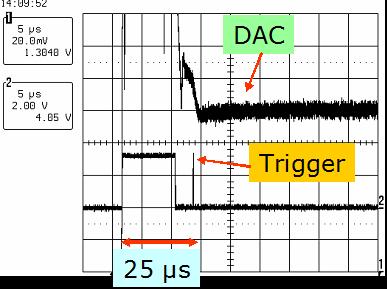}
\caption{Trigger efficiency for 100~fC input as a function of DAC
  threshold.}
\label{fig:fee_trig_eff}
\end{figure}

The trigger efficiency allows the MIPs for RPCs to be discriminated
with 100~fC threshold (10~fC for Micromegas) with a noise of 1~fC
(Fig.~\ref{fig:fee_trig_eff}). The power pulsing scheme has also been validated, also
shown in Fig.~\ref{fig:fee_trig_eff}, where 25 $\mu$s are required to start up the chip
so that it can trigger on a 10~fC input signal. Finally the readout
scheme, which is common to all the chips, has been validated on the
large square-metre board, built as a scalable technological prototype
of DHCAL that is read-out by the side.

This HARDROC chip is the first one on which large scale power-pulsing
has been tested at system level, allowing a power reduction by a
factor of 100 while keeping the detector efficiency above 95\%.

A production of 10\,000 chips started in March 2010 to equip
40 GRPC planes of a one cubic metre detector that will be tested in
2011.

\begin{figure}[h]
\centering
\includegraphics[width=0.44\textwidth]{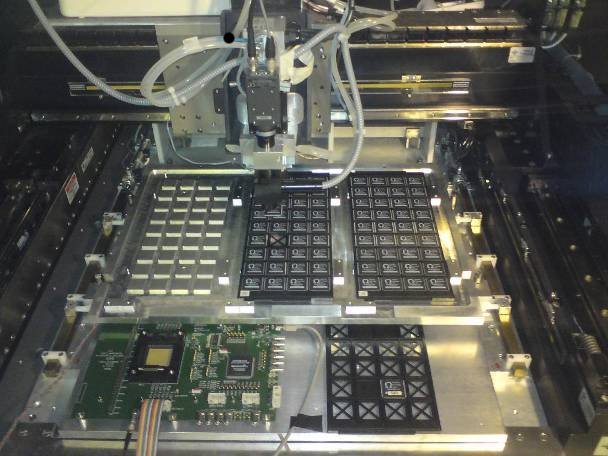}
\includegraphics[width=0.44\textwidth]{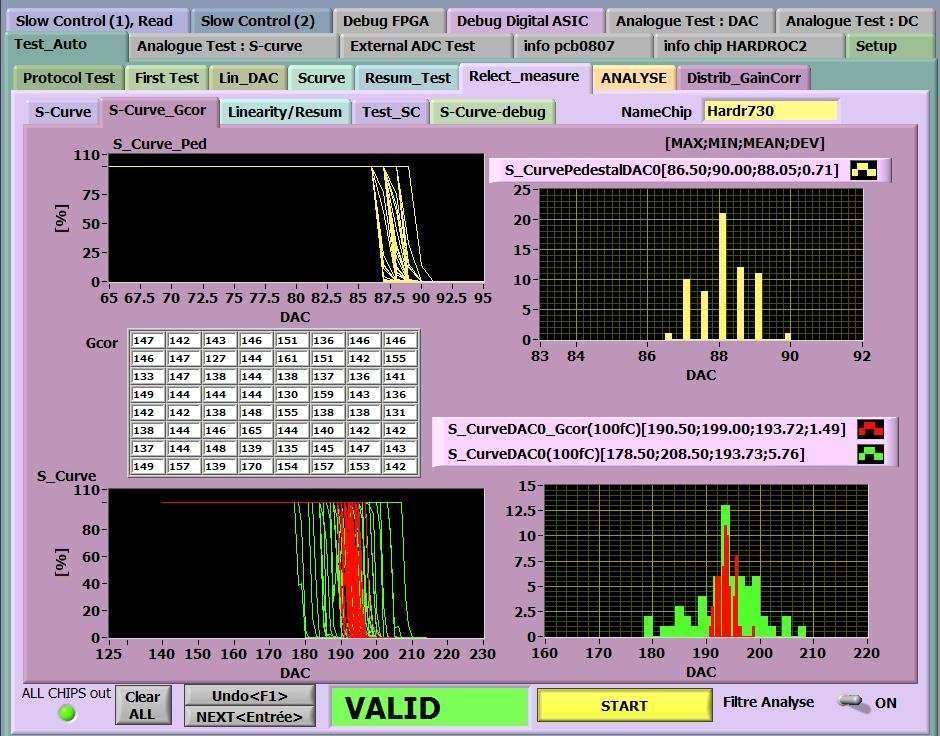}
\caption{Testbench used to test the 10\,000 HARDROC2 chips and test
  measurement results.}
\label{fig:fee_testbench}
\end{figure}

A specific test bench with a Robot, shown in
Fig.~\ref{fig:fee_testbench}, has been used at Lyon to test and
qualify the chips before mounting them on the boards. The yield is
about 93\%.

\subsubsection{MICROROC ASIC }

\begin{figure}[h!]
\centering
\includegraphics[width=0.44\textwidth]{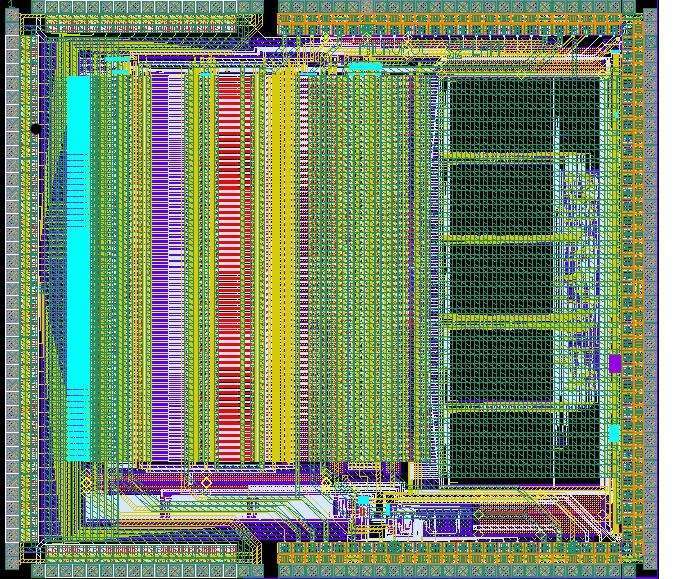}
\caption{Layout of MICROROC.}
\label{fig:fee_microroc_layout}
\end{figure}

HARDROC has also been used in test beam to readout square metre
Micromegas chambers.  However, the required robustness to HV sparks
and the abaility to detect signals down to 2~fC necessitate the design
of a more sensitive version having a low noise charge preamplifier
input stage. MICROROC is a 64 channel mixed-signal integrated circuit
based on HARDROC, designed in AMS 350~nm SiGe technology
(Fig.~\ref{fig:fee_microroc_layout}). Analogue blocks and the whole
digital part are reused from HARDROC.

Each channel of the MICROROC chip is made of a very low noise fixed
gain charge preamplifier optimised for a detector capacitance of 80~pF
and able to handle a dynamic range from 1~fC to 500~fC, two different
adjustable shapers, three comparators and a random access memory used
as a digital buffer. All these blocks are power-pulsed, thus reaching
a power consumption equal to zero in standby mode.

\begin{figure}[h!]
\centering
\includegraphics[width=0.70\textwidth]{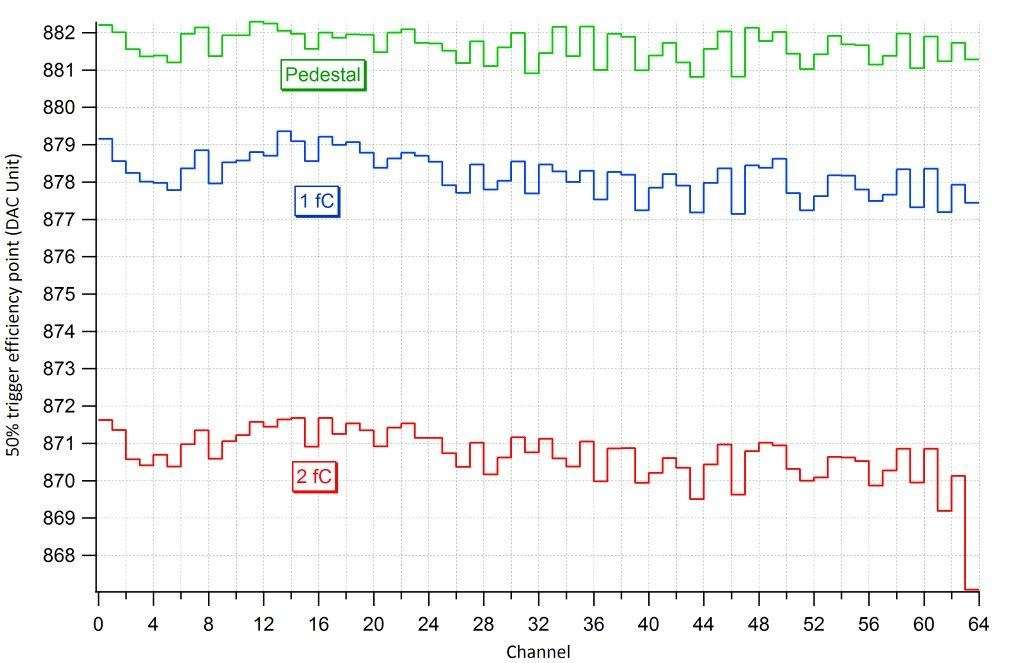}
\caption{Performance of MICROROC, threshold vs.\ channel number.}
\label{fig:fee_microroc_performance}
\end{figure}

MICROROC has been tested on test bench and exhibits very good
performance with in particular the ability to trigger down to 1~fC as
shown in Fig.~\ref{fig:fee_microroc_performance}. 300 ASICs have just
been packaged (TQFP160 plastic package) to equip and test two square
metre Micromegas chambers.

\subsection{AHCAL technological prototype}

\begin{figure}[h!]
\centering
\includegraphics[width=0.59\textwidth]{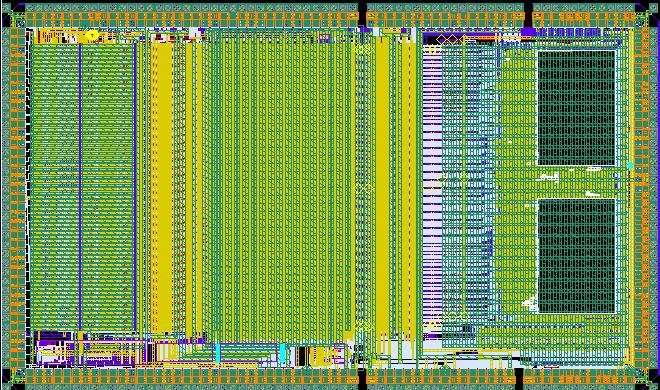}
\caption{Layout of SPIROC2.}
\label{fig:fee_spiroc2_layout}
\end{figure}

The SPIROC ASIC that reads 36 SiPMs is an evolution of the FLC\_SiPM
used in the physics prototype. The first prototype was been fabricated
in June 2007 in AMS SiGe 0.35$\mu$m. and packaged in a CQFP240
package. Similarly to HARDROC, a second version, SPIROC2, was 
realized (Fig.~\ref{fig:fee_spiroc2_layout}) in June 2008 to accommodate a thinner TQFP208 package and fix
a bug in the ADC.

\begin{figure}[h!]
\centering
\includegraphics[width=0.78\textwidth]{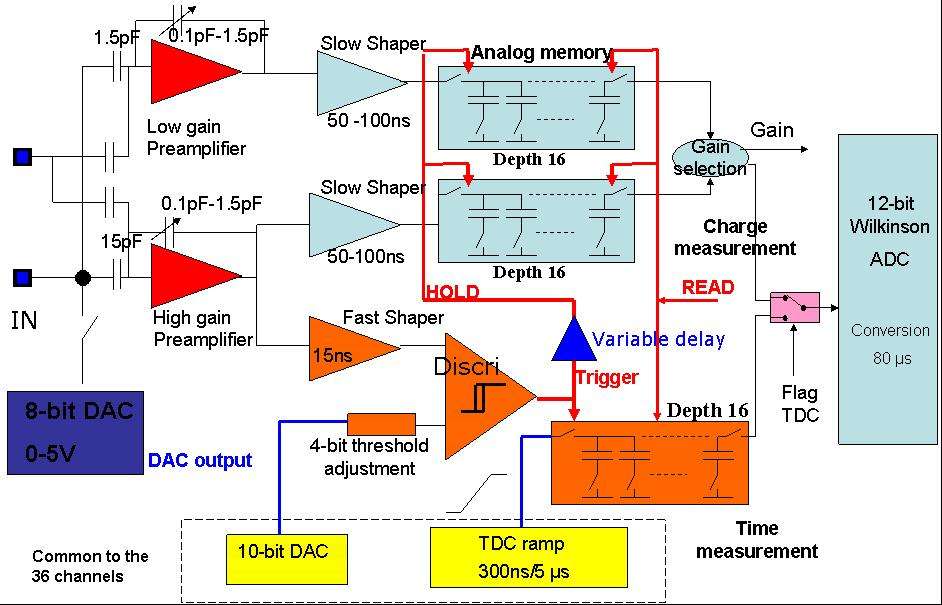}
\caption{Schematic diagram of one channel of SPIROC2.}
\label{fig:fee_spiroc2_1channel}
\end{figure}

Each channel of SPIROC2 (Fig.~\ref{fig:fee_spiroc2_1channel}) is made of:

\begin{itemize}

\item[$\bullet$] An 8-bit input DAC with a very low power of
  1$\mu$W/channel as it is not power pulsed. The DAC also has the
  particularity of being powered with 5V whereas the rest of the chip
  is powered with 3.5V.

\item[$\bullet$] A high gain and a low gain preamp in parallel on each
  input allow handling the large dynamic range. A gain adjustment over
  4 bits common for the 64 channels has been integrated in SPIROC2. A
  variant (SPIROC2B) with individual gain adjustment over 6 bits has
  also been produced.

\item[$\bullet$] The charge is measured on both gains by a ``slow''
  shaper (50--150~ns) followed by an analogue memory with a depth of 16
  capacitors.

\item[$\bullet$] The auto-trigger is taken on the high gain path with
  a high-gain fast shaper followed by a low offset discriminator. The
  discriminator output is used to generate the hold on the 36
  channels. The threshold is common to the 36 channels, given by a 10
  bit DAC similar to the one from HARDROC with a subsequent 4 bit fine
  tuning per channel.

\item[$\bullet$] The discriminator output is also used to store the
  value of a 300~ns ramp in a dedicated analogue memory to provide time
  information with an accuracy of 1~ns

\item[$\bullet$] A 12 bit Wilkinson ADC is used to digitize the data at
  the end of the acquisition period.
\end{itemize}

The digital part is complex as it must handle the SCA write and read
pointers, the ADC conversion, the data storage in a RAM and the
readout process.

\begin{figure}[h!]
\centering
\includegraphics[width=0.68\textwidth]{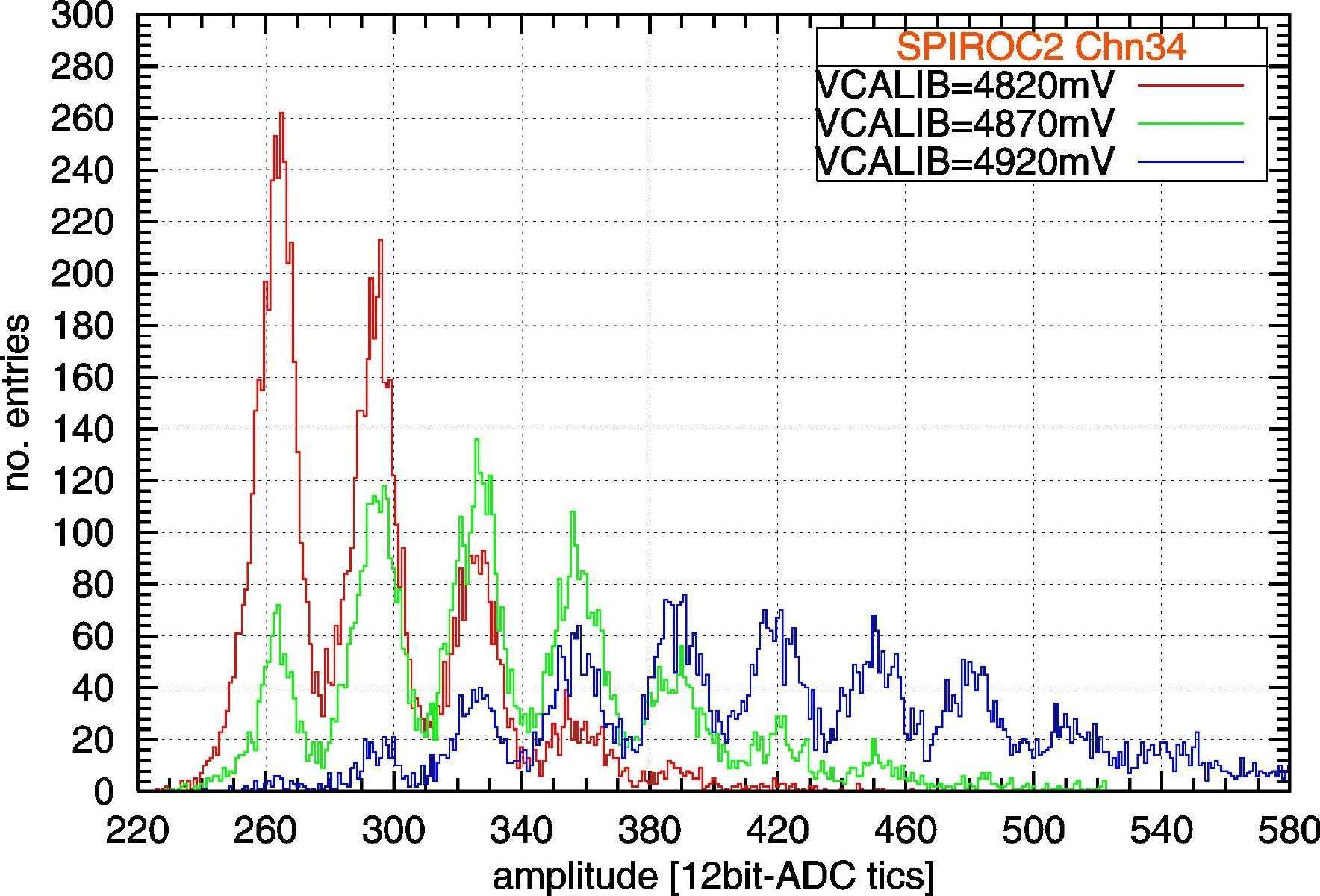}
\caption{Single photo electron spectrum.}
\label{fig:fee_single_pee}
\end{figure}

The chip has been extensively tested by many groups. The first series
of tests has been mostly devoted to characterizing the analog
performance, which meets the design specifications. A single
photoelectron spectrum using the full chain and a LED pulser is
displayed in Fig.~\ref{fig:fee_single_pee}. The one photo-electron signal to noise ratio is
around 8.

\begin{figure}[h!]
\centering
\includegraphics[width=0.8\textwidth]{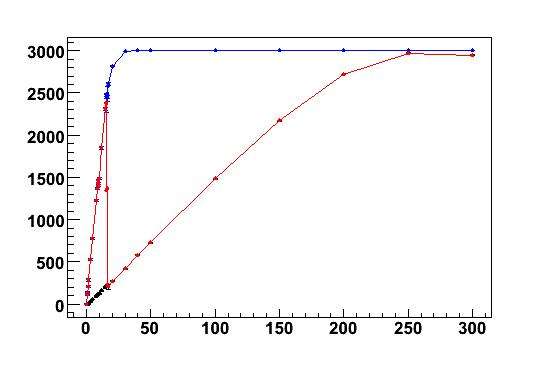}
\caption{SPIROC2B linearity using the auto gain mode.}
\label{fig:fee_spiroc2b_linearity}
\end{figure}

The linearity as a function of the input charge in the auto gain mode
and using the internal ADC is shown in
Fig.~\ref{fig:fee_spiroc2b_linearity}.

\begin{figure}[h]
\centering
\includegraphics[width=0.53\textwidth]{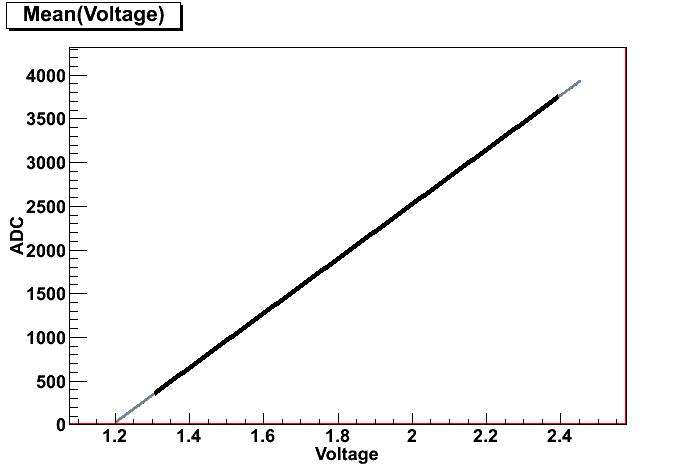}
\includegraphics[width=0.53\textwidth]{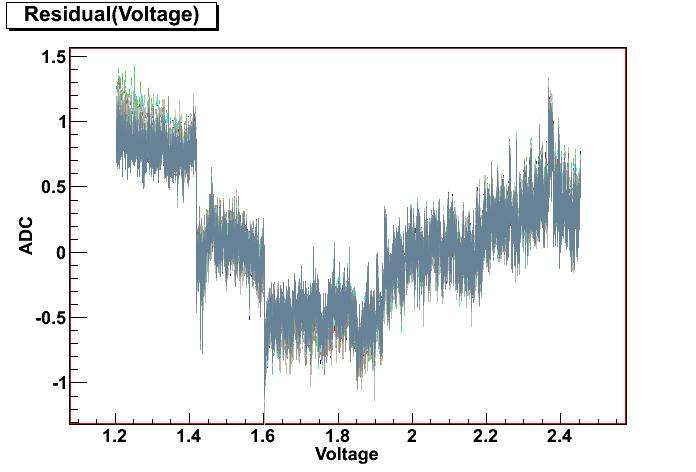}
\caption{(left) ADC response of the 36 channels over the 12 bits
  dynamic range. (right) residual to a linear fit showing an integral
  linearity better than 1 LSB.}
\label{fig:fee_adc_response}
\end{figure}

The digitization part has also been characterized and the 12 bit ADC
exhibits a very good integral non-linearity of 1 LSB and a noise
comprised between 0.5 and 1 LSB, as shown in Fig.~\ref{fig:fee_adc_response}.

The chips have been assembled on a HCAL PCB (HBU) and tested with
detector, as described in the HCAL section and are being operated with
good results.

A new, improved version of the SPIROC-Chip with increased signal-to-noise ratio is currently being developed. For this a new frontend has been designed which provides higher signal-to-noise ratio at the same time retaining a large dynamic range. A first version of a test chip (KLauS 1.0) shows a signal-to-noise ratio of greater than ten for a signal charge of 40~fC. A second version (KLauS 2.0) including the power-pulsing
option has been submitted end of 2010 and is presently tested.

\subsection{ECAL technical prototype}

\begin{figure}[h!]
\centering
\includegraphics[width=0.38\textwidth]{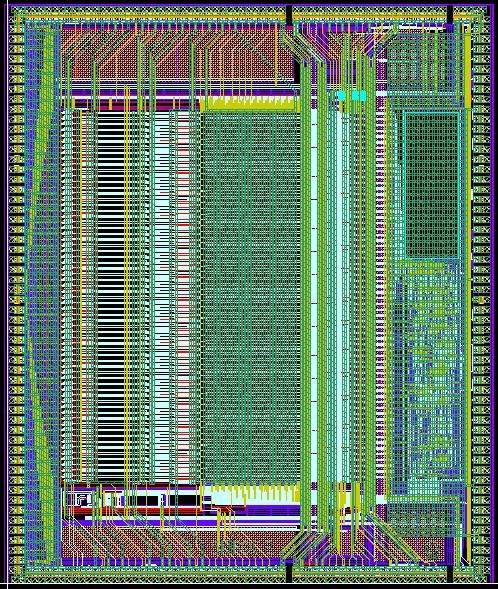}
\caption{Layout of SKIROC2 (7.2mm$\times$8.6 mm).}
\label{fig:fee_skiroc2_layout}.
\end{figure}

For the ECAL, the chip SKIROC2 has been designed
(Fig.~\ref{fig:fee_skiroc2_layout}) and submitted in the production
run of March 2010. It keeps most of the analog part of SPIROC2, except
for the preamp which is a low noise charge preamp followed by a low
gain and a high gain slow shaper to handle a large dynamic range from
0.5~MIP (~fC) up to 2500~MIPs (10~pC).

\begin{figure}[h!]
\centering
\includegraphics[width=0.58\textwidth]{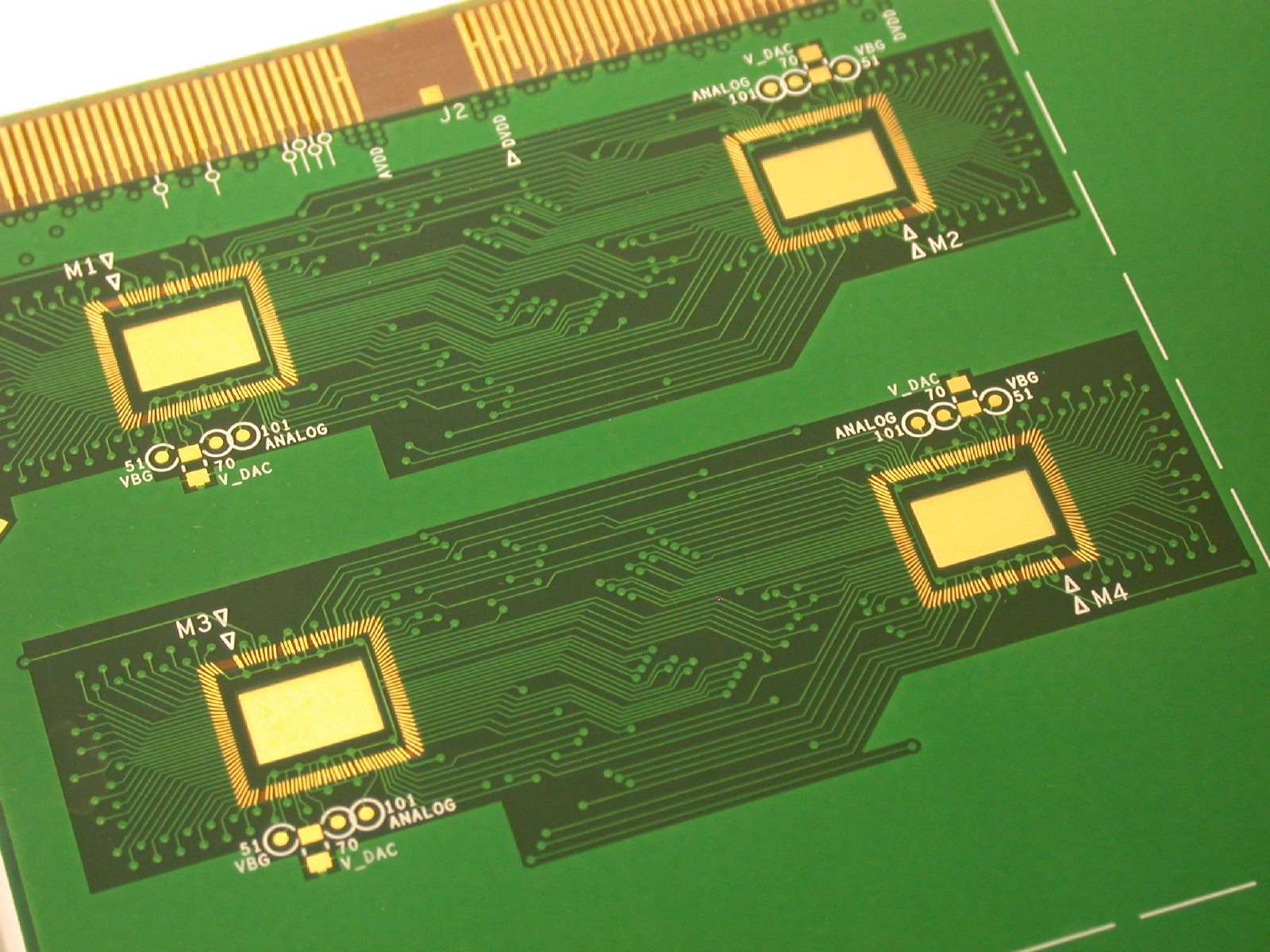}
\caption{FEV board for embedded SKIROC chips.}
\label{fig:fee_fev_skiroc}
\end{figure}

SKIROC2 chips are not packaged (except for test bench measurements) as
they must be directly bonded on the printed circuits
(Fig.~\ref{fig:fee_fev_skiroc}). The characterization is performed on
testbench using a few packaged chips but the produced chips will be
tested using a probe station. FEV boards hosting 16 SKIROC2 chips are
under design to readout 1024 channels.

\subsection{Conclusion}
The second generation ASICs for CALICE/EUDET have been produced and a
large quantity is therefore available to equip large scale detectors
to check that all key issues have been solved. The design of the third
and final generation where all channels are handled separately to
allow higher zero-suppression has just begun in the AIDA framework.

\section{Data Acquisition}
\label{sec:daq}
A second version of the CALICE DAQ was proposed \cite{daq:Eudet_mwing2006}
within the framework of EUDET to read out the various technical
prototypes being developed.  The aim was to handle the many embedded
readout chips and their direct digital output in test beams.  A second
goal was to serve as a prototype of techniques for the calorimeters of
a large detector such as the ILD.

The project is now in the latest integration phase, prior to
deployment in the CALICE labs. (DESY, IPNL, LAPP, LLR). Up-to-date
information is kept on the CALICE Twiki DAQ
pages \cite{daq:CaliceTwiki}
and regularly presented at CALICE
meetings \cite{daq:CaliceMeetings}.
It has recently triggered interest of the FCAL Collaboration for a
common interface, and will be part of the ``common DAQ'' package of the
FP7 AIDA project, together with the telescope EUDAQ.

\subsection{Functional specifications}
All the technological prototypes feature embedded electronics in the
form of ASICs with built-in memory situated in the detector volumes.
This very front-end electronics performs analogue amplification and
shaping, digitization, optionally internal triggering, and a local
storage of data in memory. The various detectors share a commonly
defined interface through custom Detector Interface (DIF) cards.

Two modes of operation are foreseen: a single event mode, in which the
data readout is triggered by an external signal (typically in beam test
operation), and an ILC-like mode which is consistent with that
expected for the ILC beam structure, i.e. the readout of full train of data
without external trigger, at a train frequency of $\sim5$~Hz.

The DAQ should handle:
\begin{itemize}
\item the loading of the configuration in the ASICs (seen as one
  stream of bits for one set of ASICs), and optionally the
  verification of the loaded data;
\item the management of the acquisition states trough commands:
 acquisition running mode [Single Event, ILC mode], StartReadout,
StopReadout, Reset, Sync mode;
\item the distribution of the fast signals in a synchronous manner:
  Clock, Trigger and the collection and ORing of the busy signals the
  other way around.  Synchronous here is detector and reconstruction
  specific: for the SDHCAL and ECAL a machine clock period (200~ns)
  should suffice, for the AHCAL the requirement is typically less than
  1~ns;
\item the data flux, which are here reasonably low: for the SDHCAL,
  with an average of 4.8 HaRDROC chips with data per plane for a
  100~GeV pion, an ASIC readout clock of 2.5~MHz, a data flux of the
  $\sim20$~MB/s is expected in beam tests allowing for a maximum allowed
  event frequency of 3.2~kHz (much higher than the working region of
  standard GRPC, limited to a few 100~Hz for example).  For the ECAL,
  equipped with 2 ASIC readout lines per DIF, 4 ASICs with data in
  each plane would generate a data flux of 113~MB/s for a maximum
  acquisition rate of 1.2~kHz. Under the same hypothesis as for the
  SDHCAL, the AHCAL data flux reaches 338~MB/s, allowing for a 4~kHz
  event rate. In all cases, the data flux is limited by the readout
  speed of the ASICs. A detailed table is available
  online \cite{daq:fluxtable}.
\item two running modes: in single event mode, an external signal
  (typically signalling the passage of a particle of a given type,
  with a combination of scintillators and Cherenkov detectors)
  triggers a suspension of the acquisition mode of the ASIC, their
  readout and the resumption of the acquisition. In ILC-like mode, the
  ASICs are switched into Acquisition mode at the start of a particle
  spill, and are read out either at the end of the spill, after a given
  time span or number of particles, or when any of the detector ASICs
  emits a RAM-full signal.  The acquisition is eventually resumed as
  soon as the readout is completed. As both modes use the
  auto-triggering capacity of the ASICs, they are very sensitive to
  noise. A careful online monitoring of the noise has to be performed,
  with an automatic correction procedure necessary to kill noisy
  channels.
\end{itemize}
\subsection{Implementation}
The overall scheme for the DAQ take the form a tree (see Fig.~\ref{seq:refIllustration0}) with :
\begin{itemize}
\item one (or more depending on data flux) DAQ PCs equipped with 1--2
  ODR (Off Detector Receiver) card, or a standard Network card, linked
  via Optical or Electrical Gigabit Ethernet to 2--4 LDAs (Link Data
  Agregator) cards;
\item each LDA card services up to 10 connections to the DIFs
  (Detector InterFace card), or (case of the SDHCAL) to the DCC cards
  (Data Concentrator Card) via HDMI cables carrying all the needed
  signals.
  \begin{itemize}
  \item each DCC is connected to up to 9 DIF cards (Detector InterFace);
  \end{itemize}
\item each DIF manages a SLAB hosting many ROCs (Read-Out Chips)
  through one or several ASIC readout lines;
\item a CCC (Clock and Control Card) dispatches the fast signal to the
  DIFs via the LDAs \& DCCs: external or internal clocks and
  Trigger/Sync signals. It collects the busy signals from the DIFs and
  blocks the external triggers accordingly. It can be controlled by
  the PC through an RS232 connection and provide simple logic and a
  50~MHz clock.
\begin{figure}
\centering
\includegraphics[width=0.75\textwidth]{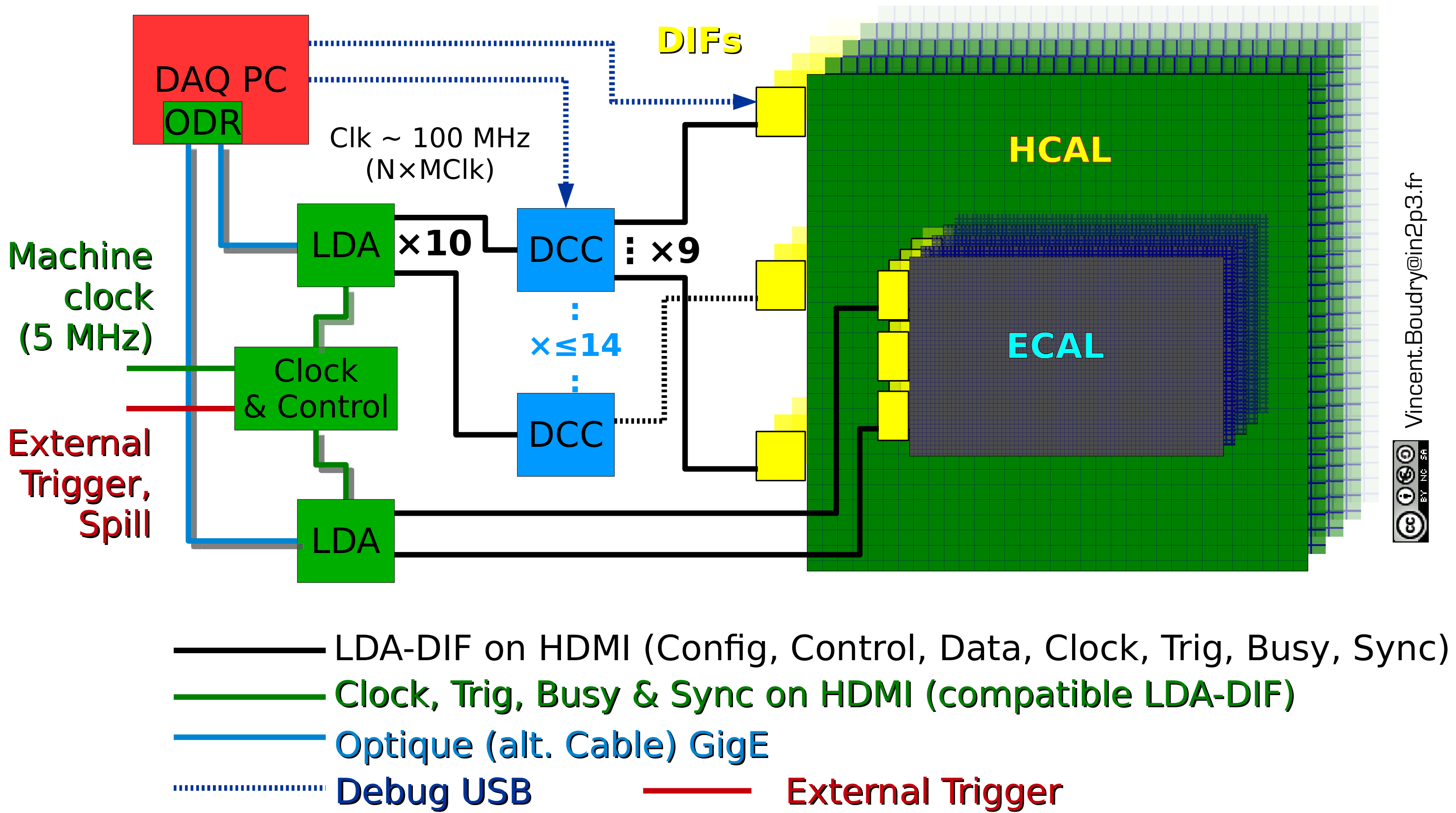}
\caption[DAQv2 scheme for the SDHCAL (with DCC cards ) and the ECAL
  (direct connection DIF--LDA)]{DAQv2 scheme for the SDHCAL (with
  DCC cards ) and the ECAL (direct connection DIF--LDA)}
\label{seq:refIllustration0}

\end{figure}
\end{itemize}
All the data on the HDMI links are transferred using 8b/10b encoding
for redundancy and passing of control signals.

The typical needs for the a complete SDHCAL (50 layers) set-up are of
150 DIFs, 17 DCCs, 2--3 LDAs; The ECAL requires 30 DIFs and 3--4 LDAs,
the AHCAL 48 DIFs and 5--7 LDAs.  Each system requires a CCC, and
1~ODR or Network card (2 for the AHCAL).

\subsection{Planning \& organisation}
The design of the DAQv2 was developed by the CALICE UK groups
(Cambridge, UCL, Manchester, RHUL) in the framework of EUDET. All the
hardware developments were fulfilled by the UK institutes, with the
exception of the DCC (LLR) and DIFs (DESY, LAPP, Cambridge). Starting in 2009, the
funding support for ILC in UK dropped, resulting in the loss of
expertise (engineers and postdocs leaving). The LLR group which
was involved in the integration for the SDHCAL gradually took over the
global integration and debugging tasks, while the IPNL group assumed
responsibility for the development of the software framework.

A DIF Task force was set up in 2008, with one expert from each of
LLR, LAPP, Cambridge and DESY, to establish
a common interface and to coordinate the firmware development. All
relevant documents and specifications (data exchange format, commands,
etc.) are available \cite{daq:DIF}.

\subsection{Hardware availability}
All the generic hardware elements (LDA and its mezzanine cards, DCC,
CCC, ODR and PC) are ready and tested.  A list of equipment is
available \cite{daq:DaqHw}. By the end of 2010 most of the equipment
was dispatched from the UK to the main testing labs
(LLR, DESY, LAPP and IPNL).

The last physical parts produced in 2010 were the DCC and LDA cards;
20 DCC cards have been produced and been tested, with almost no
failure found, they will be installed in a standard VME crate. The LDA
are physically composed of a commercial baseboard, and 3 custom
mezzanine cards respectively supporting the 10 HDMI connectors, the
connection to the CCC and the Gigabit-Ethernet connection.  After many
difficulties with the company building the card (Enterpoint Ltd) and
many delays, all the hardware parts are available; a mechanical
structure is under study to host securely the CCC and the LDAs, both
having non-standard formats, in the test beam area.

The HDMI cables' specifications are determined by two requirements:
for their length, by the physical arrangement of the detectors in the
TB hall (4-5~m are needed), and for their composition, by safety
requirements (Halogen free cables are mandatory for work in CERN).
150 cables are needed for the SDHCAL, another 30 for the ECAL, and 40
for the AHCAL. In total, 275 cables of 5~m length were manufactured
specifically for the project and are being
delivered: 100 cables are are available at LLR, the remainder are in
transit. The first tests are fully satisfactory.

\subsection[Firmware \& implemented functionalities]{Firmware \&
implemented functionalities} The major occupation of the last year has
concerned the firmware (FW) of all the cards.

First, a common FW for all DIF has been developed.  A global framework
was developed (LLR) for the interconnection of existing various parts;
an 8b/10b decoding blocks (Manchester), testing modules (pseudo random
pattern generation for high volume of data, and echo of configuration
pattern), and working implementation of the ROC chips management used
for the SDHCAL over USB (LAPP).

The DCC FW, including the 8b/10b parts and a multiplexing
engine is complete and works as expected; the main objective here being
its transparency.

The LDA FW has specifications very similar to the DCC, with the
Downstream block being replaced by a Gigabit-Ethernet block.  It is
now $\sim$98\% complete with only comfort functionality missing, such
as a soft reset of the board, and a small instability of the incoming
flux from the PC at full speed (this can easily be avoided by a small
delay in the configuration sending).

Recent performance tests using pseudo-random data have shown the
possibility to read up to 10 DIFs on 10 different DCC on one LDA or
10~DIFs on 1 LDA up to 32~MB/s almost without any failure (prelim.\
$\leq 10^{-9}$ over 1 day) using an optical link,
which is OK for the first use (20~MB/s for the SDHCAL).

The transfer of the Trigger and BUSY signal to the CCC has been much
more problematic as expected due to the use of an AC coupling between
the CCC and the LDA, leading to instabilities and stringer conditions
on the length of the signals. Special treatment (conversion of
continuous signal as clock) were implemented (UCL) and tested and
improved (LLR).  A furst working implementation was produced at the
end of March 2011 and will still need more stringent tests.

Preliminary measurements of fast signals on DIFs connected to one full
chain (CCC LDA DCC DIFs) has shown a clock
jitter of $\sim$550~ps on a single DIF and a jitter of the trigger delay
between 2 DIFs of $\sim$200~ps.  This performance is acceptable for the most
critical of the intended applications, namely the AHCAL readout.

The programming of the CCC CPLD has to be adapted for each setup
(format of input signal, ORing and blocking logic) and will have to be
performed; the FW code is fully available and expertise exits: for
exampl LAPP have modified significantly the code to allow for a direct
connection to the DIF implementing clock, trigger and (hard coded)
fast command distribution to 3 DIFs \cite{daq:prast}.

\subsection{Software}
Some early development were done in the DOOCS framework, but were
finally not pursued due to lack of manpower; in parallel, the XDAQ
framework \cite{daq:xdaq}, used by IPNL for the readout of the first SDHCAL
 prototypes using a direct USB connection to the
DIF, was also developed at IPNL to include most of the components
necessary for the TB: event building, and online reconstruction via an
interface to Marlin and root data quality histogram filling, an
interactive GUI tool manage the acquisition and display histograms,
the storage of RAW data in LCIO format.

This system has been working perfectly using the USB interface to the
DIF for more than one year and is being used as part of a several month
geological survey of a volcano near Clermont-Ferrand (by LPC colleagues),
with almost no downtime.  A stress test based on a LDA simulator
injecting pseudo random data has shown stable performances up-to 500~MB/s.

The interface to the new hardware is now being finalised.  Until recently, the
basic chain PC+LDA+DCC+DIF has been mainly been tested at LLR using ad hoc SW elements based
on python scripts \& GUI and C libraries. In early January 2011, the C
library was for the first time interfaced with the XDAQ environment as
a driver and worked immediately, allowing for intensity data flow
testing. The interface to the CCC is already done. A complete beta
version of a C++ API for the LDA is now available and is ready to be
integrated.

A first prototype of a configuration DB based on Oracle has been
designed at IPNL, with a C++ API.  In the near future, it will be
validated and thoroughly tested. It is also foreseen to have a local
file interface as a fallback solution for the scenario in which
connectivity is lost.

A GUI is being developed in parallel to ease the use of the system for
the low level electronics test in LLR.

\subsection{Summary}
The DAQ system is in good shape, without any obvious ``show stoppers''
on the horizon which could plausibly prevent it from being ready for
cosmics data taking for the SDHCAL prior to test beam data (mid-June
2011).  The other applications (ECAL and AHCAL) will follow easily.
The emphasis has now to be put software, improvement of early online
analysis (especially noise monitoring \& correction) and testbeam
interface.

\section{Test Beam Analysis Results}
\label{sec:analysis}
The aims of the analysis of the test beam data are broadly twofold.  Firstly, 
we wish to understand characterise the performance for these novel calorimeter designs, to assess their stability and to understand calibration issues.  
Secondly, we need to use these results to validate our 
Monte Carlo simulations, based on GEANT4~\cite{GEANT4}, which can then be used in the optimisation of
global detector design.  In particular, the simulation of hadronic showers is 
not theoretically well under control, especially at energies of a few GeV, characteristic of particles in jets.  New experimental input is welcomed by the GEANT developers, some of whom are now members of CALICE. 

\subsection{Si-W ECAL performance}
\label{siwperf}
\begin{figure}[btp]
\centering \includegraphics[width=0.35\textwidth]{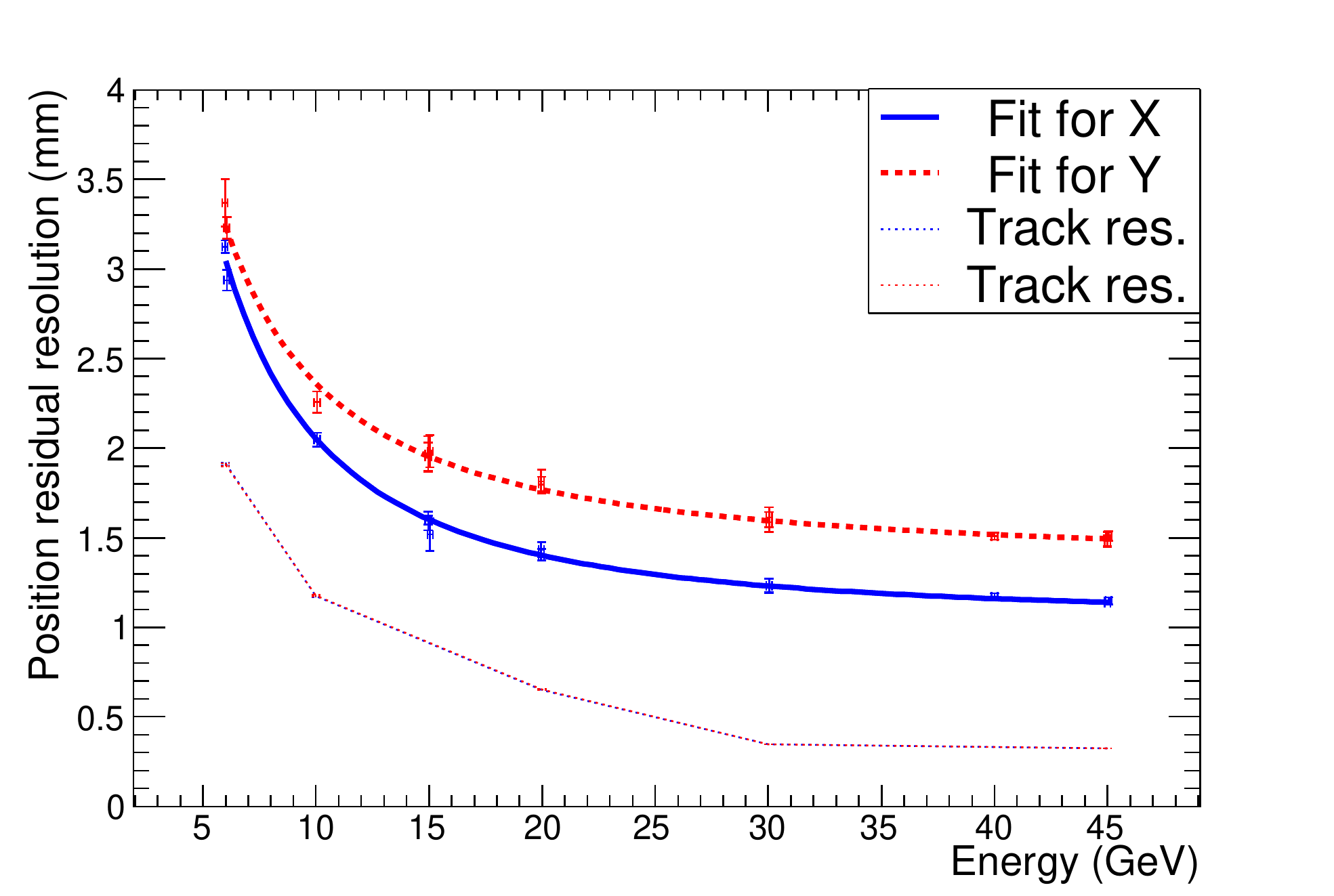}
\includegraphics[width=0.35\textwidth]{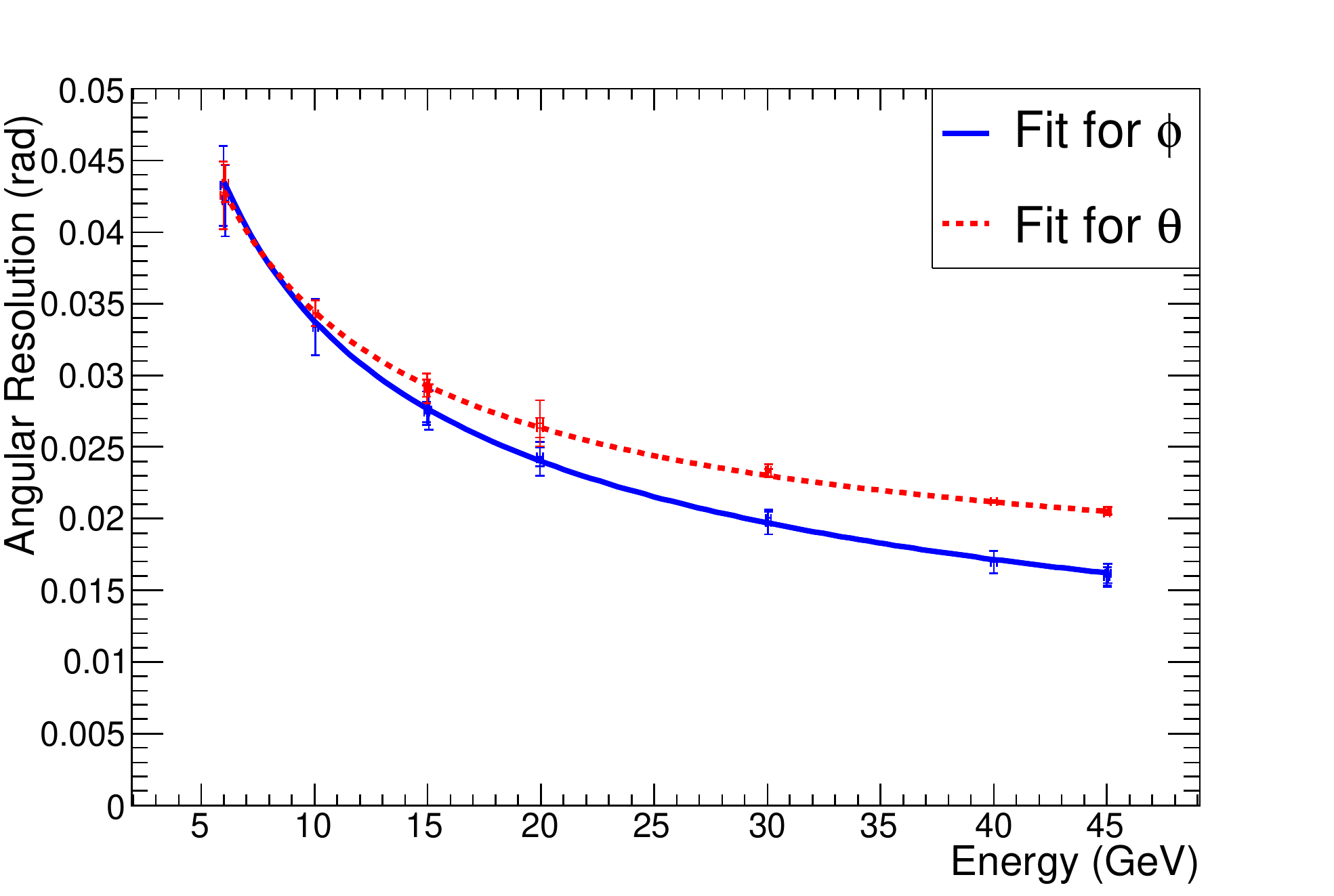}
\caption{\label{fig:EcalPosAng}\em (Left) Position resolution for electron showers in the ECAL. 
The precision of track extrapolation is also shown.
(Right) Angular resolution for electron showers. }
\end{figure}

The Si-W ECAL was operated in the CERN beam tests in 2006-7, and in the 
first stage 
of the FNAL tests in 2008.  It was exposed to beams of muons, 
electrons and hadrons.
The muon data were used as the basis of detector calibration, so that 
recorded 
signals could be converted into minimum ionising particle (MIP) 
equivalents. 
The commissioning and calibration procedures have been described in some 
detail in~\cite{ECALcomm}.  First results on the response to electrons were 
published in~\cite{ECALresp}.   More recent results~\cite{CAN-017}
have explored the position and angular resolution of electron showers,
with examples shown in Fig.~\ref{fig:EcalPosAng}.   
These results rely on a simple algorithm comparing the 
shower centroid reconstructed in the calorimeter with a track reconstructed in the 
upstream tracking chambers.  The resolutions in $x$ (horizontal) and $y$ (vertical) are different because the sensor layers are staggered in $x$.

\begin{figure}[btp]
\centering \includegraphics[width=0.68\textwidth]{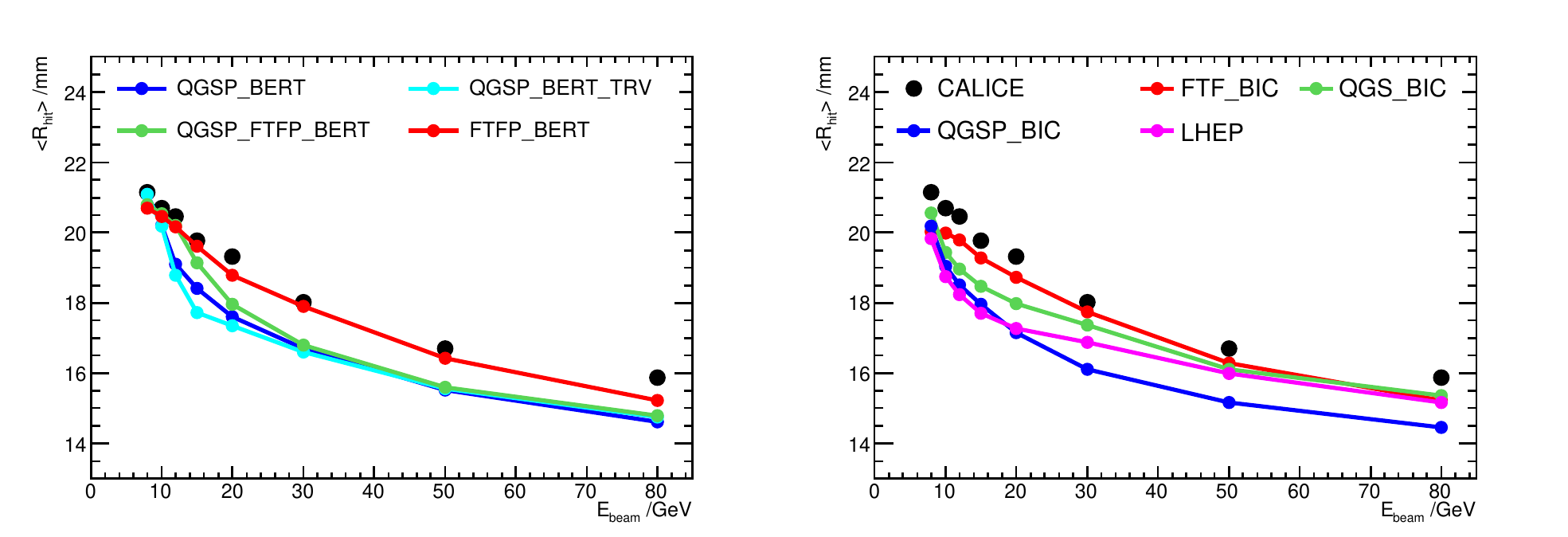}
\caption{\label{fig:EcalResp}\em Mean radius for pion showers in the ECAL as a function of energy, compared with various physics lists in GEANT4.
}
\end{figure}

\begin{figure}[btp]
\centering \includegraphics[width=0.8\textwidth]{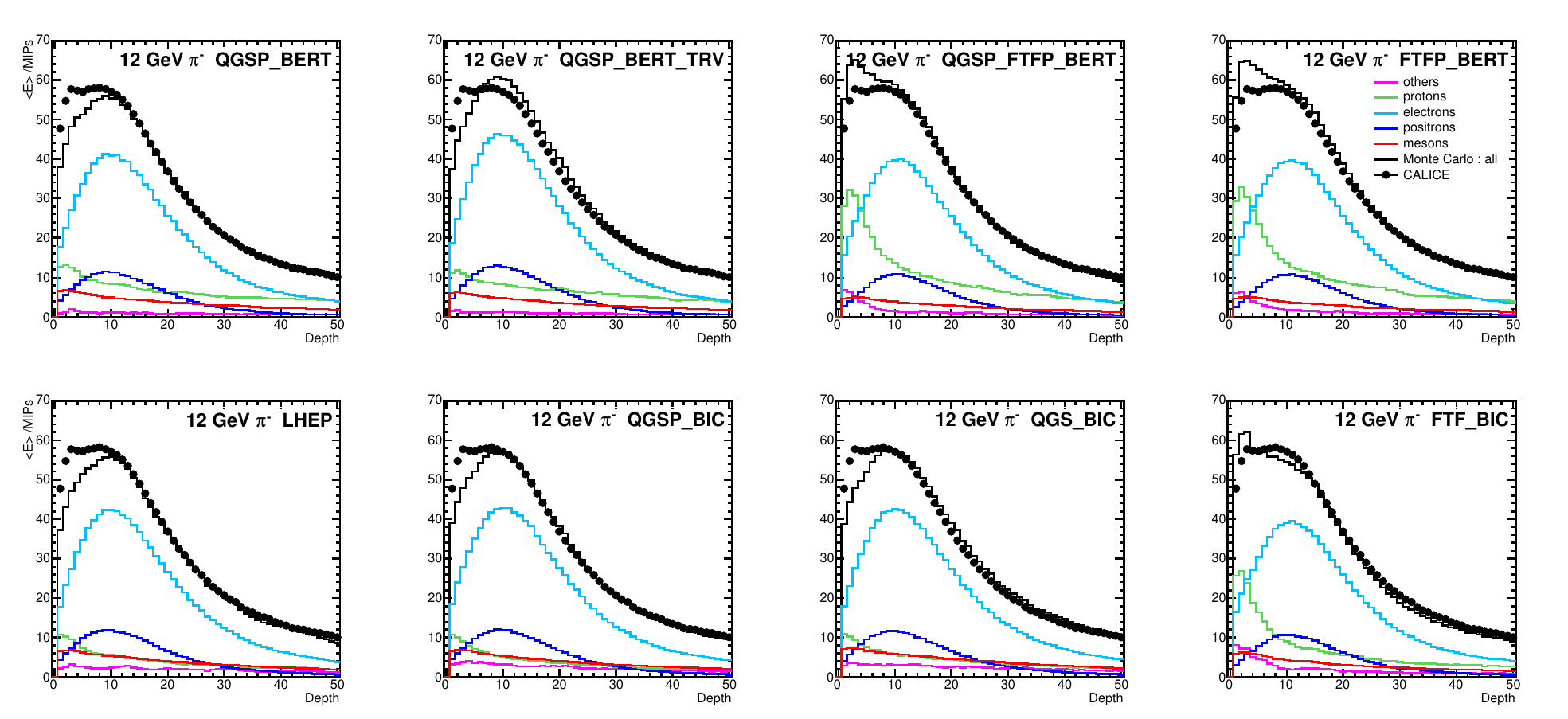}
\caption{\label{fig:EcalResp2}\em Longitudinal pion shower profiles in the SiW ECAL, 
measured with respect to the interaction point.  Data are compared with simulations, for which 
the breakdown into different particle species is illustrated.
}
\end{figure}

In a linear collider detector, some 50\% of hadronic showers will start to develop in the
ECAL, so it is of interest to study pion-induced showers in this detector.  
This allows us to probe the properties of showers in tungsten,  
and to exploit the high granularity of the ECAL to explore the 
properties of the first phase of the showers in unprecedented detail~\cite{ECALpion}. 
In Fig.~\ref{fig:EcalResp} we show the energy-weighted mean shower radius 
in the ECAL as a function of pion energy, compared with the predictions of various 
physics lists in GEANT4.  Most models underestimate the shower with, with the 
FTF models, based on the GEANT implementation of Fritjof, currently proving 
the most successful in version 4.9.3 of GEANT.
The high segmentation of the ECAL allows the start of the shower to be identified 
with precision, and the small ratio of $X_0/\lambda_{\rm int.}$ permits useful 
separation between three main components of the primary interaction --- 
the nuclear fragments produced by spallation, the electromagnetic component 
and the MIP-like relativistic hadrons.  This is illustrated in Fig.~\ref{fig:EcalResp2}, 
taken from~\cite{ECALpion},
which compares data with different physics lists.  The contribution from 
short range spallation protons is clearly visible as the shoulder in the first few layers of the detector, a feature which none of the models manages to 
reproduce very well.

\subsection{AHCAL performance}
\begin{figure}[btp]
\centering \includegraphics[width=0.68\textwidth]{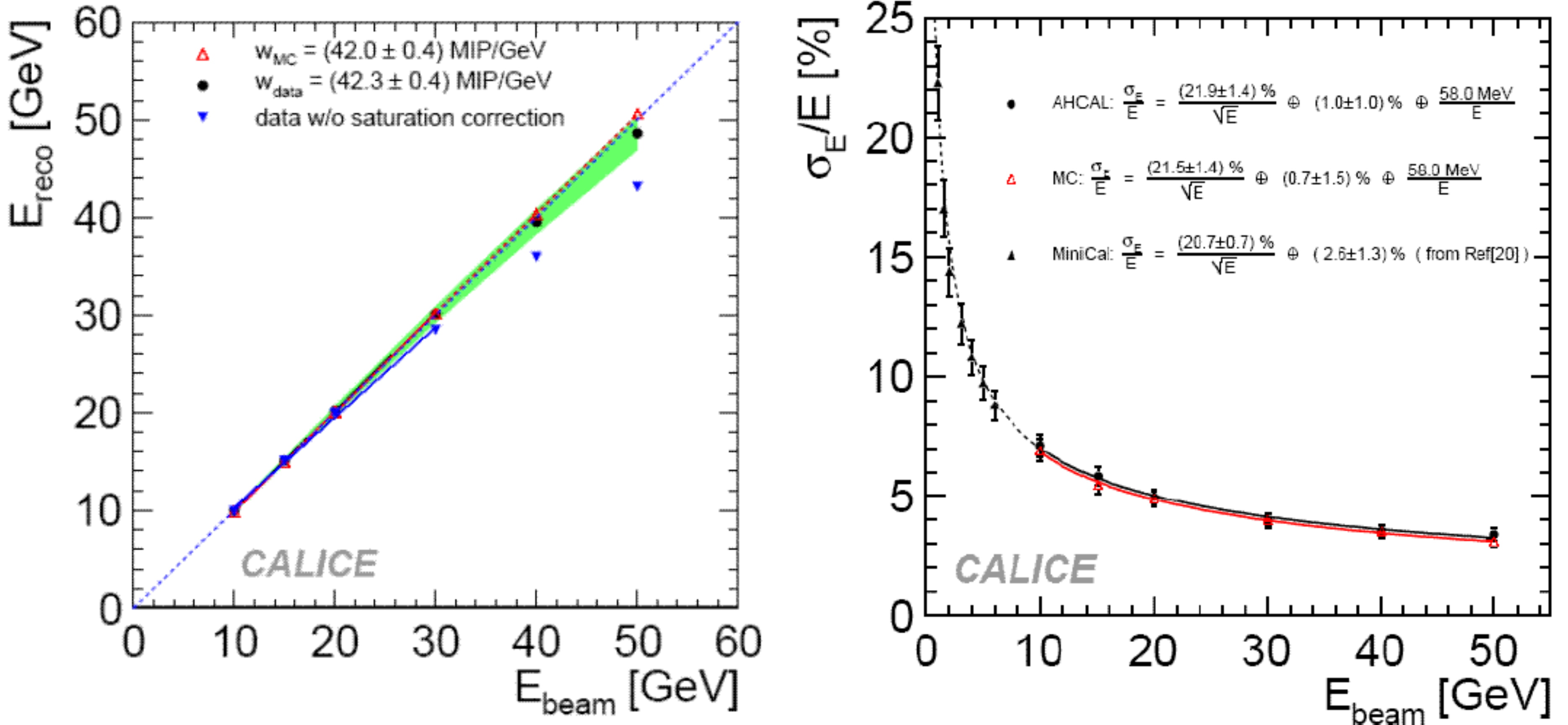}
\caption{\label{fig:AHcalPositron}\em (Left) Reconstructed energy for positrons in the AHCAL, showing that the non-linearity caused by SiPM saturation is largely recovered by the reconstruction (Right) energy resolution in data, compared with simulation.
}
\end{figure}

The completely instrumented AHCAL (described in section~\ref{sec:AHCAL}) 
was exposed to muon, electron and hadron 
beams in 2007-9, 
both with and without an ECAL in front.  Muons were used for calibration.  
An important 
test of our understanding of the calorimeter is to check the response to 
positrons with no ECAL in front of the AHCAL~\cite{AHCALpositron}.  
The energy density in electromagnetic showers is 
particularly 
large, so this is a good test of the important SiPM saturation 
corrections, and other effects.  
Typical results are shown in Fig.~\ref{fig:AHcalPositron}.  After saturation corrections are applied, satisfactory agreement between data and simulation is achieved, giving us confidence to proceed with the characterisation of hadronic showers.

\begin{figure}[btp]
\centering 
\includegraphics[width=0.32\textwidth]{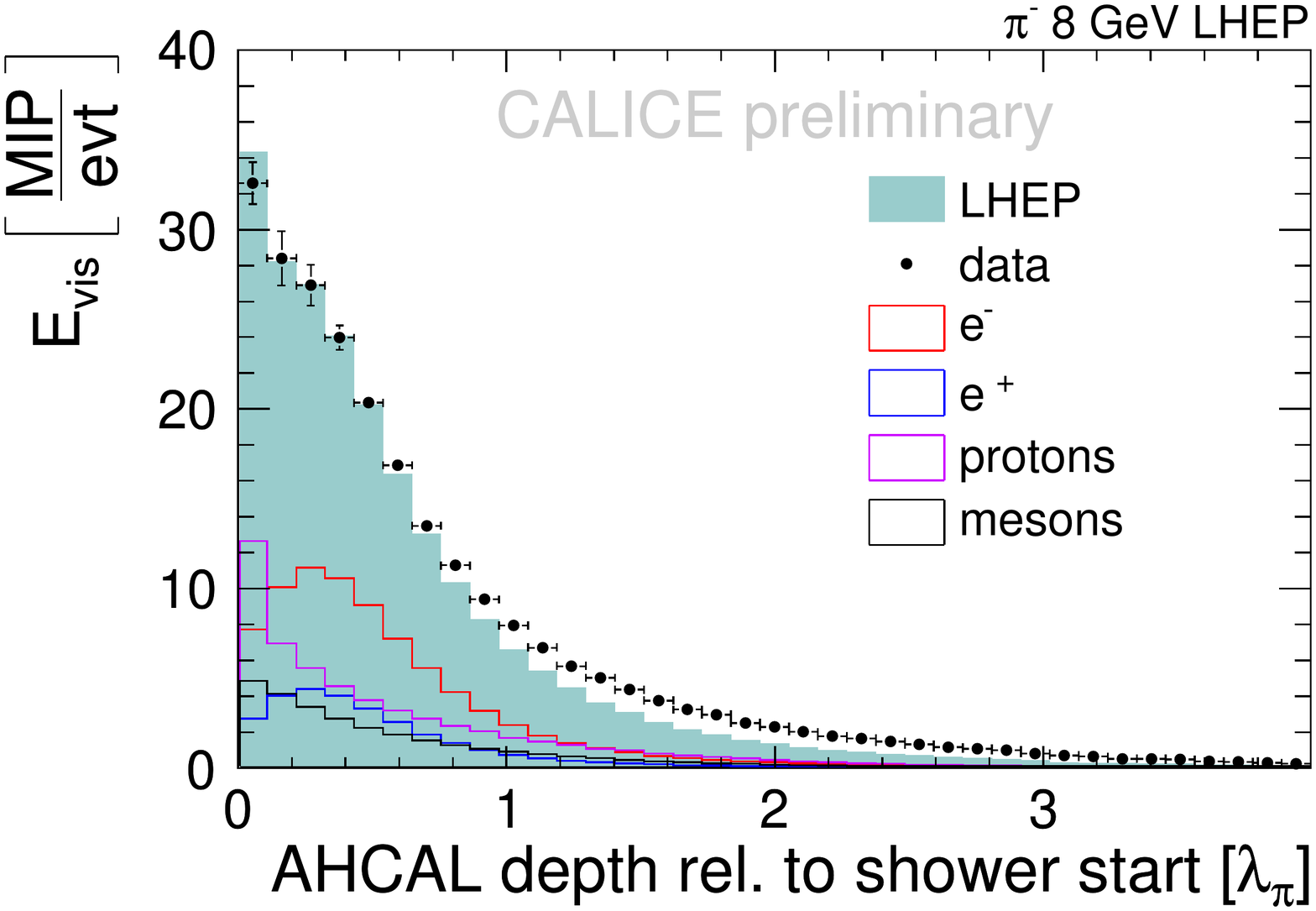}
\includegraphics[width=0.32\textwidth]{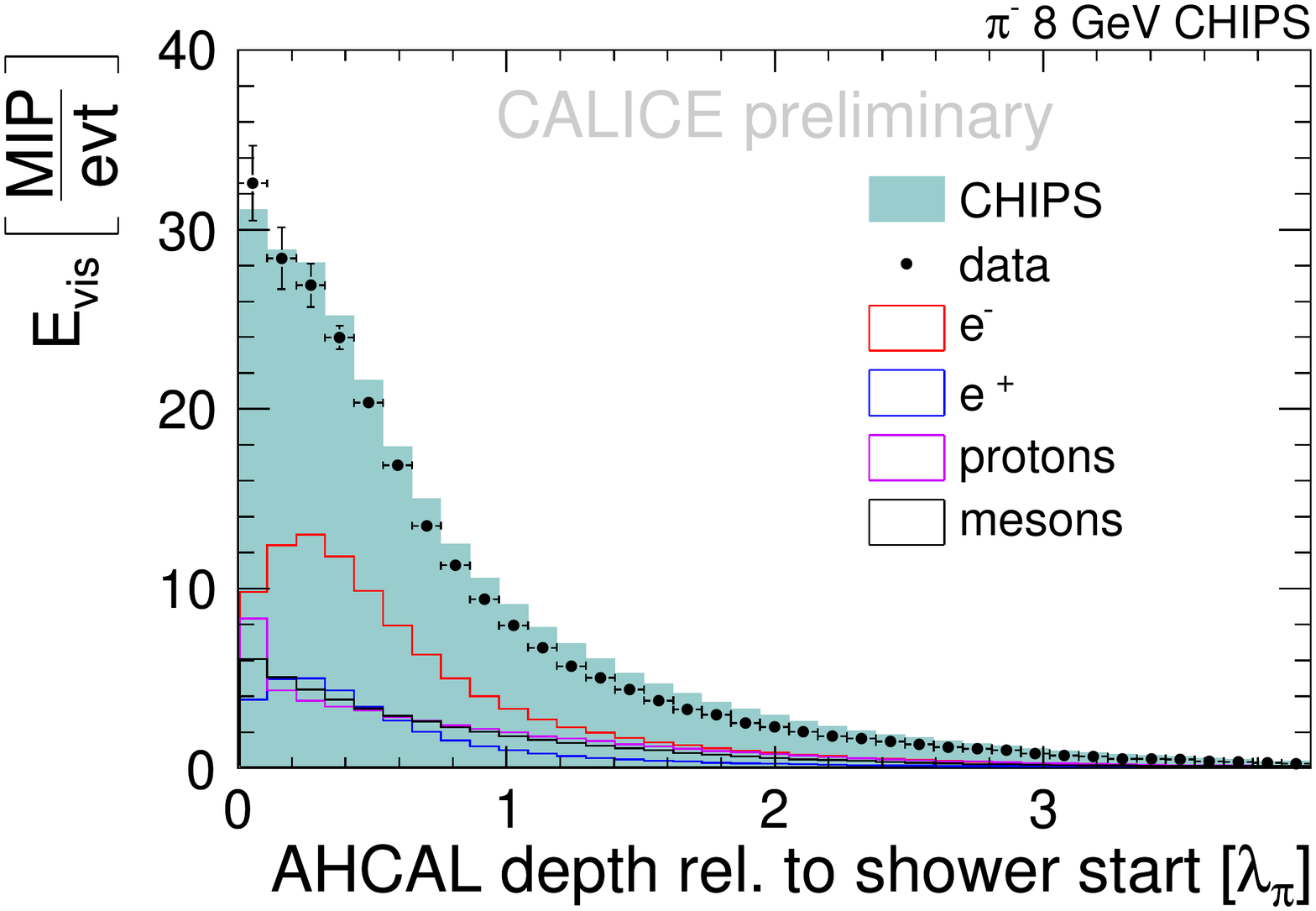}\\
\includegraphics[width=0.32\textwidth]{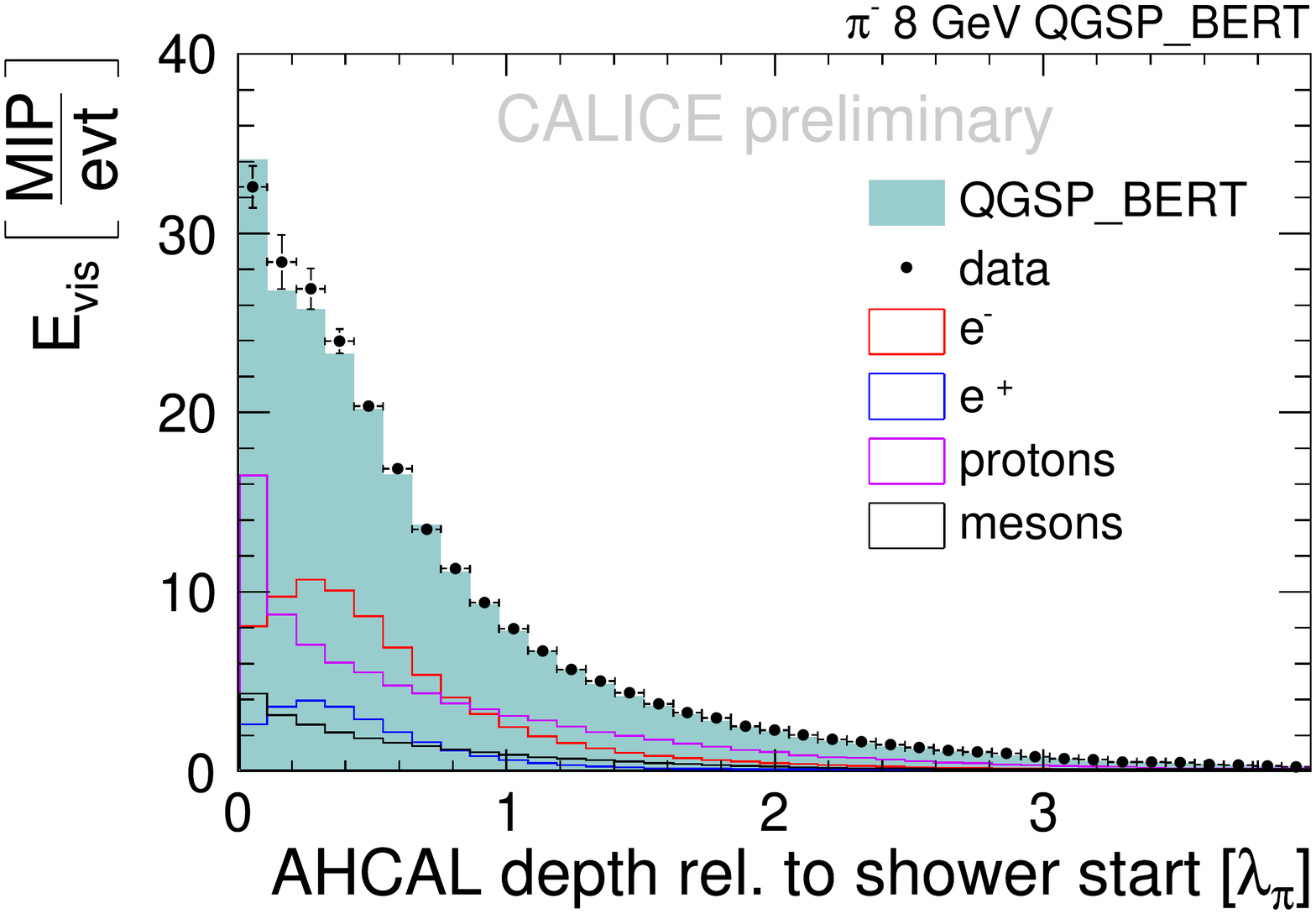}
\includegraphics[width=0.32\textwidth]{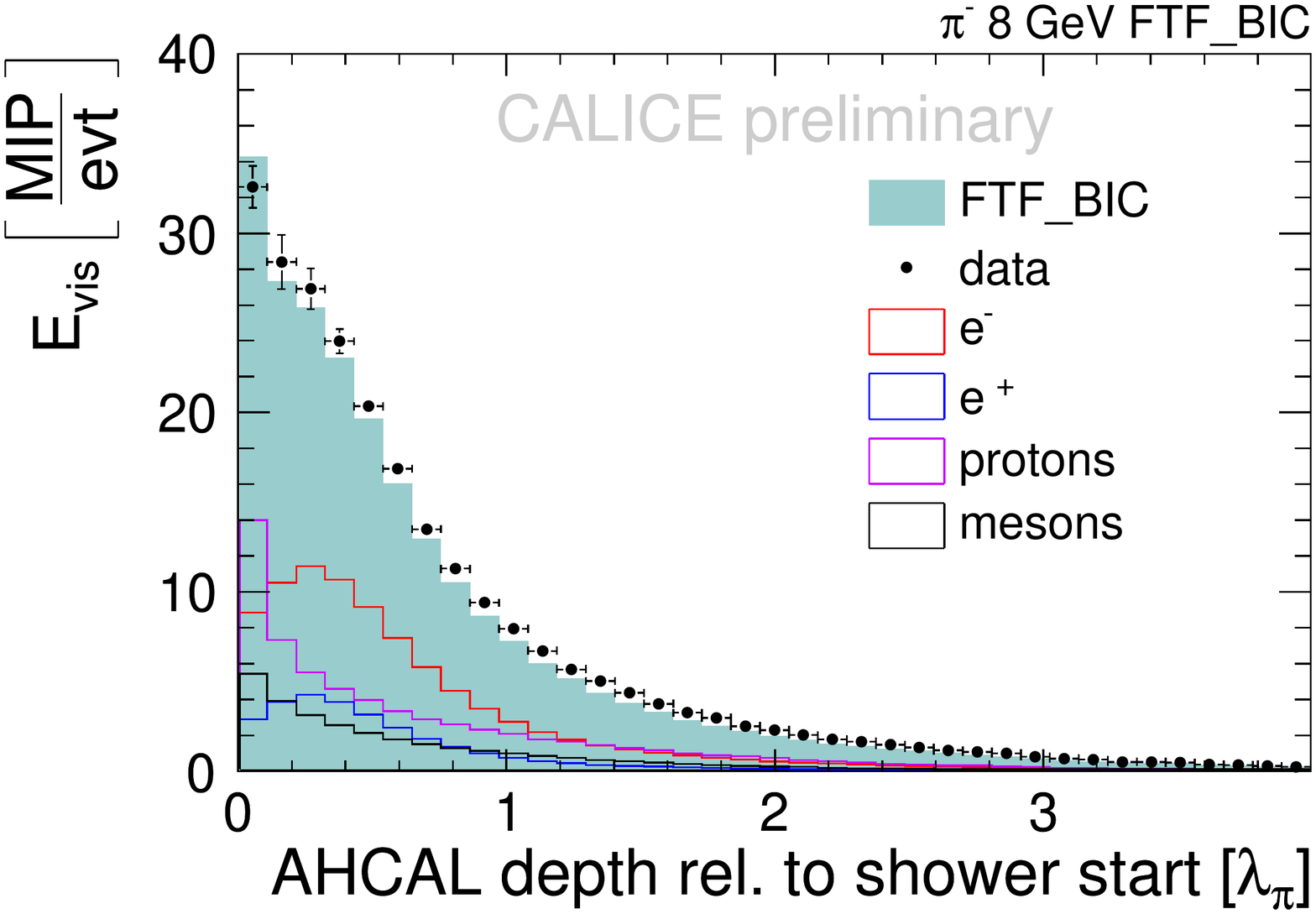}\\
\includegraphics[width=0.32\textwidth]{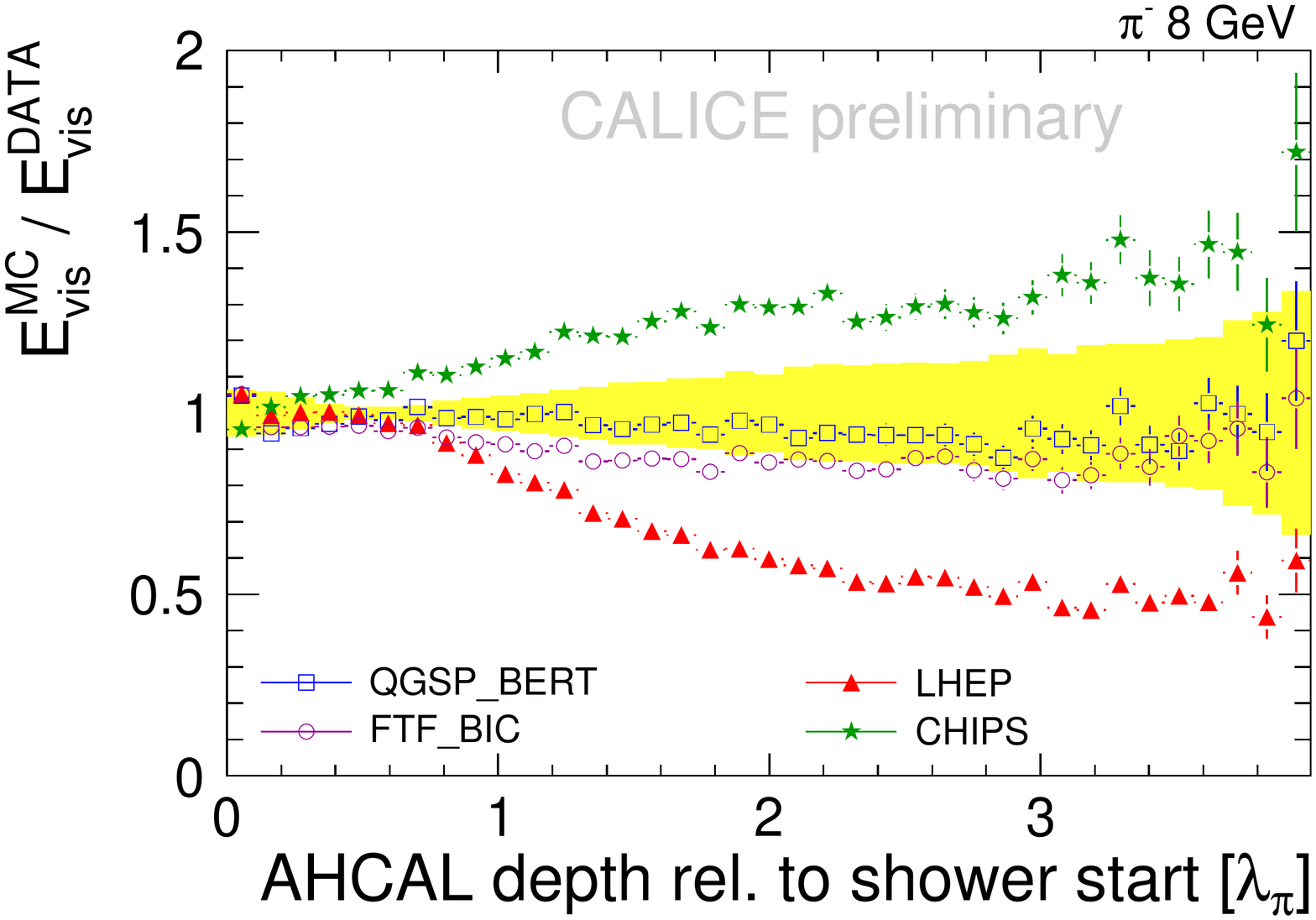}
\includegraphics[width=0.32\textwidth]{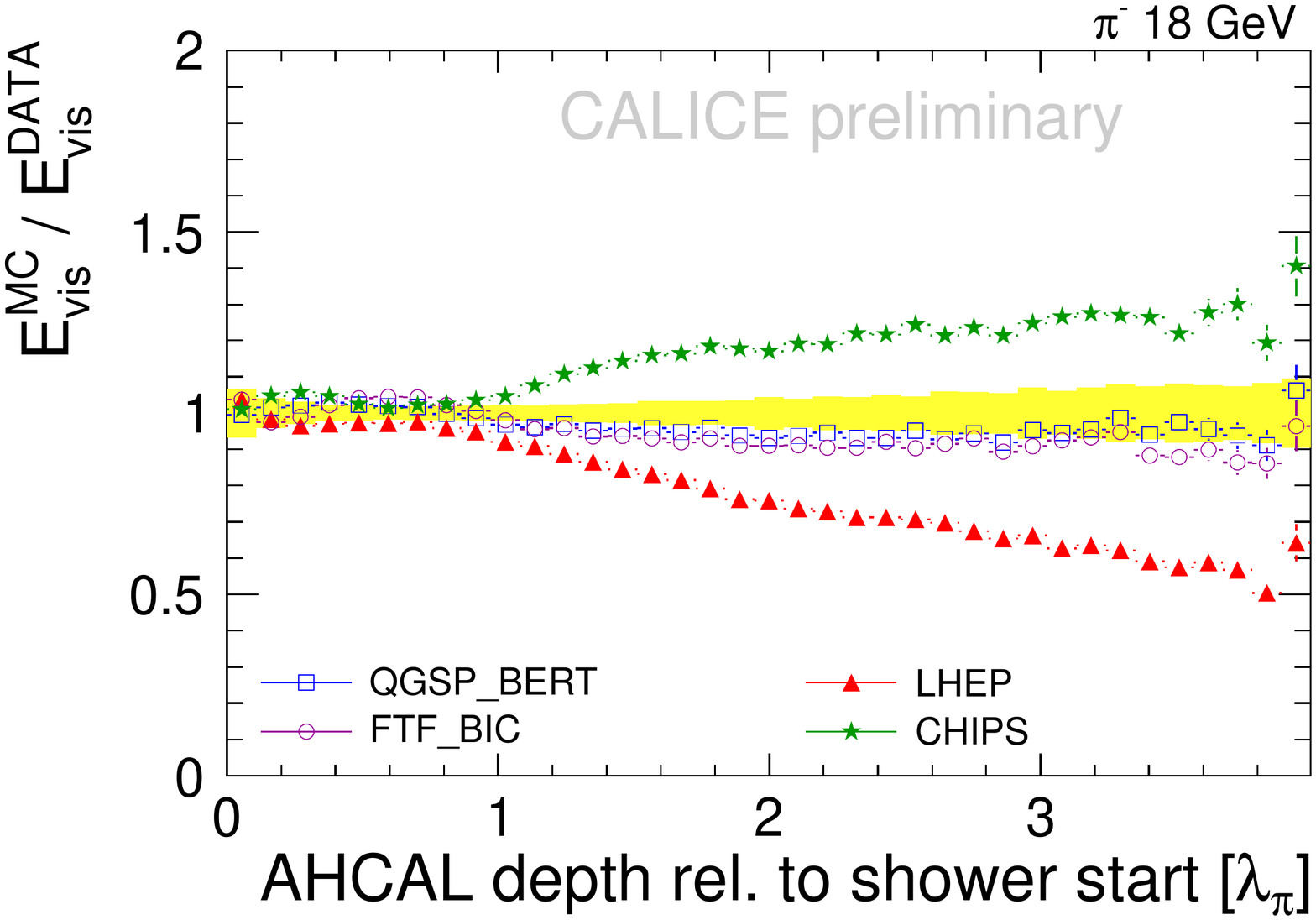}
\includegraphics[width=0.32\textwidth]{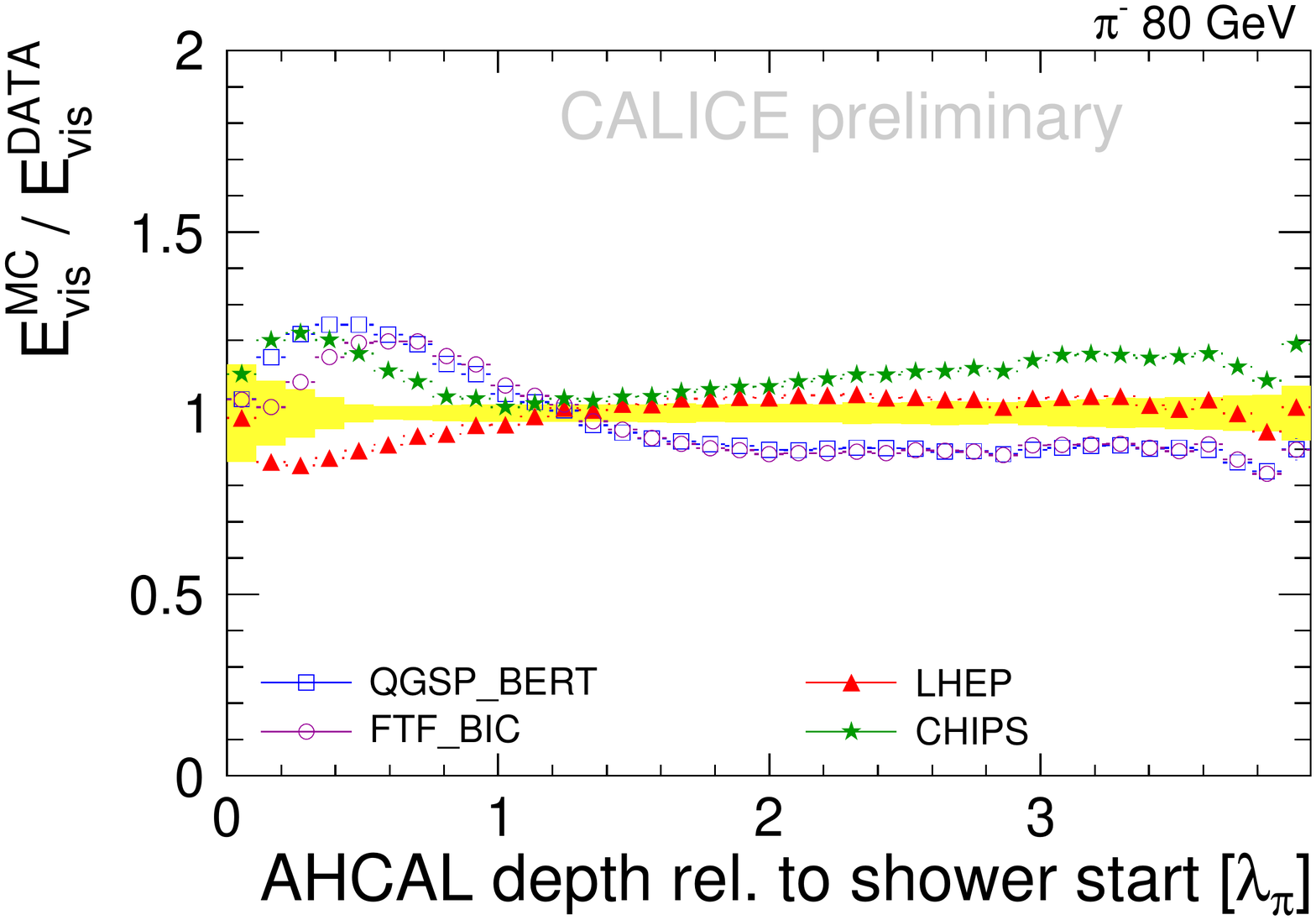}
\caption{\label{fig:AHcalProfiles}\em The upper four plots compare the longitudinal profile of pion showers in the AHCAL at 8 GeV with four GEANT4 physics lists.  Ratios of Monte Carlo to data  for three typical energies are presented in the lower plots, at 8~GeV (left), 
18~GeV and 80~GeV (right).
}
\end{figure}

Longitudinal profiles in the AHCAL have also been measured with respect to the shower starting point~\cite{CAN-026}, and can be compared with Monte Carlo simulations broken down into their separate particle constituents, in a way similar to the ECAL analysis.  In Fig.~\ref{fig:AHcalProfiles} we compare simulations to data at 8 GeV, and show ratios of Monte Carlo to data at three typical energies.  In this case, the physics lists \verb|FTF_BIC| and particularly \verb|QGSP_BERT| seem to be the most successful at modelling data.

\begin{figure}[btp]
\centering \includegraphics[width=0.6\textwidth]{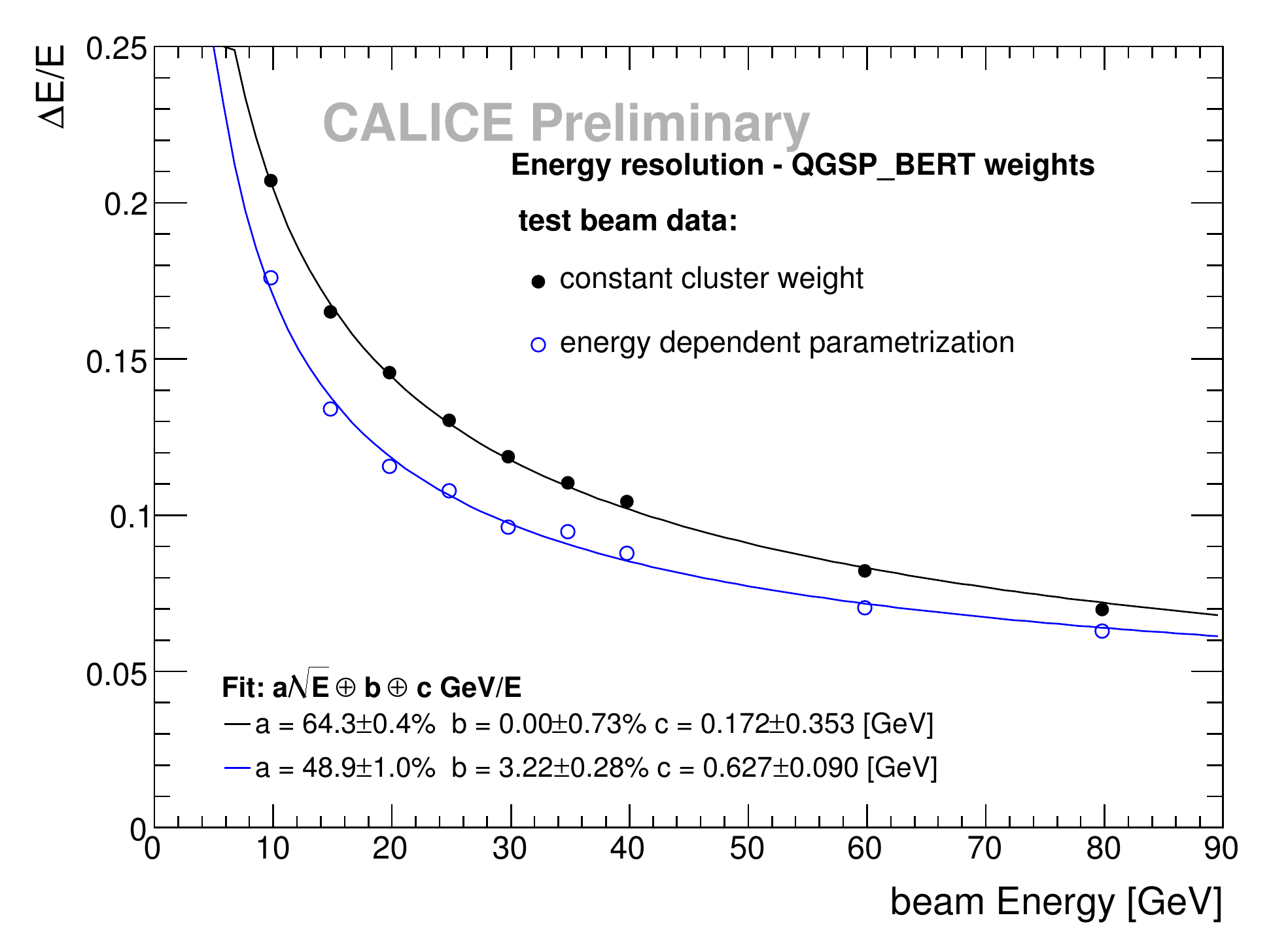}
\includegraphics[width=0.36\textwidth]{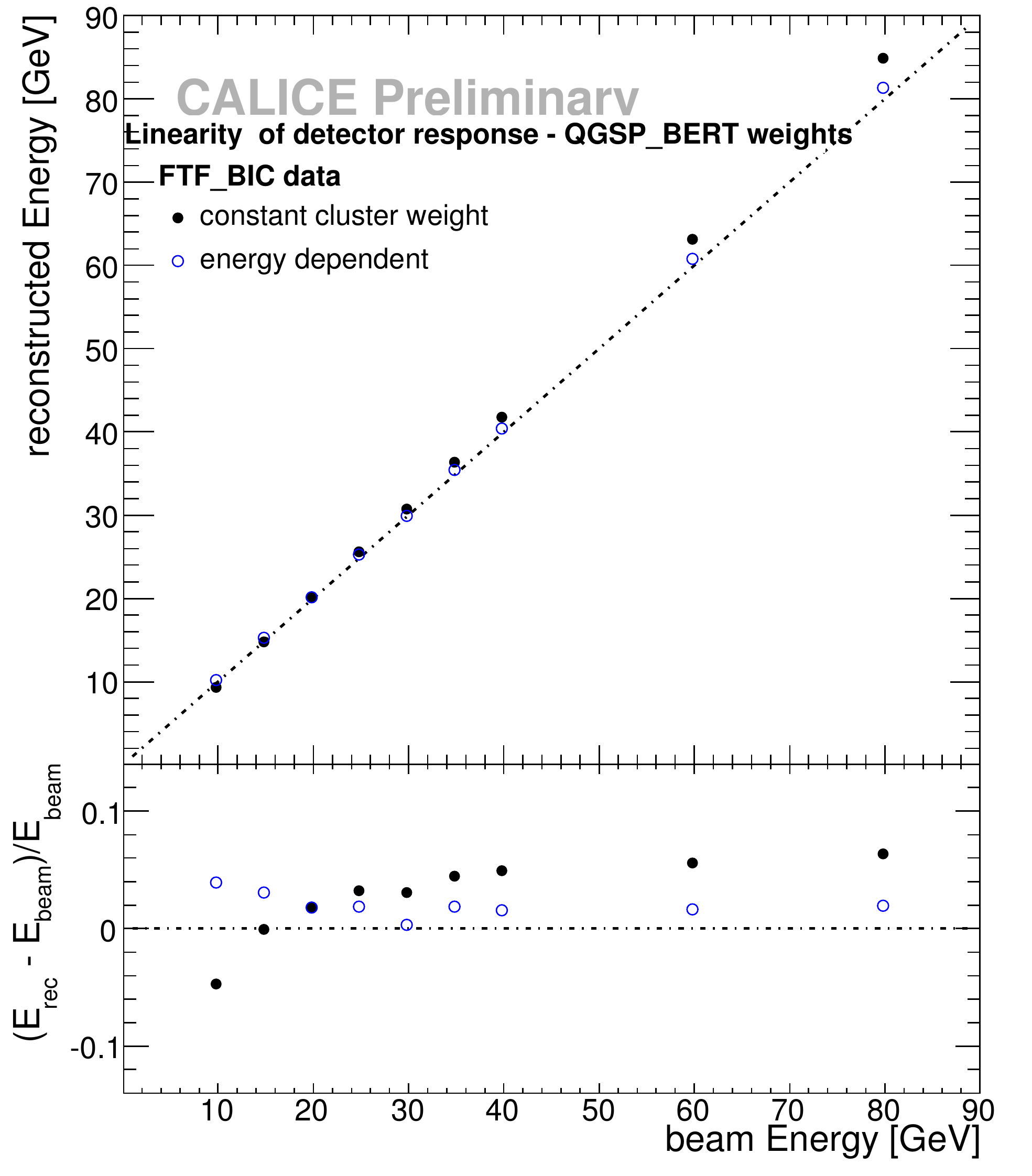}
\caption{\label{fig:AHcalSoftComp}\em (Left) Fractional energy resolution for pions with software compensation (blue open symbols) and without (black closed points).  (Right) Reconstructed vs.\ true energy,  with fractional deviations from linearity shown in the inset.  
}
\end{figure}

The CALICE calorimeters are non-compensating, i.e.\ their $\pi/e$ response ratio is not unity, and hence the hadronic energy resolution is affected by the fluctuations in the electromagnetic energy fraction in showers.  However, the high granularity of the calorimeters permits significant pattern recognition within showers, allowing some possibility of disentangling the components and weighting them differently (a procedure known as {\em software compensation}.  Several studies along these lines have been carried out exploring different techniques~\cite{CAN-015,CAN-021,CAN-028}.  The basic idea in all these approaches is that the electromagnetic parts of showers have greater energy density, so that hits of high energy, or in regions of high energy density, are weighted differently.
An example of the results is shown in Fig.~\ref{fig:AHcalSoftComp}, taken from~\cite{CAN-021}, 
and typical of all the studies.  It is found that the energy resolution can be improved by 
typically $\sim$15-20\%, achieving a stochastic term in the range 45-50\%/$\sqrt{E/{\rm GeV}}$.  In addition, the linearity of the response is improved.

\begin{figure}[btp]
\centering \includegraphics[width=0.4\textwidth]{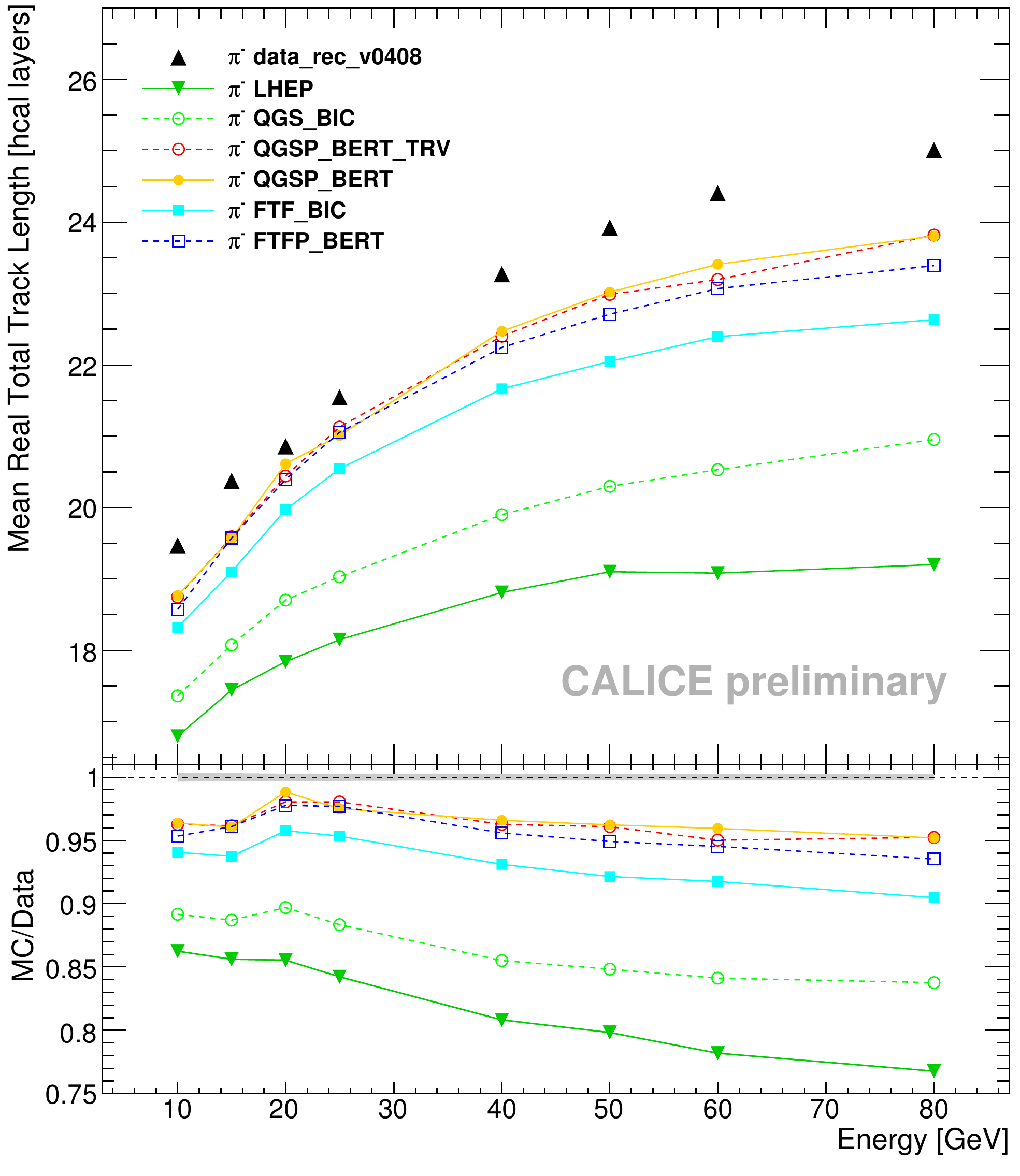}
\includegraphics[width=0.4\textwidth]{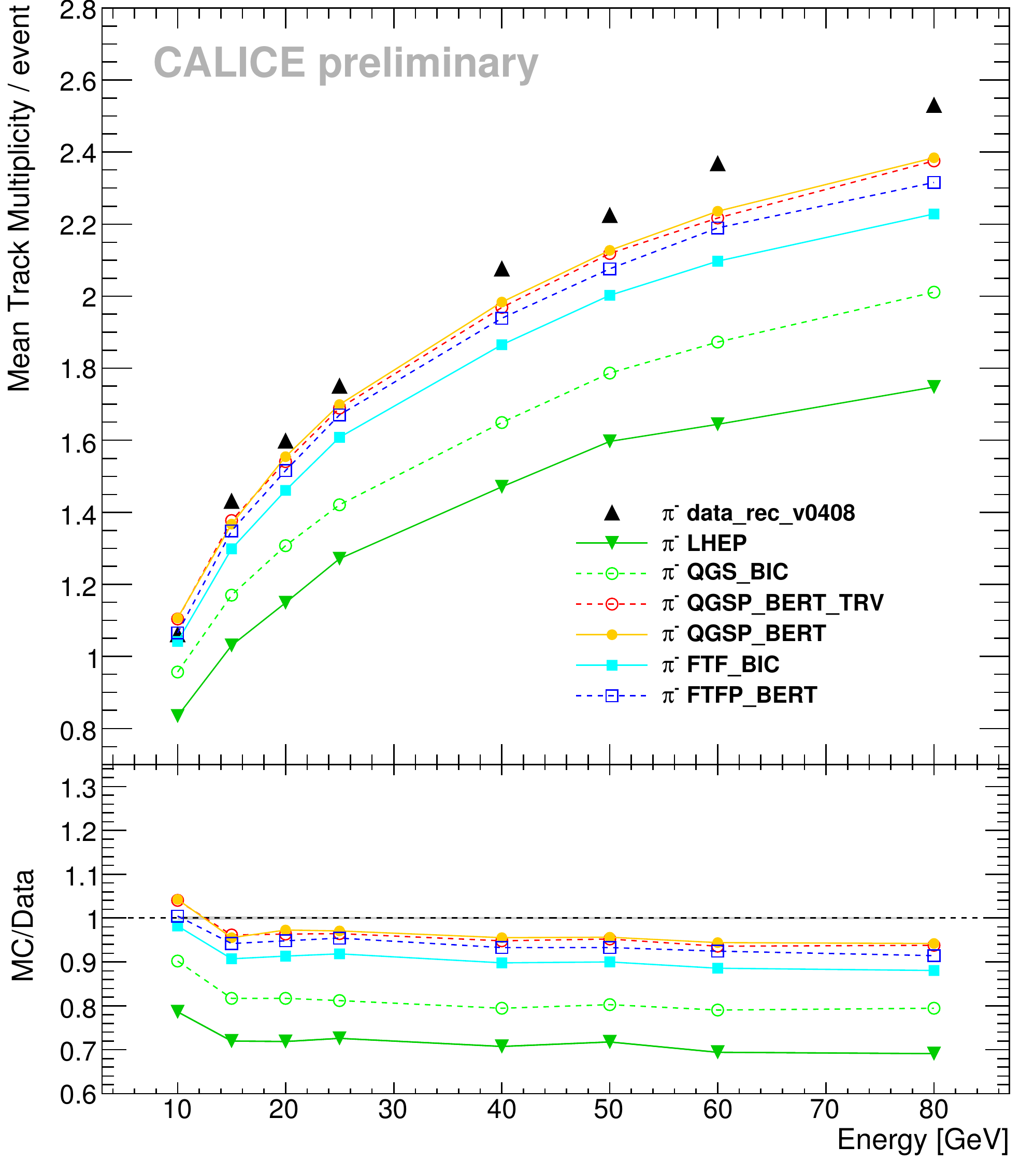}
\caption{\label{fig:AHcalSegments}\em (Left) Total track length in the AHCAL and (Right) number of track segments as a function of pion energy.  
}
\end{figure}

The high segmentation of the calorimeters also permits the identification of 
MIP-like track segments within hadronic showers.  This has been studied in detail using the 
AHCAL in~\cite{CAN-022}.  As well as providing a useful way of monitoring the calibration of the tiles, the study of these track segments permits novel tests of shower simulation models.  For example, in Fig.~\ref{fig:AHcalSegments} we show the total track length and the number of separate segments in pion-induced showers, compared with GEANT4.  We note that none of the physics lists is notably successful in modelling this aspect of the data.

\subsection{Combined calorimeter system}

The CALICE beam tests included data taken with a full system of Si-W ECAL, AHCAL and tail catcher (TCMT), in order to emulate the behaviour of a full linear collider calorimeter system.
In one of the studies of software compensation cited above~\cite{CAN-015}, the same approach were applied to showers the full system of ECAL, HCAL and TCMT, and the improvement in performance was just as good as for the HCAL alone.

\begin{figure}[btp]
\centering \includegraphics[width=0.58\textwidth]{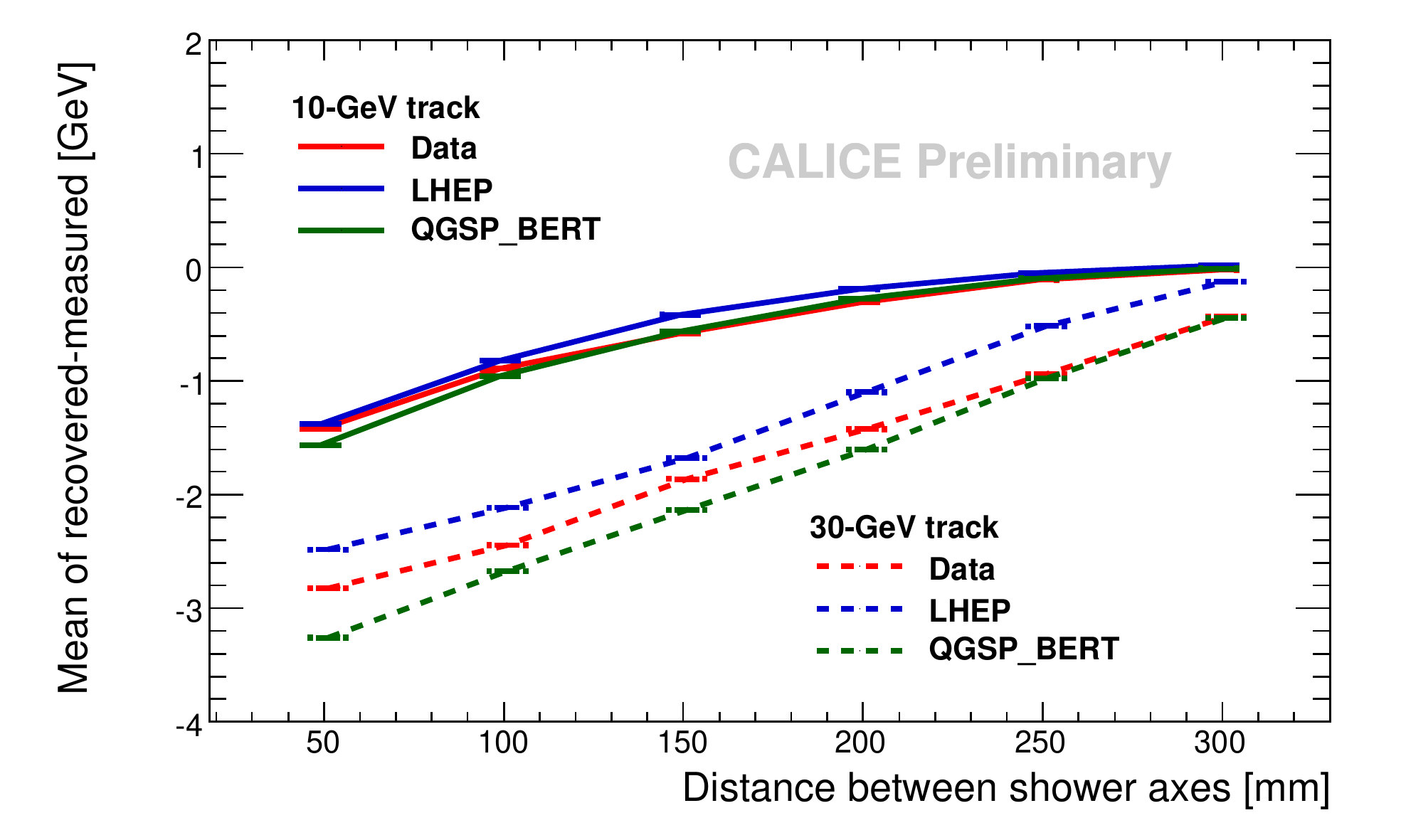}
\includegraphics[width=0.38\textwidth]{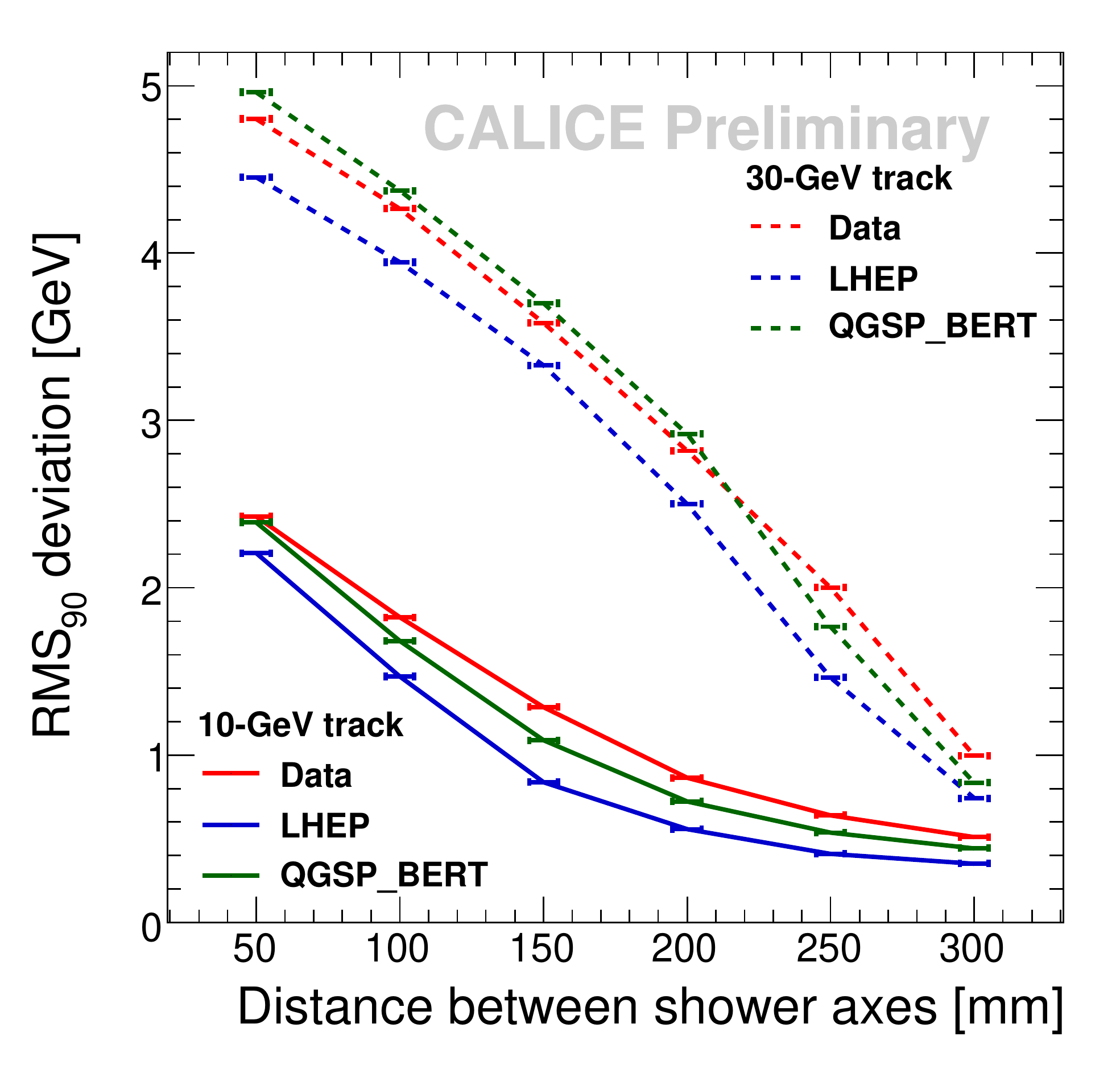}
\caption{\label{fig:AHcalPandoraTest}\em (Left) Mean difference between recovered and measured energy for a 10~GeV emulated neutral hadron in the proximity of two charged particle energies. (Right) RMS$_{90}$ deviation of the difference, providing an estimate of the ``confusion`` component of the neutral energy resolution. Note that the energies here are calibrated 
the electromagnetic energy scale, which $\sim 20\%$ underestimates the hadron energy by 
$\sim 20\%$. 
}
\end{figure}

One of the main motivating forces behind the CALICE program has been to optimise the calorimetry for particle flow reconstruction of jets.  It is therefore of obvious interest to use the test beam data in ways which directly test the features used in existing particle flow algorithms, such as Pandora-PFA~\cite{PFA}, which is currently regarded as the state-of-the-art program.  In a PFA, the calorimeter is used to measure the energy of neutral particles, and the HCAL particularly for neutral hadrons.  An important issue is the degree of ``confusion'', i.e.\ the extent to which the energy measurement of a neutral particle may be degraded by the presence of nearby charged particles. 
A first such test has been presented in~\cite{CAN-024}, in which a configuration consisting of a charged pion close to a neutral hadron shower is emulated.  This is done by 
superimposing two test beam pion-induced events, removing the incoming MIP-like track segment from one in order to emulate the neutral hadron.  These events can be overlaid using various transverse displacements, and are then presented to Pandora-PFA for reconstruction and the fidelity of the recovery of the energy of the neutral is examined.  A comparison with the measured energy before overlaying is the used to assess the degree of confusion.  In Fig.~\ref{fig:AHcalPandoraTest} we show the average difference between measured and recovered energy and its r.m.s.\ as a function of the shower separation, for two typical energies.  The data are compared with GEANT4 simulations using different physics lists.  We see that the simulation, especially using the \verb|QGSP_BERT| model, gives a reasonably faithful description of data, suggesting that the performance found for PandoraPFA in~\cite{PFA} is quite reliable.

\subsection{Sc-W ECAL performance}

\begin{figure}
\centering
    \includegraphics[angle=0,width=0.4\textwidth]{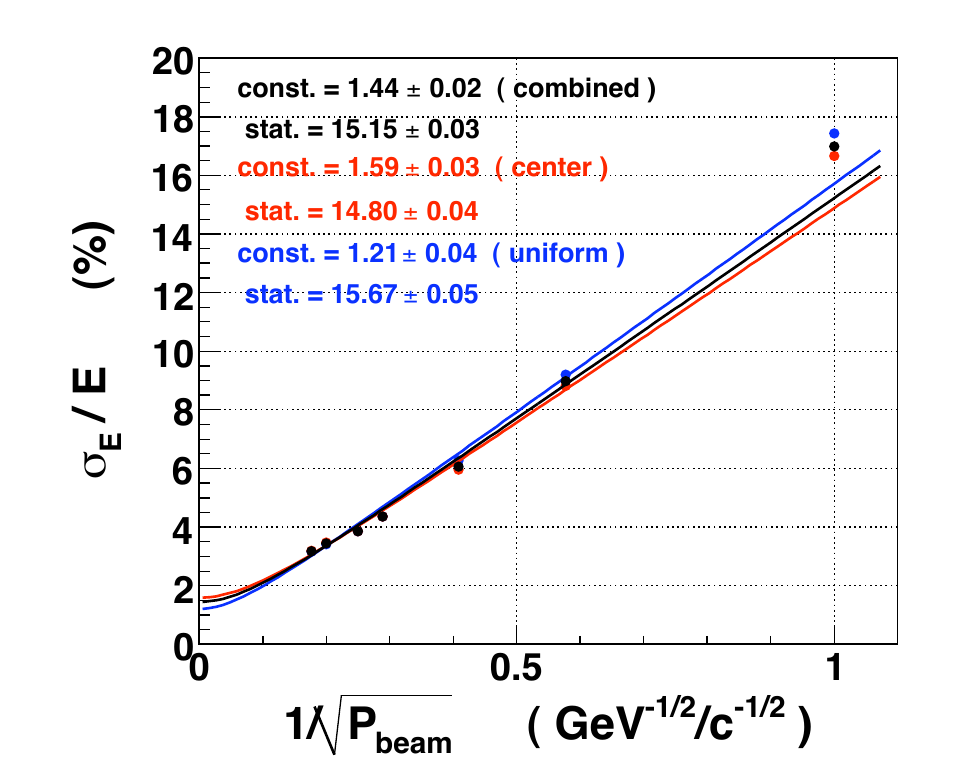}
        \includegraphics[angle=0,width=0.4\textwidth]{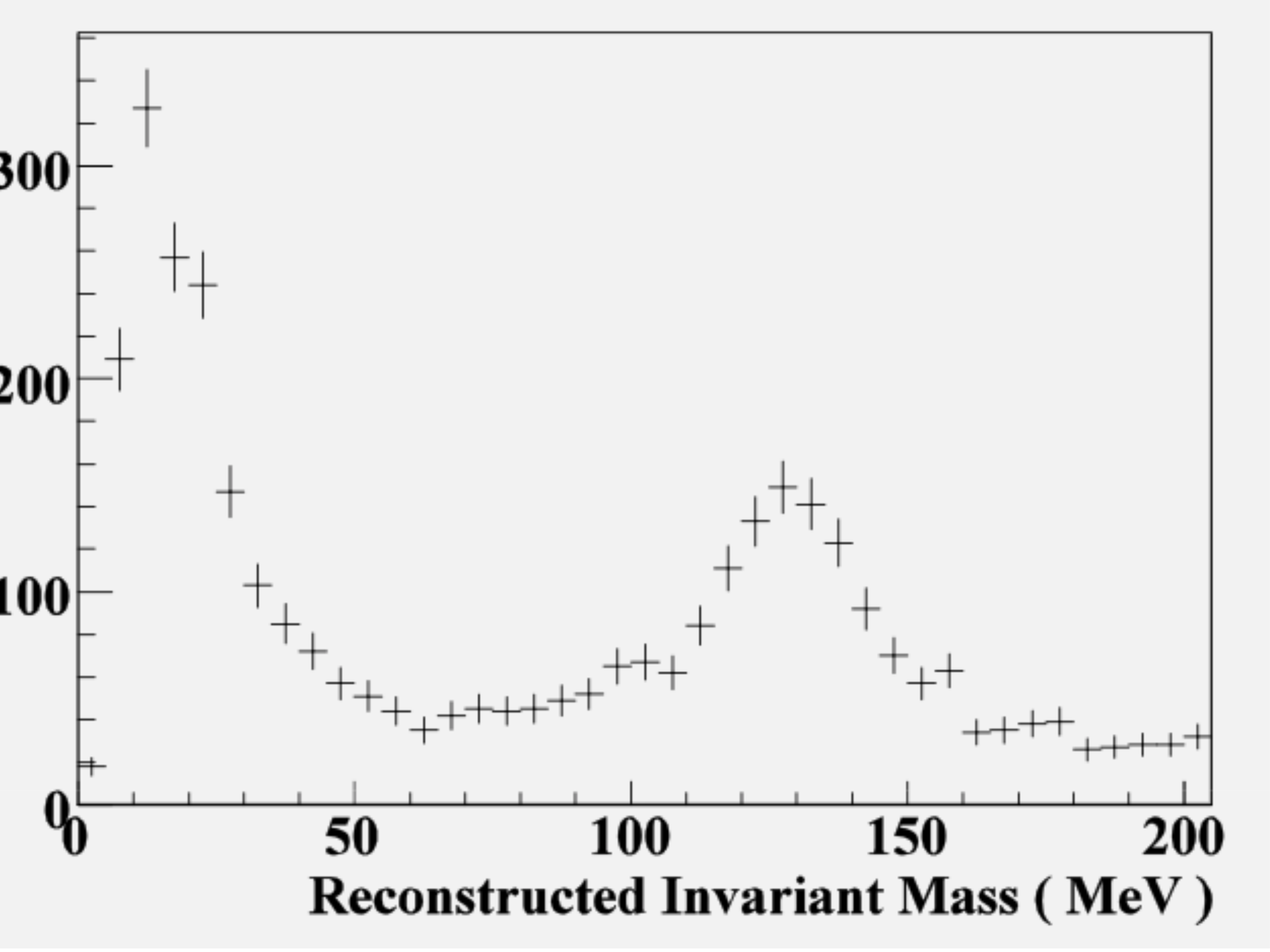}
\caption{\label{fig:scintecalresol}\em  Scintillator ECAL test beam results: 
(left) energy resolution vs.\ beam momentum, and (right) reconstructed
  di-photon invariant mass ($\pi^0$ signal). 
}
\end{figure}

A good energy resolution performance for electrons as well as a reasonable
linearity was presented in the previous report, an example being shown in
Fig.~\ref{fig:scintecalresol}. 
As part of these tests, we introduced a
target into the charged pion beam line at FNAL to produce neutral pions,
and reconstructed the invariant mass distribution of the detected
photon pairs, as shown also in Fig.~\ref{fig:scintecalresol}.

\subsection{DHCAL performance}
\begin{figure}[h]
\begin{center}
\includegraphics[width=0.8\textwidth]{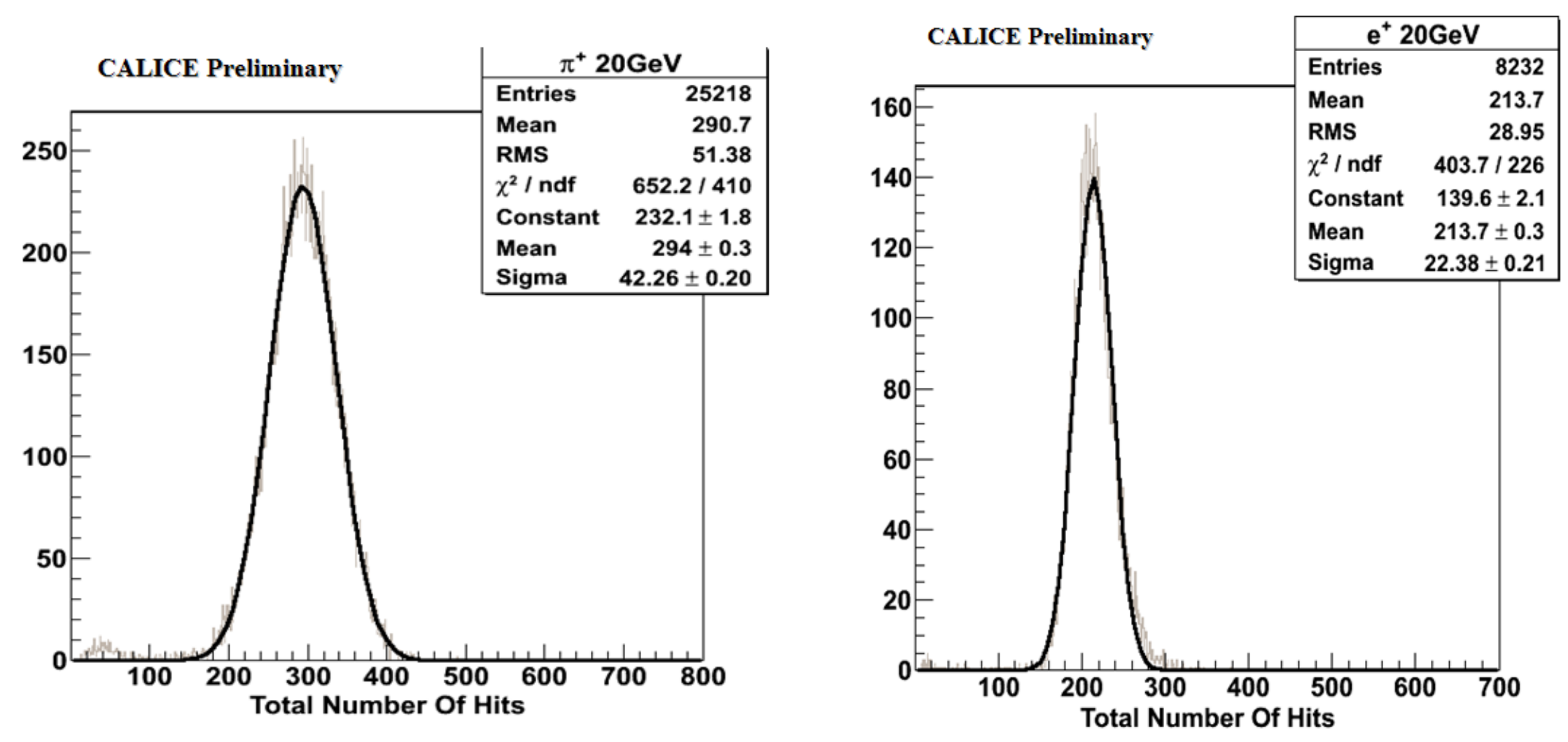}%
\caption{{\sl Response (number of hits) to 20 GeV/c pions (left)
and positrons (right)}}
\label{fig:dhcal_fig5}
\end{center}
\end{figure}

\begin{figure}[btp]
\centering 
\vspace{-1cm}
\includegraphics[width=0.78\textwidth]{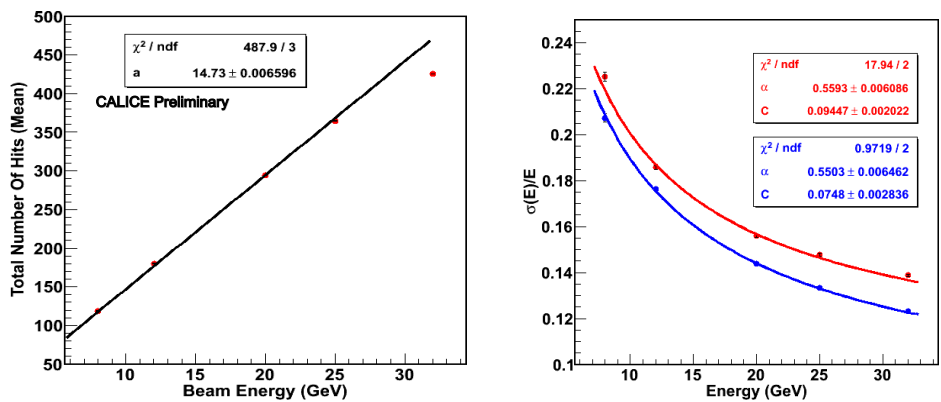}
\vspace{-2cm}
\caption{\label{fig:DHCALresult}\em (Left) DHCAL response to pions as a function of beam energy.  A linear fit is shown.
(Right) fractional energy resolution for pions.  The red curve includes all showers and 
the blue curve just those which are fully contained in the DHCAL.
}
\end{figure}

The 1~m$^3$ DHCAL prototype had its first test beam run at Fermilab in late 2010, exposed 
to beams of positrons, muons and pions.  First results were recently presented.  Clean muon tracks have been seen, and used to assess the efficiency of the RPCs and the pad multiplicity associated with single MIP hits.  Extensive studies of the noise performance have also been shown.  A start has also been made on the analysis of positron and pion shower data.  
Data were collected at various
momenta between 2 and 60 GeV/c. The content of positrons in the beam
varies from dominating at 2 to negligible at 60 GeV/c. 
First results on the response and resolution of the detector are then obtained by using the number of hits as a measure of energy.  A preliminary
analysis of the data showed the expected response for both pions and
positrons. As an example, Fig.~\ref{fig:dhcal_fig5} shows the response (number of hits)
for 20 GeV/c pions and positrons. 
Fig.~\ref{fig:DHCALresult},
taken from~\cite{CAN-032}, shows the response of the calorimeter
(as measured by the number of hits) to pions and the fractional resolution 
(estimated using the r.m.s.).  The response is seen to be well proportional to energy up to 
$\sim$24~GeV, with signs of non-linearity appearing at 32~GeV.  The resolution based on this relatively simple analysis behaves in the expected way with energy.  Clearly this calorimeter provides a great detailed information about the internal structure of the shower, which will permit improvements using software compensation techniques in a similar way to the analogue calorimeters.


\subsection{WHCAL and T3B results}
For a detector at CLIC energies, a tungsten-based HCAL is attractive because of 
its density.  The thickness of the HCAL needs to be $\sim 7.5\lambda_{\rm int.}$ thick, and the cost of immersing such a calorimeter in the magnet coil seems prohibitive if it was made of iron.  However, there is little experimental understanding of the characteristics of hadronic showers in tungsten, which could be quite different from iron; for example more energy will be expended on nuclear spallation, because of the large atomic weight.  It is therefore imporatant to validate GEANT4 for such a device.
For this reason, the CALICE AHCAL tile detectors have recently been tested 
using tungsten absorber plates in a CERN test beam, as described in Sect.~\ref{sec:WAHCAL}.
Analysis of the results is at an early stage, but present indications are that:
\begin{itemize}
\item The hadronic shower signals from pions and protons are
  comparable to the ones observed with an iron absorber.
\item The electromagnetic resolution seems to be slightly worse than
  with the iron absorber reflecting the lower sampling ratio with
  respect to $X_0$
\item The $e/\pi$ ratio seems to be stable versus energy
\end{itemize}

As described in Sect.~\ref{sec:T3B}, the T3B module was introduced for these tests in order to explore the time structure of the showers.  A measure which provides a good indication of the intrinsic time
stamping possibilities in the calorimeter is the time when a cell
which contains energy in a given event is first hit. In a first
analysis of the T3B data, this was determined from the decomposed
waveforms. Here, signals with an amplitude equivalent of at least
0.4~MIP within 9.6~ns were considered. The time of first hit was
determined from the second photon detected within the selected 9.6~ns
period (the second photon was used to reduce the impact of background
from dark noise). The same quantity was determined in simulations of
T3B with a description of the complete detector setup, including a
parametric smearing of the arrival time of photons on the SiPM to
account for the response of the scintillator and for photon travel
times. Simulations were performed with two different Geant4 physics
lists, \texttt{QGSP\_BERT} and \texttt{QGSP\_BERT\_HP}. The latter
includes high precision neutron tracking.

To provide first robust comparisons between data and simulations, the
mean time of first hit for each of the T3B cells was determined from
the distribution of the first hit. The mean was formed within a time
window of 200~ns, starting 10~ns before the maximum of the
distribution in T3B tile 0, and extending to 190~ns after the
maximum. This time window covers the time relevant for calorimetry at
CLIC, where the duration for one bunch train is expected to be 156~ns,
and is also comparable to the shaping time of 180~ns used in the front
end electronics of the CALICE analog HCAL modules.

\begin{figure}
\centering
\includegraphics[width=0.7\textwidth]{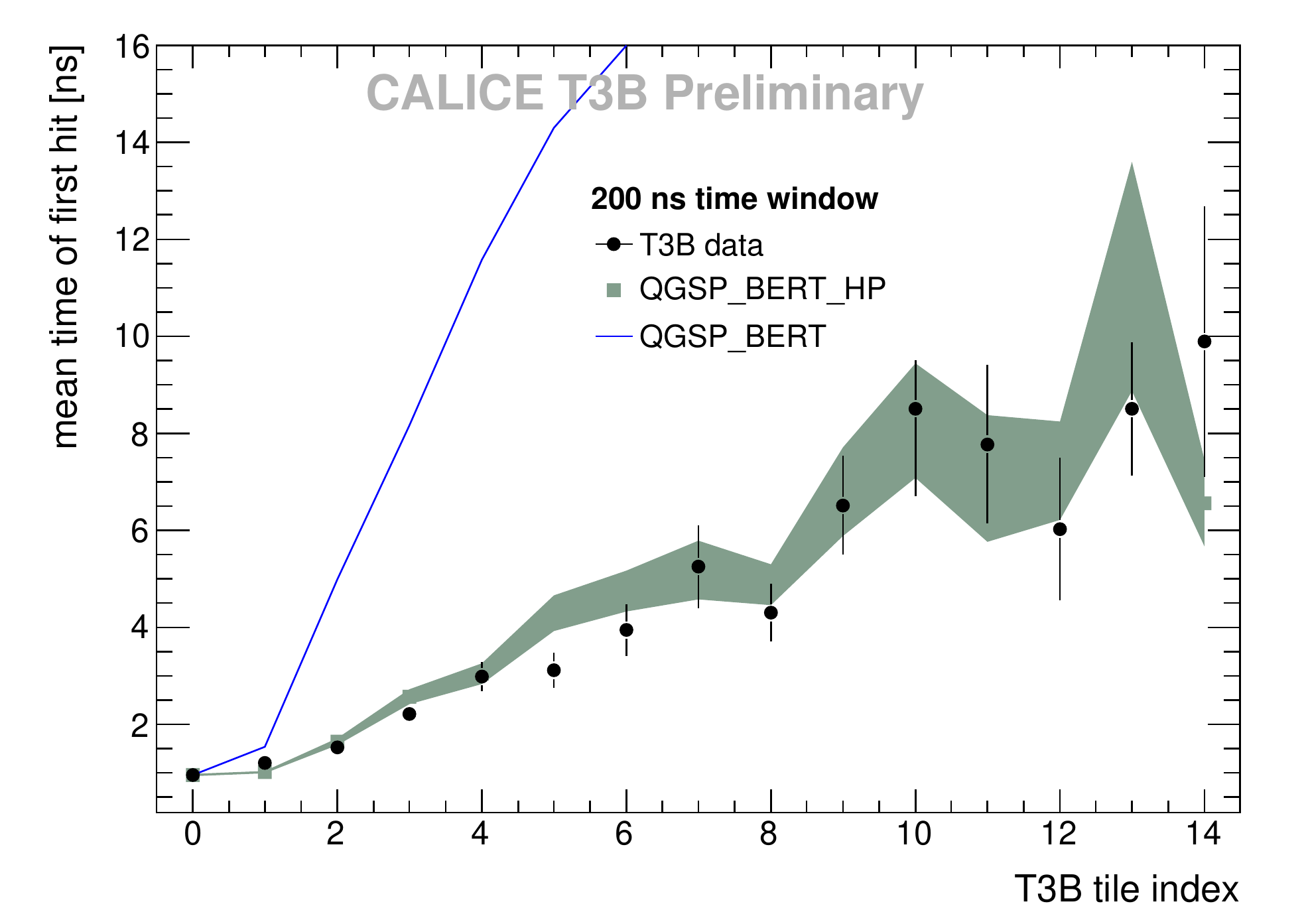}
\caption{Mean time of first hit for 10~GeV $\pi^-$ as a function of
  radial distance from the shower core (a tile index of 10 corresponds
  to approximately 30~cm). The data are compared with simulations
  using \texttt{QGSP\_BERT} and \texttt{QGSP\_BERT\_HP}. The error
  bars and the width of the area in the case of
  \texttt{QGSP\_BERT\_HP} simulations show the statistical error,
  while for \texttt{QGSP\_BERT} the errors are omitted for clarity.}
\label{fig:MeanTimeResult}
\end{figure}

Figure \ref{fig:MeanTimeResult} shows the mean time of first hit as a
function of the radial distance from the shower axis. The beam axis
passes through T3B tile 0, so that a tile index of 10 corresponds to a
distance of approximately 30~cm. The measurement is compared to the
simulations with the two physics lists, \texttt{QGSP\_BERT} and
\texttt{QGSP\_BERT\_HP}.  While \texttt{QGSP\_BERT\_HP} gives an
excellent description of the data, \texttt{QGSP\_BERT} shows very
large discrepancies, with significantly overestimated late
contributions at larger radii. This demonstrates the importance of the
high precision neutron tracking in Geant4 for a realistic reproduction
of the time evolution of hadronic showers in tungsten.

Further measurements at higher energies, measurements with steel
absorbers and an extension of the data analysis and of the simulation
studies are in preparation.



\subsection{Effects of high-energy particle showers on the embedded front-end electronics}

\begin{figure}[btp]
\centering \includegraphics[width=0.88\textwidth]{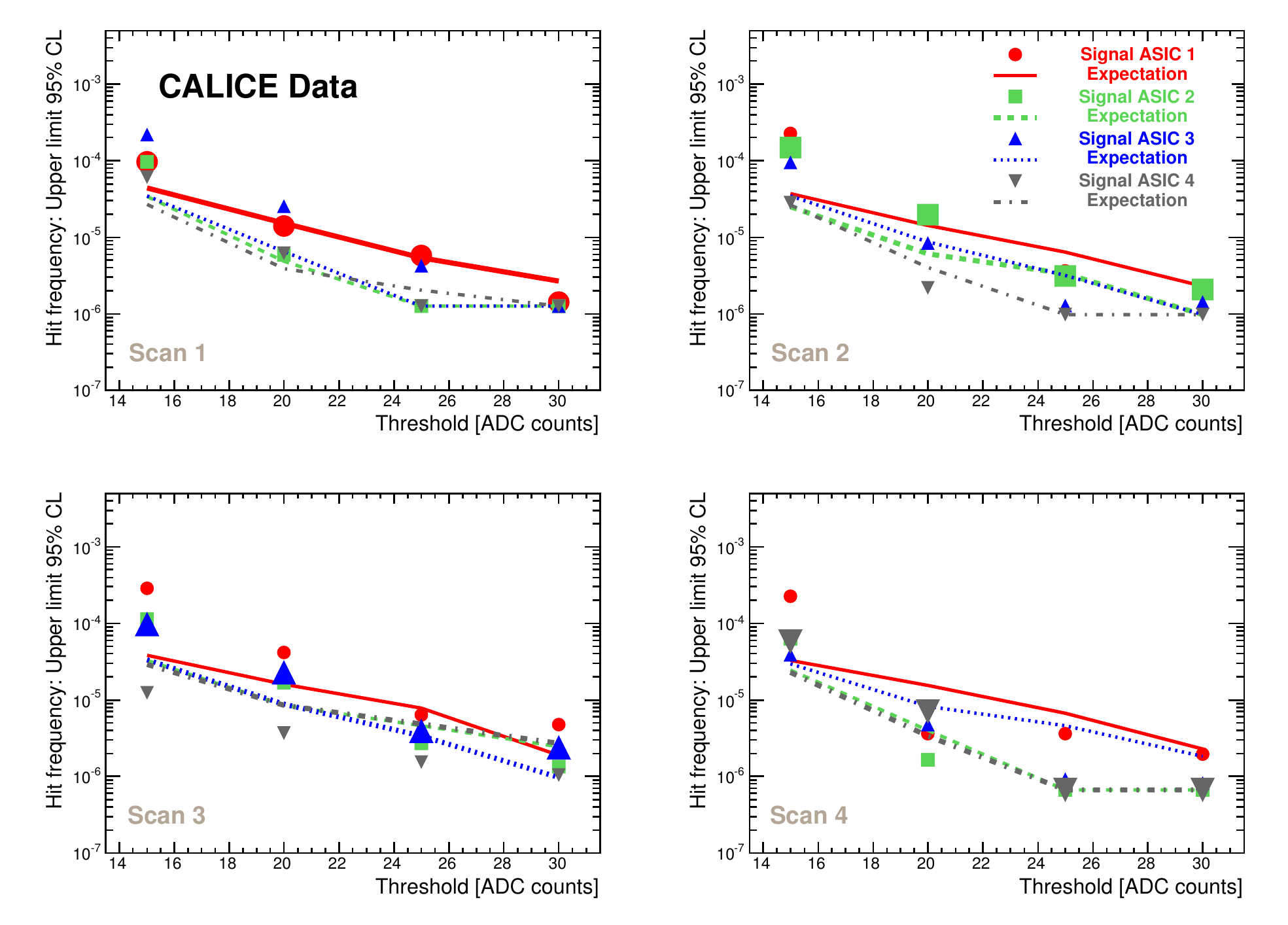}
\caption{\label{fig:exposure}\em Upper limits on frequencies at the 95\% confidence level of faked signals provoked in embedded electronics 
by high energy electromagnetic showers. The limits compared with those expected from the  pure pedestal events.  The limits are given as a function of a threshold for positive values of the ADC counts. The ASIC which is exposed to the beam is indicated by a larger symbol. For comparison the limits for ASICs outside of the beam are also shown.}
\end{figure}

As indicated above, the second generation calorimeter prototypes will have the readout  electronics embedded into the active layers of the calorimeter. The energy of electromagnetic showers produced  in the final states of $\mathrm{e}^+ \mathrm{e}^-$ collisions at a future lepton collider ranges between a few MeV up to several hundreds of GeV. A natural question arising from this design is whether the cascade particles of the  high-energy showers which penetrate through the electronics would create radiation induced effects in these circuits. 
A series of test runs has been performed and analysed in order to prove the feasibility of having embedded readout electronics 
for the calorimeters. In these  runs an ordinary calorimeter layer of the physics prototype for the silicon tungsten electromagnetic calorimeter, see Section~\ref{sec:SiWECAL}, has been replaced by a {\em special PCB} allowing for the exposure of the readout electronics to particle showers.
A detailed analysis of noise spectra of the ASICs exposed to high-energy electron beams gives no evidence that the noise pattern is altered under the influence of the electromagnetic showers. 
As shown in Figure~\ref{fig:exposure}, the probability for a faked signal is less than $10^{-5}$ for thresholds larger than 1/2 of a MIP where the MIP level is defined to be around 45\,ADC counts, see also~\cite{ECALcomm}. The probability to have faked signals above the MIP level is estimated to be smaller than $6.7\cdot 10^{-7}$. The analysis is published in~\cite{EmbedElec}.

For an event of the type $\mathrm{e}^+ \mathrm{e}^- \rightarrow \mathrm{t}\overline{\mathrm{t}}$ at $\sqrt{s}=500\,\mathrm{GeV}$ at the lepton collider about 2500 cells of dimension $1\times1\,\mathrm{cm^2}$ are expected to carry a signal above noise level which is typically defined to be (60-70)\% of a MIP. The presented results revealed no problems for the design of embedded readout electronics. It is furthermore unlikely that the residual deviations between the observed number of hits and those expected from normal noise fluctuations can be attributed to the influence of the beam but rather to an imperfect modelling of the noise spectra for signal events. In this sense, the presented results constitute a conservative upper limit.

\section{Future Beam Test Plans}
\label{sec:TB}
Beam tests a naturally vital for the success of R\&D of the detectors. Beside its large generic
approach, the CALICE efforts are targeted towards the {\em B}aseline {\em D}esign {\em D}ocument for ILC detectors, to be delivered until the end of 2012, and the {\em C}onceptual {\em D}esign {\em R}eport for CLIC detectors to be delivered until the end of 2011. The planning of beam test times is therefore very much focussed on data taking during 2011 and 2012.  This is particular true for those technologies who are supposed to be considered as base line of the detector concepts.  The beam test plans for the next years are summarised in Table~\ref{tab:testbeams}.

\subsection{Physics prototypes}
The year 2011 sees the finalisation of the main physics prototype phase. 
The physics prototype of the DHCAL based on RPCs has been completed in 
the first half of 2010 and conducted two successful beam test campaigns of four weeks each in October 2010 and January 2011. In the beam test period of October 2010 the rear part of the DHCAL stack was completed with the scintillator based TCMT while in January 2011 the scintillator layers were replaced by RPCs.
During April 2011 there is data taking in combination with  the Si-W ECAL.
The physics program to be conducted will be largely oriented on that which 
has been done in the corresponding data taking in the years 
2006-09. During 2011 small units of GEM chambers will be tested. The tests of 1\,${\rm m^2}$ chambers is planned to start around 2013.

The scintillator layers of the physics prototype of the analogue Hcal serve as sensitive devices
for a physics prototype with tungsten as absorber material. Tungsten is the most probable absorber for a hadron calorimeter for CLIC detectors. A first data taking has been realised in autumn 2010 with low energy pions. The plan for 2011 is to extend the energy range to energies
of several 100\,GeV as relevant for detectors at CLIC. Originally, 8 weeks of data taking at the SPS have been required from CERN. However, due to the high demand on the CERN beam lines and the priority to assign test beam time for LHC upgrade project, the assigned beam time by CERN will be considerably shorter. This clearly puts in jeopardy the physics goals of the test beam. 
It has to be seen to what extent the loss of beam time in 2011 can be recovered in 2012. This is also true for other projects described below.

Due to funding problems there are currently no beam test plans for the DECAL.

\subsection{Technical prototypes}
The previous sections indicate that the CALICE collaboration is entering a 
new phase of R\&D in which readout technologies and mechanical designs already meet many requirements of operation in a detector for a linear collider. 

It has to be stressed that the primary goal of these prototypes is to 
study technological solutions for the  calorimetry at a future lepton collider. 
The strategy for the coming years should take this into account. Here the 
main keywords are power pulsing and the limited depth of the 
buffers in the VFE which may require consideration of the time structure of the beams. In addition to the purely technological  issues a physics program is to be pursued. This physics program is  
derived from those of the physics prototypes, taking the new technical 
constraints into account. It requires the operators of 
testbeam sites to actively respond to the needs of the CALICE 
(LC) testbeam data taking at an very early stage. 

Two campaigns of large scale test beams with a fully equipped 1\,${\rm m^3}$ technical prototype 
of an SDHCAL equipped with GRPCs will be conducted during 2011 at CERN. As for the SDHCAL the original plan of about 8 weeks of running time has been severely cut down as indicated above.

Beyond these large scale tests, beam tests with smaller units will be conducted. Namely, these test beam efforts comprise:
\begin{itemize} 
\item Testbeams in 2011 with 1\,${\rm m^2}$ units of the technical prototype 
of the SDHCAL with Micromegas. These test will be conducted at CERN. 
\item The analogue hadron calorimeter will progressively equip the device with scintillating tiles and read out electronics.  The intermediate goal is to realise a so called vertical test towards the end of 2011 or the beginning of 2012.  This means the available equipment will be arranged to allow for the measurement of electromagnetic showers.
\item The Si-W ECAL is planning to make tests with single ASU towards the 
middle of 2011 in an electron testbeam. The detector will be progressively equipped with ASUs
to allow for larger beam test towards the end of 2012. This test may be conducted together with the
1\,${\rm m^3}$ of the SDHCAL. Tests are to be conducted at CERN, DESY or, if beam availability requires, at FNAL.
\item The Scint-W ECAL is planning to make tests with single ASU towards the 
middle of 2011 in an electron test beam. This project will benefit from the existing infrastructure for the SiW Ecal as well as from the integrated approach on front electronics and DAQ as pursued within CALICE.
\end{itemize}

It is envisaged to have a combined running of the SiW Ecal technological prototype with the
GRPC SDHCAL. The time line for this is unclear as the SiW Ecal is still going through small unit tests. In addition a number of technical questions (PCB planarity, wafer technology) need to be answerd before the start of the production of a full size prototype. Primarily these combined tests are to be conducted at CERN. The planning of the combined test needs to take into account that 
the CERN beam lines will be shut down during 2013 and 2014. An alternative would be to go Fermilab which however would require significantly more funding. In general the CALICE collaboration will also consider the option to use the new ESTB beam line at SLAC. Whether this beam line serves the CALICE needs is however not clear at the moment.   

The running of an AHCAL technical prototype alone and together with the Si-W ECAL technical prototype is planned as well but depends also on the detector readiness. Here, also the times lines of the AHCAL have to be taken into account. Therefore no plans for beam test times and locations can be given at this moment.

In any case, during the year 2011, mechanical interfaces between the different detector types will have to be defined. In general terms the realisation of a combined running is already prepared now as the DAQ system is common for all three calorimeter types. It can be expected that this DAQ system becomes fully operational during the year 2011.

\begin{table}[htdp]
\begin{center}
\begin{footnotesize}
\begin{tabular}{@{} |ccccccc| @{}}
 \hline
    Project & 2011/1 & 2011/2& 2012/1 & 2012/2 & 2013/1 & 2013/2 \\
    \hline
    Si-W ECAL/DCHAL RPC/TCMT ($\pi$)& xx & xx & xx & ? & ? & ?\\
    DHCAL GEM ($\tau$, $\pi$)& - & x & x & x & x & x\\
    W HCAL / TCMT ($\pi$)& xx & xx & xx & ? & ?  & ?\\
    GRPC SDHCAL ($\tau$)& xx & xx & xx & xx & ? & ?\\
    Mmegas SDHCAL ($\tau$)& x & x & x & x & ? & ?\\
    AHCAL ($\tau$)& x & x & x & x & ? & ?\\
    Si-W ECAL ($\tau$)& - & x & x & xx & ? & ?\\
    Sc-W ECAL ($\tau$)& - & x & x & x & ? & ?\\
    \hline
\end{tabular}
\end{footnotesize}
\end{center}
\caption{
\label{tab:testbeams}
\em The table indicate the envisaged testbeam activities until the 
end of 2013.  The symbol {\bf --} means ``No activity planned''. The symbol {\bf x} means ``Test of small units can be expected''. The symbol {\bf xx} means ``Large scale testbeam planned''. The symbol {\bf ?} means ``Activity in very early planning phase''. The symbol $\pi$ stands for physics prototype while the symbol $\tau$ stands for technological prototype. }
\end{table}%

\subsection{Concluding remarks on test beams}
A critical point is the availability of beam time. While at FNAL sufficient time could have 
been allocated in 2011, the request for CERN was, despite a good preparation and a favourable evaluation by the CERN SPS committee, severely downscaled. We would like to ask the DESY PRC to take note of this fact which might reduce proportionally the results to be expected particularly until the end of 2012.  Unfortunately, the high availability of the DESY test beam remedies the situation only by a little as most of the prototypes require high energy hadrons which after all are the largest experimental challenge for calorimeters at future linear colliders.

It needs to be stressed as well that the time plans presented here depend essentially on the available funding to complete the various prototypes. Thus, the actual running of, in particular, the test beams with fully equipped prototypes could easily be delayed by one year 
with respect to the dates indicated in Table~\ref{tab:testbeams}.






\section{Summary and Outlook}
\label{sec:summary}
The first phase of the CALICE program, the test series of physics prototypes, is nearing completion.  The silicon and scintillator based tungsten ECALs and the scintillator HCAL 
perform as expected, and the first encouraging results from the gaseous HCAL point 
into the same direction. Test beam data provide detailed feedback for the refinement 
of simulation models, and with the successful application of the PANDORA algorithm 
to real data the high granularity approach to calorimetry has been validated. The 
data presently being taken at Fermilab will for the first time allow us to assess
the potential of digital calorimetry on a firm experimental basis. 

At the same time, the second generation of highly integrated scalable prototypes enters the production phase. Power pulsing in magnetic fields and operation of the mixed circuit ASICs in dense high energy electromagnetic showers have been tested successfully in beam. With the recent advancements in front end electronics and DAQ, all components are in hand for the various ECAL and HCAL technologies, and the
semi-digital HCAL will pioneer the system integration and large scale tests. 

The multi-TeV range poses its own challenges in terms of detector density and speed. 
Tungsten as an absorber material is being explored with scintillator as active medium, 
tests with gas are aimed at and technically possible in 2012. 

In view of the timeline of the ILC technical design report and detector baseline documents, 
we expect to reach the most important goals:
\begin{itemize}
\item Studies of performance, and comparisons with Geant 4 simulations, of physics prototypes for four major technologies, the silicon and scintillator ECAL, and the scintillator and gas HCAL. All relevant ECAL HCAL combinations have been tested together in joint set-ups. 
\item A large scale system test of  second generation gaseous HCAL with power-pulsed electronics, and an experimental test of the semi-digital method for an RPC HCAL. 
\item Demonstrator tests, at least at single layer or single slab level, of the electronics integration concept for scalable prototypes of silicon and scintillator ECAL and scintillator HCAL, as well as HCALs with alternative gaseous readout (GEMs, micromegas) 
\end{itemize}

In 2012, the detector concept groups ILD and SiD are asked to establish baseline technologies for the different sub-detectors. Studies of the overall physics performance, e.g.\ in conjunction  with the tracking systems, or the integration into the overall detector engineering concept are beyond the scope of the CALICE program and need to be addressed in the context of the individual concepts. 
Based on the expected results above, 
CALICE foresees an assessment of the readiness of the different options on the technological level, and in matters of generic calorimetry. 
A number of criteria will be evaluated: 
\begin{itemize}
\item {\bf Established performance:} 
	energy resolution,
	linearity,
	uniformity,
	two particle separation
\item {\bf Validated simulation:}
	longitudinal and transverse shower profiles, response,
	linearity and resolution, for electrons and hadrons
\item {\bf Operational experience:}
	dead channels,
	noise,
	stability,
	monitoring and
	calibration
\item {\bf Scalable technology solutions:}
	power and heat reduction,
	low volume interfaces,
	data reduction,
	mechanical structures, dead spaces,
	services and supplies
\item {\bf Open R\&D issues:}
analysis and R\&D to be completed before a first pre/production prototype can be built,	
	cost reduction and industrialization issues
\end{itemize}
We expect that, due to resource limitations, not for all issues and all technologies the information will be as complete as might be desirable and as was the goal after the LOI. 
Therefore, a consensual assessment of the open issues is essential for the formulation of a coherent R\&D program for the time after 2012. 

In order to further proceed towards realistic detector proposals, the prototypes presently under construction should be extended to full module scale. The integrated electronics and associated heat dissipation, but also the on-detector zero suppression represent operational challenges which go beyond that of the first generation detectors, and performance must be re-established at large scale, including new monitoring and 
correction procedures. In addition, each of them offers added physics value for the study of hadronic showers and their reconstruction:
\begin{itemize} 
\item The ECAL will have four times finer segmentation.
\item The AHCAL has time-resolving electronics for the study of shower evolution in time and suppression of delayed neutron response. 
\item The SDHCAL combines fine segmentation with some amplitude information and should continue to take data with different absorbers, and in combination with the new ECAL. 
\end{itemize}
These issues constitute the logical next steps and need to be addressed for any future high granularity calorimeter, and the results will continue to advance our understnding of shower physics and calorimetry in general. 
 
Consequently we expect a continuing demand for extensive test beam running. In the past, our requests have been fulfilled only with compromises, and the response to our needs in the particularly intensive period towards 2012 has been plainly inadequate. Too restricted access to beam time is one of our main concerns at present and jeopardizes the timely return on the investments made into the prototypes. 

We request the PRC to endorse our planning towards 2012, to acknowledge the importance and relevance of a continuation of the studies beyond that date, and finallly to back up our test beam requests.

\section{Acknowledgments}
We would like to thank the technicians and the engineers who
contributed to the design and construction of the prototypes.  We also
gratefully acknowledge the DESY, CERN and FNAL managements for their
support and hospitality, and their accelerator staff for the reliable
and efficient beam operation.

We would like to thank the HEP group of the University of Tsukuba for
the loan of drift chambers for the DESY test beam.  The authors would
like to thank the RIMST (Zelenograd) group for their help and sensors
manufacturing.  This work was supported by the Bundesministerium
f\"{u}r Bildung und Forschung, Germany; by the the DFG cluster of
excellence `Origin and Structure of the Universe' of Germany ; by the
Helmholtz-Nachwuchsgruppen grant VH-NG-206; by the BMBF, grant
no. 05H09VHG; 
by the Alexander von Humboldt Foundation (Research Award
IV, RUS1066839 GSA);
grant HRJRG-002, Russian Agency for Atomic Energy, ISTC grant 3090; by
Russian Grants SS-1329.2008.2 and RFBR0402/17307a and by the Russian
Ministry of Education and Science; by MICINN and CPAN, Spain; by
CRI(MST) of MOST/KOSEF in Korea; by the US Department of Energy and
the US National Science Foundation; by the Ministry of Education,
Youth and Sports of the Czech Republic under the projects AV0
Z3407391, AV0 Z10100502, LC527 and LA09042 and by the Grant Agency of
the Czech Republic under the project 202/05/0653; and by the Science
and Technology Facilities Council, UK.

\newpage 

\begin{center}


\author{\centering 
\LARGE\bf The CALICE Collaboration
}

\author{\centering
C.\,Adloff, 
J.\,Blaha, 
M.\,Chefdeville, 
C.\,Drancourt,
A.\,Espargili\`{e}re, 
R.\,Gaglione, 
N.\,Geffroy, 
Y.\,Karyotakis, 
J.\,Prast,
G.\,Vouters
\\ \it
Laboratoire d'Annecy-le-Vieux de Physique des Particules, Universit\'{e} de Savoie,
CNRS/IN2P3,
9 Chemin de Bellevue BP110, F-74941 Annecy-le-Vieux CEDEX, France
}

\author{\centering
T.\,Cundiff, 
P.\,De Lurgio,
G.\,Drake,
K.\,Francis,
B.\,Haberichter, 
V.\,Guarino, 
A.\,Kreps, 
E.\,May, 
J.\,Repond, 
J.\,Schlereth, 
F.\,Skrzecz,
J.\,Smith\footnote{Also at University of Texas, Arlington},
D.\,Underwood, 
K.\,Wood, 
L.\,Xia, 
Q.\,Zhang,
A.\,Zhao
\\ \it
Argonne National Laboratory,
9700 S.\ Cass Avenue,
Argonne, IL 60439-4815,
USA}

\author{\centering
E.\,Baldolemar, 
A.\,Brandt, 
K.\,De, 
J.\,Smith, 
K.\,J.\,Park,
S.\,T.\,Park, 
M.\,Sosebee, 
A.\,White, 
J.\,Yu
\\ \it
Department of Physics, SH108, University of Texas, Arlington, TX 76019, USA
}

\author{\centering
Z.\,Deng,
 Y.\,Li,
Y.\,Wang,  
Q.\,Yue,   
 Z.\,Yang
\\ \it
Tsinghua University, Department of Engineering Physics.Beijing, 100084, P.R.
China
}

\author{\centering
G.\,Eigen, D.\,Fehlker,
H.\,Sandaker
\\ \it
University of Bergen, Inst.\, of Physics, Allegaten 55, N-5007 Bergen, Norway
}

\author{\centering
N.\,K.\,Watson 
\\ \it
University of Birmingham,
School of Physics and Astronomy,
Edgbaston, Birmingham B15 2TT, UK
}

\author{\centering 
J.\,Butler, E.\,Hazen, S.\,Wu
\\ \it
Boston University, Department of Physics, 590 Commonwealth Ave.,
Boston, MA 02215, USA
}

\author{\centering 
L.\,B.\,A.\,Hommels, 
J.\,S.\,Marshall,
M.\,A.\,Thomson, 
D.\,R.\,Ward
\\ \it
University of Cambridge, Cavendish Laboratory, J J Thomson Avenue, CB3 0HE, UK
}

\author{\centering 
D.\,Benchekroun, 
A.\,Hoummada, 
Y.\,Khoulaki
\\ \it
Universit\'{e} Hassan II A\"{\i}n Chock, Facult\'{e} des sciences.\, B.P. 5366 Maarif, Casablanca, Morocco
}

\author{\centering 
J.\,Apostolakis, 
D.\,Dannheim,
A.\,Dotti, 
K.\,Elsener,
G.\,Folger, 
C.\,Grefe,
M.\,Killenberg,
W.\,Klempt,
E.\, van der Kraaij,
L.\,Linssen,
A.\,-I.\,Lucaci-Timoce,
A.\,Muennich,
J.\,Nardulli,
D.\,Perini,
S.\,Poss,
A.\,Sailer,
D.\,Schlatter,
P.\,Speckmayer,
V.\,Uzhinskiy
\\ \it 
CERN, 1211 Gen\`{e}ve 23, Switzerland
}

\author{\centering 
M.\,Oreglia
\\ \it
University of Chicago, Dept.\, of Physics, 5720 So. Ellis Ave., KPTC 201 Chicago, 
IL 60637-1434, USA
}

\author{\centering
M.\,Benyamna, 
C.\,C\^{a}rloganu,
F.\,Fehr,
P.\,Gay, 
S.\,Manen, 
L.\,Royer
\\ \it
Clermont Univertsit\'e, Universit\'e Blaise Pascal, CNRS/IN2P3, LPC, BP
10448, F-63000 Clermont-Ferrand, France
}

\author{\centering
M.\,Tytgat,
N.\,Zaganidis
\\ \it
Ghent University, Department of Physics and Astronomy,
Proeftuinstraat 86, B-9000 Gent, Belgium
}

\author{\centering
J.\,Ha
\\ \it
Korea Atomic Energy Research Institute,
Taejon 305-600,
South Korea
}

\author{\centering
F.\,Abu-Ajamieh,
G.\,C.\,Blazey,
D.\,Chakraborty,
A.\,Dyshkant,
D.\,Hedin,
J.\,Hill,
J.\,G.\,R.\,Lima,
R.\,Salcido,
V.\,Rykalin,
V.\,Zutshi
\\ \it
NICADD, Northern  Illinois University,
Department of Physics,
DeKalb, IL 60115,
USA
}

\author{\centering 
V.\,Astakhov, V.\,A.\,Babkin, S.\,N.\,Bazylev, Yu.\,I.\,Fedotov, S.\,Golovatyuk, I.\,Golutvin, N.\,Gorbunov, 
A.\,Malakhov, S.\,Slepnev, I.\,Tyapkin, S.\,V.\,Volgin, Y.\,Zanevski, A.\,Zintchenko 
\\ \it
Joint Institute for Nuclear Research, Joliot-Curie 6,
141980, Dubna,
Moscow Region, Russia
}

\author{\centering 
D.\,Dzahini, 
L.\,Gallin-Martel, 
J.\,Giraud, 
D.\,Grondin, 
J.\,-Y.\,Hostachy, 
K.\,Krastev, 
L.\,Morin,
F-E.\,Rarbi
\\ \it
Laboratoire de Physique Subatomique et de Cosmologie - Universit\'{e} Joseph Fourier Grenoble 1 -
CNRS/IN2P3 - Institut Polytechnique de Grenoble,
53, rue des Martyrs,
38026 Grenoble CEDEX, France
}

\author{\centering 
U.\,Cornett, 
D.\,David, 
R.\,Fabbri, 
G.\,Falley, 
K.\,Gadow, 
E.\,Garutti,
P.\,G\"{o}ttlicher, 
C.\,G\"{u}nter,
S.\,Karstensen, 
F.\,Krivan,
K.\,Kschioneck, 
S.\,Lu, 
B.\,Lutz, 
I.\,Marchesini, 
N.\,Meyer,
S.\,Morozov, 
V.\,Morgunov\footnote{On leave from ITEP}, 
M.\,Reinecke, 
F.\,Sefkow, 
P.\,Smirnov,
M.\,Terwort,
A.\,Vargas-Trevino, 
N.\,Wattimena, 
O.\,Wendt
\\ \it
DESY, Notkestrasse 85,
D-22603 Hamburg, Germany
}

\author{\centering  
N.\,Feege, 
J.\,Haller, 
J.\,Samson
\\ \it
Univ. Hamburg,
Physics Department,
Institut f\"ur Experimentalphysik,
Luruper Chaussee 149,
22761 Hamburg, Germany
}

\author{\centering 
P.\,Eckert,
T.\,Harion, 
A.\,Kaplan,
 H.\,-Ch.\,Schultz-Coulon,
 W.\,Shen,
 R.\,Stamen,
 A.\,Tadday
\\ \it
 University of Heidelberg, Fakultat fur Physik und Astronomie,
Albert Uberle Str. 3-5 2.OG Ost,
D-69120 Heidelberg, Germany
}

\author{\centering 
B.\,Bilki, E.\,Norbeck, D.\,Northacker, Y.\,Onel
\\ \it
University of Iowa, Dept. of Physics and Astronomy,
203 Van Allen Hall, Iowa City, IA 52242-1479, USA
}

\author{\centering 
E.\,J.\,Kim
\\ \it
Chonbuk National University, Jeonju, 561-756, South Korea
}

\author{\centering 
G.\,Kim, D-W.\,Kim, K.\,Lee, S.\,C.\,Lee
\\ \it
Kangnung National University, HEP/PD, Kangnung, South Korea
}

\author{\centering 
B.\,van\,Doren,
G.\,W.\,Wilson
\\ \it
University of Kansas, Department of Physics and Astronomy,
Malott Hall, 1251 Wescoe Hall Drive, Lawrence, KS 66045-7582, USA
}

\author{\centering 
K.\,Kawagoe 
\\ \it
 Department of Physics, Kobe University, Kobe, 657-8501, Japan
}

\author{\centering 
P.\,D.\,Dauncey 
\\ \it
Imperial College London, Blackett Laboratory,
Department of Physics,
Prince Consort Road,
London SW7 2AZ, UK 
}

\author{\centering 
V.\,Bartsch\footnote{Now at University of Sussex, Physics and Astronomy Department, Brighton, Sussex, BN1 9QH, UK}, 
M.\,Postranecky, M.\,Warren, M.\,Wing
\\ \it
Department of Physics and Astronomy, University College London,
Gower Street,
London WC1E 6BT, UK
}

\author{\centering 
V.\, Boisvert,  
B.\,Green, 
A.\,Misiejuk, 
F.\,Salvatore\footnote{Now at University of Sussex, Physics and Astronomy Department, Brighton, Sussex, BN1 9QH, UK}
\\ \it
Royal Holloway University of London,
Dept. of Physics,
Egham, Surrey TW20 0EX, UK
}

\author{\centering 
E.\,Cortina Gil,
S.\,Mannai,
G.\,Nuessle 
\\ \it
Centre for Particle Physics and Phenomenology (CP3)
Universit\'{e} catholique de Louvain, Belgium
}

\author{\centering 
M.\,Bedjidian,    
A.\,Bonnevaux, 
C.\,Combaret, 
L.\,Caponetto, 
G.\,Grenier, 
R.\,Han, 
J.C.\,Ianigro,
R.\,Kieffer, 
I.\,Laktineh,  
N.\,Lumb, 
H.\,Mathez,
M.\,Vander\,Donckt
\\ \it
Universit\'{e} de Lyon, Universit\'{e} de Lyon 1, 
CNRS/IN2P3, IPNL 4 rue E Fermi 69622,
Villeurbanne CEDEX, France
}

\author{\centering 
J.\,Berenguer~Antequera,
E.\,Calvo~Alamillo, 
M.-C.\, Fouz, 
J.\,Marin,
J.\,Puerta-Pelayo 
\\ \it
CIEMAT, Centro de Investigaciones Energeticas, Medioambientales y Tecnologicas, Madrid, Spain 
}

\author{\centering 
D.\,S.\,Bailey, 
R.\,J.\,Barlow, 
R.\,J.\,Thompson 
\\ \it
The University of Manchester, School of Physics and Astronomy,
Schuster Laboratory,
Manchester M13 9PL,
UK
}

\author{\centering
M.\,Batouritski, 
O.\,Dvornikov, 
Yu.\,Shulhevich, 
N.\,Shumeiko, 
A.\,Solin,
P.\,Starovoitov, 
V.\,Tchekhovski, 
A.\,Terletski
\\ \it
National Centre of Particle and High Energy Physics of the
Belarusian State University, M.Bogdanovich str. 153, 220040 Minsk, Belarus
}

\author{\centering 
F.\,Corriveau, D.\,Trojand\footnote{Also at Argonne National Laboratory} 
\\ \it
Department of Physics, McGill University,
Ernest Rutherford Physics Bldg.,
3600 University Ave.,
Montr\'{e}al, Quebec,
CANADA H3A 2T8
}

\author{\centering 
V.\,Balagura,
B.\,Bobchenko, 
M.\,Chadeeva, 
M.\,Danilov, 
O.\,Markin, 
R.\,Mizuk, 
E.\,Novikov, 
V.\,Rusinov, 
E.\,Tarkovsky
\\ \it
Institute of Theoretical and Experimental Physics, B. Cheremushkinskaya ul. 25,
RU-117218 Moscow, Russia
}

\author{\centering 
V.\,Andreev, N.\,Kirikova,  A.\,Komar, V.\,Kozlov, M.\,Negodaev, P.\,Smirnov, Y.\,Soloviev, A.\,Terkulov 
\\ \it
P.\,N.\, Lebedev Physical Institute,
Russian Academy of Sciences,
117924 GSP-1 Moscow, B-333, Russia
}

\author{\centering 
P.\,Buzhan, B.\,Dolgoshein, A.\,Ilyin, V.\,Kantserov, V.\,Kaplin, A.\,Karakash, E.\,Popova, S.\,Smirnov 
\\ \it
Moscow Physical Engineering Inst., MEPhI,
Dept. of Physics,
31, Kashirskoye shosse,
115409 Moscow, Russia
}

\author{\centering 
N.\,Baranova,
E.\,Boos, 
L.\,Gladilin,
D.\,Karmanov, 
M.\,Korolev, 
M.\,Merkin,
A.\,Savin,
A.\,Voronin
\\ \it
M.V.Lomonosov Moscow State University, D.V.Skobeltsyn Institute of Nuclear
Physics (SINP MSU),
1/2 Leninskiye Gory, Moscow, 119991, Russia
}

\author{\centering 
A.\,Singh,
A.\,Topkar
\\ \it
Bhabha Atomic Research Centre,
Mumbai 400085, India
}

\author{\centering 
C.\,Kiesling,
P.\,Klenze, 
K.\,Seidel, 
F.\,Simon,
C.\,Soldner, 
M.\,Tesar,
L.\,Weuste
\\ \it
Max Planck Inst. f\"ur Physik,
F\"ohringer Ring 6,
D-80805 Munich, Germany
}

\author{\centering 
J.\,Bonis, 
B.\,Bouquet,    
S.\,Callier, 
P.\,Cornebise, 
Ph.\,Doublet,
F.\,Dulucq, 
M.\,Faucci Giannelli, 
J.\,Fleury,
T.\,Frisson,
G.\,Guilhem, 
H.\,Li,  
G.\,Martin-Chassard, 
F.\,Richard, 
Ch.\,de la Taille, 
R.\,Poeschl, 
L.\,Raux,  
N.\,Seguin-Moreau, 
F.\,Wicek, 
Z.\,Zhang
\\ \it
Laboratoire de L'acc\'elerateur Lin\'eaire,
Centre d'Orsay, Universit\'e de Paris-Sud XI,
BP 34, B\^atiment 200,
F-91898 Orsay CEDEX, France
}

\author{\centering 
M.\,Anduze,
M.\,S.\,Amjad,
K.\,Belkadhi,
M.\,Bercher, 
V.\,Boudry, 
J-C.\,Brient, 
C.\,Clerc, 
R.\,Cornat,
D.\,Decotigny,
M.\,Frotin,
F.\,Gastaldi, 
D.\,Jeans, 
A.\,Matthieu,   
P.\,Mora de Freitas, 
G.\,Musat, 
J.F.\,Roig,
M.\,Ruan,  
H.\,Videau
\\ \it
      Laboratoire Leprince-Ringuet (LLR)  -- \'{E}cole Polytechnique,
      CNRS/IN2P3,
      Palaiseau, F-91128 France
}


\author{\centering 
K-H.\,Park
\\ \it
Pohang Accelerator Laboratory, Pohang 790-784, South Korea
}

\author{\centering 
B.\,Bulanek,
J.\,Zacek 
\\ \it
Charles University, Institute of Particle \& Nuclear Physics,
V Holesovickach 2,
CZ-18000 Prague 8, Czech Republic  
}

\author{\centering 
J.\,Cvach, 
P.\,Gallus, 
M.\,Havranek, 
M.\,Janata, 
J.\,Kvasnicka,
D.\,Lednicky,
M.\,Marcisovsky, 
I.\,Polak, 
J.\,Popule, 
L.\,Tomasek, 
M.\,Tomasek, 
P.\,Ruzicka, 
P.\,Sicho, 
J.\,Smolik, 
V.\,Vrba, 
J.\,Zalesak 
\\ \it
Institute of Physics, Academy of Sciences of the Czech Republic, Na Slovance 2,
CZ-18221 Prague 8, Czech Republic
}

\author{\centering 
Yu.\,Arestov,
V.\,Ammosov, B.\,Chuiko, V.\,Gapienko,
Y.\,Gilitski,V.\,Koreshev, A.\,Semak, Yu.\,Sviridov, V.\,Zaets
\\ \it
Institute of High Energy Physics,
Moscow Region,
RU-142284 Protvino,
Russia
}

\author{\centering 
B.\,Belhorma,
H.\,Ghazlane
\\ \it
Centre National de l'Energie, des Sciences et des Techniques Nucl\'{e}aires, 
B.P. 1382, R.P. 10001, Rabat, Morocco
}

\author{\centering 
R.\,E.\,Coath,
J.\,P.\,Crooks,
M.\,Stanitzki, 
J.\,Strube, 
R.\,Turchetta, 
M.\,Tyndel,  
Z.\,Zhang 
\\ \it
Rutherford Appleton Laboratory, Chilton, Didcot,
Oxon, OX11 0QX, UK 
}


\author{\centering 
S.\,W.\,Nam, I.\,H.\,Park, J.\,Yang 
\\ \it
Ewha Womans University, Dept. of Physics,
Seoul 120,
South Korea
}

\author{\centering 
Jong-Seo Chai, Jong-Tae Kim, Geun-Bum Kim
\\ \it
Sungkyunkwan University,
300 Cheoncheon-dong, Jangan-gu, Suwon, Gyeonggi-do  440-746, South Korea
}

\author{\centering 

Y.\,Kim
\\ \it
Korea Institute of Radiological and
Medical Sciences,
215-4 Gangeung-dong,
Nowon-gu, Seoul 139-706,
South Korea
}

\author{\centering 
J.\,Kang, Y.\,-J.\,Kwon  
\\ \it
Yonsei  University, Dept. of Physics,
134 Sinchon-dong,
Sudaemoon-gu, Seoul 120-749,
South Korea
}

\author{\centering 
Ilgoo Kim, Taeyun Lee, Jaehong Park, Jinho Sung
\\ \it
School of Electric Engineering and Computing Science, Seoul National University,
Seoul 151-742, South Korea
}

\author{\centering              
K.\,Kotera, M.\, Nishiyama, T.\,Takeshita
\\ \it
Shinshu Univ.\,,
Dept. of Physics,
3-1-1 Asaki,
Matsumoto-shi, Nagano 390-861,
Japan
}

\author{\centering 
A.\,Khan, D.\,H.\,Kim, J.E.\,Kim, D.\,J.\,Kong, Y.D.\,Oh, S.\,Uozumi
\\ \it
Kyungpook National Univ., Dept. of Physics, 1370 San Kyuk-dong, Puk ku, Taegu 635, South Korea
}

\author{\centering              
H.\,Koike, 
K.\,Tanaka, 
F.\,Ukegawa
\\ \it
  University of Tsukuba, Graduate School of Pure and Applied Sciences,
  Tennoudai 1-1-1, Tsukuba, Ibaraki 305-8571, Japan
}

\author{\centering 
J.\,Sauer, 
S.\,Weber,
C.\,Zeitnitz
\\ \it
Bergische Universit\"{a}t Wuppertal
Fachbereich 8 Physik,
Gaussstrasse 20,
D-42097 Wuppertal, Germany
}

\end{center}

\renewcommand{\hepex}[1]{{\tt arXiv:#1 [hep-ex]}}
\renewcommand{\physics}[1]{{\tt arXiv:#1 [physics.ins-det]}}
\newcommand\nim[4]{{doi:10.1016/#4}{\emph{, Nucl.\ Instrum.\ Meth.} \textbf{#1} (#2) #3}}
\newcommand{\etal}{\emph{et al.}}




\end{document}